\documentclass[rmp,aps,twocolumn]{revtex4}
\usepackage{amsmath,graphics,epsfig,color,verbatim}

\newcommand{\kb}{k_{B}}

\newcommand{\trace}{\mbox{Tr}}

\renewcommand{\a}{\alpha}
\renewcommand{\b}{\beta}
\newcommand{\g}{\gamma}

\newcommand{\Fad}{\widetilde{F}^{\a\dagger}}
\newcommand{\Fa}{\widetilde{F}^{\a}}
\newcommand{\Gh}{{\cal G}}
\newcommand{\web}{http://dmftreview.rutgers.edu }
\newcommand{\iom}{i\omega}
\renewcommand{\vr}{\mathbf{r}}
\newcommand{\vk}{\mathbf{k}}

\def\todo#1{{\tt #1}}

\begin{document}

\title{Electronic Structure Calculations with Dynamical Mean--Field Theory: A
Spectral Density Functional Approach}

\author{G.~Kotliar$^{1,6}$, S.~Y.~Savrasov$^2$, K.~Haule$^{1,4}$,
V.~S.~Oudovenko$^{1,3}$, O. Parcollet$^{5}$ and C.A. Marianetti$^{1}$}

\date{\today}

\begin{abstract}
We present a review of the basic ideas and techniques of the
spectral density functional theory which are currently used in
electronic structure calculations of strongly--correlated
materials where the one--electron description breaks down. We
illustrate the method with several examples where interactions
play a dominant role: systems near metal--insulator transition,
systems near volume collapse transition, and systems with local
moments.
\end{abstract}

\address{$^1$Department of Physics and Astronomy and Center for Condensed Matter
Theory, Rutgers University, Piscataway, NJ 08854--8019}
\address{$^2$Department of Physics, University of California, Davis, CA 95616}
\address{$^3$Bogoliubov Laboratory for Theoretical Physics, Joint
Institute for Nuclear Research, 141980 Dubna, Russia}
\address{$^4$Jozef Stefan Institute, SI-1000 Ljubljana, Slovenia}
\address{$^5$ Service de Physique Theorique, CEA Saclay, 91191 Gif-Sur-Yvette, France}
\address{$^6$ Centre de Physique Theorique, Ecole Polytechnique 91128 Palaiseau Cedex, France}

\pacs{71.20.-b, 71.27.+a, 75.30.-m}

\maketitle

\tableofcontents

\section{Introduction}

\label{sec:INT}

Theoretical understanding of the behavior of materials is a great
intellectual challenge and may be the key to new technologies. We
now have a firm understanding of simple materials such as noble
metals and semiconductors. The conceptual basis characterizing
the spectrum of low--lying excitations in these systems  is well
established by the Landau Fermi liquid theory~\cite{Pines:1966}.
We also have quantitative techniques for computing ground states
properties, such as the density functional theory (DFT) in the
local density and generalized gradient approximation (LDA and GGA)
\cite{Lundqvist:1983}. These techniques also can be successfully used
as starting points for perturbative computation of one--electron
spectra, such as the GW method \cite{Aryasetiawan:1998}.

The scientific frontier that one would like to explore is a
category of materials which falls under the rubric of
strongly--correlated electron systems. These are complex
materials, with electrons occupying active $3d$-, $4f$- or
$5f$--orbitals, (and sometimes $p$- orbitals as in many organic
compounds and in Bucky--balls--based materials
\cite{Gunnarsson:1997}). The excitation spectra in these systems
cannot be described in terms of well--defined quasiparticles over
a wide range of temperatures and frequencies. In this situation
band theory concepts are not sufficient and new ideas such as those
of Hubbard bands
and narrow coherent
quasiparticle bands are needed for the
description of the electronic structure.
\cite{Georges:1996,Kotliar:2004:PT}.

Strongly correlated electron systems have frustrated interactions,
reflecting the competition between different forms of order. The
tendency towards delocalization leading to band formation and the
tendency to localization leading to atomic like behavior is
better described in real space. The competition between different
forms of long--range order (superconducting, stripe--like density
waves, complex forms of frustrated non--collinear magnetism etc.)
leads to complex phase diagrams and exotic physical properties.

Strongly correlated electron systems have many unusual
properties. They are extremely sensitive to small changes in
their control parameters resulting in large responses, tendencies
to phase separation, and formation of complex patterns in
chemically inhomogeneous situations
~\cite{Millis:2003,Mathur:2003:PT}. This makes their study
challenging, and the prospects for applications particularly
exciting.

The promise of strongly--correlated materials continues to be
realized experimentally. High superconducting transition
temperatures (above liquid Nitrogen temperatures) were totally
unexpected. They were realized in materials containing Copper and
Oxygen. A surprisingly large dielectric constant, in a wide range
of temperature was recently found in Mott insulator
CaCu$_{3}$Ti$_{4}$O$_{12}$ \cite{Lixin:2002}. Enormous mass
renormalizations are realized in systems containing rare earth
and actinide elements, the so--called heavy fermion systems
\cite{Stewart:2001}. Their large orbital degeneracy and large
effective masses give exceptionally large Seebeck coefficients,
and have the potential for being useful thermoelectrics in the
low--temperature region \cite{Sales:1996}. Colossal
magnetoresistance, a dramatic sensitivity of the resistivity to
applied magnetic fields, was discovered recently
\cite{Tokura:2000} in many materials including the prototypical
La$_{x}$Sr$_{1-x}$MnO$_{3}$. A gigantic non--linear optical
susceptibility with an ultrafast recovery time was discovered in
Mott insulating chains \cite{Ogasawara:2000}.

These non--comprehensive lists of remarkable materials and their
unusual physical properties are meant to illustrate that
discoveries in the areas of correlated materials occur
serendipitously. Unfortunately, lacking the proper theoretical
tools and daunted by the complexity of the materials, there have
not been success stories in predicting new directions for even
incremental improvement of material performance using
strongly--correlated systems.

In our view, this situation is likely to change in the
very near future as a result of the introduction of a practical
but powerful new many body method, the Dynamical Mean Field
Theory (DMFT). This method is based on a mapping of the full many
body problem of solid state physics onto  a quantum impurity
model, which is essentially a  small number of quantum degrees of freedom
embedded in a bath that obeys a self consistency condition
\cite{Georges:1992}. This approach, offers a minimal description
of the electronic structure of correlated materials, treating
both the Hubbard bands and the quasiparticle bands on the same
footing. It becomes exact in the limit of infinite lattice
coordination introduced in the pioneering work of Metzner and
Vollhardt \cite{metzner:1989}.

Recent advances
\cite{Anisimov:1997,Lichtenstein:1997,Lichtenstein:1998} have
combined dynamical mean--field theory
(DMFT)~\cite{Georges:1996,Kotliar:2004:PT} with electronic
structure techniques (for other DMFT
reviews, see \cite{Held:2001:IJMPB,Lichtenstein:2004:KLUWER,
Held:2003:PSIK,Freericks:2003,Georges:2004:CM,
Georges:2004:CM0403123,Maier:2004:CM0404055})
These developments, combined with  increasing
computational power and novel algorithms, offer the possibility
of turning DMFT into a useful method for computer aided
material design involving strongly correlated materials.

This review is an introduction to the rapidly developing field of
electronic structure calculations of strongly--correlated
materials. Our primary goal is to present some concepts and
computational tools that are allowing a first--principles
description of these systems.  We review the work of both the
many--body physics and the electronic structure communities who
are currently making important contributions in this area. For
the electronic structure community, the DMFT approach gives
access to new regimes for which traditional methods based on
extensions of DFT do not work. For the many--body community,
electronic structure calculations bring system specific
information needed to formulate interesting many--body problems
related to a given material.

The introductory section \ref{sec:INT} discusses the importance of
\textit{ab initio} description in strongly--correlated solids. We
review briefly the main concepts behind the approaches based on
model Hamiltonians and density functional theory to put in
perspective the current techniques combining DMFT with electronic
structure methods. In the last few years, the DMFT method has
reached a great degree of generality which gives the flexibility
to tackle realistic electronic structure problems, and we review
these developments in Section~\ref{sec:SDF}. This section
describes how the DMFT and electronic structure LDA theory can be
combined together. We stress the existence of new functionals for
electronic structure calculations and review applications of
these developments for calculating various properties such as
lattice dynamics, optics and transport. The heart of the
dynamical mean--field description of a system with local
interactions is the quantum impurity model. Its solution is the
bottleneck of all DMFT algorithms. In Section \ref{sec:IMP} we
review various impurity solvers which are currently in use,
ranging from the formally exact but computationally expensive
quantum Monte Carlo (QMC) method to various approximate schemes.
One of the most important developments of the past was a fully
self--consistent implementation of the LDA+DMFT approach, which
sheds new light on the mysterious properties of
Plutonium~\cite{Savrasov:2001}. Section \ref{sec:MAT} is devoted
to three typical applications of the formalism: the problem of
the electronic structure near a Mott transition, the problem of
volume collapse transitions, and the problem of the description
of systems with local moments. We conclude our review in
Section~\ref{sec:otlook}. Some technical aspects of the
implementations as well as the description of DMFT codes are
provided in the online notes to this review (see Appendix
\ref{sec:app}).

\subsection{Electronic structure of correlated systems}

\label{sec:INTels}

\begin{figure}[tbh]
\includegraphics*[height=1.8in]{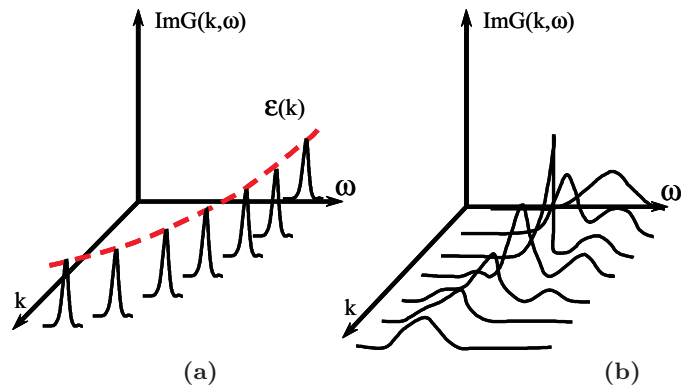} \centerline{%
\hspace*{2.cm} \it\bf (a)\hspace*{5cm} (b)  } \caption{Evolution
of the non-interacting spectrum (a) into the interacting spectrum
(b) as the Coulomb interaction increases. Panels (a) and (b)
correspond to LDA-like and DMFT-like solutions, respectively.}
\label{Fig:SpectralFunction}
\end{figure}

What do we mean by a strongly--correlated phenomenon? We can
answer this question from the perspective of electronic structure
theory, where the one--electron excitations are well--defined and
represented as delta--function--like peaks showing the locations
of quasiparticles at the energy scale of the electronic spectral
functions (Fig.~\ref{Fig:SpectralFunction}(a)). Strong
correlations would mean the breakdown of the effective
one--particle description: the wave function of the system becomes
essentially many--body--like, being represented by
combinations of Slater determinants, and the one--particle Green's
functions no longer exhibit single peaked features
(Fig.~\ref{Fig:SpectralFunction}~(b)).

The development of methods for studying strongly--correlated
materials has a long history in condensed matter physics. These
efforts have traditionally focused on model Hamiltonians using
techniques such as diagrammatic methods \cite{Bickers:1989},
quantum Monte Carlo simulations \cite{Jarrell:1996}, exact
diagonalization for finite--size clusters \cite{Dagotto:1994},
density matrix renormalization group methods
\cite{White:1992,Schollwoeck:2005:RMP} and so on. Model
Hamiltonians are usually written for a given solid--state system
based on physical grounds.  Development of LDA+U
\cite{Anisimov:1997:JPCM} and self--interaction corrected (SIC)\
\cite{Svane:1990, Szotek:1993} methods, many--body perturbative
approaches based on GW and its extensions
\cite{Aryasetiawan:1998}, as well as the time--dependent version
of the density functional theory \cite{Gross:1996} have been
carried out by the electronic structure community. Some of these
techniques are already much more complicated and time--consuming
compared to the standard LDA based algorithms, and therefore the
real exploration of materials is frequently performed by
simplified versions utilizing approximations such as the
plasmon--pole form for the dielectric function
\cite{Hybertsen:1986}, omitting the self--consistency within GW
\cite{Aryasetiawan:1998} or assuming locality of the GW
self--energy~\cite{Zein:2002}.

\begin{figure}[tbh]
\includegraphics*[height=2.4in]{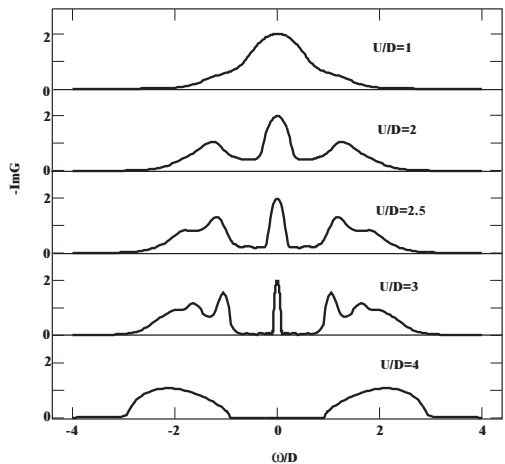}
\caption{Local spectral density at $T=0$, for several values of
$U$, obtained by the iterated perturbation theory approximation
(from \cite{Zhang:1993}).} %
\label{Fig:DMFTSpectralFunction}
\end{figure}

The one--electron densities of states of strongly correlated
systems may display both renormalized quasiparticles and atomic--like
states simultaneously~\cite{Georges:1992,Zhang:1993}. To treat
them one needs a technique which is able to treat quasi-particle
bands and Hubbard bands on the same footing, and which is able to
interpolate between atomic and band limits. Dynamical mean--field
theory \cite{Georges:1996} is the simplest approach which
captures these features; it has been extensively developed to
study model Hamiltonians. Fig.~\ref{Fig:DMFTSpectralFunction}
shows the development of the spectrum while increasing the
strength of Coulomb interaction $U$ as obtained by DMFT solution
of the Hubbard model. It illustrates the necessity to go beyond
static mean--field treatments in the situations when the on--site
Hubbard $U$ becomes comparable with the bandwidth $W$.

Model Hamiltonian based DMFT\ methods have successfully
described regimes $U/W\gtrsim 1$. However to describe strongly
correlated materials we need to incorporate realistic electronic
structure because the low--temperature physics of systems near
localization--delocalization crossover is non--universal, system
specific, and very sensitive to the lattice structure and orbital
degeneracy which is unique to each compound. We believe that
incorporating this information into the many--body treatment of
this system is a necessary first step before more general lessons
about strong--correlation phenomena can be drawn. In this
respect, we recall that DFT in its
common approximations, such as LDA or GGA,
brings a system specific description into calculations. Despite the
great success of DFT for studying weakly correlated solids, it
has  not been able thus far to address strongly--correlated
phenomena. So, we see that both density functional based and
many--body model Hamiltonian approaches are to a large extent
complementary to each other. One--electron Hamiltonians, which are
necessarily generated within density functional approaches (i.e.
the hopping terms), can be used as input for more challenging
many--body calculations. This path has been undertaken in a first
paper of Anisimov et al. \cite{Anisimov:1997} which introduced
the LDA+DMFT method of electronic structure for
strongly--correlated systems and applied it to the photoemission
spectrum of La$_{1-x}$Sr$_{x}$TiO$_{3}$. Near the Mott transition,
this system shows a number of features incompatible with the
one--electron description \cite{Fujimori:1992}. The electronic
structure of Fe has been shown to be in better agreement with
experiment within DMFT in comparison with LDA
\cite{Lichtenstein:1997,Lichtenstein:1998}.
The photoemission spectrum near the Mott
transition in V$_{2}$O$_{3}$ has been studied
\cite{Held:2001V2O3}, as well as issues connected to the finite
temperature magnetism of Fe and Ni were explored
\cite{Lichtenstein:2001}.

Despite these successful developments, we also would like to
emphasize a more ambitious goal: to build a general method which
treats all bands and all electrons on the same footing,
determines both hoppings and interactions internally using a
fully self--consistent procedure, and accesses both energetics
and spectra of correlated materials. These efforts have been
undertaken in a series of papers \cite{Chitra:2000, Chitra:2001}
which gave us a functional description of the problem in complete
analogy to the density functional theory, and its
self--consistent implementation is illustrated on
Plutonium~\cite{Savrasov:2001,Savrasov:2004:PRB}.

To summarize, we see the existence of two roads in approaching
the problem of simulating correlated materials properties, which
we illustrate in Fig.~\ref{Fig:DFT_2roads}. To describe these
efforts in a language understandable by both electronic structure
and many--body communities, and to stress qualitative differences
and great similarities between DMFT and LDA, we start our review
with discussing a general many--body framework based on the
effective action approach to strongly--correlated systems
\cite{Chitra:2001}.
\begin{figure}[tbh]
\includegraphics*[height=1.2in]{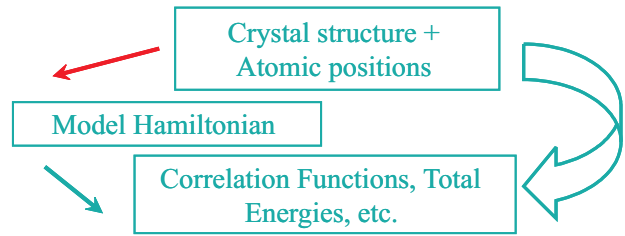}
\caption{Two roads in approaching the problem of simulating
correlated materials properties.} \label{Fig:DFT_2roads}
\end{figure}

\subsection{The effective action formalism and the constraining field}

The effective action formalism, which utilizes functional
Legendre transformations and the inversion method (for a
comprehensive review see \cite{Fukuda:1995}, also see online notes), allows us to
present a unified description of many seemingly different
approaches to electronic structure. The idea is very simple, and
has been used in other areas such as quantum field theory and
statistical mechanics of spin systems.
We begin with the free energy of the system written as a functional integral
\begin{equation}
  \exp(-F)=\int D[\psi^\dagger\psi]e^{-S}.
\label{Eq:PF}
\end{equation}
where $F$ is the free energy, $S$ is the action for a given
Hamiltonian, and $\psi$ is a Grassmann variable
\cite{Negele:1998}. One then selects an observable quantity of
interest $A$, and couples a source $J$ to the observable $A$.
This results in a modified action $S+JA$, and the free energy
$F[J]$ is now a functional of the source $J$. A Legendre
transformation is then used to eliminate the source in favor of
the observable yielding a new functional
\begin{equation}
 \Gamma \lbrack A]=F[J[A]]-{A}J[A]
 \end{equation}
$\Gamma[A]$ is useful in that the variational derivative with respect
to $A$ yields $J$. We are free to set the source to zero,
and thus the the extremum of $\Gamma[A]$ gives the free energy
of the system.


The value of the approach is that useful approximations to the
functional $\Gamma \lbrack A]$ can be constructed in practice
using the inversion method, a powerful technique introduced to
derive the TAP (Thouless, Anderson and Palmer) equations in spin
glasses by \cite{Plefka:1982} and by ~\cite{Fukuda:1988} to
investigate chiral symmetry breaking in QCD (see also Refs.
\cite{Georges:1991:PRB, Oper:2001, Yedidia:2001, Fukuda:1994}).
The approach consists in carrying out a systematic expansion of
the functional $\Gamma \lbrack A]$ to some order in a parameter
or coupling constant $\lambda $. The action is written as
$S=S_{0}+\lambda S_{1}$ and a systematic expansion is carried out
\begin{equation}
\Gamma \lbrack A]=\Gamma _{0}[A]+\lambda \Gamma _{1}[A]+ ... \ ,
\end{equation}%
\begin{equation}
J[A]=J_{0}[A]+\lambda J_{1}[A]+ ... \ .
\label{Eq:Jexpand}
\end{equation}

 A central point is that the system described by $S_0
+ A  J_{0}$ serves as a reference system for the fully
interacting problem. It is a simpler system which by
construction, reproduces the correct value of the observable
$\widehat{A}$, and when this observable is properly chosen, other
observables of the system can be obtained perturbatively from
their values in the reference system. Hence $S_0 +  A  J_{0}$ is
a simpler system which allows us to think perturbatively about
the physics of a more complex problem.  $J_{0}[A]$ is a central
quantity in this formalism and we refer to it as the
``constraining field". It is the source that needs to be added to
a reference action $S_{0}$ in order to produce a given value of
the observable $A$.

It is useful to split the functional in this way
\begin{equation}
\Gamma \lbrack A]=\Gamma _{0}[A]+\Delta \Gamma _{{}}[A]
  \label{Eq:breakUpGamma}
\end{equation}
since $\Gamma _{0}[A]=F_{0}[J_0]-A J_{0}$ we could regard

\begin{equation}
\Gamma[A,J_0]={F_0}[J_0]-A J_0+\Delta\Gamma[A]
 \label{func_twovariable}
\end{equation}
as a functional which is stationary in two variables, the
constraining field $J_0$ and A. The equation $\frac{\delta \Delta
\Gamma }{\delta A}=J_{0}[A],$ together with the definition of
$J_{0}[A]$ determines the exact constraining field for the
problem.

 One can also use the stationarity condition of the
functional  (\ref{func_twovariable}) to express A as a functional
of $J_0$ and obtain a functional of the {\it constraining field
alone} (ie. $\Gamma[J_0] = \Gamma[A[J_0], J_0] $). In the context of
the Mott transition problem, this approach allowed a clear
understanding of the analytical properties of the free energy
underlying the dynamical mean field theory \cite{Kotliar:1999}.

 $\Delta \Gamma$ can be a given a coupling constant
integration representation which is very useful, and will appear
in many guises through this review.
\begin{equation} \Delta
\Gamma \lbrack A] =
\int_{0}^{1} d \lambda \frac{\partial \Gamma}{\partial \lambda}=
\int_{0}^{1} d \lambda \langle S_1 \rangle_{J(\lambda),\lambda}
\label{Delta_Gamma}
\end{equation}

 Finally it is useful in many cases to decompose
$\Delta \Gamma =E_{H}+\Phi_{xc},$ by isolating the Hartree
contribution which can usually be evaluated explicitly. The
success of the method relies on obtaining good approximations to
the ``generalized exchange correlation" functional  $\Phi _{xc}.$

In the context of spin glasses, the parameter $\lambda $ is the
inverse temperature and this approach leads very naturally to the
TAP free energy. In the context of density functional theory,
$\lambda$ is the strength of the electron--electron interactions
as parameterized by the charge of the electron, and it can be
used to present a very transparent derivation of the density
functional approach \cite{Argaman:2000,Valiev:1997, Fukuda:1994,
Chitra:2000,Georges:2002, Savrasov:2003:CM0308053}. The central
point is that the choice of observable, and the choice of
reference system (i.e. the choice of  $S_0$ which determines
$J_0$) determine the structure of the (static or dynamic ) mean
field theory to be used.

Notice that above we coupled a source linearly to the
system of interest for the purpose of carrying out a Legendre
transformation. It should be noted that one is free to add
terms  which contain powers higher than one in the source
in order to modify the stability conditions of the functional
without changing the properties of the saddle points.
This freedom has been used to obtain functionals with better stability
properties \cite{Chitra:2001}.

We now illustrate these abstract considerations on a very
concrete example. To this end we consider the full many--body
Hamiltonian describing electrons moving in the periodic ionic
potential $V_{ext}(\mathbf{r})$ and interacting among themselves
according to the Coulomb law: $v_{C}(\mathbf{r}-\mathbf{r}^{\prime
})=e^{2}/|\mathbf{r}-\mathbf{r}^{\prime }|$. This is the formal
starting point of our all--electron first--principles
calculation. So, the ``theory of everything" is summarized in the
Hamiltonian
\begin{eqnarray}
&&H=\sum_{\sigma}\int d\mathbf{r}\psi
^{+}_{\sigma}(\mathbf{r})[-{\bigtriangledown ^{2}}+V_{ext}(
\mathbf{r})-\mu]\psi_{\sigma} (\mathbf{r})  \\
&&+{\frac{{1}}{2}}\sum_{\sigma\sigma'}\int d\mathbf{r}d\mathbf{r}^{\prime }\psi_{\sigma} ^{+}(\mathbf{r}%
)\psi ^{+}_{\sigma'}(\mathbf{r}^{\prime })v_{C}(\mathbf{r}-
\mathbf{r}^{\prime })\psi_{\sigma'} (\mathbf{r}^{\prime
})\psi_{\sigma} (\mathbf{r}).\notag  \label{INTham}
\end{eqnarray}%
Atomic Rydberg units, $\hbar =1,m_{e}=1/2$, are used throughout.
Using the functional integral formulation in the imaginary
time--frequency domain it is translated into the Euclidean
action $S$
\begin{equation}
S=\int dx\psi ^{+}(x)\partial _{\tau }\psi (x)+\int d\tau H(\tau),
\label{INTact}
\end{equation}%
where $x=(\mathbf{r}\tau \sigma )$. We will ignore relativistic
effects in this action for simplicity. In addition the position
of the atoms is taken to be fixed and we ignore the
electron--phonon interaction. We refer the reader to several
papers addressing that issue \cite{Freericks:1993,
Millis:1996:PRB}.

The effective action functional approach \cite{Chitra:2001}
allows one to obtain the free energy $F$ of a solid from a
functional $\Gamma $ evaluated at its stationary point. The main
question is the choice  of the functional variable  which is to be
extremized. This question is highly non--trivial because the
exact form of the functional is unknown and the usefulness of the
approach depends on our ability to construct good approximations
to it, which in turn depends on the choice of variables. At least
two choices are very well--known in the literature: the exact
Green's function as a variable which gives rise to the
Baym--Kadanoff (BK) theory \cite{Baym:1961, Baym:1962} and the
density as a variable which gives rise to the density functional
theory. We review both approaches using an effective action point
of view in order to highlight similarities and differences with
the spectral density functional methods which will be presented
on the same footing in Section \ref{sec:SDF}.

\subsubsection{Density functional theory}
\label{sec:DFT}

Density functional theory in the Kohn--Sham formulation is one of the
basic tools for studying weakly--interacting electronic systems and is
widely used by the electronic structure community.  We will review it
using the effective action approach, which was introduced in this
context by Fukuda \cite{Fukuda:1994, Argaman:2000, Valiev:1997}.

$\bullet $ \textit{Choice of variables. } The density of electrons
$\rho (\mathbf{r})$ is the central quantity of DFT and it is used
as a physical variable in derivation of DFT functional.

$\bullet $ \textit{Construction of exact functional. } To
construct the DFT functional we probe the system with a
time--\emph{dependent} source field $J(x)$. This modifies the
action of the system (\ref{INTact}) as follows
\begin{equation}
S^{\prime }[J]=S+\int dxJ(x)\psi ^{+}(x)\psi (x) .  \label{DFTact}
\end{equation}
The partition function $Z$ becomes a functional of the
auxiliary source field $J$
\begin{equation}
  Z[J]=\exp(-F[J])=\int D[\psi^\dagger\psi]e^{-S'[J]}.
\label{Eq:ZJDFT}
\end{equation}
The effective action for the density, i.e., the density functional, is
obtained as the Legendre transform of $F$ with respect to $\rho(x)$
\begin{equation}
\Gamma_{DFT}[\rho ]=F[J]-\mathrm{Tr}\left(J\rho\right) ,
\label{DFTfun}
\end{equation}
where trace $\mathrm{Tr}$ stands for
\begin{equation}
  \mathrm{Tr}(J\rho)=\int dx J(x)\rho(x)=T\sum_{\iom}\int d\vr
  J(\vr,\iom) \rho(\vr,\iom) .
\end{equation}
From this point forward, we shall restrict the source to be time
independent because we will only be constructing the standard
DFT. If the time dependence where retained, one could formulate
time--dependent density functional theory (TDFT). The density
appears as the variational derivative of the free energy with
respect to the source
\begin{equation}
\rho (x)=\frac{\delta F}{\delta J(x)} .  \label{DFTden}
\end{equation}

$\bullet $ \textit{The constraining field in DFT. } We shall
demonstrate below that, in the context of DFT, the constraining
field is the sum of the well known exchange--correlation potential
and the Hartree potential $V_{xc}+V_{H}$, and we refer to this
quantity as $V_{int}$. This is the potential which must be added
to the non--interacting Hamiltonian in order to yield the exact
density of the full Hamiltonian. Mathematically, $V_{int}$ is a
functional of the density which solves the equation
\begin{equation}
\rho (\mathbf{r})=T\sum_{i\omega }\langle \mathbf{r}\left\vert [
i\omega +\mu +\nabla ^{2}-V_{ext}(\mathbf{r})-V_{int}(\mathbf{r})]^{-1}\right\vert
\mathbf{r}\rangle e^{i\omega 0^{+}} .  \label{DFTror}
\end{equation}

%
The Kohn--Sham equation gives rise to a reference system of
non--interacting particles, the so called Kohn--Sham orbitals
$\psi_{\mathbf{k}j}$ which produce the interacting density
\begin{equation}
\lbrack -\nabla ^{2}+V_{KS}(\mathbf{r})]\psi _{\mathbf{k}j}(\mathbf{r}%
)=\epsilon _{\mathbf{k}j}\psi _{\mathbf{k}j}(\mathbf{r}),
\label{DFTkse}
\end{equation}
\begin{equation}
\rho (\mathbf{r})=\sum_{\mathbf{k}j}f_{\mathbf{k}j}\psi
_{\mathbf{k}j}^{\ast }(\mathbf{r})\psi _{\mathbf{k}j}(\mathbf{r})
.  \label{DFTrop}
\end{equation}
Here the Kohn--Sham potential is $V_{KS}=V_{ext}+V_{int}$,
$\epsilon _{\mathbf{k}j},$ $\psi _{\mathbf{k}j}(\mathbf{r})$ are
the Kohn--Sham energy bands and wave functions, $\mathbf{k}$ is a
wave vector which runs over the first Brillouin zone, $j$ is band
index, and $f_{\mathbf{k} j}=1/[\exp (\epsilon _{\mathbf{k}j}-\mu
)/T+1]$ is the Fermi function.

$\bullet $ \textit{Kohn--Sham Green's function. }  Alternatively,
the electron density can be obtained with the help of the
Kohn--Sham Green's function, given by
\begin{equation}
G_{KS}^{-1}(\mathbf{r},\mathbf{r}^{\prime },i\omega )=G_{0}^{-1}(\mathbf{r},%
\mathbf{r}^{\prime },i\omega )-V_{int}(\mathbf{r})\delta (\mathbf{r}-\mathbf{%
r}^{\prime }),
\label{DFTgks}
\end{equation}
where $G_0$ is the non--interacting Green's function
\begin{equation}
G_{0}^{-1}(\mathbf{r},\mathbf{r}^{\prime },i\omega )=\delta
(\mathbf{r}-\mathbf{r}^{\prime })[i\omega +\mu +\nabla
^{2}-V_{ext}(\mathbf{r})], \label{BKFzer}
\end{equation}
and the density can then be computed from
\begin{equation}
\rho (\mathbf{r})=T\sum_{i\omega
}G_{KS}(\mathbf{r},\mathbf{r},i\omega )e^{i\omega
0+}
.
\label{DFTrog}
\end{equation}
The Kohn--Sham Green's function is
defined in the entire space, where $V_{int}(\mathbf{r})$ is adjusted
in such a way that the density of the system $\rho (\mathbf{r})$ can
be found from $G_{KS}(\mathbf{r},\mathbf{r}^{\prime },i\omega )$. It
can also be expressed in terms of the Kohn--Sham particles in the
following way
\begin{equation}
G_{KS}(\mathbf{r},\mathbf{r}^{\prime },i\omega )=\sum_{\mathbf{k}j}\frac{%
\psi _{\mathbf{k}j}(\mathbf{r})\psi _{\mathbf{k}j}^{\ast }(\mathbf{r}%
^{\prime })}{i\omega +\mu -\epsilon _{\mathbf{k}j}}.
\label{DFTksg}
\end{equation}%

$\bullet $ \textit{Kohn--Sham decomposition. } Now we come to the
problem of writing exact and approximate expressions for the
functional. The strategy consists in performing an expansion of
the functional in powers of electron charge \cite{Chitra:2001,
Fukuda:1994, Valiev:1997, Plefka:1982, Georges:1991,
Georges:2002}. The Kohn--Sham decomposition consists of splitting
the functional into the zeroth order term and the remainder.
\begin{equation}
\Gamma _{DFT}(\rho )=\Gamma _{DFT}(\rho ,e^{2}=0)+\Delta \Gamma
_{DFT}(\rho ).  \label{DFTexp}
\end{equation}
This is equivalent to what Kohn and Sham did in their original
work. In the first term, $e^{2}=0$ only for the electron--electron
interactions, and not for the interaction of the electron and the
external potential. The first term consists of the kinetic energy
of the Kohn--Sham particles and the external potential. The
constraining field $J_0$ (see Eq.~(\ref{Eq:Jexpand})) is
$V_{int}$ since it generates the term that needs to be added to
the non--interacting action in order to get the exact density.
Furthermore, functional integration of the Eq.~(\ref{Eq:ZJDFT})
gives $F[V_{int}]=-\mathrm{Tr}\ln [G_0^{-1}-V_{int}]$
\cite{Negele:1998} and from Eq.~(\ref{DFTfun}) it follows that
\begin{eqnarray}  \label{DFTkin}
&&\Gamma_{DFT}(\rho,e^{2}=0)\equiv K_{DFT}[G_{KS}]= \\
&&-\mathrm{Tr}\ln(G_0^{-1}-V_{int}[G_{KS}])-\mathrm{Tr}
\left(V_{int}[G_{KS}]G_{KS}\right).  \notag
\end{eqnarray}
The remaining part $\Delta \Gamma _{DFT}(\rho )$ is the
interaction energy functional which is decomposed into the
Hartree and exchange--correlation energies in a standard way
\begin{equation}
\Delta\Gamma_{DFT}(\rho )=
E_{H}[\rho ]+\Phi
_{DFT}^{xc}[\rho ] .  \label{DFTadd}
\end{equation}
${{\Phi^{xc}}_{DFT}} [\rho]$ at zero temperature becomes  the
standard    exchange correlation energy in DFT, $E_{xc}[\rho]$.

$\bullet $ \textit{Kohn--Sham equations as saddle--point
equations. }%
The density functional $\Gamma_{DFT}(\rho)$ can be regarded as a
functional which is stationary in two variables $V_{int}$ and
$\rho$. Extremization with respect to $V_{int}$ leads to
Eq.~(\ref{DFTgks}), while stationarity with respect to $\rho$
gives $V_{int}=\delta\Delta\Gamma/\delta\rho$, or equivalently,
\begin{align}
\nonumber
 V_{KS}[\rho](\vr)&=V_{ext}(\vr)+V_{int}[\rho](\vr)
 \\
 &=V_{ext}(\vr)+V_{H}[\rho](\vr)+V_{xc}[\rho](\vr),
 \label{DFTvks}
\end{align}
where $V_{xc}(\vr)$ is the exchange--correlation potential given
by
\begin{equation}
 V_{xc}(\mathbf{r})\equiv \frac{\delta \Phi _{DFT}^{xc}}{\delta
 \rho (\mathbf{r})} .%
 \label{DFTvxc}
\end{equation}
Equations (\ref{DFTvks}) and (\ref{DFTvxc}) along with
Eqs.~(\ref{DFTrog}) and (\ref{DFTgks}) or, equivalently,
(\ref{DFTkse}) and (\ref{DFTrop}) form the system of equations of
the density functional theory. It should be noted that the Kohn-Sham
equations give the true minimum of $\Gamma_{DFT}(\rho)$, and not
only the saddle point.

$\bullet $ \textit{Exact representation for} $\Phi_{DFT}^{xc} .$
The explicit form of the interaction functional $\Phi
_{DFT}^{xc}[\rho ]$ is not available. However, it may be defined
by a power series expansion which can be constructed order by
order using the inversion method. The latter can be given, albeit
complicated, a diagrammatic interpretation. Alternatively, an
expression for it involving integration by a coupling constant
$\lambda e^{2}$  can be obtained using the Harris--Jones formula
\cite{Harris:1974, Gunnarsson:1976, Langreth:1977, Georges:2002}.
One considers $\Gamma _{DFT}[\rho ,\lambda ]$ at an arbitrary
interaction $\lambda $ and expresses it as
\begin{equation}
\Gamma _{DFT}[\rho ,e^{2}]=\Gamma _{DFT}[\rho ,0]+\int_{0}^{1}d\lambda \frac{%
\partial \Gamma _{DFT}[\rho ,\lambda ]}{\partial \lambda } .  \label{DFTint}
\end{equation}%
Here the first term is simply $K_{DFT}[G_{KS}\mathcal{]}$ as given by (\ref%
{DFTkin}) which does not depend on $\lambda $. The second part is
thus the unknown functional $\Phi _{DFT}^{xc}[\rho ].$ The
derivative with respect to the coupling constant in
(\ref{DFTint}) is given by the average $\langle \psi ^{+}(x)\psi
^{+}(x^{\prime })\psi (x^{\prime }) \psi (x)\rangle  = \Pi
_{\lambda }(x,x^{\prime },i\omega )+$ $\langle \psi ^{+}(x)\psi
(x)\rangle \langle \psi ^{+}(x^{\prime })\psi (x^{\prime
})\rangle $ where $\Pi _{\lambda }(x,x^{\prime })$ is the
density--density correlation function at a given interaction
strength $\lambda $ computed in the presence of a source which is
$\lambda $ dependent \ and chosen so that the density of the
system was $\rho $.\ Since $\langle \psi ^{+}(x)\psi (x)\rangle
=\rho (x),$ one can obtain

\begin{equation}
\Phi _{DFT}[\rho ]=E_{H}[\rho ]+\sum_{i\omega }
\int d^3 \mathbf{r} d^3 \mathbf{r}^{\prime }
\int_{0}^{1}d\lambda \frac{%
\Pi _{\lambda }(\mathbf{r},\mathbf{r}^{\prime },i\omega )}{|\mathbf{r}-%
\mathbf{r}^{\prime }|} .  \label{DFTphi}
\end{equation}

This expression has been  used to  construct more accurate
exchange correlation functionals~\cite{Dobson:1997}.

$\bullet $ \textit{Approximations. } Since $\Phi _{DFT}^{xc}[\rho
]$ is not known explicitly some approximations are needed. The LDA
assumes
\begin{equation}
\Phi _{DFT}^{xc}[\rho ]=\int \rho (\mathbf{r})\epsilon _{xc}[\rho (\mathbf{r}%
)]d\mathbf{r},  \label{DFTlda}
\end{equation}%
where $\epsilon _{xc}[\rho (\mathbf{r})]$ is the
exchange--correlation energy of the uniform electron gas, which is
easily parameterized. $V_{eff}$ is given as an explicit function
of the local density. In practice one frequently uses the
analytical formulae~\cite{Barth:1972, Gunnarsson:1976:PRA,
Moruzzi:1978, Vosko:1980, Perdew:1992}. The idea here is to fit a
functional form to quantum Monte Carlo (QMC) calculations
\cite{Ceperley:1980}. Gradient corrections to the LDA have been
worked out by Perdew and coworkers \cite{Perdew:1996}. They are
also frequently used in LDA calculations.

$\bullet $ \textit{Evaluation of the total energy. } At the saddle
point, the density functional $\Gamma_{DFT}$ delivers the total
free energy of the system
\begin{eqnarray}
F = \mathrm{Tr}\ln G_{KS}-\mathrm{Tr}\left(V_{int}\rho\right)
+E_{H}[\rho]+\Phi_{DFT}^{xc}[\rho]
,  \label{DFTtot}
\end{eqnarray}
where the trace in the second term runs only over spatial coordinates
and not over imaginary time.
If temperature goes to zero, the entropy contribution
vanishes and the total energy formulae is recovered
 \begin{equation}
E=-\mathrm{Tr}(\nabla^{2}G_{KS})+\mathrm{Tr}\left(V_{ext}\rho\right)
+E_{H}[\rho]+E_{DFT}^{xc}[\rho ].  \label{DFTene}
\end{equation}

$\bullet $ \textit{Assessment of the approach. } From a conceptual
point of view, the density functional approach is radically different
from the Green's function theory (See below). The Kohn--Sham equations
(\ref{DFTkse}), (\ref{DFTrop}) describe the Kohn--Sham quasiparticles
which are poles of $G_{KS}$ and are not rigorously identifiable with
one--electron excitations. This is very different from the Dyson
equation (see below Eq.~(\ref{BKFdys})) which determines the Green's
function $G$, which has poles at the observable one--electron
excitations. In principle the Kohn--Sham orbitals are a technical tool
for generating the total energy as they alleviate the kinetic energy
problem.
They are however not a necessary element of the
approach as DFT can be formulated without introducing the Kohn-Sham
orbitals.
In practice, they are also used as a first step in perturbative
calculations of the one--electron Green's function in powers of
screened Coulomb interaction, as e.g. the GW method. Both the LDA
and GW methods are very successful in many materials for which
the standard model of solids works. However, in correlated
electron system this is not always the case. Our view is that
this situation cannot be remedied by either using more
complicated exchange-- correlation functionals in density
functional theory or adding a finite number of diagrams in
perturbation theory. As discussed above, the spectra of
strongly--correlated electron systems have both correlated
quasiparticle bands and Hubbard bands which have no analog in
one--electron theory.

The density functional theory can also be formulated for the
model Hamiltonians, the concept of density being replaced by the
diagonal part of the density matrix in a site representation. It
was tested in the context of the Hubbard model
by~\cite{Hess:1999, Lima:2002, Schonhammer:1995}.

\subsubsection{Baym--Kadanoff functional}
\label{sec:BKfunctional}

The Baym--Kadanoff functional~\cite{Baym:1961, Baym:1962} gives
the one--particle Green's function and the total free energy at
its stationary point. It  has been derived in many papers starting
from \cite{Dominicis:1964,Dominicis:1964b} and
\cite{Cornwall:1974} (see also \cite{Chitra:2000, Chitra:2001,
Georges:2004:CM, Georges:2004:CM0403123}) using the effective
action formalism.

$\bullet $ \textit{Choice of variable. } \ The one--electron
Green's function $G(x,x^{\prime })=-\langle T_{\tau }\psi (x)\psi
^{+}(x^{\prime })\rangle $, whose poles determine the exact
spectrum of one--electron excitations, is at the center of
interest in this method and it is chosen to be the functional
variable.

$\bullet $ \textit{Construction of exact functional. }As it has
been emphasized \cite{Chitra:2001}, the Baym--Kadanoff functional
can be obtained by the Legendre transform of the action. The
electronic Green's function of a system can be obtained by
probing the system by a source field and monitoring the response.
To obtain $ \Gamma_{BK}[G]$ we probe the system with a
time--dependent two--variable source field $J(x,x^{\prime })$. %
%
Introduction of the source $J(x,x^{\prime })$ modifies the action
of the system (\ref{INTact}) in the following way
\begin{equation}
S^{\prime }[J]=S+\int dxdx^{\prime }J(x,x^{\prime })\psi
^{+}(x)\psi (x^{\prime }).  \label{BKFact}
\end{equation}%
The average of the operator $\psi^+(x)\psi(x^{\prime})$
probes the Green's function. The partition function $Z,$ or
equivalently the free energy of the system $F,$ becomes a
functional of the auxiliary source field
\begin{equation}
Z[J]=\exp (-F[J])=\int D[\psi ^{+}\psi ]e^{-S^{\prime }[J]}.
\label{BKFpar}
\end{equation}%
The effective action for the Green's function, i.e., the
Baym--Kadanoff functional, is obtained as the Legendre transform
of $F$ with respect to $G(x,x^{\prime })$
\begin{equation}
\Gamma _{BK}[G]=F[J]-\mathrm{Tr}(JG),  \label{BKFbkf}
\end{equation}
where we use the compact notation $\mathrm{Tr}(JG)$ for the
integrals
\begin{eqnarray}
&&\mathrm{Tr}(JG)=\int dxdx^{\prime }J(x,x^{\prime })G(x^{\prime },x) .
\label{BKFtra}
\end{eqnarray}
Using the condition
\begin{equation}
G(x,x^{\prime })=\frac{\delta F}{\delta J(x^{^{\prime }},x)},
\label{BKFgrn}
\end{equation}%
to eliminate $J$ in (\ref{BKFbkf}) in favor of the Green's
function, we finally obtain the functional of the Green's function
alone.

$\bullet $ \textit{Constraining field in the Baym--Kadanoff
theory. } In the context of the Baym--Kadanoff approach, the
constraining field is the familiar electron self--energy
$\Sigma_{int}(\vr,\vr^\prime,i\omega)$. This is the function
which needs to be added to the inverse of the non--interacting
Green's function to produce the inverse of the exact Green's
function, i.e.,
\begin{equation}
G^{-1}(\mathbf{r},\mathbf{r}^{\prime },i\omega )=G_{0}^{-1}(\mathbf{r},
\mathbf{r}^{\prime },i\omega )-
\Sigma_{int}(\mathbf{r},\mathbf{r}^{\prime },i\omega ) .  \label{BKFgm1}
\end{equation}
Here $G_0$ is the non--interacting Green's function given by
Eq.~(\ref{BKFzer}).  Also, if the Hartree potential is written
explicitly, the self--energy can be split into the Hartree,
$V_{H}(\mathbf{r})=\int v_{C}(\mathbf{r}-\mathbf{r}^{\prime })\rho
(\mathbf{r} ^{\prime })d\mathbf{r}^{\prime }$ and the
exchange--correlation part, $ \Sigma
_{xc}(\mathbf{r},\mathbf{r}^{\prime },i\omega).$

Ultimately, having fixed $G_{0}$ the self--energy
becomes a functional of $G$, i.e. $\Sigma_{int}[G].$

$\bullet $ \textit{Kohn--Sham decomposition. } We now come to the
problem of writing various contributions to the Baym--Kadanoff
functional. This development parallels exactly what was done in
the DFT case. The strategy consists of performing an expansion of
the\ functional $\Gamma _{BK}[G]$ in powers of the charge of
electron entering the Coulomb interaction term at fixed $G$
\cite{Chitra:2001, Fukuda:1994, Valiev:1997, Plefka:1982,
Georges:1991, Georges:2002, Georges:2004:CM,
Georges:2004:CM0403123}. The zeroth order term is denoted $K$,
and the sum of the remaining terms $\Phi $, i.e.
\begin{equation}
\Gamma _{BK}[G]=K_{BK}[G]+\Phi _{BK}[G] .
\label{BKFkpf}
\end{equation}
$K$ is the kinetic part of the action plus the energy associated
with the external potential $V_{ext}$. In the Baym--Kadanoff
theory this term has the form
\begin{eqnarray}  \label{BKFkin}
K_{BK}[G]&=&\Gamma _{BK}[G,e^{2}=0]= \\
&-&\mathrm{Tr}\ln (G_{0}^{-1}-\Sigma_{int}[G])-\mathrm{Tr}
\left(\Sigma_{int}[G]G\right).
\notag
\end{eqnarray}

$\bullet $ \textit{Saddle--point equations. }%
The functional (\ref{BKFkpf}) can again be regarded as a
functional stationary in two variables, $G$ and constraining
field $J_0$, which is $\Sigma_{int}$ in this case. Extremizing
with respect to $\Sigma_{int}$ leads to the Eq.~(\ref{BKFgm1}),
while extremizing with respect to $G$ gives the definition of the
interaction part of the electron self--energy
\begin{equation}
\Sigma_{int}(\mathbf{r},\mathbf{r}^{\prime },i\omega
)=\frac{\delta \Phi _{BK}[G]}{\delta G(\mathbf{r}^{\prime
},\mathbf{r},i\omega )}. \label{BKFsig}
\end{equation}

Using the definition for $G_0$ in Eq.~(\ref{BKFzer}), the Dyson equation
(\ref{BKFgm1}) can be written in the following way
\begin{eqnarray}
\lbrack \nabla ^{2} - V_{ext}(\mathbf{r}) + i\omega  + \mu ]G(\mathbf{r},\mathbf{r%
}^{\prime },i\omega ) &-&  \label{BKFdys} \\
\int d\mathbf{r}^{\prime \prime }\Sigma_{int}(\mathbf{\ r},\mathbf{r}%
^{\prime \prime },i\omega )G(\mathbf{r}^{\prime \prime
},\mathbf{r}^{\prime },i\omega ) &=& \delta
(\mathbf{r}-\mathbf{r}^{\prime }).  \notag
\end{eqnarray}%
The Eqs.~(\ref{BKFsig}) and (\ref{BKFdys}) constitute a system of
equations for $G$ in the Baym--Kadanoff theory.

$\bullet $ \textit{Exact representation for}\textbf{\ }$\Phi .$
Unfortunately, the interaction energy functional $\Phi _{BK}[G]$
is unknown. One can prove that it can be represented as a sum of
all two--particle irreducible diagrams constructed from the
Green's function $G$ and the bare Coulomb interaction. In
practice, we almost always can separate the Hartree diagram from
the remaining part the so called exchange--correlation
contribution
\begin{equation}
\Phi _{BK}[G]=E_{H}[\rho ]+\Phi _{BK}^{xc}[G] .
\label{BKFint}
\end{equation}

$\bullet $ \textit{Evaluation of the total energy. } At the
stationarity point, $\Gamma _{BK}[G]$ delivers the free energy
$F$ of the system
\begin{equation}
F=\mathrm{Tr}\ln G-\mathrm{Tr}
\left(\Sigma_{int}G\right)+
E_{H}[\rho ]+\Phi _{BK}^{xc}[G],
\label{BKFfre}
\end{equation}%
where the first two terms are interpreted as the kinetic energy and
the energy related to the external potential, while the last two terms
correspond to the interaction part of the free energy.  If temperature
goes to zero, the entropy part vanishes and the total energy formula
is recovered
\begin{equation}
E_{tot}= - \mathrm{Tr}(\nabla^2 G)+\mathrm{Tr}(V_{ext}G)+
E_{H}[\rho]+E_{BK}^{xc}[G],
\label{BKFtot}
\end{equation}
where $E_{BK}^{xc}=1/2\mathrm{Tr}\left(\Sigma_{xc}G\right)$
\cite{Fetter:1971} (See also online notes).

$\bullet $ \textit{Functional of the constraining
field, self-energy functional approach. }  Expressing the
functional in Eq.~(\ref{BKFkpf}) in terms of the constraining
field, (in this case $\Sigma $ rather than the observable  $G$)
recovers the self-energy functional  approach proposed by
Potthoff~\cite{Potthoff:2003, Potthoff:2003:EPJB,
Potthoff:2005:CM}.
\begin{equation}
\Gamma [ \Sigma ]
= -\mathrm{Tr} \ln[{G_{0}}^{-1}-\Sigma ]+Y[\Sigma ]
\label{double1}
\end{equation}
$Y[\Sigma ]$ is the Legendre transform with respect to $G$ of the
Baym Kadanoff functional  $\Phi_{BK}[G]$. While explicit
representations of the Baym Kadanoff functional $\Phi$ are
available for example as a sum of skeleton graphs, no equivalent
expressions have yet been obtained for $Y[\Sigma]$.

$\bullet $ \textit{Assessment of approach. } The main advantage of
the Baym--Kadanoff approach is that it delivers the full spectrum
of one--electron excitations in addition to the ground state properties.
Unfortunately, the summation of all
diagrams cannot be performed explicitly and one has to resort to
partial sets of diagrams, such as the famous GW
approximation \cite{Hedin:1965} which  has only been
useful in the weak--coupling situations.%
Resummation of diagrams to infinite order guided by
the concept of locality, which is the basis of the Dynamical
Mean Field Approximation, can be formulated neatly as truncations
of the Baym Kadanoff functional as  will be shown in the
following sections. 

\subsubsection{Formulation in terms of the screened interaction}
\label{sec:ScreenedInt}

It is sometimes useful to think of Coulomb interaction as a
screened interaction mediated by a Bose field. This allows one to
define different types of approximations. In this context, using
the locality approximation for irreducible quantities gives rise
to the so--called Extended--DMFT, as opposed to the usual DMFT.
Alternatively, the lowest order Hartree--Fock approximation in
this formulation leads to the famous GW approximation.

An independent variable of the functional is the dynamically
screened Coulomb interaction $W(\vr,\vr',i\omega)$
\cite{Almbladh:1999} see also \cite{Chitra:2001}. In the
Baym--Kadanoff theory, this is done by introducing an auxiliary
Bose variable coupled to the density, which transforms the
original problem into a problem of electrons interacting with the
Bose field. The screened interaction $W$ is the connected
correlation function of the Bose field.

By applying the Hubbard--Stratonovich transformation to the action in
Eq.~(\ref{INTact}) to decouple the quartic Coulomb interaction, one
arrives at the following action
\begin{gather}
S =\int dx\psi ^{+}(x)
\Bigl(
\partial _{\tau }-\mu-{\bigtriangledown ^{2}}
+V_{ext}(x)+V_{H}(x) \Bigr) \psi (x) \nonumber
\\
+{\frac{{1}}{2}}\int dxdx^{\prime }\phi (x)v_{C}^{-1}(x-x^{\prime
})\phi (x^{\prime })
\nonumber
\\
-ig\int dx\phi (x)
\Bigl(
\psi ^{+}(x)\psi (x)-\langle \psi ^{+}(x)\psi
(x)\rangle_S
\Bigr)
\label{RSPact}
\end{gather}
where $\phi (x)$ is a Hubbard--Stratonovich field, $V_{H}(x)$ is
the Hartree potential, $g$ is a coupling constant to be set equal
to one at the end of the calculation and the brackets denote the
average with the action $S$. In Eq.~(\ref{RSPact}), we omitted
the Hartree Coulomb energy which appears as an additive constant,
but it will be restored in the full free energy functional. The
Bose field, in this formulation has no expectation value  (since
it couples to the ``normal order" term).

{$\bullet $ }\textit{Baym--Kadanoff functional of $G$ and $W$. }
Now we have a system of interacting fermionic and bosonic fields.
By introducing two source fields $J$ and $K$ we probe the electron
Green's function $G$ defined earlier and the boson Green's
function $W=\langle T_{\tau}\phi(x)\phi(x^{\prime})\rangle$ to be
identified with the screened Coulomb interaction. The functional
is thus constructed by supplementing the action
Eq.~(\ref{RSPact}) by the following term
\begin{eqnarray}
  S'[J,K]=S&+&\int dxdx' J(x,x')\psi^\dagger(x)\psi(x')\nonumber\\
  &+&\int dxdx'  K(x,x')\phi(x)\phi(x').
\end{eqnarray}
The normal ordering of the interaction ensures that
$\langle\phi(x)\rangle=0$.
The constraining fields, which appear as the zeroth order terms in
expanding $J$ and $K$ (see Eq.~(\ref{Eq:Jexpand})), are denoted by
$\Sigma_{int}$ and $\Pi$, respectively. The zeroth order free energy
is then
\begin{equation}
F_0[\Sigma_{int},\Pi]=-\mathrm{Tr}\left(G_0^{-1}-\Sigma_{int}\right)+
\frac{1}{2}\mathrm{Tr}\left(v_C^{-1}-\Pi\right),
\end{equation}
therefore the Baym--Kadanoff functional becomes
\begin{eqnarray}
\Gamma _{BK}[G,W]=-\mathrm{Tr}\ln\left(G_0^{-1}-\Sigma_{int}\right)-
\mathrm{Tr}\left(\Sigma_{int}G\right)\\\nonumber
+\frac{1}{2}\mathrm{Tr}\ln\left(v_C^{-1}-\Pi\right)+
\frac{1}{2}\mathrm{Tr}\left(\Pi\, W\right) + \Phi_{BK}[G,W].
\label{RSPbkf}
\end{eqnarray}
Again, $\Phi_{BK}[G,W]$ can be split into Hartree contribution and
the rest
\begin{eqnarray}
\Phi _{BK}[G,W]=E_{H}[\rho ]+\Psi _{BK}[G,W] .
\label{RSPpsi}
\end{eqnarray}

The entire theory is viewed as the functional of both $G$ and
$W.$ One of the strengths of such formulation is that there is a
very simple diagrammatic interpretation for $\Psi _{BK}[G,W]$. It
is given as the sum of all two--particle irreducible diagrams constructed
from $G $ and $W$~\cite{Cornwall:1974} with the exclusion of the
Hartree term. The latter $E_{H}[\rho ]$,  is evaluated with the
bare Coulomb interaction.

{$\bullet $ }\textit{Saddle point equations. } Stationarity with
respect to $G$ and $\Sigma_{int}$ gives rise to
Eqs.~(\ref{BKFsig}) and (\ref{BKFgm1}), respectively. An
additional stationarity condition $\delta \Gamma _{BK}/\delta
W=0$ leads to equation for the screened Coulomb interaction $W$
\begin{equation}
W^{-1}(\mathbf{r},\mathbf{r}^{\prime },i\omega )=v_{C}^{-1}(\mathbf{r}-%
\mathbf{r}^{\prime })-\Pi (\mathbf{r},\mathbf{r}^{\prime
},i\omega ), \label{RSPweq}
\end{equation}%
where function $\Pi (\mathbf{r},\mathbf{r}^{\prime },i\omega
)=-2\delta \Psi _{BK}/\delta W(\mathbf{r}^{\prime
},\mathbf{r},i\omega )$ is the  susceptibility of the interacting
system.

\subsubsection{Approximations}
\label{sec:Approximations}

\begin{figure}[ht]
\includegraphics[width=0.99\linewidth]{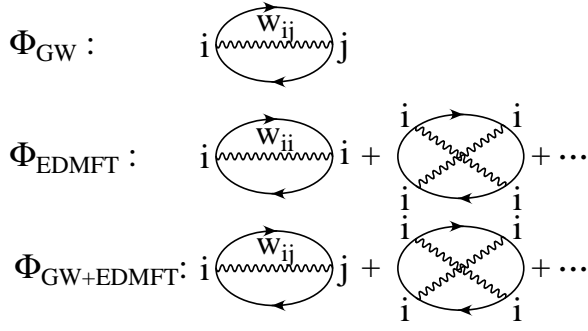}
\caption{ The Baym--Kadanoff functional $\Phi$ for various
  approximations for electron--boson action Eq.~(\ref{RSPact}). 
  In all cases, the bare Hartree diagrams have been omitted. The first
  line shows the famous GW approximation where only the lowest order
  Hartree and Fock skeleton diagrams are kept.  The second line
  corresponds to Extended--Dynamical Mean--Field Theory that sums up
  all the local graphs. Three dots represent all the remaining skeleton
  graphs which include local $G$ and local $W$ only. The combination of GW
  and EDMFT is straightforward. All lowest order Fock
  graphs are included (local and nonlocal). The higher order graphs
  are restricted to one site only (adapted from \protect\cite{Sun:2002,Sun:2004}).
}
\label{fig:cm0312303_fig1}
\end{figure}

The functional formulation in terms of a ``screened" interaction
$W$ allows one to formulate numerous approximations to the
many--body problem. The simplest approximation consists in keeping
the lowest order Hartree--Fock graph in the functional
$\Psi_{BK}[G,W]$. This is the celebrated GW approximation
\cite{Hedin:1965, Hedin:1969} (see Fig.~\ref{fig:cm0312303_fig1}).
To treat strong correlations one has to introduce
dynamical mean field ideas, which amount to a restriction of the
functionals $\Phi_{BK}, \Psi_{BK}$ to the local part of the Greens
function (see section~\ref{sec:SDF}). It is also natural to
restrict the correlation function of the Bose field $W$, which
corresponds to including information about the four point
function of the Fermion field in the self-consistency condition,
and goes under the name of the Extended Dynamical Mean--Field
Theory (EDMFT)~\cite{Bray:1980, Sachdev:1993, Sengupta:1995,
Kajueter:1996:thesis, Kajueter:Unpublished, Si:1996, Smith:2000,
Chitra:2001}.

This methodology has been useful in incorporating effects of the
long range Coulomb interactions ~\cite{Chitra:2000:PRL84} as well
as in the study of heavy fermion quantum critical
points,~\cite{Si:2001:NATURE, Si:1999} and quantum spin glasses
~\cite{Bray:1980, Sengupta:1995,Sachdev:1993}

More explicitly, in order to zero the off--diagonal Green's
functions (see Eq.~(\ref{Eq:EDMFT3})) we introduce a set of
localized orbitals $\Phi _{R\alpha }(r)$ and express $G$ and $W$
through an expansion in those orbitals.

\begin{equation}
G(r,r^{\prime},i\omega)= \sum_{RR^{\prime}\alpha\beta} G_{R\alpha,
R^{\prime}\beta}(i\omega) \Phi _{R\alpha }^{*}(r) \Phi
_{R^{\prime}\beta }(r^{\prime}),
\end{equation}
\begin{multline}
W(r,r^{\prime},i\omega)=
\sum_{R_1\alpha,R_2 \beta, R_3 \gamma, R_4 \delta}
W_{R_1\alpha,R_2 \beta, R_3 \gamma, R_4 \delta}(i\omega)
\times
\\
\Phi^* _{R_1\alpha }(r) \Phi^* _{R_2\beta }(r^{\prime}) \Phi_{R_3
\gamma}(r^{\prime}) \Phi_{R_4 \delta}(r) .
\end{multline}

The approximate EDMFT functional is obtained by \textit{restriction}
of the correlation part of the Baym--Kadanoff functional $\Psi_{BK}$ to the
diagonal parts of the $G$ and $W$ matrices:
\begin{eqnarray}
&&\Psi_{EDMFT} = \Psi_{BK}[G_{RR},W_{RRRR}] 
  \label{Eq:EDMFT3}
\end{eqnarray}
The EDMFT graphs are shown in Fig.~\ref{fig:cm0312303_fig1}.

It is straightforward to combine the GW and EDMFT approximations by
keeping the nonlocal part of the exchange graphs as well as the local
parts of the correlation graphs (see Fig.~\ref{fig:cm0312303_fig1}).

The GW approximation derived from the Baym--Kadanoff functional
is a fully self--consistent approximation which involves all
electrons. In practice sometimes two approximations are used: a)
in pseudopotential treatments only the self--energy of the valence
and conduction electrons are considered and b) instead of
evaluating $\Pi$ and $\Sigma$ self--consistently with $G$ and $W$,
one does a ``one--shot" or one iteration approximation where
$\Sigma $ and $\Pi$ are evaluated with $G_0$, the bare Green's
function which is sometimes taken as the LDA Kohn--Sham Green's
function, i.e., $\Sigma \approx \Sigma[ G_{0}, W_{0}]$ and $\Pi =
\Pi [ G_0]$. The validity of these approximations  and importance
of the self--consistency for the spectra evaluation was explored
in~\cite{Ku:2002:CM, Holm:1998, Arnaud:2000, Hybertsen:1985,
Holm:1999, Tiago:2003:CM}. The same issues arise in the context
of GW+EDMFT ~\cite{Sun:2004}.

At this point, the GW+EDMFT has been fully implemented on the
one--band model Hamiltonian level~\cite{Sun:2004, Sun:2002}. A
combination of GW and LDA+DMFT was applied to Nickel, where $W$
in the EDMFT graphs is approximated by the Hubbard $U$, in Refs.
\cite{Biermann:2003} and \cite{Biermann:2004:CM,
Aryasetiawan:2004:CM}.

\subsubsection{Model Hamiltonians and first principles approaches}%
\label{sec:INTmod}

In this section we  connect  the previous sections which were
based on real $\mathbf{r}$-space with the notation to be used
later in the review which use  local basis sets.  We perform a
transformation to a more general basis set of possibly
non--orthogonal orbitals $\chi _{\xi }(\mathbf{r})$ which can be
used to represent all the relevant quantities in our calculation.
As we wish to utilize sophisticated basis sets of modern
electronic structure calculations, we will sometimes waive the
orthogonality condition and introduce the overlap matrix
\begin{equation}
O_{\xi \xi ^{\prime }}=\langle \chi _{\xi }|\chi _{\xi ^{\prime
}}\rangle. \label{MODovr}
\end{equation}%
The field operator $\psi (x)$ becomes
\begin{equation}
\psi (x)=\sum_{\xi }c_{\xi }(\tau )\chi _{\xi }(\mathbf{r}),
\label{MODpsi}
\end{equation}%
where the coefficients $c_{\xi }$ are new operators acting in the
orbital space $\{\chi _{\xi }\}.$ The Green's function is
represented as
\begin{equation}
G(\mathbf{r},\mathbf{r}^{\prime },\tau )=\sum_{\xi \xi ^{\prime
}}\chi _{\xi }(\mathbf{r})G_{\xi \xi ^{\prime }}(\tau )\chi _{\xi
^{\prime }}^{\ast }( \mathbf{r}^{\prime }),  %
\label{MODgor}
\end{equation}%
and the free energy functional $\Gamma _{BK}$ as well as the
interaction energy $\Phi $ are now considered as functionals of
the coefficients $G_{\xi \xi ^{\prime }}$ either on the imaginary
time axis, $G_{\xi \xi ^{\prime }}(\tau )$ or imaginary frequency
axis $G_{\xi \xi ^{\prime }}(i\omega ),$ which can be
analytically continued to real times and energies.

In most cases we would like to interpret the orbital space
$\{\chi _{\xi }\}$ as a general tight--binding basis set where
the index $\xi $ combines the angular momentum index $lm$, and the
unit cell index $\mathbf{R},$ i.e., $\chi _{\xi }(\mathbf{r})=\chi
_{lm}(\mathbf{r}-\mathbf{R})=\chi _{\alpha }(\mathbf{r}-
\mathbf{R}).$ Note that we can add additional degrees of freedom
to the index $\alpha $ such as multiple kappa basis sets of the
linear muffin--tin orbital based methods \cite{Andersen:1975,
Andersen:1984, Methfessel:1988, Weyrich:1988, Bloechl:1989,
Savrasov:1992, Savrasov:1996}. If more than one atom per unit
cell is considered, index $\alpha $ should be supplemented by the
atomic basis position within the unit cell, which is currently
omitted for simplicity. For spin unrestricted calculations
$\alpha $ accumulates the spin index $\sigma $ and the orbital
space is extended to account for the eigenvectors of the Pauli
matrix.

It is useful to write down the Hamiltonian
containing the infinite space of the orbitals
\begin{equation}
\hat{H}=\sum_{\xi \xi ^{\prime }}h_{\xi \xi ^{\prime
}}^{(0)}[c_{\xi }^{+}c_{\xi ^{\prime
}}+h.c.]+\frac{1}{2}\sum_{\xi \xi ^{\prime }\xi ^{\prime \prime
}\xi ^{\prime \prime \prime }}V_{\xi \xi ^{\prime }\xi ^{\prime
\prime }\xi ^{\prime \prime \prime }}c_{\xi }^{+}c_{\xi ^{\prime
}}^{+}c_{\xi ^{\prime \prime }}c_{\xi ^{\prime \prime \prime}},
\label{MODham}
\end{equation}%
where $h_{\xi \xi ^{\prime }}^{(0)}=\langle \chi _{\xi }|-\nabla
^{2}+V_{ext}|\chi _{\xi ^{\prime }}\rangle $ is the
non--interacting Hamiltonian  and the interaction matrix element
is $V_{\xi \xi ^{\prime }\xi ^{\prime \prime }\xi ^{\prime \prime
\prime }}=\langle \chi _{\xi }(r)\chi _{\xi ^{\prime }}(r')|v_{C}|\chi
_{\xi^{\prime \prime }}(r')\chi _{\xi ^{\prime \prime \prime
}}(r)\rangle .$ Using the tight--binding interpretation this
Hamiltonian becomes
\begin{multline}  \label{MODhtb}
\hat{H}=\sum_{\alpha \beta }\sum_{RR^{\prime }}h_{\alpha R\beta
R^{\prime}}^{(0)}
\bigl(c_{\alpha R}^{+}c_{\beta R^{\prime }}+h.c. \bigr) \\
+\frac{1}{2}\sum_{\alpha \beta \gamma \delta
}\sum_{RR^{\prime}R''R^{\prime \prime \prime }}V_{\alpha \beta
\gamma \delta }^{RR^{\prime }R''R^{\prime \prime \prime
}}c_{\alpha R}^{+}c_{\beta R^{\prime }}^{+}c_{\delta R'''}c_{\gamma
R^{\prime\prime }} ,
\end{multline}%
where the diagonal elements $h_{\alpha R\beta R}^{(0)}\equiv
h_{\alpha \beta }^{(0)}$ can be interpreted as the generalized
atomic levels matrix $ \epsilon _{\alpha \beta }^{(0)}$ (which
does not depend on $R$ due to periodicity) and the off--diagonal
elements $h_{\alpha R\beta R'}^{(0)}(1-\delta _{RR^{\prime }})$ as
the generalized hopping integrals matrix $t_{\alpha R\beta
R^{\prime }}^{(0)}.$

\subsubsection{Model Hamiltonians}

Strongly correlated electron systems have been traditionally
described using model Hamiltonians. These are simplified
Hamiltonians which have the form of Eq. (\ref{MODhtb}) but
with a reduced number of band indices and sometimes assuming a
restricted form of the Coulomb interaction which is taken to be
very short ranged. The spirit of the approach is to describe a
reduced number of degrees of freedom which are active in a
restricted energy range to reduce the complexity of  a problem
and increase the accuracy of the treatment. Famous examples are
the Hubbard model (one band and multiband) and  the Anderson
lattice model.

The form of the model Hamiltonian  is  often guessed  on physical
grounds and its parameters chosen to fit a set of experiments. In
principle a more  explicit construction can be carried out using
tools such as   screening canonical transformations first used by
Bohm and Pines to eliminate the long range part of the Coulomb
interaction ~\cite{Bohm:1951,Bohm:1952,Bohm:1953}, or a Wilsonian
partial elimination (or integrating out) of the high--energy
degrees of freedom~\cite {Wilson:1975}. However,
these procedures are rarely used in practice.

One starts  from an action describing a large number of degrees of
freedom (site and orbital omitted)
\begin{equation}
S[{c}{^{+}}{c}]=\int d{x} \left({c}{^{+}}{O}%
\partial _{\tau }{c}+H[{c}{^{+}}{c}]\right) ,
\label{MODact}
\end{equation}%
where the orbital overlap $O_{\alpha R\beta R^{\prime }}$
appears  and the Hamiltonian could have the form  (\ref{MODhtb}).
Second, one divides the set of operators in the path integral in
$c_{H}$ describing the ``high--energy" orbitals which one would
like to eliminate, and $c_{L}$ describing the low--energy
orbitals that one would like to consider explicitly. The
high--energy degrees of freedom are now integrated out. This
operation  defines the effective action for the low--energy
variables~\cite{Wilson:1983}:

\begin{equation}
\frac{1}{Z_{eff}}\exp (-S_{eff}[{c}^{+}_{L}{c}_{L}]{)}=\frac{1}{Z}%
\int d{c}_{H}^{+}d{c}_{H}\exp
(-S[c_{H}^{+}c_{L}^{+}c_{L}c_{H}]).%
\label{MODrg1}
\end{equation}%
The transformation (\ref{MODrg1}) generates retarded interactions
of arbitrarily high order. If we focus on sufficiently low
energies, frequency dependence of the coupling constants beyond
linear order and non--linearities beyond quartic order can be
neglected since they are irrelevant around a Fermi liquid fixed
point \cite{Shankar:1994}. The resulting physical problem can
then be cast in the form of an effective model Hamiltonian.
Notice however that when we wish to consider a broad energy range
the full frequency dependence of the couplings has to be kept as
demonstrated in an explicit approximate calculation using the GW
method ~\cite{Aryasetiawan:2004:CM0401620}. The same ideas can be
implemented using canonical transformations and  examples of
approximate implementation of this program are provided by the
method of cell perturbation theory ~\cite{Raimondi:1996} and the
generalized tight-binding method~\cite{Ovchinnikov:1989}.

The concepts and the rational underlying the model Hamiltonian
approach are rigorous. There are very few studies of the form of
the Hamiltonians obtained by screening and elimination of
high--energy degrees of freedom, and the values of the parameters
present in those Hamiltonians. Notice  however  that if a form
for the model Hamiltonian is postulated, any technique which can
be used to treat Hamiltonians approximately, can be also used to
perform the elimination (\ref{MODrg1}).  A considerable amount of
effort has been devoted to the evaluations of the  screened
Coulomb parameter $U$ for a given material. Note that this value
is necessarily connected to the basis set representation which is
used in deriving the model Hamiltonian. It should be thought as
an effectively downfolded Hamiltonian to take into account the
fact that only the interactions at a given energy interval are
included in the description of the system. More generally, one
needs to talk about frequency--dependent interaction $W$ which
appears for example in the GW method. The outlined questions have
been addressed in many previous works \cite{Dederichs:1984,
McMahan:1988, Hybertsen:1989, Springer:1998, Kotani:2000}.
Probably, one of the most popular methods here is a constrained
density functional approach formulated with general projection
operators \cite{Dederichs:1984,Meider:1998}. First, one defines
the orbitals set which will be used to define correlated
electrons. Second, the on--site density matrix defined for these
orbitals is constrained by introducing additional constraining
fields in the density functional. Evaluating second order
derivative of the total energy with respect to the density matrix
should in principle give us the access to $U$s. The problem is
how one subtracts the kinetic energy part which appears in this
formulation of the problem. Gunnarsson \cite{Gunnarsson:1990} and
others ~\cite{McMahan:1988:book, Freeman:1987, Norman:1986} have
introduced a method which effectively cuts the hybridization of
matrix elements between correlated and uncorrelated orbitals
eliminating the kinetic contribution. This approach was used by
McMahan et al. \cite{McMahan:1988} in evaluating the Coulomb
interaction parameters in the high--temperature superconductors.
An alternative method has been used by Hybertsen et al.
\cite{Hybertsen:1989} who performed simultaneous simulations
using the LDA and solution of the model Hamiltonian at the
mean--field level. The total energy under the constraint of fixed
occupancies was evaluated within both approaches. The value of
$U$  is adjusted to make the two calculations coincide.

Much work has been done by the group of Anisimov who have
performed evaluations of the Coulomb and exchange interactions
for various systems such as NiO, MnO, CaCuO$_{2}$ and so on
\cite{Anisimov:1991}. Interestingly, the values of $U$ deduced
for such itinerant system as Fe can be as large as
6~eV~\cite{Anisimov:1991:43}. This highlights an important
problem on deciding which electrons participate in the screening
process. As a rule of thumb, one can argue that if we consider
the entire $d$-shell as a correlated set, and allow its screening
by $s$- and $p$-electrons, the values of $U$ appear to be between
5 and 10~eV on average. On the other hand, in many situations
crystal field splitting between $t_{2g}$ and $e_{g}$ levels
allows us to talk about a subset of a given crystal field symmetry
(say, $t_{2g}$), and allowing screening by another subset (say by
$e_{g}$). This usually leads to much smaller values of $U$ within
range of 1-4~eV.

It is possible to extract the value of $U$ from  GW calculations.
The simplest way to define the parameter $U=W$ ($\omega =0$).
There are also attempts to avoid the double counting inherent in
that procedure~\cite{Aryasetiawan:2004:CM0401620, Springer:1998,
Kotani:2000, Zein:2002,Zein:Unpublished}.
%
%
The values of $U$ for Ni deduced in this way appeared to be
2.2-3.3~eV which are quite reasonable. At the same time a strong
energy dependence of the interaction has been pointed out which
also addresses an important problem of treating the full
frequency--dependent interaction when  information in a broad
energy range is required.

The process of eliminating degrees of freedom with the
approximations described above gives us a physically rigorous way
of thinking about effective Hamiltonians with effective
parameters which are screened by the degrees of freedom to be
eliminated. Since we neglect retardation and terms above fourth
order, the effective Hamiltonian would have the same form
as (\ref{MODhtb}) where we only change the meaning of the
parameters. It should be regarded as the effective Hamiltonian
that one can use to treat the relevant degrees of freedom. If the
dependence on the ionic coordinates are kept, it can be used to
obtain the total energy. If the interaction matrix turns out to
be short ranged or has a simple form, this effective
Hamiltonian could be identified with the Hubbard
\cite{Hubbard:1963} or with the Anderson \cite{Anderson:1961}
Hamiltonians.

Finally we comment on  the meaning of an \textit{ab initio} or a
first--principles electronic structure  calculation. The term
implies that no empirically adjustable parameters are needed in
order to predict physical properties of compounds, only the
structure and the charges of atoms are used as an input.
First--principles does not mean exact or accurate or computationally
inexpensive. If the effective Hamiltonian is {\it derived} (i.e.
if the functional integral  or canonical transformation needed to
reduce the number of degrees of freedom is performed by a
well--defined procedure which keeps track of the energy of the
integrated out degrees of freedom as a function of the ionic
coordinates) and the consequent Hamiltonian (\ref{MODhtb}) is
solved  systematically, then we have a first--principles method. In
practice, the derivation of the effective Hamiltonian or its
solution may be inaccurate or impractical, and in this case the
\textit{ab initio} method is not very useful. Notice that
$H_{eff}$ has the form of a ``model Hamiltonian" and very often a
dichotomy between model Hamiltonians and first--principles
calculations is made. What makes a model calculation
semi--empirical is the lack of a coherent derivation of the form of
the ``model Hamiltonian" and the corresponding parameters.

\section{Spectral density functional approach}%
\label{sec:SDF}

We see that a great variety of many--body techniques developed to
attack real materials can be viewed from a unified perspective.
The energetics and excitation spectrum of the solid is deduced
within different degrees of approximation from the stationary
condition of a functional of an observable. The different
approaches differ in the choice of variable for the functional
which is to be extremized. Therefore, the choice of the variable
is a central issue since the exact form of the functional is
unknown and existing approximations entirely rely on the
given variable.

In this review we present arguments that a ``good variable" in the
functional description of a strongly--correlated material is a
``local" Green's function $G_{loc}(\mathbf{r},\mathbf{r}%
^{\prime },z).$ This is only a part of the exact electronic
Green's function, but it can be presently computed with some degree
of accuracy. Thus we would like to formulate a functional theory
where the local spectral density is the central quantity to be
computed, i.e. to develop a spectral density functional theory
(SDFT). Note that the notion of locality by itself is arbitrary
since we can probe the Green's function in a portion of a certain
space such as reciprocal space or real space. These are
the most transparent forms where the local Green's function can
be defined. We can also probe the Green's function in a portion
of the Hilbert space like Eq. (\ref{MODgor}) when the Green's
function is expanded in some basis set $ \{\chi _{\xi }\}$. Here
our interest can be associated, e.g, with diagonal elements of
the matrix $G_{\xi \xi ^{\prime }}$.

As we see, locality is a basis set dependent property.
Nevertheless, it is a very useful property because it may lead to
a very economical description of the function. The choice of the
appropriate Hilbert space is therefore crucial if we would like
to find an optimal description of the system with the accuracy
proportional to the computational cost. Therefore we always rely
on physical intuition when choosing a particular representation
which should be tailored to a specific physical problem.

\subsection{Functional of local Green's function}
\label{sec:SDFact}

We start from the Hamiltonian of the form (\ref{MODhtb}). One can
view it as the full Hamiltonian written in some complete
tight--binding basis set. Alternatively one can regard the
starting point (\ref{MODhtb}) as a model Hamiltonian, as argued
in the previous section, if an additional constant term (which
depends on the position of the atoms) is kept and (\ref{MODhtb})
is carefully derived. This can represent the full Hamiltonian in
the relevant energy range provided that one neglects higher order
interaction terms.

{$\bullet $ }\textit{Choice of variable and construction of the
exact functional. }The effective action construction of SDFT
parallels that given in Introduction. The quantity of interest is
the local (on--site) part of the one--particle Green's
function. It is generated by adding a local source $J_{loc,\alpha
\beta }(\tau ,\tau ^{\prime })$ to the action

\begin{equation}
S^{\prime }=S+\sum_{R \alpha \beta }\int J_{loc,R \alpha \beta
}(\tau ,\tau ^{\prime })c_{R \alpha }^{+}(\tau )c_{R \beta }(\tau
^{\prime })d\tau d\tau ^{\prime }.  \label{SDFacm}
\end{equation}%
The partition function $Z$, or equivalently the free energy of
the system $F$, according to (\ref{BKFpar})  becomes a functional
of the auxiliary source field  and the local Green's function is
given by the variational derivative
\begin{equation}
\frac{\delta F}{\delta J_{loc,R \beta \alpha }(\tau ^{\prime
},\tau )} =-\left\langle T_\tau c_{R\alpha }(\tau )c_{R\beta
}^{+}(\tau ^{\prime })\right\rangle =G_{loc, \alpha \beta }(\tau
,\tau ^{\prime }). \label{SDFgrn}
\end{equation}
From Eq.~(\ref{SDFgrn}) one expresses $J_{loc}$ as a functional of $%
G_{loc}$ to obtain the effective action by the standard
procedure
\begin{equation}
\Gamma _{SDFT}[G_{loc}]=F[J_{loc}]-\mathrm{Tr}\left(J_{loc}G_{loc}\right).
\label{SDFsdf}
\end{equation}
The extremum of this functional gives rise to the exact local spectral
function $G_{loc}$ and the total free energy $F$.

Below, we will introduce the Kohn--Sham representation of the
spectral density functional $\Gamma _{SDFT}$ similar to what was
done in the Baym--Kadanoff and density functional theories. A
dynamical mean--field approximation to the functional will be
introduced in order to deal with its interaction counterpart. The
theory can be developed along two alternative paths depending on
whether we stress that it is a truncation of the exact functional
when expanding $\Gamma _{SDFT}$ in powers of the hopping (atomic
expansion) or in powers of the interaction (expansion around the
band limit). The latter case is the usual situation encountered
in DFT and the Baym--Kadanoff theory, while the former has only
been applied to SDFT thus far.

\subsubsection{A non--interacting reference system: bands in a frequency--dependent potential}

{$\bullet $ }\textit{The constraining field in the context of
SDFT. } In the context of SDFT, the constraining field is defined
as $\mathcal{M}_{int,\alpha \beta }(i\omega )$. This is the
function that one needs to add to the free Hamiltonian in order
to obtain a desired spectral function:

\begin{equation}
G_{loc,\alpha \beta }(i\omega )=\sum_{\mathbf{k}}
\biggl(
(i\omega +\mu )\hat{I}-%
\hat{h}^{(0)}(\mathbf{k})-\mathcal{M}_{int}[G_{loc}](i\omega
)
\biggr)_{\alpha \beta }^{-1}, \label{SDFloc}
\end{equation}%
where $\hat{I}$ is a unit matrix, $\hat{h}^{(0)}(\mathbf{k})$ is
the Fourier transform (with respect to $R-R'$) of the bare
one--electron Hamiltonian $h_{\alpha R\beta R^{\prime }}^{(0)}$
entering (\ref{MODhtb}). The assumption that the equation
(\ref{SDFloc}) can be solved to define $\mathcal{M}_{int,\alpha
\beta }(i\omega )$ as a function of $G_{loc,\alpha \beta
}(i\omega )$, is the SDFT version of the \textit{Kohn--Sham
representability} condition of DFT. For DFT this has been proved
to exist under certain conditions, (for discussion of this
problem see \cite{Gross:1996}). The SDFT condition has not been
yet investigated in detail, but it seems to be a plausible
assumption.

$\bullet $ \textit{Significance of the constraining field in
SDFT. } If the exact self--energy of the problem is momentum
independent, then $\mathcal{M}_{int,\alpha \beta }(i\omega )$
coincides with the interaction part of the self--energy. This
statement resembles the observation in DFT: if the self--energy
of a system is momentum and frequency independent then the
self--energy coincides with the Kohn--Sham potential.

$\bullet $\textit{Analog of the Kohn--Sham Green's function. }
Having defined $\mathcal{M}_{int,\alpha \beta }(i\omega )$, we
can introduce an auxiliary Green's function $\mathcal{G}_{\alpha
R\beta R^{\prime }}(i\omega )$ connected to our new ``interacting
Kohn--Sham" particles. It is defined in the entire space by the
relationship:
\begin{equation}
\mathcal{G}_{\alpha R\beta R^{\prime }}^{-1}(i\omega
)\equiv
G_{0,\alpha R\beta R^{\prime }}^{-1}(i\omega )-\delta
_{RR^{\prime }}\mathcal{M}_{int,\alpha \beta }(i\omega ),
\label{SDFgks}
\end{equation}%
where ${G}_{0}^{-1}=(i\omega +\mu
)\hat{I}-\hat{h}^{(0)}(\mathbf{k})$ (in Fourier space). $\mathcal{M}_{int,\alpha
\beta }(i\omega )$ was defined so that $\mathcal{G} _{\alpha
R\beta R^{\prime }}(i\omega )$ coincides with the on--site Green's
function on a single site and the Kohn--Sham Green's function has
the property
\begin{equation}  \label{SDFgkl}
G_{loc,\alpha \beta }(i\omega )=\delta _{RR^{\prime
}}\mathcal{G}_{\alpha R\beta R^{\prime }}(i\omega ).
\end{equation}

Notice that $\mathcal{M}_{int}$ is a functional of $G_{loc}$ and
therefore $\mathcal{G}$ is also a function of $G_{loc}$. If this
relation can be inverted, the functionals that where previously
regarded as functionals of $G_{loc}$ can be also regarded as
functionals of the Kohn--Sham Green's function $\mathcal{G}$.

{$\bullet $ }\textit{Exact Kohn--Sham decomposition. }%
We separate the functional $\Gamma _{SDFT}[G_{loc}]$ into the
non--interacting contribution (this is the zeroth order term in
an expansion in the Coulomb interactions), $K_{SDFT}[G_{loc}]$,
and the remaining interaction contribution, $\Phi
_{SDFT}[G_{loc}]$: $\Gamma _{SDFT}[G]=K_{SDFT}[G_{loc}]+\Phi
_{SDFT}[G_{loc}]$. With the help of $\mathcal{M}_{int}$ or
equivalently the Kohn--Sham Green's function $\mathcal{G}$ the
non--interacting term in the spectral density functional theory
can be represented (compare with (\ref{DFTkin}) and
(\ref{BKFkin})) as follows
\begin{multline}
K_{SDFT}[G_{loc}]=-\mathrm{Tr}\ln
(G_{0}^{-1}-\delta _{RR^{\prime}}\mathcal{M}_{int}[G_{loc}])-
\\
\mathrm{Tr}\left(\delta _{RR^{\prime}}\mathcal{M}_{int}[G_{loc}]G_{loc}\right).
\label{SDFkin}
\end{multline}
Since $\mathcal{G}$ is a functional of $G_{loc}$, one can also
view the
entire spectral density functional $\Gamma _{SDFT}$ as a functional of
$\mathcal{G}$:
\begin{multline}
\Gamma_{SDFT}[\mathcal{G}]=
-\mathrm{Tr}\ln (G_{0}^{-1}-
\delta _{RR^{\prime}}\mathcal{M}_{int}[
\mathcal{G}])-
\\
  \mathrm{Tr}\left(\mathcal{M}_{int}[\mathcal{G}]\mathcal{G}\right)+
\Phi _{SDFT}[G_{loc}[\mathcal{G}]],%
  \label{SDFfun}
\end{multline}
where the unknown interaction part of the free energy $\Phi
_{SDFT}[G_{loc}]$ is a functional of $G_{loc}$ and
\begin{equation}
\frac{\delta G_{loc,\alpha \beta }}{\delta \mathcal{G}_{\alpha
R\beta R^{\prime }}}=\delta _{RR^{\prime }},  \label{SDFder}
\end{equation}%
according to Eq.~(\ref{SDFgkl}).

{$\bullet $ }\textit{Exact representation of} $\Phi _{SDFT}$.
Spectral density functional theory requires the interaction
functional $ \Phi _{SDFT}[G_{loc}]$. Its explicit form is
unavailable. However we can express it via an introduction of an
integral over the coupling constant $ \lambda e^{2}$ multiplying
the two--body interaction term similar to the density functional
theory \cite{Harris:1974, Gunnarsson:1976} result. Considering
$\Gamma_{SDFT}[G_{loc},\lambda ]$ at any interaction $\lambda $
(which enters $v_{C}(\mathbf{r-r^{\prime }})$) we write
\begin{equation}
\Gamma _{SDFT}[G_{loc},e^{2}]=\Gamma
_{SDFT}[G_{loc},0]+\int_{0}^{1}d\lambda \frac{\partial \Gamma
_{SDFT}[G_{loc},\lambda ]}{\partial \lambda }. \label{SDFlam}
\end{equation}%
Here the first term is simply the non--interacting part
$K_{SDFT}[G_{loc}]$ as given by (\ref{SDFkin}) which does not
depend on $\lambda $. The second part is thus the unknown
functional  (see Eq.~(\ref{Delta_Gamma}))
\begin{eqnarray}
&&\Phi _{SDFT}[G_{loc}]=\int_{0}^{1}d\lambda \frac{\partial \Gamma
_{SDFT}[G_{loc},\lambda ]}{\partial \lambda } \\
&=&{1\over 2}\int_{0}^{1}d\lambda \sum_{RR^{\prime }R^{ \prime
\prime}R^{\prime \prime \prime }}\sum_{\alpha \beta \gamma \delta
}V_{\alpha \beta \gamma \delta }^{RR^{\prime }R^{ \prime
\prime}R^{\prime \prime \prime }}\langle c_{\alpha R}^{+}c_{\beta
R^{\prime }}^{+}c_{\gamma R^{\prime \prime}}c_{\delta R^{\prime
\prime \prime }}\rangle _{\lambda }. \notag
\end{eqnarray}%
One can also further separate $\Phi _{SDFT}[G_{loc}]$ into $
E_{H}[G_{loc}]+\Phi _{SDFT}^{xc}[G_{loc}]$, where the Hartree term
is a functional of the density only.

{$\bullet $ }\textit{Exact functional as a function of two
variables. } The SDFT\ can also be viewed as a functional of two
independent variables~\cite{Kotliar:2001:Tsvelik}. This is
equivalent to what is known as Harris--Foulkes--Methfessel
functional within DFT~\cite{Harris:1985, Foulkes:1989,
Methfessel:1995}
\begin{widetext}
\begin{eqnarray}
\label{SDFtwo}
\Gamma_{SDFT}[G_{loc},\mathcal{M}_{int}]=
-\sum_{\mathbf{k}}\mathrm{Tr}\ln[(i\omega+\mu)\hat{I}-
  \hat{h}^{(0)}(\mathbf{k})-\mathcal{M}_{int}(i\omega)]-
\mathrm{Tr}\left(\mathcal{M}_{int}G_{loc}\right)+\Phi_{SDFT}[G_{loc}].
\end{eqnarray}
\end{widetext}Eq.~(\ref{SDFloc}) is a saddle point of the functional
(\ref{SDFtwo}) defining $\mathcal{M}_{int} =
\mathcal{M}_{int}[G_{loc}]$ and should be back--substituted to
obtain $\Gamma _{SDFT}[G_{loc}]$.

{$\bullet $ }\textit{Saddle point equations and their
significance. }
Differentiating the functional (\ref{SDFtwo}),
one obtains a functional equation for $G_{loc}$
\begin{equation}
\mathcal{M}_{int}[G_{loc}]=\frac{\delta \Phi
_{SDFT}[G_{loc}]}{\delta G_{loc}}. \label{SDFmas}
\end{equation}%
Combined with the definition of the constraining field (\ref{SDFloc}) it
gives the standard form of the DMFT equations. Note that thus far
{\it these are exact equations} and the constraining field $\mathcal{M}
_{int}(i\omega )$ is by definition ``local", i.e.  momentum
independent.

\subsubsection{An interacting reference system: a dressed atom}

We can obtain the spectral density functional by adopting a
different reference system, namely the atom. The starting point
of this approach is the Hamiltonian (\ref{MODhtb}) split into two
parts ~\cite{Chitra:2000, Georges:2004:CM,Georges:2004:CM0403123}:
$H=H_{0}+H_{1}$, where $H_{0}=\sum_{R}H_{at}[R]$ with $H_{at}$
defined as
\begin{eqnarray}
H_{at}[R]&=&\sum_{\alpha \beta }h_{\alpha R\beta
R}^{(0)}[c_{\alpha R}^{+}c_{\beta R}+h.c.]\\\nonumber%
&+&{1\over 2}\sum_{\alpha \beta \gamma \delta }V_{\alpha \beta
\gamma \delta }^{RRRR}c_{\alpha R}^{+}c_{\beta R}^{+}c_{\delta
R}c_{\gamma R}. \label{SDFhat}
\end{eqnarray}%
$H_1$ is the interaction term used in the inversion method done
in powers of $\lambda H_{1}$ ($\lambda $ is a new coupling
constant to be set to unity at the end of the calculation).

$\bullet $\textit{The constraining field in SDFT. }
After an unperturbed Hamiltonian is chosen the constraining field
is defined as the zeroth order term of the source in an expansion
in the coupling constant. When the reference frame is the dressed
atom, the constraining field turns out to be the hybridization
function of an Anderson impurity model (AIM) $\Delta[G_{loc}]
_{\alpha \beta }(\tau ,\tau ^{\prime })$ \cite{Anderson:1961},
which plays a central role in the dynamical mean--field theory.
It is defined as the (time dependent) field which must be added to $H_{at}$ in
order to generate the local Green's function $ G_{loc,\alpha
\beta }(\tau ,\tau ^{\prime })$
\begin{equation}
\frac{\delta F_{at}}{\delta \Delta _{\beta \alpha }(\tau ^{\prime },\tau )}%
=-\left\langle T_\tau c_{\alpha }(\tau )c_{\beta }^{+}(\tau
^{\prime })\right\rangle _{\Delta }=G_{loc,\alpha \beta }(\tau
,\tau ^{\prime }), \label{SDFvar}
\end{equation}
where
\begin{eqnarray}
\label{SDFfat}
&&F_{at}[\Delta ]= \\
&&-\ln \int dc^{+}dce^{-S_{at}[c^{+}c]-\sum_{\alpha \beta }\int
\Delta _{\alpha \beta }(\tau ,\tau ^{\prime })c_{\alpha
}^{+}(\tau )c_{\beta }(\tau ^{\prime })d\tau d\tau ^{\prime }},
\notag
\end{eqnarray}%
and the atomic action is given by
\begin{equation}
S_{at}[\Delta ]=\int d\tau\sum_{\alpha \beta} c_{\alpha}^{+}(\tau)
\left(\frac{\partial}{\partial\tau}-\mu\right)c_{\beta }
(\tau)+\int d\tau H_{at}(\tau ). \label{SDFsat}
\end{equation}%

Eq.~(\ref{SDFfat}) actually corresponds to an impurity problem
and $F_{at}[\Delta]$ can be obtained by solving an Anderson
impurity model.

$\bullet $\textit{Kohn--Sham decomposition and its significance. }
The Kohn--Sham decomposition separates the effective action into
two parts: the zeroth order part of the effective action in the
coupling constant $\Gamma _{0}[G_{loc}] \equiv\Gamma
_{SDFT}[G_{loc},\lambda=0]$ and the rest (``exchange correlation
part"). The functional corresponding to (\ref{SDFtwo}) is given by
\begin{eqnarray}
&&\Gamma _{SDFT}[G_{loc},\lambda=0]=F_{at}[\Delta \lbrack
G_{loc}]]-
    \mathrm{Tr}\left(\Delta[G_{loc}]G_{loc}\right)=  \notag \\
&&\mathrm{Tr}\ln G_{loc}-\mathrm{Tr}\left(G_{at}^{-1}G_{loc}\right)+
    \Phi_{at}[G_{loc}], \label{SDFga0}
\end{eqnarray}%
with the $G_{at,\alpha \beta }^{-1}(i\omega )=(i\omega +\mu
)\delta _{\alpha \beta }-h_{\alpha \beta }^{(0)}$. $F_{at}$ is
the free energy when $\lambda =0$ and $\Phi _{at}$ is the sum of
all two--particle irreducible diagrams constructed with the local
vertex $V_{\alpha \beta \gamma \delta }^{RRRR}$ and $G_{loc}$.

{$\bullet $}\textit{\ Saddle point equations and their
significance. }%
The saddle point equations determine the exact spectral function
(and the exact Weiss field). They have the form
\begin{equation}
-\left\langle T_\tau c_{\alpha }(\tau )c_{\beta }^{+}(\tau ^{\prime
})\right\rangle _{\Delta }=G_{loc,\alpha \beta }(\tau ,\tau
^{\prime }),  \label{SDFavr}
\end{equation}%
\begin{equation}
\Delta _{\alpha \beta }(\tau ,\tau ^{\prime
})=\frac{\delta\Delta\Gamma}
 {\delta G_{loc,\beta \alpha }(\tau ^{\prime },\tau )},
\label{SDFdel}
\end{equation}
where $\Delta\Gamma$ can be expressed using coupling constant
integration as is in
Eq.~(\ref{Eq:breakUpGamma})~\cite{Georges:2004:CM0403123,Georges:2004:CM}.
This set of equations describes an atom or a set of atoms in the
unit cell embedded in the medium. $\Delta $ is the exact Weiss
field (with respect to the expansion around the atomic limit)
which is defined from the equation for the local Green's function
$G_{loc}$ (see Eq. (\ref{SDFvar})). The general Weiss source
$\Delta $ in this case should be identified with the
hybridization of the Anderson impurity model.

When the system is adequately represented as a collection of
paramagnetic atoms, the Weiss field is a weak perturbation
representing the environment to which it is weakly coupled. Since
this is an exact construction, it can also describe the band
limit when the hybridization becomes large.

\subsubsection{Construction of approximations: dynamical
mean--field theory as an approximation. }

The SDFT should be viewed as a separate exact theory whose
manifestly local constraining field is an auxiliary mass operator
introduced to reproduce the local part of the Green's function of
the system, exactly like the Kohn--Sham potential is an auxiliary
operator introduced to reproduce the density of the electrons in
DFT. However, to obtain practical results, we need practical
approximations. The dynamical mean--field theory can be thought of
as an approximation to the exact SDFT functional in the same
spirit as LDA appears as an approximation to the exact DFT
functional.

The diagrammatic rules for the exact SDFT functional can be
developed but they are more complicated than in the
Baym--Kadanoff theory as discussed in (\onlinecite{Chitra:2000}).
The single--site DMFT approximation to this functional consists of
taking $\Phi_{SDFT}[G_{loc}]$ to be a sum of all graphs (on a
single site $R$), constructed with $V_{\alpha \beta \gamma \delta
}^{RRRR}$ as a vertex and $G_{loc}$ as a propagator, which are
two--particle irreducible, namely $\Phi _{DMFT}[G_{loc}]=\Phi
_{at}{[}G_{loc}] $. This together with Eq.~(\ref{SDFtwo}) defines
the \ DMFT\ approximation to the exact spectral density
functional.

It is  possible to arrive at this functional by summing up
diagrams~\cite{Chitra:2000} or using the coupling constant
integration trick ~\cite{Georges:2004:CM,Georges:2004:CM0403123}
(see Eq.~(\ref{Delta_Gamma})) with a coupling dependent Greens
function having the DMFT form, namely with a local self-energy.
This results in
\begin{eqnarray}
\label{2variable}
&&\Gamma _{DMFT}(G_{{loc}\ {ii}}) = \sum_i F_{at} [\Delta
(G_{{loc}\ {ii}}) ]\\\nonumber
&&-\sum_{\mathbf{k}}\mathrm{Tr}\ln \Bigl( (i\omega +\mu )\hat{I}-
\hat{h}^{(0)}(\mathbf{k})-
\mathcal{M}_{int}(G_{{loc}\ {ii}} ) \Bigr)\\\nonumber%
&&+\mathrm{Tr} \ln  \bigl(-\mathcal{M}_{int}(G_{{loc}\ {ii}} )
+i\omega+\mu-h^{(0)} - \Delta(i \omega)
\bigr)  .%
\end{eqnarray}
with $\mathcal{M}_{int}(G_{{loc}\ {ii}} )$ in Eq.
(\ref{2variable}) the self-energy of the Anderson impurity model.
It is useful to have a formulation of this DMFT functional
as a function of \emph{three} variables,
\cite{Kotliar:2001:Tsvelik} namely combining the hybridization
with that atomic Greens function to form the Weiss function
$\mathcal{G}_{0}^{-1}=G_{at}^{-1} -\Delta $, one can obtain the
DMFT equations from the stationary point of a functional of
 $G_{loc}$, $\mathcal{M}_{int}$ and the Weiss field
$\mathcal{G}_{0}$:
%
\begin{eqnarray}
\label{SDFdmf} %
\Gamma \lbrack G_{loc},\mathcal{M}_{int}, {\mathcal{G}_{0}}] &=&
F_{imp}[{\mathcal{G}_{0}^{-1}} ]- \mathrm{Tr} \ln [G_{loc}] -\\\nonumber %
&& \mathrm{Tr} \ln(i\omega +\mu -{h^{0}}[k]-\mathcal{M}_{int} )+\\\nonumber %
&& \mathrm{Tr}[({\mathcal{G}_{0}} ^{-1}- \mathcal{M}_{int}
-G_{loc}^{-1}) G_{loc}].
\end{eqnarray}

One can eliminate $G_{loc}$ and $\mathcal{M}_{int}$ from
(\ref{SDFdmf})  using the stationary conditions and recover a
functional of the  Weiss field function only.  This form of the
functional, applied to the Hubbard model, allowed the analytical
determination of the nature of the transition and the
characterization of the zero temperature critical points
\cite{Kotliar:1999}.
Alternatively eliminating  $\mathcal{G}_{0}$ and    $ G_{loc} $
in favor of $\mathcal{M}_{int}$  one obtains the DMFT
approximation to the self-energy functional discussed in section
~\ref{sec:BKfunctional}.

\subsubsection{Cavity construction}

An alternative view to
derive the DMFT approximation is by means of the cavity
construction. This approach gives complementary insights to the
nature of the DMFT and its extensions. It is remarkable that the
summation over all local diagrams can be performed exactly via
introduction of an auxiliary quantum impurity model subjected to
a self--consistency condition \cite{Georges:1992, Georges:1996}.
If this impurity is considered as a cluster $C$, either a
dynamical cluster approximation or cellular DMFT technique can be
used. In single--site DMFT, considering the effective action $S$
in Eq.~(\ref{MODact}), the integration volume is separated into
the impurity $V_{imp}$ and the remaining volume is referred to as
the bath: $V-V_{imp}=V_{bath}.$ The action is now represented as
the action of the cluster cell, $V_{imp}$ plus the action of the
bath,$\ V_{bath},\ $plus the interaction between those two. We are
interested in the local effective action $S_{imp}$ of the cluster
degrees of freedom only, which is obtained conceptually by
integrating out the bath in the functional integral
\begin{equation}
\frac{1}{Z_{imp}}\exp
[-S_{imp}]=\frac{1}{Z}\int_{V_{bath}}D[c^{\dagger}c]\exp [-S] ,
\label{SDFint}
\end{equation}%
where $Z_{imp}$ and $Z$\ are the corresponding partition
functions. Carrying out this integration and neglecting all
quartic and higher order terms
(which is correct in the infinite dimension limit)
we arrive to the result \cite{Georges:1992}
\begin{eqnarray}
\label{eq:SDFimp} &&S_{imp}=-\sum_{\alpha \beta }\int d\tau d\tau
^{\prime }c_{\alpha }^{+}(\tau )\mathcal{G}_{0,\alpha \beta
}^{-1}(\tau ,\tau ^{\prime
})c_{\beta }(\tau ^{\prime }) \\
 &&+{\frac{{1}}{2}}\sum_{\alpha \beta \gamma
\delta }\int d\tau d\tau ^{\prime }c_{\alpha }^{+}(\tau )c_{\beta
}^{+}(\tau ^{\prime })V_{\alpha \beta \gamma \delta }(\tau ,\tau
^{\prime })c_{\gamma}(\tau^\prime)c_{\delta }(\tau) .
\notag
\end{eqnarray}
Here $\mathcal{G}_{0,\alpha \beta }(\tau ,\tau ^{\prime })$ or
its Fourier transform $\mathcal{G}_{0,\alpha \beta }(i\omega )$
is identified as the bath Green's function which appeared in the
Dyson equation for $\mathcal{M}_{int,\alpha \beta }(i\omega )$
and for the local Green's function $G_{loc,\alpha \beta }(i\omega
)$ of the impurity, i.e.
\begin{equation}  \label{SDFgm1}
\mathcal{G}_{0,\alpha \beta }^{-1}(i\omega )=G_{loc,\alpha \beta
}^{-1}(i\omega )+\mathcal{M}_{int,\alpha \beta }(i\omega ).
\end{equation}%
Note that $\mathcal{G}_{0}$ cannot be associated with
non--interacting $G_{0} $.

\begin{figure}[tbh]
\includegraphics*[height=2.0in,angle=0]{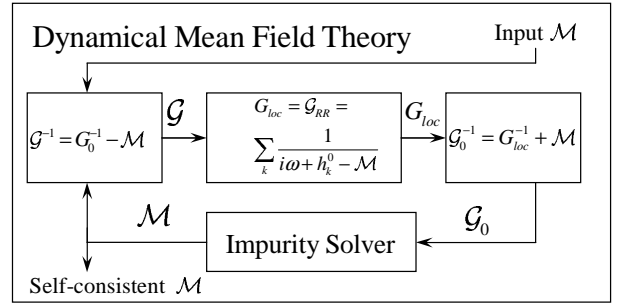}
\caption{Illustration of the self--consistent cycle in DMFT. }
\label{Fig:DMFT}
\end{figure}

The impurity action (\ref{eq:SDFimp}), the Dyson equation
(\ref{SDFgm1}), connecting local and bath quantities as well as
the original Dyson equation (\ref{SDFgks}), constitute the
self--consistent set of equations of the spectral density
functional theory. They are obtained as the saddle--point
conditions extremizing the spectral density functional $\Gamma
_{SDFT}(\mathcal{G})$. Since $\mathcal{M}_{int}$ is not known at
the beginning, the solution of these equations requires an
iterative procedure. First, assuming some initial
$\mathcal{M}_{int},$ the original Dyson equation (\ref{SDFgks})
is used to find Green's function $\mathcal{G}.$ Second, the Dyson
equation for the local quantity (\ref{SDFgm1}) is used to find
$\mathcal{G}_{0}.$ Third, quantum impurity model with the
impurity action $S_{imp}$ after (\ref{eq:SDFimp}) is solved by
available many--body techniques to give a new local
$\mathcal{M}_{int}$. The process is repeated until
self--consistency is reached. We illustrate this loop in
Fig.~\ref{Fig:DMFT}.

\subsubsection{Practical implementation of the self--consistency
condition in DMFT. } %
In many practical calculations, the local Green's function can be
evaluated via Fourier transform. First, given the
non--interacting Hamiltonian $h_{\alpha \beta
}^{(0)}(\mathbf{k})$, we define the Green's function in the
$\mathbf{k}$--space
\begin{equation}
\mathcal{G}_{\alpha \beta }(\mathbf{k},i\omega )=
    \left\{[(i\omega +\mu )\hat{O}(
\mathbf{k})-\hat{h}^{(0)}(\mathbf{k})-\mathcal{M}_{int}(i\omega)]^{-1}
    \right\}_{\a\b} ,  \label{SDFgkw}
\end{equation}%
where the overlap matrix $O_{\alpha \beta }(\mathbf{k})$ replaces
the unitary matrix $\hat{I}$ introduced earlier in (\ref{SDFloc})
if one takes into account possible non--orthogonality between the
basis functions \cite{Wegner:2000}. Second, the local Green's function is evaluated
as the average in the momentum space
\begin{equation}
G_{loc,\alpha \beta }(i\omega )=\sum_{\mathbf{k}}\mathcal{G}_{\alpha \beta }(%
\mathbf{k},i\omega ) ,  \label{SDFsum}
\end{equation}%
which can then be used in Eq.~(\ref{SDFgm1}) to determine the bath
Green's function $\mathcal{G}_{0,\alpha \beta }(i\omega).$

The self--consistency condition in the dynamical mean--field
theory requires the inversion of the matrix, Eq.~(\ref{SDFgkw})
and the summation over $\mathbf{k}$ of an integrand,
(\ref{SDFsum}), which in some cases has a pole singularity. This
problem is handled by introducing left and right eigenvectors of
the inverse of the Kohn--Sham Green's function
\begin{eqnarray}
\sum_{\beta }\left[ h_{\alpha \beta
}^{(0)}(\mathbf{k})+\mathcal{M} _{int,\alpha \beta }(i\omega
)-\epsilon _{\mathbf{k}j\omega} O_{\alpha \beta
}(\mathbf{k})\right]\psi^{R}_{\mathbf{k}j\omega,\beta}=0,\ \ &&
\label{SDFaar} \\
\sum_{\alpha }\psi^{L}_{\mathbf{k}j\omega,\a}\left[ h_{\alpha
\beta
}^{(0)}(\mathbf{k})+\mathcal{M}_{int,\alpha \beta }(i\omega )-
\epsilon _{\mathbf{k}j\omega }O_{\alpha \beta }(\mathbf{k})\right]
=0.\ \ && \label{SDFaal}
\end{eqnarray}
This is a non--hermitian eigenvalue problem solved by standard
numerical methods. The orthogonality condition involving the
overlap matrix is
\begin{equation}
\sum_{\alpha \beta}\psi^{L}_{\mathbf{k}j\omega,\a}
O_{\alpha\beta}(\mathbf{k})
\psi^{R}_{\mathbf{k}j^{\prime}\omega,\b}=\delta_{jj^{\prime }}. \label{SDFort}
\end{equation}
Note that the present algorithm just inverts the matrix
(\ref{SDFgkw}) with help of the ``right" and ``left" eigenvectors.
The Green's function (\ref{SDFgkw}) in the basis of its
eigenvectors becomes
\begin{equation}
\mathcal{G}_{\alpha \beta }(\mathbf{k},i\omega )=\sum_{j}
\frac{\psi^{R}_{\mathbf{k}j\omega,\a}\psi^{L}_{\mathbf{k}j\omega,\b}}
{i\omega+\mu -\epsilon _{\mathbf{k}j\omega }} .
\label{SDFeig}
\end{equation}
This representation generalizes the orthogonal case in the
original LDA+DMFT paper \cite{Anisimov:1997}. The formula
(\ref{SDFeig}) can be safely used to compute the Green's function
as the integral over the Brillouin zone, because the energy
denominator can be integrated analytically using the tetrahedron
method \cite{Lambin:1984}.


The self-consistency condition becomes computationally very
expensive when many atoms need to be considered in a unit cell,
as for example in compounds or complicated crystal structures. A
computationally efficient approach was proposed in
Ref.~\onlinecite{Savrasov:2005:CM0507552}. If the self-energy is
expressed by the rational interpolation in the form
\begin{equation}
  {\cal M}_{\a\b}(i\omega)={\cal
  M}_\a(\infty)\delta_{\a\b}+\sum_i\frac{w^i_{\a\b}}{i\omega-P_i},
  \label{M_SS_idea}
\end{equation}
where $w^i$ are weights and $P_i$ are poles of the self-energy
matrix. The non-linear Dyson equation (\ref{SDFaar}),
(\ref{SDFaal}) can be replaced by a linear Schroedinger-like
equation in an extended subset of auxiliary states. This is clear
due to mathematical identity
\begin{eqnarray}
  \sum_{\vk}\left[
    \begin{array}{rr}
      (i\omega+\mu)\hat{O}_\vk-\hat{h}^0(\vk)-{\cal M}(\infty), &\sqrt{W}\\
      \sqrt{W}^\dagger,& i\omega-P
      \end{array}
    \right]^{-1}=\\
  \left[
  \begin{array}{rr}
    \sum_{\vk}\left[{(i\omega+\mu)\hat{O}_\vk-\hat{h}^0(\vk)-
    {\cal M}(i\omega)}\right]^{-1}, &\cdots\\
      \cdots, & \cdots
      \end{array}
  \right].
  \notag
\end{eqnarray}
where ${\cal M}(i\omega)$ is given by Eq.~(\ref{M_SS_idea}). Since the
matrix $P$ can always be chosen to be a diagonal matrix, we have
$w^i_{\a\b} = \sqrt{W}_{\a i}\sqrt{W}_{\b i}^*$.

The most important advantage of this method is that the eigenvalue
problem Eq.~(\ref{SDFaar}), (\ref{SDFaal}) does not need to be
solved for each frequency separately but only one inversion is
required in the extended space including ``pole states".
In many applications, only a small number of poles is necessary to
reproduce the overall structure of the self-energy matrix (see
section \ref{new_ipt}). In this case, the DMFT self-consistency
condition can be computed as fast as solving the usual Kohn-Sham
equations.

The situation is even simpler in some symmetry cases.  For
example, if Hamiltonian is diagonal $h_{\alpha \beta
}^{(0)}(\mathbf{k})=\delta _{\alpha \beta }h_{\alpha
}^{(0)}(\mathbf{k})$ and the self--energy
$\mathcal{M}_{int,\alpha \beta }(i\omega )=\delta _{\alpha \beta}
\mathcal{M}_{int,\alpha }(i\omega )$, the inversion in the above
equations is trivial and the summation over $\mathbf{k}$ is
performed by introducing the non--interacting density of states
$N_{\alpha }(\epsilon )$
\begin{equation}
N_{\alpha }(\epsilon )=\sum_{\mathbf{k}}\delta \lbrack \epsilon
-h_{\alpha }^{(0)}(\mathbf{k})] .  \label{SDFdos}
\end{equation}%
The resulting equation for the bath Green's function becomes%
\begin{equation}
\mathcal{G}_{0,\alpha }^{-1}(i\omega )=
\left(\int d\epsilon
\frac{N_{\alpha }(\epsilon )}{i\omega +\mu -\epsilon
-\mathcal{M}_{int,\alpha }(i\omega )}\right)^{-1}+
\mathcal{M}_{int,\alpha}(i\omega ).
\label{SDFds1}
\end{equation}

{$\bullet $ }\textit{Assessing the DMFT approximation. }  %
Both the dressed atom and the dressed band viewpoint indicate that
$\Gamma_{DMFT}$ is going to be a poor approximation to
$\Gamma_{SDFT}(G_{loc})$ when interactions are highly nonlocal.
However, extensions of the DMFT formalism allow us to tackle this
problem. The EDMFT~\cite{Kajueter:1996, Kajueter:Unpublished,
Si:1996} allows us the introduction of long--range Coulomb
interactions in the formalism. The short--range Coulomb
interaction is more local in the non--orthogonal basis set and can
be incorporated using the CDMFT \cite{Kotliar:2001:CDMFT}.

\subsection{Extension to clusters}
\label{subsec:CDMFT}

The notion of locality is not restricted to a single site or a
single unit cell, and it is easily extended to a cluster of sites
or supercells. We explain the ideas in the context of model
Hamiltonians written in an orthogonal basis set
to keep the presentation and the notation simple.
The extension to general basis sets ~\cite{Kotliar:2001:CDMFT,
Savrasov:2003:CM0308053} is straightforward.

The motivations for cluster extension of DMFT are multiple: {\it
i)} Clusters are necessary to study some ordered states like
$d$-wave superconductivity which can not  be described by a
single--site method (\cite{Lichtenstein:2000,
Maier:2005:CM0504529,Macridin:2005, Maier:2004,
Maier:2003,Macridin:2004, Maier:2002:PRB66,
Maier:2000:PRL,Maier:2000:PRL,Maier:2000}) 
{\it
ii)} In cluster methods the lattice self--energies have some $k$
dependence (contrary to single--site DMFT) which is clearly an
important ingredient of any theory of the high--$T_c$ cuprates
for example. Cluster methods may then explain variations of the
quasiparticle residue or lifetime along the Fermi surface
\cite{Parcollet:2004,Civelli:2005} {\it iii)} Having a cluster of
sites allows the description of non--magnetic insulators (eg.
valence bond solids) instead of the trivial non--magnetic
insulator of the single--site approach. Similarly, a cluster is
needed when Mott and Peierls correlations are simultaneously
present leading to dimerization
\cite{Poteryaev:2004:PRL,Biermann:2004:CM:VO2} in which case a
correlated  link is the appropriate reference frame. {\it iv)}
The effect of nonlocal interactions within the cluster ({\it e.g.
} next neighbor Coulomb repulsion) can be investigated
\cite{Bolech:2003}. {\it v)} Since cluster methods interpolate
between the single--site DMFT  and the full problem on the
lattice when the size of the cluster increases from one to
infinity,  they resum $1/d$ corrections to DMFT in a
non-perturbative way. Therefore they constitute a systematic way
to assert the validity of and improve the DMFT calculations.

Many cluster methods have been studied in the literature. They
differ both in the self--consistency condition (how to compute
the Weiss bath from the cluster quantities) and on the
parameterization of the momentum dependence of the self--energy
on the lattice. Different perspectives on single-site DMFT lead
to different cluster generalizations : analogy with classical
spin systems lead to the Bethe-Peierls approximation
\cite{Georges:1996}, short range approximations of the
Baym-Kadanoff functional lead to the ``pair
scheme''\cite{Georges:1996}  and its nested cluster
generalizations (which reduces to the Cluster Variation Method in
the classical limit)\cite{Biroli:2004}, approximating the
self-energy by a piecewise constant function of momentum lead to
the dynamical cluster approximation (DCA)
\cite{Hettler:1998,Hettler:2000,Maier:2000}, approximating the
self-energy by the lower harmonics lead to the work of Katsnelson
and Lichtenstein ~\cite{Lichtenstein:2000:R}, and a real space
perspective lead to Cellular DMFT (CDMFT).
\cite{Kotliar:2001:CDMFT}. In this review, we focus mainly on the
CDMFT method, since it has been used more in the context of
realistic computations. For a detailed review of DCA, CDMFT  and
other schemes and their applications to model Hamiltonians see
\cite{Maier:2004:CM0404055}.

$\bullet $ \textit{Cellular dynamical mean--field theory : definition}

 The construction of an
exact functional of a ``local" Green's function in
Eqs.~(\ref{SDFacm}), (\ref{SDFgrn}),(\ref{SDFsdf}) is unchanged,
except that the labels $\alpha, \beta$ denotes orbitals and sites
within the chosen cluster. The cluster DMFT equations have the
form (\ref{SDFloc}), (\ref{SDFgm1}),  where
$\hat{h}^{0}[\mathbf{k}]$ is now replaced by $\hat{t}(K)$ the
matrix of hoppings in supercell notation  and we use the notation
$\Sigma ^{\text{C} }(i\omega_{n})$ \ for the cluster self--energy
(note that the notation $\mathcal{M}_{int}$ was used for this
quantity in the preceding sections).

\begin{multline}
\label{eq:CDMFT_sigma}
G_{0}^{-1}(i\omega _{n})=
\left( \sum_{K\in RBZ}
\Bigl(
i\omega _{n}+\mu - \hat{t}(K)-\Sigma ^{\text{C}}(i\omega _{n})
\Bigr)^{-1}
\right) ^{-1}\\
+ \Sigma
^{\text{C}}(i\omega _{n}), %
\end{multline}%
where the sum over $K$ is taken over the Reduced Brillouin Zone
(RBZ) of the superlattice and normalized.

Just like single--site DMFT, one can either view CDMFT as an
approximation to an exact functional to compute the cluster
Green's function, or as an approximation to the exact
Baym--Kadanoff functional obtained by restricting it to the
Green's functions on the sites restricted to a cluster and its
translation by a supercell lattice vector (see Eq.
(\ref{eq:phi_CDMFT}) below) ~\cite{Maier:2002,Georges:2002}.
From a spectral density functional point of view, Eqs.
(\ref{SDFgks}), (\ref{SDFgkl}), and the equation
$\mathcal{M}_{int}[G]={\delta\Phi_{SDFT}}/{\delta G_{loc}}$ can be
viewed as the exact equations provided that the exact functional
$\Phi _{SDFT}$ is known.

\begin{figure}
\begin{center}
\includegraphics[height=5cm,angle=-0] {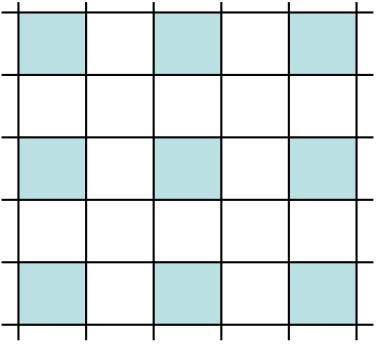}\\
\end{center}
\caption{Example of a $2\times 2$ superlattice construction to
define CDMFT on a plaquette. Notice that the definition is
dependent on the tiling of space by supercells and is therefore
not unique} \label{superlattice}
\end{figure}

A good approximation to the ``exact functional", whose knowledge
would deliver us the exact cluster Green's function, is obtained
by restricting the exact Baym--Kadanoff functional. In this case,
it is restricted to a cluster and all its translations by a
supercell vector. Denoting by $C$ the set of couples $(i,j)$
where $i$ and $j$ belong to the same cluster  (see Fig.
\ref{superlattice} for an example),
\begin{equation}\label{eq:phi_CDMFT}
\Phi^{SDFT}_{CDMFT}=\Phi_{BK}|_{G_{ij}=0\ \mbox{if}\ (i,j)\notin C}
\end{equation}

Alternatively the CDMFT equations can be derived from the point
of view of a functional of the Weiss field generalizing
Eq.~(\ref{2variable}) from single sites to supercells as shown in
Fig.~\ref{superlattice}. A fundamental concept in DMFT is that of
a Weiss field. This is the function describing the environment
that one needs to add to an \textit{interacting but local}
problem to obtain the correct local Greens function of an
extended system. Now expressed in terms of the Weiss field of the
cluster $ \mathcal{G}_{0}^{-1}=G_{at}^{-1} -\Delta
$.
This concept can be used to
highlight the connection of the mean field theory of lattice
systems with impurity models and the relation of their free
energies~\cite{Georges:1996}. For this purpose it is useful to
define the DMFT functional of \textit{three} variables,
\cite{Kotliar:2001:Tsvelik} $G_{loc}$, $\Sigma $ and the Weiss
field $\mathcal{G}_{0}$:
\begin{eqnarray}
\label{eq:triple_functionnal}
&&\Gamma_{CDMFT} [G,\Sigma ,{\mathcal{G}_{0}}]=\Sigma_{cells}
F_{cell}[{\mathcal{G}_{0}^{-1}}] -\mathrm{Tr}\ln[G] \\\nonumber%
&& -\mathrm{Tr}\ln[G]\left(i\omega +\mu -{\hat{t}}[k]
-\Sigma \right)+\mathrm{Tr}[({\mathcal{G}_{0}}%
^{-1}- \Sigma -G^{-1}) G]\
\end{eqnarray}
Extremizing this functional gives again the standard CDMFT
equations.

$\bullet $ \textit{CDMFT : approximation of lattice quantities}

The impurity model delivers cluster quantities. In order to make a
connection with the original lattice problem,  we need to
formulate estimates for the lattice Green's function. A natural
way to produce these estimates is by considering the superlattice
(SL) (see Fig.~\ref{superlattice})
and  constructing  lattice objects from superlattice objects by
averaging the relevant quantities to restore periodicity, namely
\begin{equation}
W_{latt} (i-j) \approx
\frac{1}{N_s}\sum_{k}~W^{SL}_{k,k+i-j},        \label{averageW}
\end{equation}
where $N_s$ represents the total number of sites, and $i,j,k$ are
site indices. Notice that Eq. (\ref{averageW}) represents a
super-lattice average, not a cluster average. In particular, if
$W$ is the cluster irreducible cumulant, $M_c \equiv G_c^{-1}  +
\Delta $ (where $\Delta$ is the hybridization) or the cluster
self-energy $\Sigma$ all the contributions with $k$ and $k+i-j$
belonging to different cells are zero by construction.
The lattice  Greens function can then be reconstructed  from the
lattice cumulant ~\cite{Stanescu:Unpublished} or the lattice
self-energy ~\cite{Kotliar:2001:CDMFT, Biroli:2002}. Namely
$G_{latt}(k,\omega) \approx (\omega - t(k) +\mu
-\Sigma_{latt}(k,\omega))^{-1}$ or $G_{latt}(k,\omega) \approx (-
t(k) +{M_{latt}(k,\omega)}^{-1})^{-1}  $ . Criteria for more
general periodizations respecting causality were derived in ref
\cite{Biroli:2002}.
Another alternative suggested originally by S\'en\'echal and
Tremblay is to  directly  periodize the Green's function, i.e.
apply Eq.~(\ref{averageW}) with $W=G$ to obtain
~\cite{Senechal:2000}
\begin{eqnarray}
{G}({k},\omega) = \frac{1}{N_c}\sum_{a,b\in\mathcal{C}}
[\hat{M}_c^{-1} - \hat{t}_{k}]^{-1}_{ab}~e^{i{k}({r}_a-{r}_b)},
\label{Gk}
\end{eqnarray}
where $\hat{t}_{k}$ is the Fourier transform of the ``hopping''
on the super-lattice, and $N_c$ the number of sites in the
cluster.

For example, consider the two-dimensional Hubbard model
on a square lattice within a four-site approximation (plaquette)
in which square symmetry is preserved. After performing the average
(\ref{averageW}) and then taking the Fourier transform, we obtain
the following expressions for the self-energy and the irreducible cumulant 
of the lattice problem.
\begin{eqnarray*}
{\Sigma}({k},\omega) &=& {\Sigma}_0(\omega) +
{\Sigma}_1(\omega)~\alpha({k}) + {\Sigma}_2(\omega)~\beta({k}),
\label{sigk} \\
{M}({k},\omega) &=& {M}_0(\omega) + {M}_1(\omega)~\alpha({k}) +
{M}_2(\omega)~\beta({k}), \label{mk}
\end{eqnarray*}
where for the cluster quantities $W^{(c)}_{ab}$ we used the
notations $W_0$ for the on-site values ($a=b$), $W_1$ if a and b
are nearest neighbors (on a link) and $W_2$ if a and b are
next-nearest neighbors (on the diagonal),
 and 
$\alpha({k}) = \cos(k_x)+\cos(k_y)$ and $\beta({k}) =
\cos(k_x)\cos(k_y)$.

It is important to notice that it is better to reconstruct on
site quantities from the cluster Greens function
\cite{Capone:2004} and non local quantities from the lattice
quantities \cite{Stanescu:Unpublished}. Using cumulants there is
not much difference between estimates from the lattice or the
local green function for either one of these quantities, and the
same is true about the lattice periodization. Alternatively,
periodizing the self-energy has the drawback that the local
quantities inferred from $G_{latt}$ differ from $G_c$ near the
Mott transition.

$\bullet $ \textit{Other cluster schemes}

We briefly comment on various other cluster schemes
mentioned in the introduction (see also \cite{Maier:2004:CM0404055}).
Nested cluster schemes are defined by another truncation of the Baym-Kadanoff
functional :
\begin{equation}\label{eq:phi_CDMFT}
\Phi^{SDFT}_{Nested}=\Phi_{BK}|_{G_{ij}=0\ \mbox{if}\ (i,j)\notin C}
\end{equation}
where $C$ is the set of couples $(i,j)$ with $|i-j| \leq L$ with
$L$ is the size of the cluster and we use the Manhattan distance
on the lattice. Those schemes  combine information from various
cluster sizes, and can give very accurate determination of
critical temperatures using small cluster sizes, but they are not
causal when the range of the self-energy exceeds the size of the
truncation \cite{Biroli:2004} (See also \cite{Okamoto:2004}).

There is a class of cluster schemes which are guaranteed to be
causal and which requires the solution of one impurity problem :
DCA, CDMFT, and PCDMFT (periodized cluster cellular dynamical mean--field theory).
The self--consistency condition of all three
schemes can be summarized into the same matrix equation
\begin{multline}
\label{eq:cluster_sc}
G_{0}^{-1}(i\omega _{n})=\Sigma^{\text{C}}(i\omega _{n})+ \\
\left( \sum_{K\in RBZ}
\Bigl(
i\omega _{n}+\mu - \hat{t}_S(K)- \Sigma_S(K,i\omega_n)
\Bigr)^{-1}
\right) ^{-1},
\end{multline}
where the difference between the three schemes is enclosed in the
value of $t_{S}$ and of $\Sigma_{S}$ that enter in the
self--consistency condition. Namely, if $\hat t_S(K)=\hat t(K)$
and $\Sigma_S (K,i\omega_n ) = \Sigma^{C}(i\omega_n )$ we have
the CDMFT case, $\hat t_S(K) = \hat t(K)$ and $\Sigma_S =
\Sigma_{latt}$ corresponds to PCDMFT case, and DCA is realized
when $t_S(K) = t^{DCA}\equiv t_{\mu\nu}(K)\exp[-iK(\mu-\nu)]$ and
$\Sigma_S (K,i\omega_n ) = \Sigma^{C}(i\omega_n
)$~\cite{Biroli:2004}. 
PCDMFT uses the lattice self--energy in the
sum over the reduced Brillouin zone in the self--consistency
Eq.~(\ref{eq:CDMFT_sigma}). It is similar to the scheme proposed
by Katsnelson and Lichtenstein ~\cite{Lichtenstein:2000:R}, but
can be proven to be explicitly causal. The dynamical cluster
approximation or DCA \cite{Hettler:1998,Hettler:2000,Maier:2000}
derives cluster equations starting from momentum space. Its real
space formulation of Eq.\ref{eq:cluster_sc} was introduced in
\cite{Biroli:2002}. While in CDMFT  (or PCDMFT) the lattice
self--energy is expanded on the lowest harmonics in $k$, in DCA
the self--energy is taken piecewise constant in the Brillouin
Zone.

Simpler approximations, such as cluster perturbation theory (CPT) and 
variational cluster perturbation theory (VCPT), can also be fruitfully viewed
as limiting cases of cluster DMFT. Indeed CPT is obtained by
setting  the DMFT hybridization equal to zero. The self--energy
then becomes the atomic self--energy of the cluster. The lattice
self--energy is then obtained by restoring the periodicity in the
Green's function \cite{Senechal:2000, Senechal:2002, Gros:1993,
Zacher:2000, Zacher:2002, Dahnken:2002}. The restriction of the
functional (\ref{eq:triple_functionnal}) to a {\it non zero but
static} Weiss field, gives rise to the variational cluster
perturbation theory (VCPT) introduced in
Refs.~\cite{Potthoff:2003:PRL, Senechal:2004, Dahnken:2004}. An
extensions of these ideas in the context of EDMFT has been
recently been carried out by Tong ~\cite{Tong:2005:CM0504778}.

$\bullet $ \textit{Hartree--Fock terms}

In realistic computations, it is natural to separate the
Hartree--Fock term, which can be treated easily from the more
complex ``exchange'' contributions to the Baym--Kadanoff
functional $\Phi$. This idea can also be extended to CDMFT, in
the case of nonlocal interactions connecting different clusters
({\it e.g. } spin--spin interactions). The Hartree--Fock
contribution to the Baym--Kadanoff functional induces a
self--energy which is frequency independent and therefore does not
cause problems with causality and can be evaluated with little
computational cost. So it is convenient to separate $\Phi=
\Phi_{HF} + \Phi_{int}$, and apply the cluster DMFT truncation
only to $\Phi_{int}$, and to the self--energy it generates while
treating the Hartree contributions exactly \cite{Biroli:2004}.
More precisely, one can treat with Hartree--Fock terms that
connect the cluster to the exterior only, to avoid a double
counting problem. This observation is particularly relevant in
the treatment of broken symmetries induced by nonlocal
interactions as exemplified in the study of the transition to a
charge density wave in the extended Hubbard model in one
dimension which was studied in \cite{Bolech:2003}.

$\bullet $ \textit{Cluster size dependence}

\begin{figure}[!htb]
\begin{center}
\includegraphics[height=7cm,angle=-0] {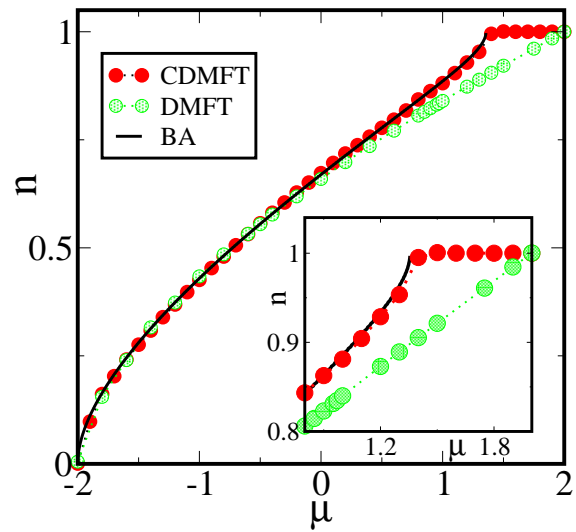}\\
\end{center}
\caption{\label{fig:0401060} Density $n$ as a function of
$\protect\mu$ for the one--dimensional Hubbard model with on--site
repulsion strength  $U/t=4$, number of cluster sites $N_c=2$
within single--site DMFT, two--site CDMFT, two--site PCDMFT and
two--site DCA compared with the exact solution by Bethe Ansatz
(BA). The lower--right inset shows a region near the Mott
transition (adapted from \cite{Capone:2004})}.
\end{figure}

\begin{figure}[!htb]
\begin{center}
\includegraphics[height=6cm,angle=-0] {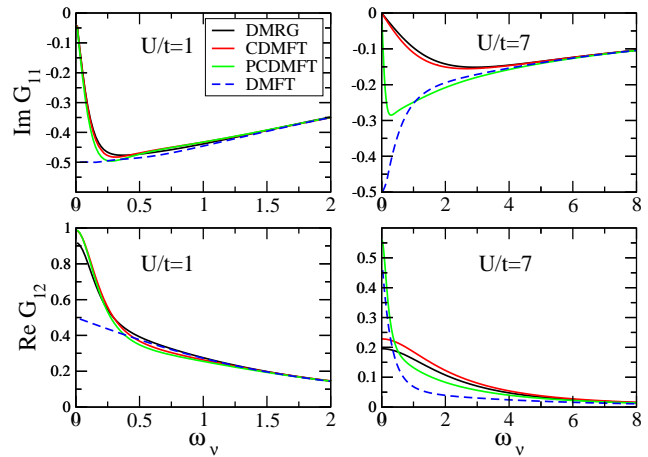}\\
\end{center}
\caption{\label{fig:0401060_DYN} $\Im G_{11}$ and $\Re G_{12}$
 for the one--dimensional Hubbard model with on--site
repulsion strength  $U/t=1$ and $U/t=7$, number of cluster sites $N_c=2$
within single--site DMFT, two--site CDMFT, two--site PCDMFT
compared with the numerically exact DMRG calculation.
(from \cite{Capone:2004}).}
\end{figure}

The cluster DMFT methods are in an early stage of development but
a few investigations of the performance of the methods for
different sizes have already appeared (See
\cite{Maier:2004:CM0404055} and references below). There are two
distinct issues to consider. The first issue is what can be
achieved with very small clusters (e.g. 2 sites in one dimension
or 2x2 plaquette in two dimensions). Cluster studies have
demonstrated that in a broad region of parameter space,
single-site DMFT is quantitatively quite accurate. Similarly, one
would like to know what are the minimal clusters needed to capture
e.g. the physics of the cuprates.
The one--dimensional Hubbard model is a very
challenging case to study this question. Application of DMFT and
cluster methods to this problem was carried out in
\cite{Capone:2004} and is reproduced in Figures
~\ref{fig:0401060} and ~\ref{fig:0401060_DYN}. Let us note that
{\it i)} far from the transition, single-site DMFT is quite
accurate, {\it ii)} a cluster of 2 sites is already very close to
the exact solution (obtained by Bethe Ansatz for thermodynamics
quantities or DMRG for the dynamical ones). Even though no
mean--field approach can produce a Luttinger liquid (a very large
cluster would be necessary) CDMFT is shown to perform remarkably
well when considering quantities related to intermediate or high
energies or associated with the total energy, even near the Mott
transition. 

The second issue is the asymptotic convergence of the methods
to the exact solution of the problem (in the infinite cluster limit).
At present, this is still a somewhat academic issue because
large cluster can not be studied for large $U$ or at low temperature,
but algorithmic advances and increase of computer power may
change the situation in a near future.
 The convergence properties of the CDMFT method for
large cluster size can be easily improved. Away from critical
points, local quantities in CDMFT converge exponentially to their
bulk value when measured at the core of the cluster. However,
averages over the cluster converge like $1/L$ where $L$ is the
linear size of the cluster \cite{BiroliKotliar2004} (see also
\cite{AryanpourMaierJarrell2003,BiroliKotliar2005}), because in
CDMFT the cluster is defined in real space, and the error is
maximal and of order 1 in $L$ at the boundary. Therefore, to
estimate the value of a local quantity, one should preferentially
use the core of the cluster ({\it i.e. } giving a lower weight to
the boundary) assuming of course that the cluster is large enough
to distinguish between a core and a boundary. Failure to do so in
CDMFT can lead to non--physical results, as illustrated in the
one--dimensional Hubbard model. In this case, the critical
temperature for the N\'eel order does not go to zero when the
size of the cluster increases \cite{MaierGonzalez:CM0205460}. In
fact, the boundary of a large (chain) cluster sees an effective
field given by the other boundary, not by the sites at the center
of the cluster, which leads to spurious ordering. It is however
possible to greatly improve the convergence properties of CDMFT
in ordered phases by weighting the self--energy more at the core
of the cluster, a cluster scheme called weighted-CDMFT
\cite{ParcolletKotliar2005}:  in the self--consistency condition
(\ref{eq:CDMFT_sigma}), we replace the self--energy
$\Sigma^{\text{C}}$ by  $\Sigma^{\text{w}}$
\begin{align}\label{eq:defwCDMFT}
\Sigma^{\text{w}}_{\alpha\beta} &= \sum_{\alpha'\beta'} w^{\alpha'\beta'}_{\alpha\beta}
\Sigma^{\text{C}}_{\alpha'\beta'}
\\
w_{\gamma\delta}^{\alpha \beta }&=\delta _{\alpha -\beta ,\gamma-\delta}f_{c}(\alpha )f_{c}(\beta)
\end{align}
where $f_c$ is a normalized function that decays exponentially
from the center of the cluster towards the boundaries and satisfy
$\sum_\alpha f_c(\alpha)^2 =1$. This scheme is causal, it does
not present the spurious ordering in one dimension, and it can be
shown to have faster convergence of the critical temperature in
the classical limit of the Falikov--Kimball model
\cite{ParcolletKotliar2005}. Therefore,
for applications to  (quasi)-1d systems (chains or ladders)
where relatively large cluster size can be reached,
weighted-CDMFT should be preferred to CDMFT.

$\bullet $ \textit{Numerical solutions}

Since the impurity model to be solved for a cluster method is
formally a multi-orbital problem, most the  solvers used for
single-site DMFT can be extended cluster methods, at the expense
of an increase of computational cost (See section \ref{sec:IMP}).
The computational cost of  solving the  impurity model entering
the CDMFT equations  using QMC is the same as that of an isolated
system of the same cluster, or sometimes even less since it has been
found that the presence of the bath reduces  the sign problem. To
solve the  CDMFT equations with exact diagonalization, the bath
needs to be discretized and represented by free fermions. This
results in an increase in the size of the Hilbert space.
$\Delta(\omega)$ has to be represented by a discrete set of poles
and, as in single--site DMFT, there are various approaches for
choosing a  parameterization \cite{Georges:1996}. A
modification  of the original procedure of Caffarel and Krauth
\cite{Caffarel:1994} which gives stronger weight to the
low--frequency part of the Weiss field has been
suggested~\cite{Capone:2004}.

Another possibility for parameterizing the bath is to simply
insert a discretized form of the Weiss field into the CDMFT
functional, which is  viewed as a function of three variables
(the obvious generalization of Eq.~(\ref{SDFdmf}) to clusters),
and varying the functional with respect to the free parameters
parameterizing the Weiss field. An alternative choice of the bath
parameters can be obtained by inserting approximate expressions
of the self-energy parameterized by a few set of parameters into
the self-energy functional~\cite{Potthoff:2003}.

$\bullet $ \textit{Application to realistic calculations}

In the context of realistic studies
of materials, the applications of cluster methods are only
beginning. An interesting class of problems are
posed by materials
with dimerization or charge--charge correlations in the
paramagnetic phasei, such as
{NaV}$_{2}$O$_{5}$. This  compound, where the Vanadium atoms
are arranged to form  two leg ladders which are quarter filled,
served as a first application of LDA+DMFT cluster methods. At
low temperatures the system is a charge ordered insulator, a
situation that is well described by the LDA+U
method~\cite{Yaresko:2000}.  Above the charge ordering
temperature the insulating  gap persist, and cluster DMFT is
required to describe this unusual insulating
state~\cite{Mazurenko:2002}.
The  second application of this approach focused on the  
interplay of Pauling- Peierls distortions and Mott correlations
\cite{Poteryaev:2004:PRL} that occur in
$Ti_2 O_3$. 
 Titanium sesquioxide, ${Ti}_2 O_3$, is
isostructural  to Vanadium Sesquioxide, $V_2 O_3$,  the
prototypical Mott-Hubbard system.  In the  corundum structure
the  pair of Titanium atoms form a structural motif.
Titanium sesquioxide
displays a rapid crossover from a bad metal regime at high
temperatures, to an insulating regime  at low temperatures.
Standard first principles electronic structure methods have
failed to account for  this crossover.
While  single site DMFT was quite succesful describing the high
temperature physics of $V_2 O_3$ it  cannot account for the
observed temperature driven crossover in $Ti_2 O_3$ with a
reasonable set of parameters. A two site  CDMFT calculation with a
very reasonable set of onsite interactions and an intersite
Coulomb repulsion successfully describes the observed
crossover. A surprising  result of the cluster
calculations\cite{Poteryaev:2004:PRL}, was the strong frequency
dependence of the  inter-site Titanium self energy  which can be
interpret as a scale dependent modification of the bare bonding
antibonding splitting.
The link reference frame provides an intuitive picture of
the synergistic interplay of the  lattice distortion (ie. the Pauling
-Goodenough-Peierls  mechanism \cite{pauling} 
which decreases the
distance between the Ti atoms) and the Hubbard-Mott mechanism  in
correlated materials having dimers in the unit cell.
The bare (high frequency) parameters of the problem are such
that a static mean field calculation yields a metal. However as
temperature and frequency are lowered, important correlation
effects  develop. The bandwidth of the a1g and eg bands  is
reduced by the correlations while the crystal field splitting
between the bonding and the antibonding orbital increases in such
a way
that  the low energy renormalized  parameters result in   a band
insulator.
We have a case where the Coulomb interactions enhance
the crystal field splitting and reduce the bandwidth, in a
synergistic cooperation with the lattice distortions to drive the
system thru a metal to insulator crossover. 

Another example of the interplay of the Peierls and the Mott
mechanism is provided by Vanadium dioxide $V O_2$ .  This
material undergoes a first-order transition from a
high-temperature metallic phase to a low-temperature insulating
phase near room- temperature.  The
resistivity jumps by several orders of magnitude through this
transition, and the crystal structure changes from rutile
at high-temperature to monoclinic
at low-temperature. The latter is characterized by a dimerization
of the vanadium atoms into pairs, as well as a tilting of
these pairs with respect to the c-axis. CDMFT studies of this
material 
\cite{Biermann:2004:CM:VO2}
can account for the metallic and the
insulating phase with reasonable interaction parameters.

\subsection{LDA+U method. }

\label{sec:SDFldu}

We now start the discussion on how the ideas of spectral density
functional theory and conventional electronic structure
calculations can be bridged together. In many materials, the
comparison of LDA calculations with experiment demonstrates that
delocalized $s$- and $p$--states are satisfactorily described by
local and frequency independent potentials. This leads to the
introduction of hybrid methods which separate the electrons into
light and heavy. Treating the light electrons using LDA and the
heavy electrons using many--body techniques, such as DMFT (see
\ref{sec:SDFlda}), has already proven to be effective.

As a first illustration, we consider the LDA+U method of Anisimov
and co--workers \cite{Anisimov:1991}. Historically, this was
introduced as an extension of the local spin density
approximation (LSDA) to treat the ordered phases of the Mott
insulating solids. In this respect, the method can be seen as a
natural extension of LSDA. However, this method was the first to
recognize that a better energy functional can be constructed if
not only the density, but also the density matrix of correlated
orbitals is brought into the density functional. In this sense,
the LDA+U approach is the
Hartree--Fock approximation for the spectral density functional
within LDA+DMFT, which is discussed in
the following section.

{$\bullet $ }\textit{Motivation and choice of variables. }%
From the effective action point of view, the LDA+U constructs a
functional of the density $\rho (\mathbf{r}),$ magnetization
$\mathbf{m}(\mathbf{r})$, and the occupancy matrix of the
correlated orbitals. The latter is defined by projecting the
electron creation and destruction operators on a set of local
orbitals, $c_{aR}=\int \chi _{a}^{\ast }(\mathbf{r-R})\psi
(\mathbf{r})d\mathbf{r},$ i.e. by constructing the occupancy
matrix from the local Green's function
\begin{equation}
n_{ab}=T\sum_{i\omega }e^{i\omega 0^{+}}G_{loc,ab}(i\omega ).
\label{LDUnab}
\end{equation}%
In principle, an exact functional of the spin density and the
occupancy matrix can be constructed, so as to give the total free
energy at the stationary point using the techniques described in
previous sections. The LDA+U approximation is an approximate
functional of these variables which can be written down
explicitly. In the context of LDA+U, the constraining field is
designated as $\lambda_{ab}$.

{$\bullet $ }\textit{Form of the functional. }%
The total free energy now is represented as a functional of $\rho
(\mathbf{r})$, $\mathbf{m}(\mathbf{r}),$  $n_{ab},$ $\lambda
_{ab}, $ the Kohn--Sham potential $V_{KS}(\mathbf{r})$ and
Kohn--Sham magnetic field, $\mathbf{B}_{KS}(\mathbf{r})$. This
representation parallels the Harris--Methfessel form (see
Eq.~\ref{SDFtwo}). The LDA+U functional is a sum of the kinetic
energy, energy related to the external potential and possible
external magnetic field, $K_{LDA+U},$ as well as the interaction
energy $\Phi _{LDA+U}[\rho ,\mathbf{m},n_{ab}]$ (see
\cite{Kotliar:2001:Tsvelik} for more details),
i.e.
\begin{eqnarray}
&&\Gamma_{LDA+U}[\rho ,\mathbf{m},n_{ab},\lambda_{ab}] =\\\nonumber%
&& K_{LDA+U}[\rho ,\mathbf{m} ,n_{ab}]-\lambda_{ab}n_{ab}+\Phi
_{LDA+U}[\rho ,\mathbf{m},n_{ab}] . \label{LDUfun}
\end{eqnarray}%
The form of the functional $K_{LDA+U}$ is analogous to the
discussed equations (\ref{DFTkin}), (\ref{BKFkin}),
(\ref{SDFkin}) for the DFT, BK
and SDFT theories. The interaction energy $\Phi _{LDA+U}[\rho ,\mathbf{m}%
,n_{ab}]$ is represented as follows
\begin{eqnarray}
&&\Phi_{LDA+U}[\rho ,\mathbf{m},n_{ab}]= \\
&&E_{H}[\rho ]+E_{xc}^{LDA}[\rho , \mathbf{m}]+\Phi
_{U}^{Model}[n_{ab}]-\Phi _{DC}^{Model}[n_{ab}],  \notag
\label{LDUint}
\end{eqnarray}
This is the LDA interaction energy to which we have added a
contribution from the on-site Coulomb energy in the shell of correlated
electrons evaluated in the Hartree--Fock approximation
\begin{equation}
\Phi_{U}^{Model}[n_{ab}]=\frac{1}{2}
\sum_{abcd\in l_c}(U_{acdb}-U_{acbd})n_{ab}n_{cd} .  \label{LDUmod}
\end{equation}%
Here, indexes $a,b,c,d$ involve fixed angular momentum $l_{c}$ of
the correlated orbitals and run over magnetic $m$ and spin
$\sigma $ quantum numbers. The on--site Coulomb interaction
matrix $U_{abcd}$ is the on--site Coulomb interaction matrix
element $V_{\alpha =a\beta =b\gamma =c\delta =d}^{RRRR}$ from
(\ref{MODhtb}) taken for the sub--block of the correlated
orbitals. Since the on-site Coulomb interaction is already
approximately accounted for within LDA, the LDA contribution to
the on-site interaction needs to be removed. This quantity is
referred to as the ``double counting" term, and is denoted by
$\Phi_{DC}^{Model}[n_{ab}]$.
Various forms of the double--counting functional have
been proposed in the past. In particular, one of the popular
choices is given by \cite{Anisimov:1997:JPCM}
\begin{equation}
\Phi_{DC}^{Model}[n_{ab}]=\frac{1}{2}\bar{U}\bar{n}_{c}(\bar{n}_{c}-1)-%
\frac{1}{2}\bar{J}[\bar{n}_{c}^{\uparrow }(\bar{n}_{c}^{\uparrow }-1)+\bar{n}%
_{c}^{\downarrow }(\bar{n}_{c}^{\downarrow }-1)],
\label{LDUedc}
\end{equation}
where $\bar{n}_{c}^{\sigma }=\sum_{a\in l_{c}}n_{aa}\delta
_{\sigma _{a}\sigma },$ $\bar{n}_{c}=\bar{n}_{c}^{\uparrow
}+\bar{n}_{c}^{\downarrow
} $ and where $\bar{U}=\frac{1}{(2l_{c}+1)^{2}}\sum_{ab\in
  l_c}U_{abba}$ ,
$\bar{J}=\bar{U}-\frac{1}{2l_{c}(2l_{c}+1)}\sum_{ab\in l_c}(U_{abba}-U_{abab})$.

{$\bullet $ }\textit{\ Saddle point equations. }%
The Kohn--Sham equations are now obtained by the standard
procedure which gives the definitions for the Kohn--Sham potential
$V_{KS}(\mathbf{r})$, the effective magnetic field
$\mathbf{B}_{KS}(\mathbf{r})$, and the constraining field matrix
$\lambda _{ab}.$ The latter is the difference between the
orbital--dependent potential $\mathcal{M}_{ab}$ and the
contribution due to double counting, $V_{ab}^{DC}, $ i.e.
\begin{eqnarray}
\lambda _{ab} &=&\frac{\delta \Phi _{U}^{Model}}{\delta
n_{ab}}-\frac{\delta \Phi _{DC}^{Model}}{\delta
n_{ab}}=\mathcal{M}_{ab}-V_{ab}^{DC},
\label{LDUlam} \\
\mathcal{M}_{ab} &=&\sum_{cd}(U_{acdb}-U_{acbd})n_{cd},  \label{LDUmas} \\
V_{ab}^{DC} &=&\delta _{ab}[\bar{U}(\bar{n}_{c}-\frac{1}{2})-\bar{J}(\bar{n}%
_{c}^{\sigma }-\frac{1}{2})] .  \label{LDUvdc}
\end{eqnarray}

{$\bullet $ }\textit{Comments on the parameterization of the
functional. }%
(i) The LDA+U functional and the LDA+U equations are defined once
a set of projectors and a matrix of interactions $U_{abcd}$ are
prescribed. In practice, one can express these matrices via a set
of Slater integrals which, for example, for $d$--electrons are
given by three constants $F^{(0)},F^{(2)},$ and $F^{(4)}.$ These
can be computed from constrained LDA calculations as discussed in
Section \ref{sec:INTmod} or taken to be adjustable parameters. An
important question is the form of the double counting term
$\Phi_{DC}^{Model}$ in Eq.~(\ref{LDUedc}).
The question arises whether
double--counting term should include self--interaction effects or
not. In principle, if the total--energy functional contains this
spurious term, the same should be taken into account in the
double--counting expression. Judged by the experience that the
LDA total energy is essentially free of self--interaction (total
energy of the hydrogen atom is, for example, very close to
--1~Ry, while the Kohn--Sham eigenvalue is only about --0.5~Ry),
the construction $\Phi _{DC}^{Model}$ is made such that it is free
of the self--interaction. However given the unclear nature of the
procedure, alternative forms of the double counting may include
the effects of self--interaction. This issue has been
reconsidered recently by Petukhov et al. \cite{Petukhov:2003} who
proposed more general expressions of double counting corrections.

{$\bullet $ }\textit{Assessment of the method. }%
Introducing additional variables into the energy functional
allowed for better approximations to the ground--state energy in
strongly--correlated situations. This turned out to be a major
advance over LDA in situations where orbital order is present.
The density matrix for the correlated orbitals is the order
parameter for orbital ordering, and its introduction into the
functional resembles the introduction of the spin density when
going from the LDA to the LSDA.

Unfortunately it suffers from some obvious drawbacks. The most
noticeable one is that it only describes spectra which has
Hubbard bands when the system is orbitally ordered. We have
argued in the previous sections that a correct treatment of the
electronic structure of strongly--correlated electron systems has
to treat both Hubbard bands and quasiparticle bands on the same
footing.  Another problem occurs in the paramagnetic phase of
Mott insulators: in the absence of  broken orbital  symmetry, the
LDA+U results are very close to the  LDA-like solution, and the
gap collapses. In systems like NiO where the gap is of the order
of several eV, but the N\'eel temperature is a few hundred
Kelvin, it is unphysical to assume that the gap and the magnetic
ordering are related.

The drawbacks of the LDA+U method are the same as those of the
static Hartree Fock approximation on which it is based. It
improves substantially the energetics in situations where a
symmetry is broken, but it cannot predict reliably the breaking
of a symmetry in some situations. This is clearly illustrated in
the context of the Hubbard model where correlation effects reduce
the double occupancy, and Hartree Fock can only achieve this
effect by breaking the  spin system which results in magnetic
ordering. For this reason, the LDA+U predicts magnetic order in
cases where it is not observed, as, e.g., in the case of Pu
\cite{Savrasov:2000, Bouchet:2000}.

Finally, notice that LDA+U can be viewed as an approximation to
the more sophisticated LDA+DMFT  treatment  consisting  of taking
the Hartree--Fock approximation for the exchange--correlation
functional $\Phi_{DMFT}$ (see (\ref{DMFfun})), which results in a
\emph{static} self--energy. Even in the limit of large interaction
$U$, LDA+DMFT does \emph{not} reduce to LDA+U. For example, LDA+U
will incorrectly predict spin ordering temperatures to be on the
scale of $U$, while LDA+DMFT correctly predicts them to be on the
order of $J$, the exchange interaction. Hence LDA+DMFT captures
the local moment regime of various materials (see
\ref{subsec:MATloc}), while LDA+U does not.

\subsection{LDA+DMFT theory}

\label{sec:SDFlda}

{$\bullet $ }\textit{Motivation and choice of variables. } %
We now turn to the LDA+DMFT method \cite{Anisimov:1997,
Lichtenstein:1998}. This approach can be motivated from different
perspectives. It can be viewed as a natural evolution of the
LDA+U method to eliminate some of its difficulties. It can also be
viewed as a way to upgrade the DMFT approach, which so far has
been applied to model Hamiltonians, in order to bring in realistic
microscopic details.

To compute the energy in a combination of LDA and DMFT one can
use an approximate formula to avoid the overcounting of the free
energy $F_{tot}=F_{LDA}+F_{DMFT}-F_{mLDA}$ where $F_{mLDA}$ is a
mean--field treatment of the LDA Hamiltonian. This procedure was
used by Held et al. in their work on Cerium \cite{Held:2001}.
Alternatively, the approach proposed in this section uses an
effective action construction and obtains an approximate
functional merging LDA and DMFT. This has the advantage of
offering in principle stationarity in the computation of the
energy.

In this review we have built a hierarchy of theories, which focus
on more refined observables (see Table \ref{tab:PQCF}). At the
bottom of the hierarchy we have the density functional theory
which focuses on the density, and at the top of the hierarchy we
have a Baym--Kadanoff approach which focuses on the full
electronic Green's function. The LDA+DMFT is seen as an
intermediate theory, which focuses on two variables, the
\textit{density} and the \textit{local Green's function} of the
heavy electrons. It can be justified by reducing theories
containing additional variables, a point of view put forward
recently in Ref.~\onlinecite{Savrasov:2003:CM0308053}.
\begin{table}[tbp]
\centering
\begin{tabular}{lll}
Method & Physical Quantity & Constraining Field
\\ \hline
Baym--Kadanoff & $G_{\alpha\beta}(\vk,i\omega)$ & $\Sigma_{int,\alpha\beta}(\vk,i\omega)$ \\
DMFT (BL)     & $G_{loc,\alpha\beta}(i\omega)$ & $\mathcal{M}_{int,\alpha\beta}(i\omega )$\\
DMFT (AL)     & $G_{loc,\alpha\beta}(i\omega)$ & $\Delta_{\alpha\beta}(i\omega )$ \\
LDA+DMFT (BL) & $\rho(r),G_{loc,ab}(i\omega)$  & $V_{int}(r),\mathcal{M}_{int,ab}(i\omega)$ \\
LDA+DMFT (AL) & $\rho(r),G_{loc,ab}(i\omega)$  & $V_{int}(r),\Delta_{ab}(i\omega)$\\
LDA+U         & $\rho(r),n_{ab}$               & $V_{int}(r),\lambda_{ab}$ \\
LDA           & $\rho(r)$                      & $V_{int}(r)$
\end{tabular}
\caption{ Parallel between the different approaches, indicating
the physical quantity which has to be extremized, and the field
which is introduced to impose a constraint (constraining field).
(BL) means band limit and (AL) corresponds to atomic limit.}
\label{tab:PQCF}
\end{table}

{$\bullet $ }\textit{Construction of the exact functional. } %
We derive the equations following the effective action point of
view \cite{Chitra:2001}. To facilitate the comparison between the
approaches discussed in the earlier sections we have tabulated
(see Table~\ref{tab:PQCF}) the central quantities which have to be
minimized, and the fields which are introduced to impose a
constraint in the effective action method \cite{Fukuda:1994}. As
in the LDA+U method one introduces a set of correlated orbitals
$\chi _{a}(\mathbf{r}-\mathbf{R}).$ One defines an exact
functional of the total density $\rho (x)$ and of the local
spectral function of the correlated orbitals discussed before:
\begin{equation}
G_{loc,ab}(\tau ,\tau ^{\prime })=-\int \sum_R \chi _{Ra}(\mathbf{r}%
)\left\langle \psi (x)\psi ^{+}(x^{\prime })\right\rangle \chi _{Rb}^{\ast }(\mathbf{%
r }^{\prime })d\mathbf{r}d\mathbf{r}^{\prime }  \label{DMFloc}
\end{equation}%
where indexes $a,b$ refer exclusively to the correlated orbitals,
and $c_{Rb}^{+}$ creates $\chi _{b}(\mathbf{r}-\mathbf{R}).$.

We now introduce the sources for the density, $L(x)$, and for the
local spectral function of the correlated orbitals,
$J_{loc,Rab}(\tau ,\tau ^{\prime }).$ These two sources modify the
action as follows
\begin{eqnarray}
&&S^{\prime }= S +\int L(x)\psi ^{+}(x)\psi (x) dx + \\\nonumber%
&&\frac{1}{N}\sum_{Rab}\int J_{loc,Rab}(\tau \mathbf{,}\tau
^{\prime })c_{Ra}^{+}(\tau )c_{Rb}(\tau ^{\prime })d\tau d\tau
^{\prime } . \label{DMFact}
\end{eqnarray}%
This defines the free energy of the system as a functional of the
source fields after (\ref{BKFpar}). Both density and local
Green's function can be
calculated as follows
\begin{equation}
\frac{\delta F}{\delta L(x)}=\rho (x) ,  \label{DMFvar}
\end{equation}
\begin{equation}
\frac{\delta F}{\delta J_{loc,Rba}(\tau^{\prime } \mathbf{,}\tau
)\ } =G_{loc,ab}(\tau \mathbf{,}\tau ^{\prime }) .  \label{DMFvaw}
\end{equation}%
Then, the functional of both density and the spectral function is
constructed by the Legendre transform. This is an exact
functional of the density and the local\ Green's function,
$\Gamma(\rho ,G_{loc})$, which gives the exact total
free energy, the exact density, and the exact local Green's
function of the heavy electrons at the stationary point.

{$\bullet $ }\textit{Exact representations of the constraining
field. } %
A perturbative construction can be carried out either around the
atomic limit or around the band limit following the inversion
method. Unfortunately the latter is very involved and has not
been yet evaluated, except for the lowest order (so--called
``tree") level neglecting nonlocal interactions. One can also
perform a decomposition into the lowest order term (consisting of
``kinetic energy") and the rest (with an exchange and correlation
energy).

{$\bullet $}\textit{\ Constructing approximations. }%
Given that DMFT has proven to accurately describe many systems at
the level of model Hamiltonians, and that LDA has a long history
of success in treating weakly correlated materials, LDA+DMFT is
obviously a reasonable choice for an approximation to the exact
functional.

The functional implementation corresponding to this approximation
is given by $\Gamma_{LDA+DMFT}$ and has the form
\begin{widetext}
\begin{eqnarray}\nonumber
&&\Gamma_{LDA+DMFT}[\rho,G_{loc},V_{int},\mathcal{M}_{int}] =\\
&&-\mathrm{Tr}\ln
[i\omega+\mu+\nabla^{2}-V_{ext}-V_{int}-\sum_{abR}
  [\mathcal{M}_{int,ab}(i\omega)-\mathcal{M}_{DC,ab}]\chi_{a}(\mathbf{r-R})\chi _{b}^{\ast}
  (\mathbf{r}^{\prime}-\mathbf{R})]\\
&&- \int V_{int}(\mathbf{r})\rho(\mathbf{r})d\mathbf{r}
-T\sum_{i\omega}\sum_{ab}
[\mathcal{M}_{int,ab}(i\omega)-\mathcal{M}_{DC,ab}]G_{loc,ba}(i\omega)
+ E_{H}[\rho]+E_{xc}^{LDA}[\rho ]+\Phi
_{DMFT}[G_{loc,ab}]-\Phi_{DC}[n_{ab}].\nonumber
 \label{DMFfun}
\end{eqnarray}%
\end{widetext} %
$\Phi _{DMFT}[G_{loc,ab}]$ is the sum of all two--particle
irreducible graphs constructed with the local part of the
interaction and the local Green's function, and $ \Phi
_{DC}[n_{ab}]$ is taken to have the same form as in the LDA+U
method, Eq. (\ref{LDUedc}). In a fixed tight--binding basis,
$-\nabla ^{2}+V_{ext}$ reduces to $h_{ab}^{(0)}(\mathbf{k})$ and
the functional $\Gamma _{LDA+DMFT}$ Eq.~(\ref{DMFfun}),  for a
fixed density truncated to a finite basis set, takes a form
identical to the DMFT functional which was discussed in Section
\ref{sec:SDFact}.

\begin{figure}[tbh]
\input{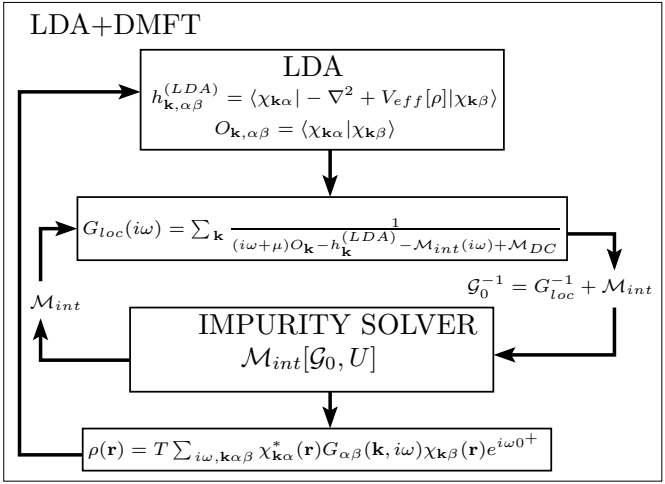}
\caption{Illustration of self--consistent cycle in spectral
density functional theory with LDA+DMFT approximation: double
iteration cycle consists of the inner DMFT loop and outer
(density plus total energy) loop. } \label{FigLDA+DMFT}
\end{figure}

{$\bullet $ }\textit{Saddle point equations. }%
Minimization of the functional leads to the set of equations with the
Kohn--Sham potential and
\begin{eqnarray}
\mathcal{M}_{int,ab}(i\omega ) &=&\frac{\delta \Phi
_{DMFT}}{\delta
G_{loc,ba}(i\omega )},  \label{DMFsig} \\
\mathcal{M}_{DC,ab} &=&\frac{\delta \Phi _{DC}}{\delta n_{ba}},
\label{DMFdoc}
\end{eqnarray}%
which identifies matrix $\mathcal{M}_{int,ab}(i\omega )$ as the
self--energy of the generalized Anderson impurity model in a bath
characterized by a hybridization function $\Delta _{ab}(i\omega
)$ obeying the
self--consistency condition
\begin{eqnarray}
\label{DMFdel}
&&(i\omega +\mu )\bar{O}_{ab}-\bar{\epsilon}_{ab}-\Delta _{ab}(i\omega )- \mathcal{M}_{int,ab}(i\omega ) =\\
&&\left[ \sum_{\mathbf{k}}[(i\omega +\mu )\hat{O}(\mathbf{k})-\hat{h}%
^{(LDA)}(\mathbf{k})-\mathcal{M}_{int}(i\omega )+\mathcal{M}_{DC}]^{-1}%
\right] _{ab}^{-1} .
  \notag
\end{eqnarray}%
By examining the limiting behavior $i\omega \rightarrow \infty$,
we get the definition of the average overlap matrix
$\bar{O}_{ab}$ for the impurity levels as inverse of the average
inverse overlap, i.e.
\begin{equation}
\bar{O}_{ab}=\left[
\sum_{\mathbf{k}}\hat{O}^{-1}(\mathbf{k})\right] _{ab}^{-1}.
\label{DMFovr}
\end{equation}
Similarly, the matrix of the impurity levels has the following form%
\begin{widetext}
\begin{equation}
\bar{\epsilon}_{ab}=\sum_{cd}\bar{O}_{ac}\left[ \sum_{\mathbf{k}}\hat{O}%
^{-1}(\mathbf{k})[h^{(LDA)}(\mathbf{k})+\mathcal{M}_{int}(i\infty )-\mathcal{%
M}_{DC}]\hat{O}^{-1}(\mathbf{k})\right] _{cd}\bar{O}_{db} -\mathcal{M}_{int,ab}(i\infty ) .%
\label{DMFimp}
\end{equation}
Finally, minimization of Eq. (\ref{DMFfun}) with respect to
$V_{eff}$
indicates that $\rho (\mathbf{r})$ should be computed as follows%
\begin{equation}
\rho(\mathbf{r})=T\sum_{i\omega }\left\langle
\mathbf{r}\left\vert [i\omega
+\mu +\nabla ^{2}-V_{eff}-\sum_{abR}[\mathcal{M}_{int,ab}(i\omega )-\mathcal{%
M}_{DC,ab}]\chi _{a}(\mathbf{r-R})\chi _{b}^{\ast }(\mathbf{r}^{\prime }-%
\mathbf{R})]^{-1}\right\vert \mathbf{r}\right\rangle e^{i\omega 0^{+}}.%
\label{DMFrho}
\end{equation}
\end{widetext}
The self--consistency in the LDA+DMFT theory is performed as a
double iteration loop, the inside loop is over the DMFT cycle and
the outside loop is over the electron density, which modifies the
one--electron LDA Hamiltonian. The self--consistent cycle is
illustrated in Fig.~\ref{FigLDA+DMFT}.

\textbf{$\bullet$}\textit{Evaluation of the total energy. }%
In general, the free energy is $F_{tot}=E_{tot}-TS,$ where
$E_{tot}$ is the total energy and $S$ is the entropy. Both energy
and entropy terms exist in the kinetic and interaction
functionals. The kinetic energy part of the functional is given by
$K_{SDFT}[\mathcal{G}]=\mathrm{Tr(} -\nabla
^{2}+V_{ext})\mathcal{G}$ while the potential energy part is
$\frac{1}{2}\mathrm{Tr}\mathcal{M}_{int}G_{loc}$ therefore the
total energy within LDA+DMFT becomes
\begin{widetext}
\begin{eqnarray}
E_{tot}&=&T\sum_{\vk j}\sum_{i\omega}g_{\vk j\omega}\epsilon_{\vk j\omega}
-\int V_{int}(\vr)\rho(\vr)d\vr-
T\sum_{i\omega}\sum_{ab}[\mathcal{M}_{int,ab}(i\omega)-\mathcal{M}_{DC,ab}]
G_{loc,ba}(i\omega )+ \notag \\
&&
+E_{H}[\rho]+E_{xc}^{LDA}[\rho]+
\frac{1}{2}T\sum_{i\omega }\sum_{ab}\mathcal{M}%
_{int,ab}(i\omega )G_{loc,ba}(i\omega )-\Phi _{DC}[n_{ab}],%
\label{DMFtot}
\end{eqnarray}
\end{widetext} where the frequency--dependent eigenvalues $\epsilon _{\mathbf{%
k}j\omega }$ come as a result of diagonalizing the following
non--hermitian
eigenvalue problem similar to (\ref{SDFaar}):%
\begin{eqnarray}
&&\sum_{\beta }\left[ h_{\alpha \beta
}^{(LDA)}(\mathbf{k})+\delta _{\alpha
a}\delta _{\beta b}(\mathcal{M}_{int,ab}(i\omega )-\mathcal{M}%
_{DC,ab})-\right.  \notag \\
&&\left.\epsilon _{\mathbf{k}j\omega }O_{\alpha \beta
}(\mathbf{k})\right] \psi^{R}_{\mathbf{k}j\omega,\b}=0 .
\label{DMFekj}
\end{eqnarray}%
Also,%
\begin{equation}
g_{\mathbf{k}j\omega }=\frac{1}{i\omega +\mu -\epsilon _{\mathbf{k}j\omega }}%
,  \label{DMFgkj}
\end{equation}%
is the Green's function in the orthogonal left/right
representation which plays a role of a ``frequency--dependent
occupation number".

Evaluation of the entropy contribution to the free energy can be
performed by finding the total energy at several temperatures and
then taking the integral \cite{Georges:1996}
\begin{equation}
S(T)=S(\infty )-\int_{T}^{\infty }dT^{\prime }\frac{1}{T^{\prime }}\frac{%
dE_{LDA+DMFT}}{dT^{\prime }} .  \label{DMFent}
\end{equation}%
The infinite temperature limit $S(\infty )$ for a well defined
model Hamiltonian can be worked out. This program was implemented
for the Hubbard model \cite{Rozenberg:1994} and for\ Ce
\cite{Held:2001}. If we are not dealing with a model Hamiltonian
construction \ one has to take instead of infinity a sufficiently
high temperature so that the entropy $S(\infty )$ can be
evaluated by semiclassical considerations.

\textbf{$\bullet $ }\textit{Choice of basis and double counting. }%
The basis can be gradually refined so as to obtain more accurate
solutions in certain energy range. In principle this improvement
is done by changing the linearization energies, and the
experience from density functional implementations could be
carried over to the DMFT case.

Notice that the rational for the double counting term described
in Eq.~(\ref{LDUedc}) was chosen empirically to fit the
one--particle spectra of Mott insulator (for further discussion
see \cite{Petukhov:2003}) and deserves further investigations. All
the discussion of double counting terms in LDA+U literature can
be extended to LDA+DMFT. Notice that as long as the equations are
derivable from a functional, \textit{the Luttinger theorem is
satisfied} (in the single--site DMFT case).

In addition to the forms of double counting terms discussed in
Section \ref{sec:SDFldu}, it has been proposed to use the DMFT
self--energy at zero or at infinity for the double counting. One
possibility

\begin{equation}
\mathcal{M}_{DC,ab}=\frac{1}{N_{\deg }^{}}\delta _{ab}\sum_{a'}
\mathcal{M}_{int,a'a'}(0),  \label{DMFmdc}
\end{equation}
was suggested and implemented by Lichtenstein et al. in their
work on Fe and Ni \cite{Lichtenstein:2001}. The spin polarized
version of this term, has
been recently applied to Iron with encouraging results \cite%
{Lichtenstein:2000}.

\textbf{$\bullet $ }\textit{Assessment of the LDA+DMFT method. }%
The addition of a realistic band theory to the DMFT treatment of
models of correlated electron systems has opened a new area of
investigations. To the many--body theorist, the infusion of a
realistic band theory allows one to make system--specific studies.
Some of them are listed in Section~\ref{sec:MAT} on materials.
For the electronic structure community, the LDA+DMFT method allows
the treatment of a variety of materials which are not well treated
by the LDA or the LDA+U method, such as correlated metals and
systems with paramagnetic local moments. The main shortcoming is
in the arbitrariness in the choice of the correlated orbitals, in
the estimation of the $U$, and the ambiguity in the choice of
double counting correction. This may turn out to be hard to
resolve within this formalism. The ideas described in the
following section formulate the many--body problem in terms
of fluctuating electric fields and electrons, treat all the
electrons on the same footing, provide an internally consistent
evaluation of the interaction, and eliminate the need for the
double counting correction.

\subsection{Equations in real space}

\label{sec:SDFrsp}

$\bullet ${\ }\textit{Functional of the local Green's function in
real space. }%
The success of the dynamical mean--field approximations is
related to the notion that the local approximation is good in
many situations. Thus far, the notion of locality has only been
explored after choosing a set of tight--binding orbitals, but it
can also be formulated directly in real space, as stressed
recently in Ref. \cite{Savrasov:2003:CM0308053,Chitra:2000}. This
is necessary in order to make direct contact with theories such as
density functional theory, which is formulated directly in the
continuum without resorting to a choice of orbitals or preferred
basis set. The theory is formulated by defining the local Green's
function to be the exact Green's function
$G(\mathbf{r},\mathbf{r}^{\prime },z)$ within a given volume
$\Omega _{loc}$
and zero outside. In other words,%
\begin{equation}
G_{loc}(\mathbf{r},\mathbf{r}^{\prime
},z)=G(\mathbf{r},\mathbf{r}^{\prime },z)\theta
_{loc}(\mathbf{r},\mathbf{r}^{\prime }) ,  \label{RSPloc}
\end{equation}%
where $\mathbf{r}$ is within a primitive unit cell $\Omega _{c}$
positioned at $\mathbf{R}=0$ while $\mathbf{r}^{\prime }$ travels
within some volume $\Omega _{loc}$ centered at $\mathbf{R}=0$.
Theta function is unity
when $\mathbf{r}\in \Omega _{c},\mathbf{r}^{\prime }\in \Omega _{loc}$%
\ and zero otherwise. This construction can be translationally
continued onto entire lattice by enforcing the property $\theta
_{loc}(\mathbf{r+R}, \mathbf{r}^{\prime }+\mathbf{R})=\theta
_{loc}(\mathbf{r},\mathbf{r}^{\prime }).$

The procedures outlined in the previous sections can be applied
to the continuum in order to construct an exact functional which
gives the exact free energy, the local Green's function (in real
space), its Kohn--Sham \ formulation, and its dynamical
mean--field approximation (by restricting the interaction in the
full Baym--Kadanoff functional to the local Green's function).

This approach has the advantage that the density is naturally
contained in this definition of a local Green's function, and
therefore the density functional theory is naturally embedded in
this formalism. Another advantage is that the approach contains
the bare Coulomb interaction, and therefore is free from
phenomenological parameters such as the Hubbard $U$. However, this
may create problems since it is well--known that significant
screening of the interactions occurs within real materials.
Therefore, it is useful to incorporate the effects of screening
at a level of functional description of the system.

{$\bullet $ }\textit{Motivation and choice of variables: spectral
density functional of the local Green's functions and of the
local interaction. }%
We introduce two local source fields $J_{loc}$ and $K_{loc}$
which probe the local electron Green's function $G_{loc}$ defined
earlier and the local part of the boson Green's function
$W_{loc}(x,x^{\prime })=\langle T_{\tau }\phi (x)\phi (x^{\prime
})\rangle \theta _{loc}(\mathbf{r},\mathbf{r}^{\prime })$ being
the screened interaction (see Section \ref{sec:ScreenedInt}). This
generalization represents the ideas of the extended dynamical
mean--field theory \cite{Si:1996}, now viewed as an exact theory.
Note that formally the cluster for the interaction can be
different from the one considered to define the local Green's
function (\ref{RSPloc}) but we will not distinguish between them
for simplicity. The auxiliary Green's function
$\mathcal{G}(\mathbf{r},\mathbf{r}^{\prime },i\omega )$ as well
as the auxiliary interaction $
\mathcal{W}(\mathbf{r},\mathbf{r}^{\prime },i\omega )$ are
introduced which are the same as the local functions within
non--zero volume of $\theta _{loc}(\mathbf{r},\mathbf{r}^{\prime
})$
\begin{eqnarray}
G_{loc}(\mathbf{r},\mathbf{r}^{\prime },i\omega ) &=&\mathcal{G}(\mathbf{r},%
\mathbf{r}^{\prime },i\omega )\theta
_{loc}(\mathbf{r},\mathbf{r}^{\prime }),
\label{RSPgrn} \\
W_{loc}(\mathbf{r},\mathbf{r}^{\prime },i\omega ) &=&\mathcal{W}(\mathbf{r},%
\mathbf{r}^{\prime },i\omega )\theta
_{loc}(\mathbf{r},\mathbf{r}^{\prime }). \label{RSPint}
\end{eqnarray}%
The spectral density functional is represented in the form
\begin{widetext}
\begin{equation}
\Gamma_{SDFT}[G_{loc},W_{loc}]=Tr\ln \mathcal{G}-Tr[G_{0}^{-1}-\mathcal{G}%
^{-1}]\mathcal{G+}E_{H}[\rho ]-\frac{1}{2}\mathrm{Tr}\ln \mathcal{W}+\frac{1%
}{2}\mathrm{Tr}[v_{C}^{-1}-\mathcal{W}^{-1}]\mathcal{W}+
\Psi_{SDFT}[G_{loc},W_{loc}]  .%
\label{RSPfun}
\end{equation}%
\end{widetext}It can be viewed as a functional $\Gamma
_{SDFT}[G_{loc},W_{loc}]$\ or alternatively as a functional
$\Gamma _{SDFT}[\mathcal{G},\mathcal{W}]$. $\Psi
_{SDFT}[G_{loc},W_{loc}]$ is formally \emph{ not} a sum of
two--particle diagrams constructed with $G_{loc}\ $and $ W_{loc}
$, but in principle a more complicated diagrammatic expression
can be derived following Refs. \cite{Fukuda:1994, Valiev:1997,
Chitra:2001}. A more explicit expression involving a coupling
constant integration can be given. Examining stationarity of
$\Gamma _{SDFT}$ yields saddle--point equations for
$\mathcal{G}(\mathbf{r},\mathbf{r}^{\prime },i\omega )$ and for
$\mathcal{W}(\mathbf{r},\mathbf{r}^{\prime },i\omega )$
\begin{eqnarray}
\mathcal{G}^{-1}(\mathbf{r},\mathbf{r}^{\prime },i\omega ) &=&G_{0}^{-1}(%
\mathbf{r},\mathbf{r}^{\prime },i\omega )-\mathcal{M}_{int}(\mathbf{r},%
\mathbf{r}^{\prime },i\omega ),  \label{RSPdeg}\ \ \ \ \ \ \\
\mathcal{W}^{-1}(\mathbf{r},\mathbf{r}^{\prime },i\omega ) &=&v_{C}^{-1}(%
\mathbf{r}-\mathbf{r}^{\prime
})-\mathcal{P}(\mathbf{r},\mathbf{r}^{\prime },i\omega ),
\label{RSPdew}
\end{eqnarray}%
where $\mathcal{M}_{int}(\mathbf{r},\mathbf{r}^{\prime },i\omega
)$ is the auxiliary local mass operator
defined as the variational derivative of the interaction functional:%
\begin{equation}
\mathcal{M}_{int}(\mathbf{r},\mathbf{r}^{\prime },i\omega
)=\frac{\delta
\Phi _{SDFT}[G_{loc}]}{\delta \mathcal{G}(\mathbf{r}^{\prime },\mathbf{r}%
,i\omega )}=\frac{\delta \Phi _{SDFT}[G_{loc}]}{\delta G_{loc}(\mathbf{r}%
^{\prime },\mathbf{r},i\omega )}\theta
_{loc}(\mathbf{r},\mathbf{r}^{\prime }),  \label{RSPsig}
\end{equation}%
$\mathcal{P}(\mathbf{r},\mathbf{r}^{\prime },i\omega )$ is the
effective susceptibility of the system defined as the variational
derivative
\begin{equation}
\mathcal{P}(\mathbf{r},\mathbf{r}^{\prime },i\omega
)=\frac{-2\delta \Psi
_{SDFT}}{\delta \mathcal{W}(\mathbf{r}^{\prime },\mathbf{r},i\omega )}=\frac{%
-2\delta \Psi _{SDFT}}{\delta W_{loc}(\mathbf{r}^{\prime
},\mathbf{r},i\omega )}\theta
_{loc}(\mathbf{r},\mathbf{r}^{\prime }).  \label{RSPchi}
\end{equation}%
Notice again a set of parallel observations for $\mathcal{P}$ as
for $\mathcal{M}_{int}$. Both $\mathcal{P}$ and
$\mathcal{M}_{int}$ are local by construction, i.e. these are
non--zero only within the cluster restricted by $\theta
_{loc}(\mathbf{r}, \mathbf{r}^{\prime }).$ Formally, they are
auxiliary objects and cannot be identified with the exact
self--energy and susceptibility of the electronic system.
However, if the exact self--energy and susceptibility are
sufficiently localized, this identification becomes possible. If
the cluster $\Omega _{loc}$ includes the physical area of localization,
we can immediately identify $\mathcal{M}_{int}(r,r^{\prime
},i\omega )$ with $\Sigma _{int}( \mathbf{r},\mathbf{r}^{\prime
},i\omega ),$ $\mathcal{P}(\mathbf{r},\mathbf{r}^{\prime
},i\omega )$ with $\Pi (\mathbf{r},\mathbf{r}^{\prime },i\omega
)$ in all space. However, both
$\mathcal{G}(\mathbf{r},\mathbf{r}^{\prime },i\omega )$ and
$G(\mathbf{r},\mathbf{r}^{\prime },i\omega )$ as well as $
\mathcal{W}$ and $W$ are always the same within $\Omega _{loc}$
regardless its size, as it is seen from (\ref{RSPgrn}) and
(\ref{RSPint}).

\textbf{$\bullet $ }\textit{Practical implementation and
Kohn--Sham representation. }%
The Kohn--Sham Green's function can be calculated using the
following representation
\begin{equation}
\mathcal{G}(\mathbf{r},\mathbf{r}^{\prime },i\omega )=\sum_{\mathbf{k}j}%
\frac{\psi _{\mathbf{k}j\omega }^{R}(\mathbf{r})\psi
_{\mathbf{k}j\omega }^{L}(\mathbf{r}^{\prime })}{i\omega +\mu
-\epsilon _{\mathbf{k}j\omega }}, \label{RSPgk1}
\end{equation}%
where the left $\psi _{\mathbf{k}j\omega }^{L}(\mathbf{r})$ and
right $\psi _{\mathbf{k}j\omega }^{R}(\mathbf{r})$ states satisfy
the following Dyson equations
\begin{widetext}
\begin{eqnarray}
\lbrack -\nabla ^{2}+V_{ext}(\mathbf{r})+V_{H}(\mathbf{r})]\psi _{\mathbf{k}%
j\omega }^{R}(\mathbf{r})+\int \mathcal{M}_{xc}(\mathbf{r},\mathbf{r}%
^{\prime },i\omega )\psi _{\mathbf{k}j\omega }^{R}(\mathbf{r}^{\prime })d%
\mathbf{r}^{\prime } &=&\epsilon _{\mathbf{k}j\omega }\psi _{\mathbf{k}%
j\omega }^{R}(\mathbf{r}) , \label{RSPder} \\%
\lbrack -\nabla ^{2}+V_{ext}(\mathbf{r}^{\prime
})+V_{H}(\mathbf{r}^{\prime
})]\psi _{\mathbf{k}j\omega }^{L}(\mathbf{r}^{\prime })+\int \psi _{\mathbf{k%
}j\omega
}^{L}(\mathbf{r})\mathcal{M}_{xc}(\mathbf{r},\mathbf{r}^{\prime
},i\omega )d\mathbf{r} &=&\epsilon _{\mathbf{k}j\omega }\psi _{\mathbf{k}%
j\omega }^{L}(\mathbf{r}^{\prime })  .%
\label{RSPdel}
\end{eqnarray}%
\end{widetext}These equations should be considered as eigenvalue
problems with a complex, non--hermitian self--energy. As a result,
the eigenvalues $\epsilon _{\mathbf{k}j\omega }$ are complex in
general, and the same for both equations. The explicit dependence
on the frequency $ i\omega $ of both the eigenvectors and
eigenvalues comes from the self--energy. Note that left and right
eigenfunctions are orthonormal
\begin{equation}
\int d\mathbf{r}\psi _{\mathbf{k}j\omega }^{L}(\mathbf{r})\psi
_{\mathbf{k} j^{\prime }\omega }^{R}(\mathbf{r})\mathbf{=}\delta
_{jj^{\prime }}, \label{RSPort}
\end{equation}%
and can be used to evaluate the charge density of a given system
using the
Matsubara sum and the integral over the momentum space
\begin{equation}
\rho(\mathbf{r})=T\sum_{i\omega}\sum_{\mathbf{k}j}
\frac{
\psi_{\mathbf{k}j\omega}^{R}(\mathbf{r})
\psi_{\mathbf{k}j\omega }^{L}(\mathbf{r})
}{i\omega+\mu-\epsilon_{\mathbf{k}j\omega}}
e^{i\omega 0^{+}} .  \label{RSPrho}
\end{equation}
It can be shown \cite{Savrasov:2003:CM0308053} that this system
of equations reduces to the Kohn--Sham eigensystem when the
self--energy is frequency independent.

Note that the frequency--dependent energy bands $\epsilon
_{\mathbf{k}j\omega }$ represent an auxiliary set of complex
eigenvalues. These are not the true poles of the exact
one--electron Green's function
$G(\mathbf{r},\mathbf{r}^{\prime},z)$. However, they are designed
to reproduce the local spectral density of the system. Note also
that these bands $\epsilon_{\mathbf{k}jz}$ are not the true poles
of the auxiliary Green's function
$\mathcal{G}(\vr,\vr^{\prime},z)$. Only in the situation when
$\mathcal{G}$ is a good approximation to $G$, the solution of the
equation $z+\mu-\epsilon_{\vk jz}=0$ gives a good approximation
for the quasiparticle energies

\textbf{$\bullet $ }\textit{Evaluation of the total energy. }%
The energy--dependent representation allows one to obtain a very
compact expression for the total energy. As we have argued, the
entropy terms are more difficult to evaluate. However, as long as
we stay at low temperatures, these contributions are small and
the total energy approach is valid. In this respect, the SDFT
total energy formula is obtained by utilizing the relationship
$\epsilon _{\mathbf{k}j\omega }=\langle \psi _{\mathbf{k} j\omega
}^{L}|-\nabla ^{2}+\mathcal{M}_{eff}|\psi _{\mathbf{k}j\omega
}^{R}\rangle =\langle \psi _{\mathbf{k}j\omega }^{L}|-\nabla
^{2}+V_{ext}+V_{H}+\mathcal{M}_{xc}|\psi _{\mathbf{k}j\omega
}^{R}\rangle $
\begin{widetext}
\begin{eqnarray}
&&E_{SDFT}=T\sum_{i\omega }e^{i\omega 0^{+}}\sum_{\mathbf{k}j}g_{\mathbf{k}%
j\omega }\epsilon _{\mathbf{k}j\omega }-T\sum_{i\omega }\int d\mathbf{r}d%
\mathbf{r}^{\prime }\mathcal{M}_{eff}(\mathbf{r,r}^{\prime },i\omega )%
\mathcal{G}(\mathbf{r}^{\prime },\mathbf{r},i\omega )+  \notag \\
&&+\int d\mathbf{r}V_{ext}(\mathbf{r})\rho (\mathbf{r})+E_{H}[\rho ]+\frac{1%
}{2}T\sum_{i\omega }\int d\mathbf{r}d\mathbf{r}^{\prime }\mathcal{M}_{xc}(%
\mathbf{r,r}^{\prime },i\omega )G_{loc}(\mathbf{r}^{\prime },\mathbf{r}%
,i\omega ),
\label{RSPtot}
\end{eqnarray}
\end{widetext}
where $\mathcal{M}_{eff}=\mathcal{M}_{int}+V_{ext}$ and
$g_{\mathbf{k}j\omega}=1/(i\omega+\mu-\epsilon_{\mathbf{k}j\omega})$.
For the same reason as in DFT, this expression should be evaluated
with the value of the  the self-energy $\mathcal{M}_{eff}$ which
is used as  input to  the routine performing the inversion of the
Dyson equation,  and with the value of  the Green's function $\cal
G$ which is the output of that inversion.

{$\bullet $ }\textit{Constructions of approximations. }%
The dynamical mean--field approximation to the exact spectral
density functional is defined by restricting the interaction part
of Baym--Kadanoff functional $\Psi_{SDFT}[G_{loc},W_{loc}]$ to
$G_{loc}(\mathbf{r},\mathbf{r}^{\prime },z)$ and
$W_{loc}(\mathbf{r},\mathbf{r}^{\prime },i\omega ).$ The sum over
all the diagrams, constrained to a given site, together with the
Dyson equations can be formulated in terms of the solution of an
auxiliary Anderson impurity model, after the introduction of a
basis set. We introduce a bath Green's function
$\mathcal{G}_{0}(\mathbf{r}, \mathbf{r}^{\prime },i\omega )$ and
a \textquotedblleft bath interaction"
$\mathcal{V}_{0}(\mathbf{r},\mathbf{r}^{\prime },i\omega )$
defined by the following Dyson equations
\begin{eqnarray}
\mathcal{G}_{0}^{-1}(\mathbf{r},\mathbf{r}^{\prime },i\omega )
&=&G_{loc}^{-1}(\mathbf{r},\mathbf{r}^{\prime },i\omega )+\mathcal{M}_{int}(
\mathbf{r},\mathbf{r}^{\prime },i\omega ),  \ \ \ \ \ \label{RSPbat} \\
\mathcal{V}_{0}^{-1}(\mathbf{r},\mathbf{r}^{\prime },i\omega )
&=&W_{loc}^{-1}(\mathbf{r},\mathbf{r}^{\prime },i\omega )+\mathcal{P}(
\mathbf{r},\mathbf{r}^{\prime },i\omega ).  \label{RSPv01}
\end{eqnarray}
Note that formally neither $\mathcal{G}_{0}$ nor
$\mathcal{V}_{0}$ can be associated with non--interacting $G_{0}$
and the bare interaction $v_{C},$ respectively. These two
functions are to be considered as an input to the auxiliary
impurity model which delivers new $\mathcal{M}_{int}(\mathbf{r},
\mathbf{r}^{\prime },i\omega )$ and
$\mathcal{P}(\mathbf{r},\mathbf{r}^{\prime },i\omega ).$

\begin{figure}[tbh]
\includegraphics*[height=2.4in]{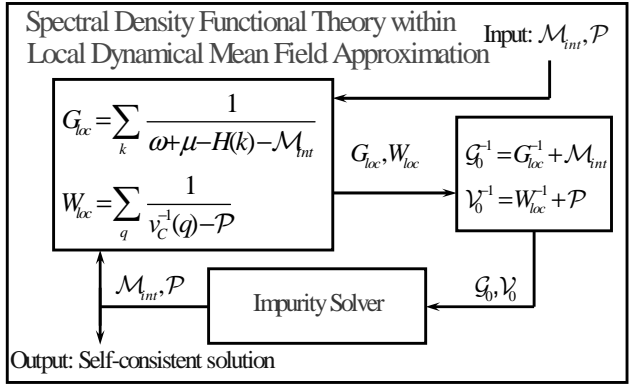}
\caption{
  Illustration of the self--consistent cycle in spectral
density functional theory within the local dynamical mean--field
approximation: both local Green's function $G_{loc}$ and local
Coulomb interaction $W_{loc}$ are iterated. Here we illustrate
one possible explicit realization of the abstract general SDFT
construction. This requires an explicit  definition of
 $G_{loc}$, which for the purpose of this figure is
done   by means of the use of a tight binding basis set.}
\label{FigSDFT}
\end{figure}

To summarize, the effective impurity action, the Dyson equations
(\ref{RSPbat}), (\ref{RSPv01}) connecting local and bath
quantities as well as the original Dyson equations (\ref{RSPdeg}),
(\ref{RSPdew}) constitute a self--consistent set of equations as
the saddle--point conditions extremizing the spectral density
functional $\Gamma _{SDFT}(\mathcal{G}, \mathcal{W})$. They
combine cellular and extended versions of DMFT and represent our
philosophy in the \textit{ab initio} simulation of a
strongly--correlated system. Since $\mathcal{M}_{int}$ and
$\mathcal{P}$ are unknown at the beginning, the solution of these
equations assumes self--consistency. First, assuming some initial
$\mathcal{M}_{int}$ and $\mathcal{P}$, the original Dyson
equations (\ref{RSPdeg}), (\ref{RSPdew}) are used to find Green's
function $\mathcal{G}$ and screened interaction $\mathcal{W}.$
Second, the Dyson equations for the local quantities
(\ref{RSPbat}), (\ref{RSPv01}) are used to find
$\mathcal{G}_{0}$, $\mathcal{V}_{0}.$ Third, quantum impurity
model with input $\mathcal{G}_{0}$, $\mathcal{V}_{0}$ is solved
by available many--body technique to give new local
$\mathcal{M}_{int}$ and $\mathcal{P}$ : this is a
much more challenging task than purely fermionic calculations
(e.g. cluster DMFT in Hubbard model), which can only be addressed
at present with Quantum Monte Carlo methods using continuous
Hubbard-Stratonovich fields \cite{Sun:2002} or possibly with
continuous Quantum Monte Carlo studied in
\cite{Rubtsov:2004:CM0411344}.  The process is
repeated until self--consistency is reached. This is
schematically illustrated in Fig.~\ref{FigSDFT}. Note here that
while single--site impurity problem has a well--defined algorithm
to extract the lattice self--energy, this is not generally true
for the cluster impurity models \cite{Biroli:2004}. The latter
provides the self--energy of the cluster, and an additional
prescription such as implemented within cellular DMFT or DCA
should be given to construct the self--energy of the lattice.
An interesting observation can be made on the role of the
impurity model which in the present context appeared as an
approximate way to extract the self--energy of the lattice using
input bath Green's function and bath interaction. Alternatively,
the impurity problem can be thought of itself as the model which
delivers the exact mass operator of the spectral density
functional \cite{Chitra:2001}. If the latter is known, there
should exist a bath Green's function and a bath interaction which
can be used to reproduce it. In this respect, the local
interaction $W_{loc}$ appearing in our formulation can be thought
of as an exact way to define the local Coulomb repulsion ``$U$",
i.e. the interaction which delivers exact local self--energy.

\textbf{$\bullet $ }\textit{Local GW. }%
A simplified version of the described construction
\cite{Kotliar:2001:Tsvelik,Zein:2002} is known
as a local version of the GW method (LGW). Within the spectral
density functional theory, this approximation appears as an
approximation to the functional $\Psi
_{SDFT}[G_{loc},W_{loc}]$ taken in the form%
\begin{equation}
\Psi_{LGW}[G_{loc},W_{loc}]=-\frac{1}{2}\mathrm{Tr}G_{loc}W_{loc}G_{loc}.
\label{RSPlgw}
\end{equation}%
As a result, the susceptibility
$\mathcal{P}(\mathbf{r},\mathbf{r}^{\prime },i\omega )$ is
approximated by the product of two local Green's functions, i.e.
$\mathcal{P}=-2\delta \Psi _{LGW}/\delta W_{loc}=G_{loc}G_{loc},$
and the exchange--correlation part of our mass operator is
approximated by the local GW diagram, i.e.
$\mathcal{M}_{xc}=\delta \Psi _{LGW}/\delta
G_{loc}=-G_{loc}W_{loc}$. Note that since the local GW
approximation (\ref{RSPlgw}) is relatively cheap from a
computational point of view, its implementation for all orbitals
within a cluster is feasible. The results of the single--site
approximation for the local quantities were already reported in
the literature~\cite{Zein:2002}.

Note finally that the local GW approximation is just one of the
possible impurity solvers to be used in this context. For
example, another popular approximation known as the fluctuation
exchange approximation (FLEX) \cite{Bickers:1989} can be worked
out along the same lines.

{$\bullet $ }\textit{Assessment of the method. }%
The described algorithm is quite general, totally \textit{ab
initio}, and allows the determination of various quantities, such
as the local one--electron Green's functions $G_{loc}$ and the
dynamically screened local interactions $W_{loc}$. This
challenging project so far has only been carried out on the level
of a model Hamiltonian \cite{Sun:2002}. On the other hand, one
can view the LDA+DMFT method as an approximate implementation of
this program, as discussed in Ref.
\cite{Savrasov:2003:CM0308053}. Note also that the combination of
the DMFT and full GW algorithm has been recently proposed and
applied to Ni \cite{Biermann:2003}. This in principle shows the
way to incorporate full $\mathbf{k}$--dependence of the
self--energy known diagrammatically within GW.
The first implementation of a fully self-consistent spectral
density functional calculation within the LDA+DMFT approximation was
carried out in~\cite{Savrasov:2001} using the full potential LMTO
basis set (for details see \cite{Savrasov:2004:PRB}).  Since then
the method has been implemented in the exact muffin tin orbital
basis set ~\cite{Chioncel:2003:PRB67} as well as in a fully KKR
implementation ~\cite{Minar:2005:CM0504760}.

The spectral density functional theory contains the local or
cluster GW diagrams together with all higher order local
corrections to construct an approximation to the exact
$\mathcal{M}_{xc}$. Just like the Kohn--Sham spectra were a good
starting point for constructing the quasiparticle spectra for
weakly correlated electron systems, we expect that
$\mathcal{M}_{xc}$ will be a good approximation for
strongly--correlated electron systems. This is a hypothesis that
can be checked by carrying out the perturbation expansion in
nonlocal corrections.

\subsection{Application to lattice dynamics}

\label{sec:SDFphn}

Computational studies of lattice dynamics and structural
stability in strongly--correlated situations is another
challenging theoretical problem which has been recently addressed
in Refs. \cite{Savrasov:2003, Dai:2003}. LDA has delivered the
full lattice dynamical\ information and electron--phonon related
properties of a variety of simple metals, transition metals, as
well as semiconductors with exceptional \
accuracy~\cite{Baroni:2001}. This is mainly due to an
introduction of a linear response approach \cite{Baroni:1987,
Zein:1984}. This method overcame the problems of traditional
techniques based on static susceptibility calculations which
generally fail to reproduce lattice dynamical properties of real
materials due to difficulties connected with the summations in
high--energy states and the inversion of very large dielectric
matrix \cite{Devreese:1983}.

Despite these impressive successes, there is by now clear
evidence that the present methodology fails when applied to
strongly--correlated materials. For example, the local density
predictions for such properties as bulk modulus and elastic
constants in metallic Plutonium are approximately one order of
magnitude off from experiment~\cite{Bouchet:2000}; the phonon
spectrum of Mott insulators such as MnO is not predicted
correctly by LDA \cite{Massidda:1999}.

Recently, a linear response method to study the lattice dynamics
of correlated materials has been developed \cite{Savrasov:2003,
Dai:2003}. The dynamical matrix being the second order derivative
of the energy can be computed using spectral density functional
theory. As with the ordinary density functional formulation of
the problem~\cite{Savrasov:1996}, we deal with the first order
corrections to the charge density, $\delta \rho ,$ as well as the
first order correction to the Green's function $\delta
\mathcal{G} (i\omega )$ which should be considered as two
independent variables in the functional of the dynamical matrix.
To find the extremum, a set of the linearized Dyson equations has
to be solved self--consistently
\begin{equation}
\lbrack -\nabla ^{2}+\mathcal{\hat{M}}_{eff}(i\omega )-\epsilon _{\mathbf{k}%
j\omega }]\delta \psi _{\mathbf{k}j\omega }^{R}+[\delta \mathcal{\hat{M}}%
_{eff}(i\omega )-\delta \epsilon _{\mathbf{k}j\omega }]\psi _{\mathbf{k}%
j\omega }^{R}=0,  \label{PHNleq}
\end{equation}%
which leads us to consider the first order changes in the local
mass operator $\mathcal{\hat{M}}_{eff}(i\omega ).$ Here and in
the following we will assume that the phonon wave vector of the
perturbation $\mathbf{q}$ is different from zero, and, therefore,
the first order changes in the eigenvalues $\delta \epsilon
_{\mathbf{k}j\omega }$ drop out. The quantity $\delta
\mathcal{\hat{M}}_{eff}(i\omega )$ is a functional of $\delta
\mathcal{G}(i\omega )$ and should be found self--consistently. In
particular, the change in the self--energy $\delta
\mathcal{\hat{M}}_{eff}(i\omega )$  needs a solution of an AIM
linearized with respect to the atomic displacement, which  in
practice requires the computation of a two-particle vertex function
$\Gamma=\delta^2\Phi_{SDFT}(G_{loc})/(\delta G_{loc}\delta G_{loc})$.

In practice, change in the eigenvector $\delta \psi
_{\mathbf{k}j\omega }$ has to be expanded in some basis set.
Previous linear response schemes were based on tight--binding
methods \cite{Varma:1977}, plane wave pseudopotentials
\cite{Baroni:1987, Zein:1984, Gonze:1992, Quong:1992}, linear
augmented plane waves\cite{Yu:1994}, mixed
orbitals~\cite{Heid:1999} and linear muffin--tin orbitals
\cite{Savrasov:1992}. Due to explicit dependence on the atomic
positions of local orbital basis sets both Hellmann--Feynman
contributions and incomplete basis set corrections appear in the
expression for the dynamical matrix \cite{Savrasov:1996}. The
functions $\delta \psi _{\mathbf{k}j\omega }$ are represented as
follows
\begin{equation}
\delta \psi _{\mathbf{k}j\omega }=\sum_{\alpha }\{\delta A_{\alpha }^{%
\mathbf{k}j\omega }\chi _{\alpha }^{\mathbf{k+q}}+A_{\alpha }^{\mathbf{k}%
j\omega }\delta \chi _{\alpha }^{\mathbf{k}}\} ,  \label{PHNpsi}
\end{equation}%
where we introduced both changes in the frequency--dependent
variational coefficients $\delta A_{\alpha }^{\mathbf{k}j\omega
}$ as well as changes in the basis functions $\delta \chi
_{\alpha }^{\mathbf{k}}.$ The latter helps us to reach fast
convergence in the entire expression (\ref{PHNpsi}) with respect
to the number of the basis functions \{$\alpha \}$  since the
contribution with $\delta \chi _{\alpha }^{\mathbf{k}}$ takes
into account all rigid movements of the localized orbitals
\cite{Savrasov:1992}.

\begin{figure}[tbh]
\includegraphics*[height=2.4in]{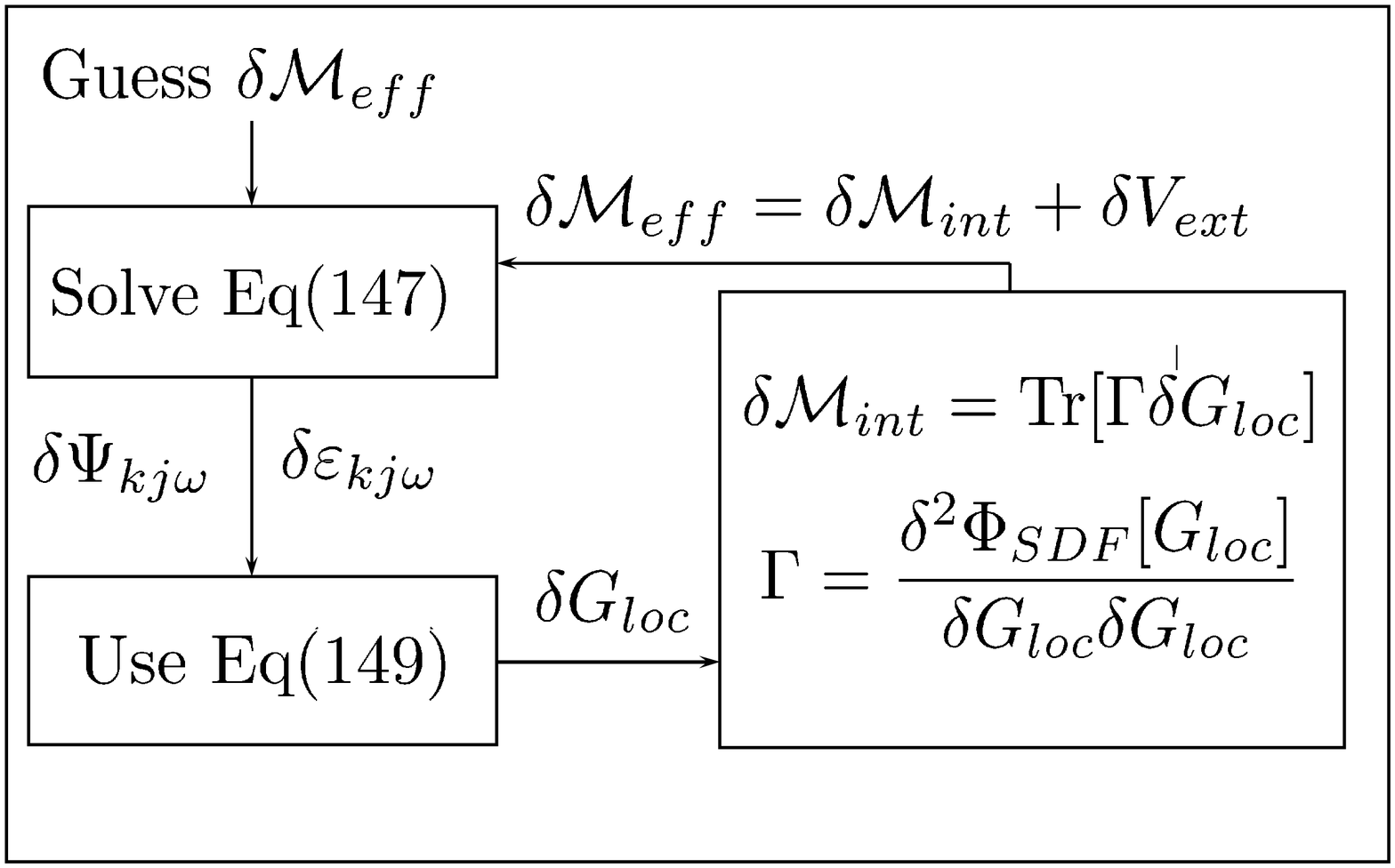}
\caption{
  Illustration of the self--consistent cycle to calculate lattice
  dynamics within the spectral density functional theory.}
\label{FigPhonon}
\end{figure}
The first--order changes in the Green's function can be found as
follows
\begin{equation}
\delta G_{loc}(i\omega )=\sum_{\mathbf{k}j}\frac{\delta \psi _{\mathbf{k}%
j\omega }^{L}\psi _{\mathbf{k}j\omega }^{R}+\psi
_{\mathbf{k}j\omega }^{L}\delta \psi _{\mathbf{k}j\omega
}^{R}}{i\omega+\mu-\epsilon _{\mathbf{k}j\omega } } ,
\label{PHNdgw}
\end{equation}%
which should be used to evaluate the first order change in the
charge density and the dynamical matrix itself (see Fig.
\ref{FigPhonon}).

A simplified version of the approach, neglecting the impurity vertex
function, was successfully applied to the
paramagnetic phases of Mott insulators \cite{Savrasov:2003} as
well as to high--temperature phases of Plutonium \cite{Dai:2003}.
We will describe these applications in Section \ref{sec:MAT}.

\subsection{Application to optics and transport}

\label{sec:SDFopt}

Optical spectral functions such as conductivity or reflectivity
are very important characteristics of solids and give us a direct
probe of the electronic structure.

Here we outline an approach which allows to calculate the optical
properties of a strongly--correlated material within the spectral
density functional framework
\cite{Oudovenko:2004:PRB,Perlov:2004:CM}. This work extends the
methodology in use for weak correlated systems (see
\cite{Maksimov:1988}) to correlated materials. The optical
conductivity can be expressed via equilibrium state
current--current correlation function \cite{Mahan:1993} and is
given by
\begin{eqnarray}
&&\sigma _{\mu \nu }(\omega )=\\\nonumber %
&&\pi e^{2}\int\limits_{-\infty }^{+\infty }d\varepsilon \phi
_{\mu \nu }(\varepsilon +\omega /2,\varepsilon -\omega
/2) \frac{f(\varepsilon -\omega /2{)-}f(\varepsilon +\omega /2{)}}{%
\omega },  \label{OPTsig}
\end{eqnarray}%
where $f(\varepsilon )$ is the Fermi function, and the transport
function $\phi _{\mu \nu }(\varepsilon ,\varepsilon ^{\prime })$
is
\begin{equation}
\phi _{\mu \nu }(\varepsilon ,\varepsilon ^{\prime })=\frac{1}{\Omega _{c}}%
\sum\limits_{\mathbf{k}jj^{\prime }}\mathrm{Tr}\left\{ \nabla _{\mu }\rho _{%
\mathbf{k}j}{(\varepsilon )\nabla _{\nu }}\rho _{\mathbf{k}j^{\prime }}{%
(\varepsilon }^{\prime }{)}\right\} ,  \label{OPTtra}
\end{equation}%
where $\Omega _{c}$ is the unit cell volume and
\begin{equation}
\hat{\rho}_{\mathbf{k}j}(\varepsilon )=-\frac{1}{2\pi i}\left( \mathcal{G}_{%
\mathbf{k}j}(\varepsilon )-\mathcal{G}_{\mathbf{k}j}^{\dagger
}(\varepsilon )\right) ,  \label{OPTrho}
\end{equation}%
is expressed via retarded one--particle Green's function,
$\mathcal{G}_{\mathbf{k}j}(\varepsilon )$, of the system. Taking
limit of zero temperature and using the solutions $\epsilon
_{\mathbf{k}j\omega }$ and $\psi _{\mathbf{k}j\omega }^{R,L}$ of
the Dyson equation (\ref{RSPder}), (\ref{RSPdel}) on the real
frequency axis we express the optical conductivity in the form
\begin{widetext}
\begin{equation}
\label{OPTcon}%
\sigma _{\mu \nu }(\omega )={\pi e^{2}\over
\omega}\sum_{ss^{\prime }=\pm }ss^{\prime
}\sum_{\mathbf{k}jj^{\prime }}\int\limits_{-\omega /2}^{+\omega
/2}d\varepsilon \frac{M_{\mathbf{k}jj^{\prime }}^{ss^{\prime
},\mu \nu
}(\varepsilon ^{-},\varepsilon ^{+})}{\omega +\epsilon _{\mathbf{k}%
j\varepsilon ^{-}}^{s}-\epsilon _{\mathbf{k}j^{\prime }\varepsilon
^{+}}^{s^{\prime }}} \left[ \frac{1}{\varepsilon ^{-}+\mu -\epsilon _{%
\mathbf{k}j\varepsilon ^{-}}^{s}}{-}\frac{1}{\varepsilon ^{+}+\mu
-\epsilon
_{\mathbf{k}j^{\prime }\varepsilon ^{+}}^{s^{\prime }}}\right] ,%
\end{equation}%
\end{widetext}where we have denoted $\varepsilon ^{\pm }=\varepsilon \pm
\omega /2$, and used the abbreviated notations $\epsilon _{\mathbf{k}
j\varepsilon }^{+}\equiv \epsilon _{\mathbf{k}j\varepsilon }$,
$\epsilon _{\mathbf{k}j\varepsilon }^{-}=\epsilon
_{\mathbf{k}j\varepsilon }^{\ast }$.

The matrix elements $M_{\mathbf{k}jj^{\prime }}$ are
generalizations of the standard dipole allowed transition
probabilities which are now defined with the right and left
solutions $\psi ^{R}$and $\psi ^{L}$ of the Dyson equation
\begin{eqnarray}
&&M_{\mathbf{k}jj^{\prime }}^{ss^{\prime },\mu \nu
}(\varepsilon,\varepsilon
^{\prime })= \\
&&\int (\psi _{\mathbf{k}j\varepsilon }^{s})^{s}\nabla _{\mu }(\psi _{%
\mathbf{k}j\varepsilon ^{\prime }}^{-s^{\prime }})^{s^{\prime }}d\mathbf{r}%
\int (\psi _{\mathbf{k}j^{\prime }\varepsilon ^{\prime
}}^{s^{\prime
}})^{s^{\prime }}\nabla _{\nu }(\psi _{\mathbf{k}j\varepsilon }^{-s})^{s}d%
\mathbf{r},  \notag  \label{OPTmat}
\end{eqnarray}%
where we have denoted $\psi _{\mathbf{k}j\varepsilon }^{+}=\psi
_{\mathbf{k} j\varepsilon }^{L}$ , $\psi _{\mathbf{k}j\varepsilon
}^{-}=\psi _{\mathbf{k} j\varepsilon }^{R}$ and assumed that
$(\psi _{\mathbf{k}j\varepsilon }^{s})^{+}\equiv \psi
_{\mathbf{k}j\varepsilon }^{s}$ while $(\psi
_{\mathbf{k}j\varepsilon }^{s})^{-}=\psi _{\mathbf{k}j\varepsilon
}^{s\ast }$. The expressions (\ref{OPTcon}), (\ref{OPTmat})
represent generalizations of the formulae for optical
conductivity for a strongly--correlated system, and involve the
extra internal frequency integral in Eq.~(\ref{OPTcon}).

Let us consider the non--interacting limit when
$\mathcal{\hat{M}}_{xc}(\omega )\rightarrow i\gamma \rightarrow
0.$ In this case, the eigenvalues $\epsilon
_{\mathbf{k}j\varepsilon }=\epsilon _{\mathbf{k} j}+i\gamma ,\psi
_{\mathbf{k}j\varepsilon }^{R}\equiv |\mathbf{k}j\rangle ,$ $\psi
_{\mathbf{k}j\varepsilon }^{L}\equiv |\mathbf{k}j\rangle ^{\ast
}\equiv \langle \mathbf{k}j|$ and the matrix elements
$M_{\mathbf{k} jj^{\prime }}^{ss^{\prime },\mu \nu}(\varepsilon
,\varepsilon ^{\prime })$ are all expressed via the standard
dipole transitions $|\langle \mathbf{k} j|\nabla
|\mathbf{k}j^{\prime }\rangle |^{2}.$ Working out the energy
denominators in the expression (\ref{OPTcon}) in the limit
$i\gamma \rightarrow 0$ and for $\omega \neq 0$ leads us to the
usual form for the conductivity which for its interband
contribution has the form
\begin{eqnarray}
\sigma _{\mu \nu }(\omega ) &=&\frac{\pi e^{2}}{\omega }\sum_{\mathbf{k}%
,j^{\prime }\neq j}\langle \mathbf{k}j|\nabla _{\mu
}|\mathbf{k}j^{\prime }\rangle \langle \mathbf{k}j^{\prime
}|\nabla _{\nu }|\mathbf{k}j\rangle
\times  \notag \\
&&[f(\epsilon _{\mathbf{k}j})-f(\epsilon _{\mathbf{k}j^{\prime
}})]\delta (\epsilon _{\mathbf{k}j}-\epsilon
_{\mathbf{k}j^{\prime }}+\omega ). \label{OPTfer}
\end{eqnarray}%

To evaluate the expression $\sigma _{\mu \nu }(\omega )$
numerically, one needs to pay special attention to the energy
denominator $1/(\omega +\epsilon _{\mathbf{k}j\varepsilon
^{-}}^{s}-\epsilon _{\mathbf{k}j^{\prime }\varepsilon
^{+}}^{s^{\prime }})$ in (\ref{OPTcon}). Due to its strong
$\mathbf{k}$-dependence the tetrahedron method of Lambin and
Vigneron \cite{Lambin:1984} should be used. On the other hand,
the difference in the square brackets of Eq.~(\ref{OPTcon}) is a
smooth function of $\mathbf{k}$ and one can evaluate it using
linear interpolation. This allows one to calculate the integral
over $\varepsilon $ by dividing the interval $-\omega
/2<\varepsilon <+\omega /2$ into discrete set of points
$\varepsilon _{i}$ and assuming that the eigenvalues $\epsilon
_{\mathbf{k}j\varepsilon }$ and eigenvectors $\psi
_{\mathbf{k}j\varepsilon }$ can to zeroth order be approximated
by their values at the middle between each pair of points i.e.
$\bar{\varepsilon}_{i}^{\pm }=\varepsilon _{i}\pm \omega
/2+(\varepsilon _{i+1}-\varepsilon _{i})/2$. In this way,\ the
integral is replaced by the discrete sum over internal grid
$\varepsilon _{i}$ defined for each frequency $\omega ,$ and the
Dyson equation needs to be solved twice for the energy
$\bar{\varepsilon}_{i}^{+}$ and for the energy
$\bar{\varepsilon}_{i}^{-}$ The described procedure produces fast
and accurate algorithm for evaluating the optical response
functions of a strongly--correlated material.

Similar developments can be applied to calculate the transport
properties such as $dc$-resistivity. The transport
parameters of the system are expressed in terms of so called
kinetic coefficient, denoted here by $A_{m}$.  The equation for
the electrical resistivity is given by
\begin{equation}
\label{eq:resistivity} \rho = \frac{\kb T}{e^{2}}\frac{1}{A_{0}},
\end{equation}
and the thermopower and the thermal conductivity are given by
\begin{equation}
  \label{eq:thermalparameters}
  S = \frac{-\kb}{|e|}\frac{A_{1}}{A_{0}},
  \qquad
  \kappa = \kb\left(A_{2}-\frac{A_{1}^{2}}{A_{0}}\right).
\end{equation}
Within the Kubo formalism \cite{Mahan:1993} the kinetic
coefficients are given in terms of equilibrium state
current--current correlation functions of the particle and the
heat current in the system. To evaluate these correlation
functions an expression for electric and heat currents are
needed. Once those currents are evaluated, transport with DMFT
reduces to the evaluation of the transport function
\begin{equation}
  \label{eq:transportfunction}
  \phi^{xx}(\epsilon) = \frac{1}{\Omega_c}\sum_{k}
  \trace
  \left\{
    \hat{v}^{x}_{k}(\epsilon)\hat{\rho}_{k}(\epsilon)
    \hat{v}^{x}_{k}(\epsilon)\hat{\rho}_{k}(\epsilon)
  \right\},
\end{equation}
and the transport coefficients
\begin{eqnarray}
  \label{eq:A_coefficients2}
  A_{m} =  \pi
  \int_{-\infty}^{\infty}d\epsilon\phi^{xx}(\epsilon)
  f(\epsilon)f(-\epsilon)(\beta\epsilon)^{m},
\end{eqnarray}
The described methodology has been applied to calculate the
optical conductivity~\cite{Oudovenko:2004:PRB}, the
thermopower~\cite{Palsson:1998},  the DC--resistivity, and the thermal
conductivity for LaTiO$_3$~\cite{Oudovenko:2004}.


\section{Techniques for solving the impurity model}
\label{sec:IMP}

In practice the solution of the dynamical mean--field (DMFT)
equations is more involved than the solution of the Kohn--Sham
equations, which now appear as static analogs. There are two
central elements in DMFT: the self--consistency condition and the
impurity problem (see Fig.~\ref{Fig:DMFT}). The first step is
trivial for model calculations but becomes time consuming when
realistic band structures are considered. Usually it is done
using the tetrahedron method (see, e.g., \cite{Anisimov:1997},
programs and algorithms for caring out this step are described at
$\web$).

The second step in the DMFT algorithm, i.e. the solution of the
impurity problem, is normally the most difficult task.
Fortunately, we can now rely on many years of experience to
devise reasonable approximations for carrying out this step. At
the present time, there is no universal impurity solver that
works efficiently and produces accurate solutions for the Green's
function in all
regimes of parameters. Instead what we have is a large number of
techniques, which are good in some regions of parameters. In many
cases when there are various methods that can be applied, there
is a conflict between accuracy and computational cost, and in many
instances one has to make a compromise between efficiency and
accuracy to carry out the exploration of new complex materials.
It should be noted that the impurity solver is one component of
the various algorithms discussed, and that for a given material
or series of materials, one should strive to use comparable
realism and accuracy in the various stages of the solution of a
specific problem.

For space limitations, we have not covered all the methods that
are available for studying impurity models, but we simply chose a
few illustrative methods which have been useful in the study of
correlated materials using DMFT. In this introductory section, we
give an overview of some of the methods, pointing out
the strengths and limitations of them and we expand on the
technical details in the following subsections.

There are two exactly soluble limits of the multiorbital Anderson
impurity model, for a general bath. The atomic limit when the
hybridization vanishes and the band limit when the interaction
matrix $U$ is zero. There are methods which are tied to expansions
around each of these limits. The perturbative expansion in the
interactions is described in section~\ref{sec:expandinU}. It is
straightforward to construct the perturbative expansion of the
self--energy in powers of $U$ up to second order, and resum
certain classes of diagrams such as ring diagrams and ladder
diagrams. This is an approach known as the fluctuation exchange
approximation (FLEX), and it is certainly reliable when $U$ is
less than the half--bandwidth, $D$. These impurity solvers are
very fast since they only involve matrix multiplications and
inversions. They also have good scaling, going as $N^{3}$ where
$N$ is  the number of orbitals or the cluster size.

The expansion around the atomic limit is more complicated. A
hybridization function with spectral weight at low frequencies is
a singular perturbation at zero temperature. Nevertheless
approaches based on expansion around the atomic limit are
suitable for describing materials where there is a gap in the
one--particle spectra, or when the temperature is sufficiently
high that one can neglect the Kondo effect. This includes Mott
insulating states at finite temperatures, and the incoherent
regime of many transition metal oxides and heavy fermion
systems.  Many approaches that go beyond the atomic limit exist:
direct perturbation theory in the hybridization, resummations
based on equation of motion methods, such as the Hubbard
approximations, resolvent methods, and slave particle techniques
such as the non--crossing approximation (NCA) and their
extensions. We describe them in sections~\ref{sec:expandinV} and
\ref{sec:IMPatm}.

There are methods, such as the quantum Monte Carlo method (QMC),
or functional integral methods which are not perturbative in
either $U$ or in the bandwidth, $W$. In the QMC method one
introduces a Hubbard--Stratonovich field and averages over this
field using Monte Carlo sampling. This is a controlled
approximation using a different expansion parameter, the size of
the mesh for the imaginary time discretization. Unfortunately, it
is computationally very expensive as the number of time slices and
the number of Hubbard--Stratonovich fields increases. The QMC
method is described in section \ref{sec:IMPqmc}. It also has a
poor scaling with the orbital degeneracy, since the number of
Hubbard--Stratonovich fields increases as the square of the
orbital degeneracy. Mean--field methods are based on a functional
integral representation of the partition function, and the
introduction of auxiliary slave bosons \cite{Barnes:1976,
Barnes:1977, Coleman:1984}. The saddle point approximation
\cite{KotliarR:1986,Rasul:1988} gives results which are very
similar to those of the Gutzwiller method, and corrections to the
saddle point can be carried out by a loop expansion
\cite{Li:1989}. Unfortunately the perturbative corrections to the
saddle point are complicated and have not been evaluated in many
cases. We review the mean--field theory in section
\ref{sec:IMPsbmf}.

Interpolative methods bear some resemblance to the analytic
parameterizations of $V_{xc}$ in LDA. One uses different
approximations to the self--energy of the impurity model, viewed
as a functional of $\Delta (i\omega )$, in different regions of
frequency. The idea is to construct interpolative formulae that
become exact in various limits, such as zero frequency where the
value of the Green's function is dictated by Luttinger theorem,
high frequencies where the limiting behavior is controlled by
some low--order moments, and in weak and strong coupling limits
where one can apply some form of perturbation theory. This
approach has been very successful in unraveling the Mott
transition problem in the context of model Hamiltonians, and it
is beginning to be used for more realistic studies. We review
some of these ideas in section \ref{sec:IMPipt}.

In this review, we have not covered techniques based on exact
diagonalization methods, and their improvements such as Wilson
renormalization group (RG) techniques and density matrix
renormalization group methods. These are very powerful
techniques, but due to the exponential growth of the Hilbert
space, they need to be tailored to the application at hand. In
the context of model Hamiltonians, it is worth mentioning that
the exact solution for the critical properties of the Mott
transition was obtained with the projective self--consistent
method \cite{Goetz:1995}, which is an adaptation of Wilson RG
ideas to the specifics of the DMFT study of the Mott transition.
This method sets up a Landau theory and justifies the use of
exact diagonalization for small systems to determine the critical
properties near the transition. Further simplifications of these
ideas, which in practice amounts to the exact diagonalization
methods with one site, have been used by Potthoff and coworkers
\cite{Potthoff:2001}. The flow equation method of Wilson and
Glazek and Wegner \cite{Glazek:1993,Glazek:1994,Wegner:1994} is
another adaptive technique for diagonalizing large systems, and
it has been applied to the impurity model. Clearly, the
renormalization group approach in the cluster DMFT context is
necessary to attack complex problems. Some ideas for combining
cellular DMFT with RG formalism were put forward in the context
of model Hamiltonians \cite{Bolech:2003}.

Finally, we point out that a much insight is gained when
numerical methods are combined with analytic studies.
As in the previous applications of DMFT to model Hamiltonians, fast
approximate techniques and algorithms are needed to make progress
in the exploration of complex problems, but they should be used
with care and tested with more exact methods.

\subsection{Perturbation expansion in Coulomb interaction}
\label{sec:expandinU}

The application of perturbation theory in the interaction $U$ has
a long history in many--body theory. For DMFT applications, we
consider here a general multiorbital Anderson impurity model
(AIM) given by
\begin{eqnarray}
H &=& \sum_{\a\b}\epsilon
_{\a\b}d_{\a}^{\dagger}d_{\b}+\frac{1}{2}
\sum_{\a\b\g\delta}U_{\a\b\g\delta}d_\a^{\dagger}d_\b^{\dagger}d_\g
d_\delta\\\nonumber &+& \sum_{k\a\b}(V_{k\a\b}^* d^{\dagger}_\a
c_{k\b}+H.c.)+\sum_{k\a}\epsilon_{k\alpha}c^{\dagger}_{k\a}c_{k\a},
\label{eq:atom}
\end{eqnarray}
where $U_{\a\b\g\delta}$ is the most general interaction matrix
and $\a$ combines spin and orbital index (or position of an atom
in the unit cell or cluster, in cellular DMFT applications).

The lowest order term is the Hartree--Fock formulae
\begin{equation}
  \Sigma_{12}^{(HF)} = \sum_{34}(U_{1342}-U_{1324})n_{43}.
\end{equation}
The second order term is given by
\begin{widetext}
\begin{equation}
\Sigma_{12}^{(2)}(i\omega)=\sum_{\{3-8\}} U_{1456}U_{7832}
\int\int\int d\epsilon d\epsilon^{\prime}d\epsilon^{\prime\prime}
\rho_{67}(\epsilon)\rho_{58}(\epsilon^{\prime})\rho_{34}(\epsilon^{\prime\prime})
{\frac{{f(\epsilon)f(\epsilon^\prime)f(-\epsilon^{\prime\prime})+
      f(-\epsilon)f(-\epsilon^\prime)f(\epsilon^{\prime\prime})}}
  {i\omega-\epsilon+\epsilon^{\prime\prime}-\epsilon^{\prime}}} ,
\label{eq:impurity2OSE}
\end{equation}
\end{widetext}
where $f$ is the Fermi function, ${\rho}_{12} $ is the spectral
function of the impurity Green's function.

\begin{figure}
  \includegraphics[width=0.99\linewidth,angle=0]{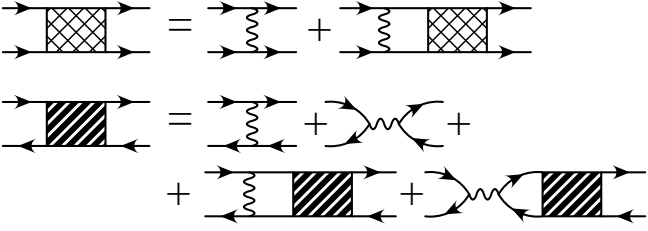}
\caption{
  Particle--particle (top row) and particle--hole (bottom row)
  $T$--matrices which appear in the FLEX approximation. Full lines
  correspond to electron propagators and wiggly lines stand for the
  bare interaction $U$.
  \label{fig:TM_FLEX}}
\end{figure}
\begin{figure}
  \includegraphics[width=0.99\linewidth,angle=0]{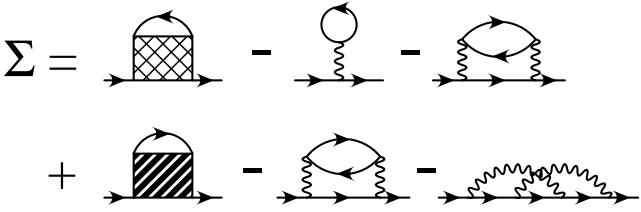}
\caption{
  Definition of the FLEX self--energy constructed with the help of
  particle--particle and particle--hole $T$--matrices. Note that
  some lower order terms appear many times and need to be subtracted
  to avoid double counting.
\label{fig:SE_FLEX}}
\end{figure}
Higher order terms in perturbation theory that can be easily
summed up are those in the form of a ladder or, equivalently,
$T$--matrix. There are two distinct types of ladder graphs, the
particle--particle type \cite{Galitskii:1958} shown in the top
row of Fig.~\ref{fig:TM_FLEX} and particle--hole $T$--matrix
depicted in the bottom row of Fig.~\ref{fig:TM_FLEX}. The
one--particle self--energy can then be constructed using those two
building blocks in the way shown in Fig.~\ref{fig:SE_FLEX}.
Although we did not plot the generating functional, which in the
general case is somewhat more involved, it can be constructed
order by order from the above definition of the self--energy.
Hence, the approximation is conserving if the propagators are
fully dressed, and therefore automatically obeys certain
microscopic conservation laws as well as the Friedel sum rule.
The method was first proposed by Bickers and Scalapino in the
context of lattice models \cite{Bickers:1989} under the name
fluctuation exchange approximation (FLEX).  It is the minimal set
of graphs describing the interaction of quasiparticles with
collective modes (pairs, spin and charge fluctuations).

Particle--particle and particle--hole $T$--matrices correspond
algebraically to
\begin{eqnarray}
  \hat{T}^{pp}(i\Omega) &=&
  (1-\hat{U}\hat{\chi}^{pp}_{i\Omega})^{-1}\hat{U}
  \hat{\chi}^{pp}_{i\Omega}\hat{U}\hat{\chi}^{pp}_{i\Omega}\hat{U},
  \label{eq:flex_Tpp}\\\nonumber
  \hat{T}^{ph}(i\Omega) &=&
 (1-(\hat{V}+\hat{W})\hat{\chi}^{ph}_{i\Omega})^{-1}(\hat{V}+\hat{W})\\
  &-&(\hat{V}+\hat{W})\hat{\chi}^{ph}_{i\Omega}\hat{V},
  \label{eq:flex_Tph}
\end{eqnarray}
where $V_{1234} = U_{1324}$,  $W_{1234} = -U_{1342}$ and
\begin{eqnarray}
  \chi_{1234}^{pp}(i\Omega) = -T \sum_{i\omega^\prime}
  G_{23}(i\omega^\prime)G_{14}(i\Omega-i\omega^\prime),\\
  \chi_{1234}^{ph}(i\Omega) = -T \sum_{i\omega^\prime}
  G_{23}(i\omega^\prime)G_{41}(i\Omega+i\omega^\prime).
\end{eqnarray}
We assumed here a product of the form $(\hat A \hat
B)_{1234} = \sum_{56}A_{1256} B_{5634}$. With these building
blocks one can construct the self--energy of the form
\begin{eqnarray}\nonumber
  \Sigma_{12}^{(FLEX)}(i\omega) = %
  T \sum_{i\Omega34}\left(
  T^{pp}_{1432}(i\Omega)G_{34}(i\Omega-i\omega) \right.\\%
  +   \left. T^{ph}_{1432}(i\Omega)G_{43}(i\Omega+i\omega)
  \right) .
\end{eqnarray}

The Feynman graphs in perturbation theory can be evaluated
self--consistently (namely in terms of fully dressed Green's
function, $G$, including only skeleton graphs) or
non--self--consistently (namely using $G_0$). In practice the
results start being different once $U$ is comparable to the half
bandwidth. The skeleton perturbation theory in $G$ sums more
graphs than the bare perturbation theory, but in many--body theory
more does not necessarily imply better. In the context of the
single--band AIM model, the perturbative approach in powers of the
Hartree--Fock Green's function $G_0$ was pioneered by Yamada and
Yoshida \cite{Yosida:1970, Yamada:1975a, Yamada:1975b,
Yamada:1975c}.  These ideas were crucial for the first
implementation of DMFT \cite{Georges:1992a} for the one--band
Hubbard model, where the expansion in $G_0$ proved to be
qualitatively and quantitatively superior to the expansion in
$G$. In the multiorbital case, the situation is far less clear as
discussed recently in Ref.~\onlinecite{Janis:2003}.

\begin{figure}
  \includegraphics[width=0.85\linewidth,angle=0]{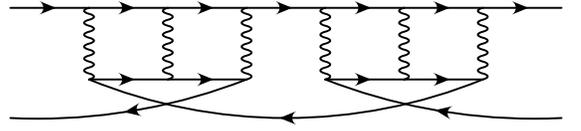}
\caption{
  Particle hole $T_{ph}$ ladder contribution with screened effective
  interaction $U_{eff}$ mediated by $T_{pp}$.
  \label{fig:renormalizedU}}
\end{figure}
Bulut and Scalapino \cite{Bulut:1993}, tested Kanamori's
\cite{Kanamori:1963} observation that the particle--hole bubbles
should interact not with the bare interaction matrix, $U$ in
Eq.~\ref{eq:flex_Tph}, but with an effective interaction screened
by the particle--particle ladder (see
Fig~\ref{fig:renormalizedU}). This can be approximated by
replacing $U$ by $U_{eff} = T_{pp}(\omega=0)$ in
Eq.~\ref{eq:flex_Tph}.  Notice that those diagrams are a subset
of the parquet graphs, recently implemented by Bickers
\cite{Bickers:1991}.  It is also worth noticing that the FLEX
approach is exact to order $U^3$.

Within the realistic DMFT, the FLEX method was implemented for
Iron and Nickel (in its non--self--consistent form) by Drchal et
al.~\cite{Drchal:1999} and Katsnelson et al.~\cite{Chioncel:2003}
(these authors used the expansion in $G_0$ where an additional
shift of the impurity level is implemented to satisfy Luttinger's
theorem following Ref.~\onlinecite{Kajueter:1996b}, and the
screened interaction in the particle--hole channel was assumed).

\begin{figure}[tbh]
\center{\epsfig{file=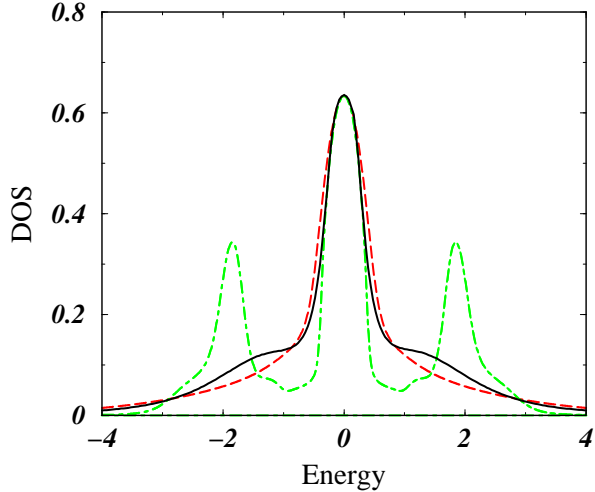,width=0.9\linewidth}}
\caption{
  Comparison of FLEX density of states (using $T_{pp}$
  graphs only) with QMC results (full line). Dashed line corresponds to FLEX
  approximation with fully dressed propagators and dashed--dotted line
  stands for the same approximation with undressed propagators. Calculation was
  performed for a two--band Hubbard model for semicircular density of
  states with $U=2D$ and $T=1/16 D$ \cite{Janis:2003}.
\label{fig:zflex}  }
\end{figure}
When the interaction is much less than the half--bandwidth the
perturbative corrections are small and all the approaches
(self--consistent, non--self--consistent, screened or unscreened)
are equivalent to the second order graph. However, when $U$
becomes comparable to the half bandwidth differences appear, and
we highlight some qualitative insights gained from a comparison
of the various methods~\cite{Janis:2003,Putz:1996}. The
perturbation theory in $G_0$ tends to overestimate $Z^{-1}$ and
overemphasize the weight of the satellites. On the other hand the
skeleton perturbation theory tends to underestimate the effects
of the correlations and suppress the satellites. This is clearly
seen in Fig.~\ref{fig:zflex} where perturbative results are
compared to QMC data, analytically continued to real axis with
the maximum entropy method (MEM)~\cite{Jarrell:1996}.

\begin{figure}[tbh]
\center{\epsfig{file=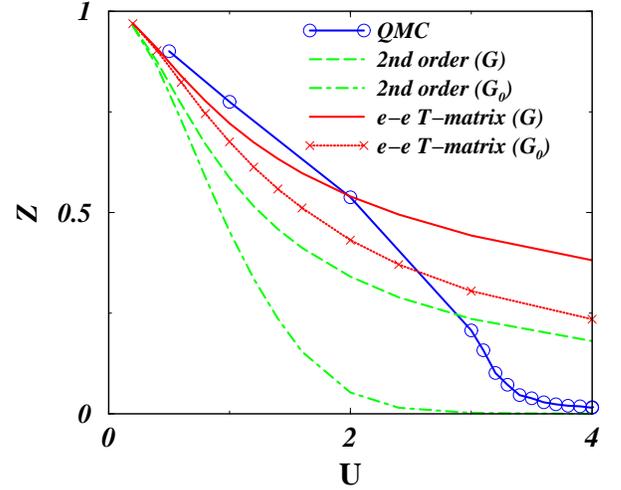,width=0.9\linewidth}}
\caption{
  Variation of quasiparticle residue with interaction strength of the
  interaction
  for two--band half--filled Hubbard model with a semicircular density
  of states of bandwidth $D$.
  Schemes presented are: second order perturbation with fully
  dressed propagators(dashed curve), second order with Hartree--Fock
  dressing of propagators (dot--dashed),
  FLEX with electron--electron $T$--matrix only but fully
  self--consistent (solid line), FLEX with electron--electron $T$--matrix only and
  Hartree--Fock dressing of propagators (line with stars) \cite{Janis:2003}.
\label{fig:Zflex2} }
\end{figure}
To gauge the region in which the approach is applicable, we
compare the quasiparticle weight from various perturbative
approaches to QMC data in Fig.~\ref{fig:Zflex2}.  All
self--consistent approaches miss the existence of the Mott
transition, while its presence or at least a clear hint of its
existence appears in the second order non--self--consistent
approach.

\begin{figure}[tbh]
\center{\epsfig{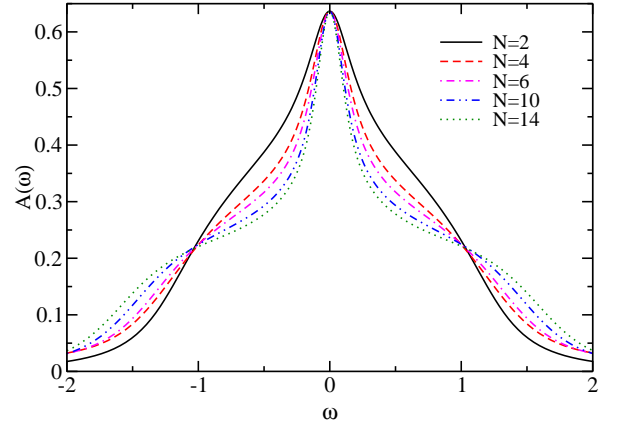}}
\caption{
  Variation of FLEX spectra with increasing number of bands
  for $U=D$ and $T=0.001$ on the Bethe lattice. The self--consistency is
  obtained by fully dressed propagators including all three FLEX
  channels \cite{Janis:2003}.
\label{fig:ANflex}}
\end{figure}
While FLEX performs reasonably well in the case of two- and
three--band models, it is important to stress that this can not
persist to very large degeneracy. With increasing number of
bands, the quasiparticle residue must increase due to enhancement
of screening effect and therefore $Z$ must grow and eventually
approach unity. This remarkable screening effect is not captured
by FLEX which displays the opposite trend as shown in
Fig.~\ref{fig:ANflex}.

It is worth noticing that for Ni the full $d$--bandwidth is
approximately 4.5~eV and $U$ is estimated to be around 3~eV so the
approach is near the boundary of its applicability.

\subsection{Perturbation expansion in the hybridization strength}
\label{sec:expandinV}

The perturbation expansion with respect to the hybridization
strength can be derived with the help of resolvent techniques or
by decoupling of Hubbard operators in terms of slave particles.
In the latter case, an auxiliary operator $a_n$ is assigned to
each state of the local Hilbert space, such that a slave particle
creates an atomic state out of the new vacuum
\begin{equation}
\left|n\right> = a_n^\dagger\left|vac\right>,
\end{equation}
where $\left|vac\right>$ is a new vacuum state. The Hubbard
operators are easily expressed in terms of this auxiliary
particles, $X_{n,m}= a_n^\dagger a_m$. The creation operator of
an electron is expressed by
\begin{equation}
d^{\dagger}_{\a} = \sum_{nm} (F^{\a \dagger})_{nm} a_n^{\dagger}
a_m, \label{eq:PH:breakup}
\end{equation}
where $F_{nm}^{\a}=\langle n|d_{\a}|m\rangle$ are matrix elements
of a destruction operator. In terms of pseudo particles, the
general Anderson impurity model reads
\begin{eqnarray}
H=\sum_{mn} E_{mn} a_n^\dagger a_m + \sum_{k\g}
\varepsilon_{k\g}c_{k\g}^\dagger c_{k\g} + \nonumber\\
\sum_{k,mn,\a\b}\left(V^*_{k\a\b} (F^{\a\dagger})_{nm} a_n^\dagger
a_{m} c_{k\b}+H.c.\right),%
\label{eq:H_AIM}
\end{eqnarray}
where $c_{k\g}^{\dagger}$ creates an electron in the bath and
$\gamma$ stands for the spin and band index.

In order that electrons are faithfully represented by the
auxiliary particles, i.e.
$\{d_{\a},d_{\b}^\dagger\}=\delta_{\a\b}$, the auxiliary particle
$a_n$ must be boson (fermion), if the state $\left|n\right>$
contains even (odd) number of electrons, and the constraint:
\begin{equation}
Q \equiv \sum_n a_n^{\dagger}a_n = 1 ,
\end{equation}
must be imposed at all times. This condition merely expresses the
completeness relation for the local states $\sum_n
\left|n\right>\left<n\right|=1$. The constraint is imposed by
adding a Lagrange multiplier $\lambda Q$ to the Hamiltonian and
the limit $\lambda\rightarrow\infty$ is carried out.

The physical local Green's function (electron Green's function in
$Q=1$ subspace) and other observables are calculated with the
help of the Abrikosov trick \cite{Abrikosov:1965} which states
that the average of any local operator that vanishes in the $Q=0$
subspace is proportional to the grand--canonical (all $Q$ values
allowed) average of the same operator
\begin{equation}
\langle A \rangle_{Q=1} =
\lim_{\lambda\rightarrow\infty}{\frac{\langle A\rangle_G
}{\langle Q\rangle_G}}.
\end{equation}

The advantage of pseudo--particle  approach is that standard
diagrammatic perturbation theory techniques such as Wicks theorem
can be applied. The limit $\lambda\rightarrow\infty$ is to be
taken after the analytic continuation to the real frequency axes
is performed. Taking this limit, actually leads to a substantial
simplification of the analytic continuation \cite{Haule:2001}.

A different approach, is to ``soften" the constraint $Q=1$, and
replace it by
\begin{equation}
\sum_n a^{\dagger}_n a_n = q_0 N .
\end{equation}
The original problem corresponds to taking $q_0 = 1/ N $ but one
can obtain a saddle point by keeping $q_0$ of order of one while
allowing $Q$ to be large. This approach, was studied in
Ref.~\cite{Parcollet:1997,Parcollet:1998}. While the standard NCA
approach suffers from exceeding the unitary limit leading to
causality problems, the soft NCA's do not suffer from that
problem. Other sub--unitary impurity solvers were developed
recently based on slave rotor methods
\cite{Florens:2002b,Florens:2004} and on a decoupling schemes
\cite{Costi:1986, Jeschke:2003}.

The perturbation expansion in the hybridization strength can be easily
carried out in the pseudo--particle representation. The desired
quantity of the expansion is the local Green's function which is
proportional bath electron $T$--matrix, therefore we have
\begin{equation}
G_{loc} = \lim_{\lambda\rightarrow\infty}{\frac{1}{V^2 \langle Q
\rangle_G}} \Sigma_c, \label{eq:locG}
\end{equation}
where $\Sigma_c$ is the bath electron self--energy calculated in
the grand--canonical ensemble. The latter quantity has a simple
diagrammatic interpretation.

\begin{figure}[tbp]
\includegraphics[angle=-90,width=0.9\linewidth]{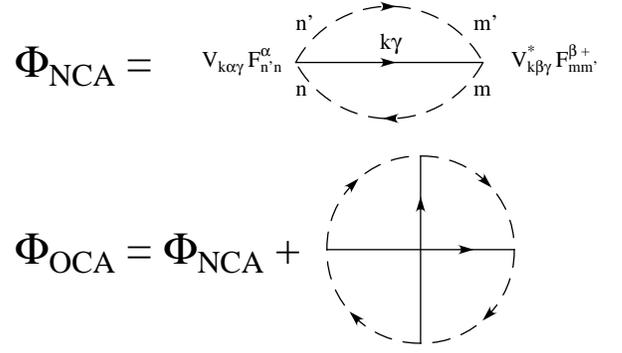}
\caption{ Diagrammatic representation of the  non--crossing
  approximation (NCA) and one--crossing approximation (OCA) functional
  for the Anderson impurity model.\label{fig:LuttWard_NCA}}
\end{figure}

The selection of diagrams, is best illustrated using the
Baym--Kadanoff functional $\Phi$.  The building blocks of $\Phi$
are dressed Green's functions of pseudo--particles $G_{mn}$
(depicted as dashed lines), and bath electrons $G_{\a\b}$ (solid
lines). Due to the exact projection, only pseudo--particles are
fully dressed while bath electron Green's functions are
non--dressed because the bath self--energy vanishes as
$exp(-\beta\lambda)$ with $\lambda\rightarrow\infty$. The bare
vertex $V_{k\a\b}$, when combined with the conduction electron
propagator, can be expressed in terms of the bath spectral
function $A_{\a\b}(\omega)=-{1\over 2\pi
i}[\Delta_{\a\b}(\omega+i0^+)-\Delta_{\a\b}(\omega-i0^+)]$.
Because propagators are fully dressed, only skeleton diagrams
need to be considered in the expansion.

The lowest order contribution, depicted in the first line of
Fig.~\ref{fig:LuttWard_NCA} is known in literature as
non--crossing approximation (NCA). Pseudo--particle self--energies
$\Sigma_{mn}$, defined through $(G^{-1})_{mn} =
(\omega-\lambda)\delta_{mn}-E_{mn}-\Sigma_{mn}$, are obtained by
taking the functional derivative of $\Phi$ with respect to the
corresponding Green's function, i.e.
$\Sigma_{mn}=\delta\Phi/\delta G_{nm}$. After analytic
continuation and exact projection, the self--energies obey the
following coupled equations
\begin{widetext}
\begin{eqnarray}
  \Sigma_{mn}(\omega+i0^+) &=&
  \sum_{\a\b,m^{\prime}n^{\prime}}\left[ F^{\b}_{mm^{\prime}}(F^{\a\dagger})_{n^\prime n}\int d\xi
  f(\xi) A_{\a\b}(\xi)G_{m^\prime n^\prime}(\omega+\xi+i 0^+)\right.
  \nonumber\\
  &&\left.\qquad+
  (F^{\b\dagger})_{mm^{\prime}}F^{\a}_{n^\prime n}\int d\xi
  f(-\xi) A_{\b\a}(\xi)G_{m^\prime n^\prime}(\omega-\xi+i
  0^+)\right].\ \ \ \ \
\end{eqnarray}
The local electron Green's function, obtained by functional
derivative of $\Phi$ with respect to the bath Green's function,
becomes
\begin{eqnarray}
  G_{\a\b}(\omega+i0^+)=\sum_{mnm^\prime n^\prime}F^{\a}_{n^\prime
    n}(F^{\b\dagger})_{m m^{\prime}} {1\over Q}\int d\xi\exp(-\beta\xi)\left[
    \rho_{m^\prime n^\prime}(\xi)G_{nm}(\xi+\omega+i0^+)-G_{m^\prime
  n^\prime}(\xi-\omega-i0^+)\rho_{nm}(\xi)\right], \ \ \ \ \ \
\end{eqnarray}
\end{widetext}
where $A_{\a\b}$ is the bath spectral function defined above,
$Q=\int d\xi\exp(-\beta\xi)\sum_m \rho_{mm}(\xi)$ is the grand
canonical expectation value of charge $Q$ and $\rho_{mn}=-{1\over
2\pi i}[G_{mn}(\omega+i0^+)-G_{mn}(\omega-i0^+)]$ is the
pseudo--particle density of states. Note that equations are
invariant with respect to shift of frequency in the pseudo
particle quantities due to local gauge symmetry, therefore
$\lambda$ that appears in the definition of the pseudo Green's
functions can be an arbitrary number. In numerical evaluation,
one can use this to our advantage and choose zero frequency at
the point where the pseudo particle spectral functions diverge.

The NCA has many virtues, it is very simple, it captures the
atomic limit, it contains the Kondo energy as a non--perturbative
scale and describes the incipient formation of the Kondo
resonance. However, it has several pathologies that can be
examined analytically by considering the pseudo--particle
threshold exponents  at zero temperature \cite{MHartman:1984}.
Within NCA, the infrared exponents are independent of doping and
follow the exact non--Fermi liquid exponents in the multichannel
Kondo problem \cite{Cox:1993}. From the Friedel sum rule,
however, it follows that fully screened local moment leads to a
doping--dependent Fermi liquid exponents that differ substantially
from the NCA exponents \cite{Costi:1996}. When calculating the local
spectral function within NCA, this leads to a spurious peak at
zero frequency with the Abrikosov--Suhl resonance exceeding the
unitary limit. At the temperature at which the Kondo resonance
exceeds the unitary limit, the approximation breaks down when
combined with DMFT self--consistency conditions, because it
causes the spectral function to become negative hence violating
causality.

\begin{figure}[tbp]
\includegraphics[angle=-90,width=0.9\linewidth]{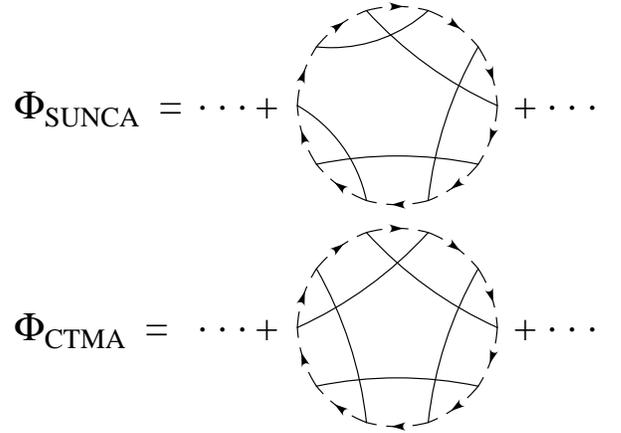}
\caption{ Diagrammatic representation of the two--crossing
  approximation (TCA) generating functional. It consists of all
  skeleton diagrams (infinite number) where conduction lines cross at
  most twice.  Conserving $T$--matrix approximation (CTMA) is a subset of
  diagrams where all conduction lines cross exactly twice and have
  either clockwise or counter--clockwise direction. Symmetrized U--NCA
  (SUNCA) is a subset of TCA where a conduction line exists that
  crosses only once. Conduction lines can have either clockwise or
  counter--clockwise direction.  \label{fig:LutWard_SUNCA}}
\end{figure}

At finite $U$, the NCA has a new problem, namely, it severely
underestimates the width of the Kondo resonance and hence the
Kondo temperature.  This problem is partly corrected by the vertex
corrections shown in the second row of
Fig.~\ref{fig:LuttWard_NCA}, which was introduced by Pruschke and
Grewe and is called one--crossing approximation (OCA). This
diagram is a natural generalization of NCA, namely, including all
diagrams with a single line crossing. It is also the lowest order
self--consistent approximation exact up to $V^2$. Although the
pseudo particle self--energies within NCA are calculated up to
$V^2$, the local Green's function is not. Only the conduction
electron self--energy is exact up to $V^2$ and from
Eq.~(\ref{eq:locG}) it follows that the physical spectral
function is not calculated to this order within NCA.

Several attempts were made to circumvent the shortcomings of NCA
by summing up certain types of diagrams with ultimate goal to
recover the correct infrared exponents and satisfy the unitarity
limit given by the Friedel sum rule. It can also be analytically
shown, that an infinite resummation of skeleton diagrams is
necessary to change the infrared exponents from their NCA values.
The natural choice is to consider the ladder type of scatterings
between the pseudo--particle and the bath electrons which leads to
crossings of conduction lines in the Luttinger--Ward functional.
In the infinite $U$ limit, only diagrams where all conduction
electrons cross at least twice, are possible. Note that due to
the projection, any contribution to the Luttinger--Ward functional
consists of a single ring of pseudo--particles since at any moment
in time there must be exactly one pseudo--particle in the system.
The diagrams where all conduction electrons cross exactly twice
is called CTMA \cite{Kroha:1997} and has not yet been implemented
in the context of the DMFT. A typical contribution to the CTMA
Luttinger Ward functional is shown in the second line of
Fig.~\ref{fig:LutWard_SUNCA}. In the impurity context, this
approximation recovers correct Fermi liquid infrared exponents in
the whole doping range and it is believed to restore Fermi liquid
behavior at low temperature and low frequency.

At finite $U$, however, skeleton diagrams with less crossings
exist. Namely, a ladder ring where conduction lines cross exactly
twice can be closed such that two conduction lines cross only
ones. This approximation, depicted in the first row of
Fig.~\ref{fig:LutWard_SUNCA}, is called SUNCA. It has been shown
in the context of single impurity calculation \cite{Haule:2001},
that this approximation further improves the Kondo scale bringing
it to the Bethe ansatz value and also restores Fermi liquid
exponents in the strict Kondo regime. Unfortunately, this is not
enough to restore Friedel sum rule in the local spectral function
at low temperatures and to apply the approach to this regime, it
has to be combined with other approaches such as renormalized
perturbation theory or ideas in the spirit of the interpolative
methods discussed in \ref{sec:IMPipt}.

The results of the computationally  less expensive SUNCA
calculation agree very well with the considerably more time
consuming QMC calculation. The agreement is especially accurate
around the most interesting region of the Mott transition while the
half--integer filling shows some discrepancy due to the
restriction to a small number of valences in the SUNCA
calculation.

\begin{figure}[tbh]
\includegraphics[angle=0,width=0.9\linewidth]{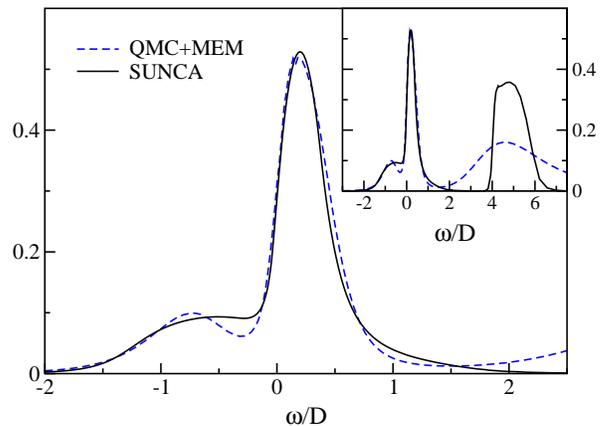}
\caption{
  Comparison between SUNCA (full line) and QMC (dashed line) density
  of states for the three--band Hubbard model on the Bethe lattice for
  $n_d=0.8$ at $U=5D$ and $T=D/16$.
 \label{fig:DOS_SUNCA}}
\end{figure}
We illustrate the agreement between these methods by means of the
real axis data of QMC+MEM and SUNCA where the latter results are
obtained on the real axis. As shown in Fig.~\ref{fig:DOS_SUNCA},
both calculations produce almost identical quasiparticle peak,
while some discrepancy can be observed in the shape of Hubbard
bands. We believe that this is mostly due to analytic
continuation of QMC data which do not contain much high--frequency
information. Notice that the width of the upper Hubbard band is
correctly obtained within SUNCA while QMC results show
redistribution of the weight in much broader region. Namely, in
the large $U$ limit, i.e. when a band is separated from the
quasiparticle peak, its width has to approach the width of the
non--interacting density of states, in this case $2D$.

\begin{figure}[tbh]
\includegraphics[angle=0,width=0.99\linewidth]{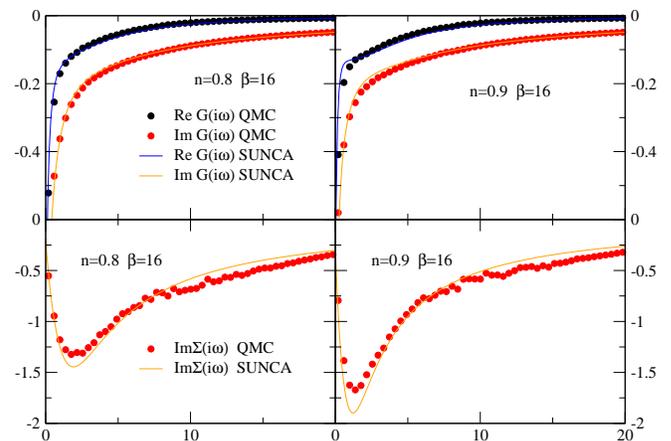}
\caption{
  Imaginary axis QMC data (dots) and SUNCA results (full--lines) are
  compared for three--band Hubbard model on Bethe lattice for
  $\beta=16$ and $U=5D$. Left panel shows results for doping levels
  $n_d=0.8$ right panel corresponds to doping $n_d=0.9$.
 \label{fig:IAD_SUNCA}}
\end{figure}
Finally, we compare the imaginary axis data in
Fig.~\ref{fig:IAD_SUNCA} for two doping levels of $n_d=0.8$ and
$n_d=0.9$. Notice that the results are practically identical with
discrepancy smaller that the error of the QMC data.

The NCA was used in the DMFT context to study Cerium
\cite{Zolfl:2001}, ${\rm La}_{1-x}{\rm Sr}_x{\rm TiO}_3$
\cite{Zolfl:2000b} and ${\rm Ca}_{2-x}{\rm Sr}_x{\rm RuO}_4$
\cite{Anisimov:2002}. In the case of Cerium, it does capture the
most essential differences of the alpha and gamma phases, and
compares very favorably to the quantum Monte Carlo results
\cite{Held:2001} and to experiments.  Using NCA in the context of
the SrRuO$_3$, \onlinecite{Anisimov:2002} was the first to
predict the so-called orbitally selective Mott transition, the
phase where one--band is in a Mott insulting state while the rest
are metallic. These results are discussed further within
(see section \ref{PDMIT}). The SUNCA approach was first tested in
the LaSrTiO$_3$ where it is in a good agreement with QMC results
~\cite{Oudovenko:2004}.

\subsection{Approaching the atomic limit: decoupling scheme, Hubbard I
and lowest order perturbation theory}
\label{sec:IMPatm}

The AIM Hamiltonian can also be expressed in terms of Hubbard
operators $X_{n m}$ by replacing $a_n^\dagger a_m$ in
Eq.~(\ref{eq:H_AIM}) with $X_{n m}$:
\begin{eqnarray}
H=\sum_{m} E_{m} X_{m m} + \sum_{k\g}
\varepsilon_{k\g}c_{k\g}^\dagger c_{k\g} + \nonumber\\
\sum_{k,mn,\a\b}\left(V^*_{k\a\b} (F^{\a\dagger})_{nm} X_{n m}
c_{k\b}+H.c.\right). \label{eq:H_atm}
\end{eqnarray}
For convenience, we choose here the local base to be the atomic
eigenbase, i.e., the atomic Hamiltonian is diagonal.

The atomic Green's function can be most simply deduced from the
Lehmann representation of the Green's function
\begin{equation}
  G^{(at)}_{\a\b}(i\omega) = {1\over Z}\sum_{nm} {F^{\a}_{nm}
  (F^{\b\dagger})_{mn} \left(e^{-\beta E_n}+e^{-\beta E_m}\right)\over
  i\omega -E_m+E_n},
  \label{eq:atmgf}
\end{equation}
where $F$'s are given by $F^{\a}_{nm}=\langle n|d_\a|m\rangle$ as
in section \ref{sec:expandinV} and $Z=\sum_n e^{-\beta E_n}$ is
the partition function.  The atomic Green's function has a
discrete number of poles, at energies corresponding to the atomic
excitations, weighted with the appropriate factors $e^{-\beta
E_n}/Z$ that can be interpreted as probabilities to find an atom
in the atomic configuration $|n\rangle$.

One can compute corrections to the Green's function
Eq.~(\ref{eq:atmgf}) by expanding around the atomic limit, using
the technique of cumulants \cite{Metzner:1991}. However there are
many resummations of these expansions, and no extensive test of
this problem has been carried out. Various methods start from the
equations of motion for the Green's functions of the Hubbard
operators, for the Green's functions of the conduction electrons,
and for the mixed Green's functions of the conduction electrons
and the Hubbard operators.

Once the Green's functions for the Hubbard operators
\begin{equation}
\Gh_{n_1 n_2 n_3 n_4}(\tau) = -\langle T_\tau X_{n_1 n_2}(\tau)
X_{n_3 n_4}(0)\rangle,
\end{equation}
are determined, the local Green's function $G_{\a\b}$ can be
deduced by the following linear combination of $\Gh$'s:
\begin{equation}
G_{\a\b}(i\omega) = \sum_{n_1 n_2 n_3 n_4} F^\a_{n_1 n_2}
\Gh_{n_1 n_2 n_3 n_4}(i\omega) (F^{\b\dagger})_{n_3 n_4} .
\end{equation}

In the decoupling method of L. Roth~\cite{Roth:1969}, one replaces
commutator $[H_{hyb}, X]$ by a linear combination of the
operators $c$ and $X$, namely
\begin{eqnarray}
[H_{hyb}, X_{n_1 n_2}]&=& \sum_{n_3 n_4} A_{n_1 n_2 n_3 n_4}X_{n_3 n_4}\\\nonumber %
& +& \sum_{k\a}B^{k\a}_{n_1 n_2}c_{k\a} + \sum_{k\a}C^{k\a}_{n_1
n_2}c_{k\a}^\dagger,%
 \label{eq:decouple}
\end{eqnarray}
and the coefficients $A$ $B$ and $C$ are determined by projecting
onto the basis set of $X$ and $c$, by means of a scalar product
defined by the anticommutator. This leads to a set of closed
equations for the coefficients $A$ $B$ and $C$. The Green's
function for the Hubbard operators can then be deduced from the
following matrix equation
\begin{equation}
 \Gh^{-1} = {\Gh^{(at)}}^{-1} - \widetilde{\Delta} - Y .
 \label{eq:atm_me}
\end{equation}
where the effective hybridization function $\widetilde{\Delta}$
and atomic Green's function for the Hubbard operators
$\Gh^{(at)}$ are
\begin{eqnarray}
&&  \widetilde{\Delta}_{n_1 n_2 n_3 n_4} = \sum_{\a\b}
  (F^{\a\dagger})_{n_1 n_2}\Delta_{\a \b}F^{\b}_{n_3 n_4},\\
&&  \Gh^{(at)}_{n_1 n_2 n_3 n_4}(i\omega) = {\delta_{n_1 n_4}
   \rho_{n_3 n_2} + \delta_{n_2 n_3}\rho_{n_1
   n_4} \over {(i\omega+E_{n_1}-E_{n_2})}},\ \ \ \
 \end{eqnarray}
and $\rho$ is the ``density matrix". The equations close once the
density matrix $\rho$ is computed from an equation such as
$\rho_{n_1 n_2} =  -{1\over\beta}\sum_{i\omega} e^{-i\omega
0^+}\Gh_{n_1 n^\prime n^\prime n_2}(i\omega)$ or $\rho_{n_1 n_2}
=  {1\over\beta}\sum_{i\omega} e^{i\omega
  0^+}\Gh_{n^\prime n_1 n_2 n^\prime}(i\omega)$ or any
combination. The result is not unique.

Finally, the matrix $Y$, which
is proportional to the coefficient $A$ introduced in
Eq.~(\ref{eq:decouple}), becomes
\begin{widetext}
\begin{equation}
  Y_{n_1 n_2 n_3 n_4} = {1\over(\rho_{n_1
  n_1}+\rho_{n_2 n_2})(\rho_{n_3
  n_3}+\rho_{n_4 n_4})}\sum_m (Z_{m n_2 m n_3}\delta_{n_1
  n_4}-Z_{n_1 m n_4 m}\delta_{n_2 n_3}),
\end{equation}
with
\begin{equation}
  Z_{n_1 n_2 n_3 n_4} = -T \sum_{n_5 n_6}\sum_{i\omega}
  \left(\widetilde{\Delta}_{n_1 n_2 n_5 n_6}(i\omega) \Gh_{n_5 n_5 n_3  n_4}%
  (i\omega)+\Gh_{n_3 n_4 n_5 n_6}(i\omega)\widetilde{\Delta}_{n_5 n_6 n_1 n_2}(i\omega)\right).
\end{equation}
\end{widetext}

The equation for the Green's function (\ref{eq:atm_me}) is non--linear
because of the coefficients $\overline{X}$ and $Y$, and has to be
solved iteratively.
Neglecting $Y$ results in the famous Hubbard I approximation
\begin{equation}
  G^{-1} = {G^{(at)}}^{-1} - \Delta.
\end{equation}

Perhaps, the best approximate method for the system in the
Mott--insulating state is the straightforward perturbation
expansion in hybridization strength to the lowest order. By
expanding the $S$ matrix
\begin{equation}
  \exp{\left(-\sum_{\a\b}\int_0^\beta\int_0^\beta d\tau_1 d\tau_2
  \; d_\a(\tau_1)\Delta_{\a\b}(\tau_1,\tau_2)d_{\b}(\tau_2)\right)},
\end{equation}
to the lowest order, one immediately obtains the following correction to the
Green's function
\begin{widetext}
\begin{eqnarray}
  && G_{\a\b}(\iom)-G^{(at)}_{\a\b}(\iom)=G_{\a\b}^{(at)}(\iom)\sum_{\g,\delta}\sum_{\iom'}
  G^{(at)}_{\delta \g}(\iom')\Delta_{\g \delta}(i\omega')\nonumber\\
  &+&
  \int_0^\beta d\tau\int_0^\beta d\tau_1\int_0^\beta d\tau_2 e^{i\omega\tau}\sum_{\g,\delta}
  \langle T_\tau d_\a(\tau)d_{\b}^\dagger(0)d_{\g}^\dagger(\tau_1)d_{\delta}(\tau_2)\rangle_0
  \Delta_{\g \delta}(\tau_1-\tau_2).
  \label{Eq:SOPTV}
\end{eqnarray}
\end{widetext}

It is straightforward to evaluate the two--particle Green's
function for the atom in Eq.~(\ref{Eq:SOPTV}). One can insert the
identity $|m\rangle\langle m|$ between any pair of creation and
destruction operators and then integrate over time the resulting
exponential factors. The resulting six terms, due to six
different time orderings of the product, can also be drawn by
Feynman diagrams and evaluated by the straightforward
non--self--consistent expansion along the lines of section
\ref{sec:expandinV}. To make the method exact also in the band
$U=0$ limit, we calculate the lowest order correction to the
self--energy rather than to the Green's function. The correction
is
\begin{eqnarray}
  \Sigma = {G^{(at)}}^{-1}(G-G^{(at)}){G^{(at)}}^{-1}-\Delta .
\end{eqnarray}
This self--energy is exact up to second order in hybridization $V$
and also in the non--interacting $U=0$ case. It also gives correct
width of the Hubbard bands which is underestimated by a factor of
$2$ by the Hubbard I approximation in the large $U$ limit.
This method has recently been tested for the
Hubbard model where it performed very satisfactorily whenever the system
has a finite gap in the one electron spectrum \cite{Dai:2004}.

Many other approaches were recently used to develop
impurity solvers. The local moment method
\cite{Vidhyadhiraja:2004,Vidhyadhiraja:2005,Logan:2005} has been
successfully applied to the periodic Anderson model.  It would be
interesting to extend it to a full multiorbital case.
Also decoupling technique, or mode coupling technique, the
factorization technique, the alloy analogy, the modified alloy
analogy, and the methods of moments were used.  These approaches can
be applied directly to the lattice and simplified using the DMFT
locality ansatz, or applied directly to the AIM. For a recent review
with a DMFT perspective see \cite{Shvaika:2000}.

\subsection{Quantum Monte Carlo: Hirsch--Fye method}
\label{sec:IMPqmc}


\def\hc{{\it h.c.}}
\def\dst{\hat \rho}
\def\dstBare{ \rho}

\def\todo#1{{\bf #1}}
\def\GroupeEquations#1{\begin{subequations}  #1  \end{subequations}}
\def\moy#1{\left\langle #1 \right\rangle}
\def\dmoy#1{\left\langle \! \moy{#1} \! \right\rangle}
\def\dkmoy#1{\left\langle \! \moy{#1} \! \right\rangle_{{\cal K}}}
\def\kmoy#1{\moy{#1}_{{\cal K}}}
\def\bigmoy#1{\bigl\langle #1 \bigr\rangle}
\def\Bigmoy#1{\Bigl\langle #1 \Bigr\rangle}
\def\Im{\hbox{Im}}
\def\Tr{\text{Tr}}

\def\sgn{\text{sgn}\,}
\def\parent#1{\left(#1\right)}
\def\fig#1#2{\includegraphics[height=#1]{#2}}
\def\figx#1#2{\includegraphics[width=#1]{#2}}

\def\ket#1{\left\vert #1 \right\rangle}
\def\bra#1{\left\langle #1 \right\vert}
\def\matel#1#2#3{\langle #1 \vert #2 \vert #3 \rangle}
\def\inprod#1#2{\langle #1 \vert #2 \rangle}
\def\parent#1{\left(#1\right)}
\newcommand{\empile}[2]{\genfrac{}{}{0pt}{}{#1}{#2}}
\def\BigOinv#1#2{O \parent{\frac{1}{#1^{#2}}}}
\def\matrice#1{{\begin{pmatrix}#1\end{pmatrix}}}

\def\bsig{\mbox{\boldmath{$\sigma$}}}
\def\cc{{\it c.c.}}

\def\anticommutator#1#2{\left\{ #1,#2 \right\}}
\def\commutator#1#2{\left[ #1,#2 \right]}

\def\htilde{\widetilde{h}}
\def\SigmaSkel{{ \Sigma \kern -5.5pt \raise 1pt \hbox{/}}}
\def\Mdot{M_{\text{dot}}}

\def\bk{{\bf k}}
\def\bq{{\bf q}}
\def\bsig{\mbox{\boldmath{$\sigma$}}}

\def\Spdemi{S+\frac{1}{2}}
\def\Bigparent#1{\Bigl (#1\Bigr)}
\def\biggparent#1{\biggl (#1\biggr)}
\def\biggbra#1{\biggl [#1\biggr]}
\def\Sm{{\cal  S}_{m}}
\def\order#1{\BigOinv{\Sm}{#1}}

\def\var#1{{\tt #1}}

\def \QMC{{\Delta\tau}}
\def\caln{{\cal  N}}
\def\calnsig{{\caln_\sigma}}

The general idea underlying the Hirsch--Fye determinantal QMC
method is to discretize the path integrals representing the
partition function and the Green's function of an interacting
problem. These discretized path integrals are then converted,
using a Hubbard--Stratonovich transformation, into a statistical
average over a set of non--interacting Green's functions in a
time--dependent field, which can be either continuous or discrete.
The sum over the auxiliary fields is done using Monte Carlo
sampling methods. The QMC algorithm for the solution of the
Anderson impurity model was introduced  in ~\cite{Hirsch:1986,Fye:1989},
and generalized to the multiorbital case in
\cite{Takegahara:1993} and \cite{Bonca:1993}.
Applications to the solution of the lattice models via DMFT was
introduced in \cite{Jarrell:1992}, see also \cite{Rozenberg:1992,
Georges:1992b, Jarrell:1993:book} in the single--orbital case. In
the multiorbital context it was implemented  in
\cite{Rozenberg:1997}, see also \cite{Held:1998}. Some DMFT
applications, such as the study of the electron--phonon
interactions within single--site DMFT, require a QMC
implementation using continuous Hubbard--Stratonovich
fields~\cite{Jarrell:1993:book}. This is also essential for the
implementation of Extended Dynamical Mean--Field Theory (EDMFT)
(see for example \cite{Motome:2000}, \cite{Pankov:2002} and for
the combination of EDMFT and GW method ~\cite{Sun:2002}). An
alternative algorithm for EDMFT using discrete spins was
introduced in \cite{Grempel:1998}. Zero temperature QMC
algorithms, which are closely related to determinantal
algorithms, have been extensively developed for lattice models.
DMFT applications have been recently introduced in
\cite{Feldbacher:2004}. There are alternative methods of
evaluating the partition function and the correlation functions,
which are amenable to Monte Carlo methods \cite{Rombouts:1999}.
These are free of discretization errors and have been introduced
in
\cite{Rubtsov:2003:CM0302228,Rubtsov:2004:CM0411344,Savkin:2005}
in the DMFT context.

QMC has been used extensively in DMFT calculations. Due to space
limitations, only a few illustrative examples shall be described.
The QMC method has been applied to the study of Cerium by McMahan
and collaborators ~\cite {McMahan:2003, Held:2001} and to Iron
and Nickel by Lichtenstein and collaborators
\cite{Lichtenstein:2000, Lichtenstein:2001}. DMFT with QMC as an
impurity solver, has been applied to many other $d$--electron
systems. These include perovskites with a $d^1$ configuration
such as LaTiO$_3$ ~\cite{ Nekrasov:2000} SrVO3 and CaVO3 \cite
{Sekiyama:2004, Pavarini:2004:PRL}, ruthenates  such as RuSrO$_4$
\cite{Liebsch:2000, Liebsch:2003:CM0301536} and vanadates such
as  V$_2$O$_3$ ~\cite{Held:2001V2O3} and VO$_2$
\cite{Biermann:2004:CM:VO2}.

Since a detailed review~\cite{Georges:1996} for the
single--orbital case is already available, we focus here on the
generalization of the QMC method for multiorbital or cluster
problems (for the impurity solver, ``cluster DMFT''  is a
particular case of multiorbital DMFT where the cluster index
plays the role of an orbital). The emphasis of this section is on
generality; for a more pedagogical introduction to the method in
a simple case, see \cite{Georges:1996}. This section is organized
as follows. First in \ref{sec:QMC:quantumimpurity}, we present
the general impurity problem to be solved by QMC. In
\ref{sec:QMC:algo} we present the Hirsch--Fye algorithm, where we
discuss the time discretization, derive the discrete Dyson
equation, and present the algorithm. In \ref{sec:QMC:dens}, we
present in more detail the case of density--density interactions,
which has been the most widely used. Details of the derivations
are provided in Appendix \ref{sec:QMC:Appendix} for completeness.

\subsubsection{ A generic quantum impurity problem}
\label{sec:QMC:quantumimpurity}
\paragraph{Definitions}

We will focus on the solution of a generic quantum impurity
problem like (\ref{eq:SDFimp}) defined  by the following action
\GroupeEquations{
\begin{align}\label{eq:qmc_S}
S_{\text{eff}} =& - \iint_{0}^{\beta } d \tau  d \tau'
\sum_{\empile{1 \leq \mu,\nu \leq \caln}{1 \leq \sigma \leq \calnsig}}
d^{\dagger }_{\mu \sigma } (\tau )
\mathcal{G}_{0\sigma \mu \nu }^{-1} (\tau ,\tau')
d^{ }_{\nu \sigma } (\tau' )
\nonumber \\
& +  \int_{0}^{\beta } d \tau   H_{\text{int}} (\tau )
\\
H_{\text{int}} \equiv&
U_{\mu_1 \mu_2 \mu_3 \mu_4}^{\sigma_1 \sigma_2 \sigma_3 \sigma_4}
d^{\dagger }_{\mu_1\sigma_1 } d_{\mu_2\sigma_2 } d^{\dagger }_{\mu_3\sigma_3  } d_{\mu_4\sigma_4  }
\end{align}
}
where $\mathcal{G}_{0\sigma\mu\nu}^{-1}$ is the Weiss function,
$1\leq \sigma  \leq \caln_{\sigma }$ are indices in which the
Green's functions are diagonal (conserved quantum numbers),
$1\leq \mu,\nu \leq \caln$ are indices in which the Green's
functions are not diagonal, and repeated indices are summed over.
The value of $\caln$ and
$\caln_{\sigma }$ depends on the problem (see
\ref{eq:QMC:examples}). 
$H_{int}$ is the interaction part of the action.
It should be noted that we have defined a
completely general $H_{int}$, which is necessary to capture
multiplets which occur in real materials.
The purpose of the impurity solver is to compute the
Green's function
\begin{equation}
  \label{eq:QMC:defGreenFunction}
  G_{\sigma\mu \nu }( \tau ) = \moy{T d_{\mu \sigma}(\tau) d_{\nu \sigma}^{\dagger }(0)}_{S_{\text{eff}}}
\end{equation}
and higher order correlation functions. {\sl In this section, we
will use a different convention for the sign of the Green's
function than in the rest of this review}: in accordance with
\cite{Georges:1996} and the QMC literature, we define the Green's
functions without the minus sign. 

\paragraph{Generalized Hubbard--Stratonovich decoupling}
Hirsch--Fye QMC can only solve impurity problems where the
interaction have a decoupling formula of the following form
\begin{subequations}
  \label{eq:qmc:DefDecouplingFormula}
\begin{align}
H_{\text{int}}&= H_{1}+\dots +H_{n}
\label{eq:qmc:decomposeH}
\\
e^{- \Delta \tau  H_{i} }
 &= \sum_{S_{i}\in {\cal S}_{i}} w_{i} (S_{i})
\exp \biggl (
\sum_{\sigma\mu\nu}
d^{\dagger }_{\mu \sigma }
V_{\mu \nu }^{i\sigma } (S_{i})
d_{\nu \sigma }
\biggr)
\end{align}
\end{subequations}
where $S_{i}$ is a index (referred to in the following as a
``QMC--spin'') in a set ${\cal S}_{i}$ (discrete or continuous),
$w_{i} (S_{i}) >0$ is a positive weight, $V=V^{\dagger}$, and
$H_i=H_i^{\dagger}$. Approximate decouplings,  where
(\ref{eq:qmc:DefDecouplingFormula}) holds only up to $O (\Delta
\tau^{m}), m\geq 3$, are discussed below (see also
\cite{Gunnarsson:1997:PLA}).
Equation (\ref{eq:qmc:decomposeH}) is a generalized form of the
familiar Hubbard--Stratonovich transformation. Multiple
Hubbard--Stratonovich fields per time slice allow the decoupling
of more general interactions, as exemplified below.

\paragraph{Examples}
\label{eq:QMC:examples}

Let us consider first a multiorbital or cluster DMFT solution of
the Hubbard model in the normal phase (non--superconducting). In
this case, $\caln$ is the number of impurity sites or orbitals,
and $\calnsig = 2$: $\sigma$ is the spin index which is
conserved. The interaction term  $H_{int}$ is given by
 \begin{equation}\label{eq:qmc_Hint_diag}
 H_{int}=\sum_{(\mu,\sigma) < (\nu, \sigma') }
U_{\mu\nu}^{\sigma\sigma'}
n_{\mu}^\sigma n_{\nu}^{\sigma'} 
\end{equation}
where we use the lexicographic order: $(\mu,\sigma) <
(\nu,\sigma')$ if $\mu<\nu$ or $\mu=\nu$ and $\sigma < \sigma'$.
In this case, the decoupling formula uses the discrete
Hubbard--Stratonovich transformation using Ising spins introduced
by Hirsch \cite{Hirsch:1983} (see also ~\cite{Takegahara:1993}
and \cite{Bonca:1993})
\begin{align}
\label{eq:QMC:expHdiag}
\nonumber
e^{-\Delta \tau H_{\text{int}}}&=
\frac{1}{2} \sum_{\{S_{\mu\nu}^{\sigma\sigma'}=\pm 1 \} }
\\
\exp \biggl[ \sum_{(\mu,\sigma) < (\nu, \sigma') } \!\!\! \!\!\!
&
\,\,\left(  
\lambda_{\mu\nu}^{\sigma\sigma'} S_{\mu\nu}^{\sigma\sigma'}
 ( n_{\mu}^\sigma - n_{\nu}^{\sigma'})
- \Delta \tau \frac{U_{\mu\nu}^{\sigma\sigma'}}{2}(
n_{\mu}^\sigma + n_{\nu}^{\sigma'}) 
\right)  
\biggr]
\\
\lambda_{\mu\nu}^{\sigma\sigma'} &\equiv\mathrm{arccosh} \left(
  \exp
\biggl( \frac{\Delta \tau }{2}U_{\mu\nu}^{\sigma\sigma'} \biggr)
\right)
 \end{align}
The weight $w(S)=\frac{1}{2}$ is independent of the
auxiliary Ising fields $S_{\mu\nu}^{\sigma\sigma'} $ defined for each  $U$ term.
The matrix $V$ of Eq. (\ref{eq:qmc:DefDecouplingFormula}) is
diagonal and reads
\begin{align}
  \label{eq:QMC:V}
\nonumber
V_{\mu\nu}^\sigma (\{  S\})=&
\delta_{\mu\nu}
  \sum_{\empile{\rho,\sigma'}{(\mu,\sigma) <(\rho,\sigma')}}
\!\!\!
\!\!\!
( 
\lambda_{\mu\rho}^{\sigma\sigma'} S_{\mu\rho}^{\sigma \sigma'}
- \frac{\Delta \tau}{2} U_{\mu\rho}^{\sigma\sigma'} ) 
-\\ &
\delta_{\mu\nu}
  \sum_{\empile{\rho,\sigma'}{(\mu,\sigma) > (\rho,\sigma')}}
\!\!\!
\!\!\!
( 
\lambda_{\rho\mu}^{\sigma'\sigma} S_{\rho\mu}^{\sigma' \sigma}
+\frac{\Delta \tau}{2} U_{\rho\mu}^{\sigma'\sigma} ) 
\end{align}
A general goal of the Hirsch--Fye algorithm is to minimize the
number of decoupling fields, to reduce the size of the
configuration space where the Monte Carlo is done (see below). In
this respect, decoupling each $U$ term in the interaction with a
different field is not optimal, especially when there is a
symmetry between orbitals. However, there is currently no
efficient solution to this problem.

A second example is the study of a superconducting phase. We
restrict our discussion to the Hubbard model for simplicity, but
the generalization to more realistic models is straightforward.
For the study of superconductivity in a two--band model see
\cite{Georges:1993}. For the QMC calculation of $d$--wave
superconductivity in a cluster see \cite{Lichtenstein:2000:R,
MaierJarrel2000}. We restrict ourselves to a case where a
possible antiferromagnetic order and the superconducting order
are collinear.  In this case, we introduce the Nambu spinor
notation at each site $i$: $ \psi^{\dagger} \equiv
(d_{\uparrow}^{\dagger},d_{\downarrow})$ so that the Green's
function is
\begin{equation}\label{eq:QMC:GreenNambu}
G (\tau) \equiv  \moy{T_\tau \psi (\tau) \psi^{\dagger}(0)} =
\begin{pmatrix} G_{\uparrow} (\tau)& F (\tau)\\
F^{*} (\tau) & G_{\downarrow} (\beta -\tau)\end{pmatrix}
\end{equation}
where $F$ is the anomalous Green's function $ F (\tau) =
\moy{T_\tau d_{\uparrow} (\tau) d_{\downarrow} (0)} $. We denote
by $+$ and $-$ the Nambu indices, and $ n_{+} = n_{\uparrow} $
and $ n_{-} = 1 - n_{\downarrow} $. In this case, we take
$\caln_{\sigma } = 1$ (spin is not conserved) and $\caln= 2
\caln_\text{normal case}$, twice the number of sites. The index
$\mu$ is a double index $(i,\pm)$, where $i$ is a site index.
For simplicity, we take a local $U$ interaction, which is then
decoupled as (for each cluster site)
\begin{equation*}
\exp
\biggl(
-\Delta \tau U
 n_{i\uparrow} n_{i\downarrow}
\biggr)
\propto \sum_{S_i =\pm
\lambda }  \frac{e^{- S_i }}{2}
 e^{ S_i (n_{i+} + n_{i-}) - \frac{\Delta\tau U}{2} (n_{i+} - n_{i-})}
\end{equation*}
with $\lambda = \mathrm{arccosh}\bigl(  \exp (
{\Delta \tau }U/2 \bigr )$  
(we drop a constant since it cancels in the
algorithm).
%

A third example is a further generalization of the Hirsch--Fye
formula to decouple the square of some operator. For example, if
$M$ has a spectrum contained in $\{0,\pm 1,\pm 2 \}$, one can use
\begin{subequations}
  \label{eq:3spin}
  \begin{align}
e^{\alpha M^2} &= \sum_{\sigma = 0,\pm S} w_\sigma e^{\sigma M} \\
S &= \cosh^{-1} \parent{\frac{e^{3\alpha} + e^{2 \alpha} + e^\alpha -1 }{2}}\\
w_S &= w_{-S} = \frac{e^\alpha -1 }{e^{3\alpha} + e^{2 \alpha} + e^\alpha - 3}\\
w_0 &= 1- 2w_S
 \end{align}
\end{subequations}
This can be used to decouple a nearest neighbor density--density
interaction in a Hubbard model (as an alternative to the more
commonly used method that splits this interaction  into four terms
using $n = n_{\uparrow} + n_{\downarrow}$ and the Hirsch--Fye
formula).

In the case of a quantum impurity problem formulated in a general
non--orthogonal basis, we can orthogonalize within the impurity
degree of freedom (however, there may still be an overlap between
different unit cells), in order to reduce the problem to the case
where the $c$ basis is orthogonal. The $V$ matrix transforms as
$V' = (P^{ \dagger})^{-1} V  P^{-1}$, where $P$ is the matrix
that transform into the orthogonal basis. Note however that  in
general  a diagonal $V$ will transform into a non--diagonal $V'$,
which will make the QMC more costly.

\subsubsection{Hirsch--Fye algorithm}
\label{sec:QMC:algo}
\paragraph{Time discretization}
\label{sec:QMC:time}

We start by  writing a Hamiltonian form $H=H_{0} + H_{int}$ of
the action using an effective generalized Anderson model with
$n_{s}$ bath sites
\begin{align}\label{Def H0}
H_{0}
=&
\sum_{p=1}^{n_{s}} \sum_{\mu \nu\sigma } \epsilon^{0}_{p\sigma\mu\nu
} a^{\dagger}_{p\mu \sigma}a_{p\nu  \sigma }
+ \sum_{p\mu\nu \sigma } V^{0}_{p\mu \nu \sigma }
(a^{\dagger}_{p\nu  \sigma}
d_{\mu \sigma } +  h.c.)
\nonumber\\
& + \epsilon_{\mu \nu \sigma }  d^{\dagger
}_{\mu \sigma }d_{\nu \sigma }
\nonumber\\
\end{align}
${\cal K}^{\sigma }_{p \mu, p'\nu }$ is defined as follows
\begin{align}
H_0^{\sigma} \equiv& \sum_{p \mu p'\nu} a^{\dagger }_{p\mu \sigma }
{\cal K}^{\sigma }_{p \mu, p'\nu }
a_{p'\nu \sigma }
\end{align}
where $H_0 \equiv \sum_{\sigma}H_0^{\sigma}$, and
$a_{p\mu\sigma}$ is the annihilation operator of the electron on
the bath site for $p>0$ and for $p=0$ we identify $ d=a_{p=0}$
which corresponds to the impurity site.
In the Hirsch--Fye algorithm, the imaginary time is discretized
with $L$ discrete times $\tau_l = (l-1)\beta/L$, with $1\leq l
\leq L$. Using the Trotter formula, we approximate the partition
function by $ Z \approx Z^{\Delta\tau}$ with
\begin{widetext}
\begin{align}
  \label{eq:QMC:trotter}
Z^{\Delta\tau}
\equiv&
\Tr \prod_{l=1}^L
\left(
  \exp^{-\Delta \tau H_0} \prod_{i=1}^n
\exp^{-\Delta \tau H_i}
\right)
\\
Z^{\Delta\tau}
=&  \sum_{\{S_{i}^{l} \}}
\left(
\prod_{l=1}^{L}\prod_{i=1}^{n} w_{i} (S_{i}^{l})
 \right)
\prod_\sigma
\Tr \prod_{l=1}^{L}
\Biggl[
e^{-\Delta \tau H_{0} }
 \prod_{i=1}^{n}
\exp \biggl (
\sum_{\mu\nu}
{d^{\dagger }_{\mu \sigma } V_{\mu \nu }^{i\sigma } (S_{i}^{l}) d_{\nu \sigma }  }
 \biggr)
\Biggr]
\end{align}
\end{widetext}
where $S_{i}^{l}$ are $L$ copies of the decoupling QMC--spins. The
Green's function defined in (\ref{eq:QMC:defGreenFunction}) at
time $\tau_l$
\begin{subequations}
  \label{eq:QMC:defDiscreteG}
\begin{align}
G_{\sigma;\mu\nu} (\tau_{l_{1}},\tau_{l_{2}} ) &= \frac{1}{Z}
\Tr \left( U^{L-l_{1}} d_{\mu\sigma} U^{l_{1} - l_{2}}
d_{\nu\sigma}^{\dagger} U^{l_{2}} \right)
\\
U &\equiv e^{-\Delta \tau H}
\end{align}
\end{subequations}
for $1\leq \mu,\nu \leq \caln$  and $l_1 \geq l_2$, is replaced
by its discretized version
\begin{subequations}
  \label{eq:QMC:G_qmc_comp}
\begin{gather}
  \label{eq:QMC:G_qmc_comp1}
{G}_{\sigma;(l_1,\mu),(l_2,\nu)}^{\QMC}  \equiv \frac{1}{Z^\QMC} \Tr \left(
\widetilde{U^{}}^{L-l_{1}+1} d_{\mu\sigma} \widetilde{U^{}}^{l_{1} - l_{2}}
d_{\nu\sigma}^{\dagger} \widetilde{U^{}}^{l_{2}-1} \right)
\\
\nonumber
\widetilde{U^{}}
\equiv  \prod_{0\leq i\leq  n} \sum_{S_{i}\in {\cal S}_{i}} w_{i} (S_{i})
\exp \biggl ({d^{\dagger }_{\rho \sigma } V_{\rho \lambda }^{i\sigma }
(S_{i}) d_{\lambda \sigma }  } \biggr) \\
\widetilde{U^{}}
= \prod_{0\leq i\leq  n} e^{-
\Delta \tau H_{i}} + O (\Delta \tau^{m})
\end{gather}
\end{subequations}

Since the Trotter formula is an approximation controlled by
$\Delta \tau$, one may use {\it approximate} decoupling formulas,
up to order $O (\Delta \tau^{m} )$ ($m\geq 3$), that would not
introduce a priori a bigger error than the Trotter formula itself.
For density--density interactions exact formula are available (as
described above) but for a more general interactions this may not
be the case. A priori, $\widetilde{U^{}} = U + O (\Delta
\tau^{2})$ thus $G^\QMC =G + O (\Delta \tau)$ (since $L\times O
(\Delta \tau^{2}) = O (\Delta \tau) $). However, given that $H$
is hermitian, we see that $G$ is hermitian
$G_{\sigma;\mu\nu}(\tau_{l_1},\tau_{l_2}) =
\bigl(G_{\sigma;\nu\mu}(\tau_{l_1},\tau_{l_2})\bigr)^{*}$. Using
$\widetilde{U} = U\Bigl( 1 - (\Delta \tau)^{2}/2 \sum_{i<j}
[H_{i},H_{j}] + O (\Delta \tau^{3})\Bigr )$, the fact that the
commutator is anti--hermitian when all the $H_{i}$ are hermitian,
and $U^{\dagger}=U$, we get $Z^{\Delta\tau} = Z + O (\Delta
\tau^{2})$ and the  stronger result
\begin{gather}\label{eq:QMC:symm_sep}
G_{\sigma;\mu\nu}(\tau_{l_1},\tau_{l_2})
=
\frac{G^\QMC_{\sigma;(l_1,\mu),(l_2,\nu)} + \Bigl(G^{\QMC }_{\sigma;(l_1,\nu),(l_2,\mu)}\Bigr)^* }{2} + O (\Delta \tau^{2})
\end{gather}
Eq. (\ref{eq:QMC:symm_sep}) shows that {\it i)} we gain one order
in $\Delta\tau$ with symmetrization, {\it ii)} various hermitian
$H_i$ can be decoupled separately  to the same order, and {\it
iii)} we only need a decoupling formula that is correct  up to
order ($\Delta \tau ^{2}$) included.

\paragraph{The Dyson equation}
\label{sec:QMC:Dyson}

Let us introduce a matrix of size ${\cal  N} n_{s}$ defined by
\begin{equation}\label{DefBigV}
{\cal  V}^{i\sigma} (S)_{|p\mu,p'\nu } \equiv  \delta_{p
p'}\delta_{p 0} V_{\mu \nu }^{i\sigma } (S)
\end{equation}
and the notation  $\{S \} \equiv \{S_{i}^{l}, 1 \leq i \leq n;
1\leq l \leq L \}$ for a configuration of the QMC--spin,
 we have immediately from (\ref{eq:QMC:trotter})
\begin{align}\label{}
Z^{\Delta \tau} =&   \sum_{\{S \}}
\left(\prod_{l=1}^{L}\prod_{i=1}^{n} w_{i} (S_{i}^{l})
 \right) Z[\{S \}] \\
Z[\{S \}] \equiv&
\prod_{\sigma }
\Biggl[
\Tr \prod_{l=L}^{1}
\Biggl(
\exp \biggl (
-\Delta \tau
a^{\dagger }_{p\mu \sigma }
{\cal K}^{\sigma }_{p \mu, p'\nu }
a_{p'\nu \sigma }
\biggr)
\times\nonumber
\\
&\prod_{i=1}^{n}
\exp \biggl (
{a^{\dagger }_{p\mu \sigma }
{\cal V}^{i\sigma } (S_{i}^{l})_{|p\mu,p'\nu } a_{p'\nu \sigma }  }
\biggr)
 \Biggr)
\Biggr]
\end{align}
\def \ak {A^{(k)}}
Introducing, ${\cal  N}n_{s}\times {\cal  N}n_{s}$ matrices
 $B_{l} (S)$ defined by
\begin{equation}
  B_{l} (S) \equiv
\exp \biggl (
-\Delta \tau
{\cal K}^{\sigma }
\biggr)
\prod_{i=1}^{n}
\exp \biggl (
{\cal V}^{i\sigma } (S)
\biggr)
\end{equation}
we can rewrite the partition function as (see
Appendix~\ref{sec:QMC:Appendix}):
\begin{subequations}
\label{eq:QMC:ZasdetO}
  \begin{align}
Z[\{S \}] &= \prod_{\sigma }\det {\cal O_{\sigma }} (\{S \})\\
{\cal O}_{\sigma } (\{S \}) &\equiv
\begin{pmatrix}
1      
 &\vdots  &   0   &   B_{L}^\sigma (S_{i}^{L})
\\
-B_{1}^\sigma (S_{i}^{1}) 
    & \vdots   &  \dots&   0
\\
0     
 & \vdots     & \dots     &   \dots
\\
\dots  
 &\vdots  & 1 & 0
\\
\dots  
 &\vdots  & -B_{L-1}^\sigma (S_{i}^{L-1}) & 1
\end{pmatrix}
\end{align}
\end{subequations}
Note that ${\cal O_{\sigma }}$ has size $L{\cal  N}n_{s}$.
Moreover, the Green's function for a fixed QMC--spins
configuration, defined as in Eq. (\ref{eq:QMC:G_qmc_comp1}) can
be shown to be (see \ref{sec:QMC:Appendix})
\begin{equation}
\label{eq:QMC:gisOinverse}
g_{{\{S \}}}^\sigma = {\cal O}^{-1}_\sigma (\{S \})
\end{equation}
The formula for the partition function can be generalized to the
average of any operator $M$
\begin{equation}\label{eq:QMC:AverageOperator}
\moy{M} = \frac{\sum_{\{ S  \}}
\moy{M}_{\{S \}}
\left(\prod_{i,l} w_{i} (S_{i}^{l})
 \right)\prod_{\sigma }\det {\cal O_{\sigma }} (\{S \})
}{\sum_{\{S \}}
\left(\prod_{i,l} w_{i} (S_{i}^{l})
 \right)\prod_{\sigma }\det {\cal O_{\sigma }} (\{S \})
}
\end{equation}
where $\moy{M}_{\{S \}}$ is the average of the operator at fixed
configuration ${\{S \}}$. In particular, the Green's function is
given by averaging $g_{{\{S \}}}^\sigma $. Moreover, for a fixed
QMC--spins configuration, the action is Gaussian, allowing to
compute any correlation functions with Wick theorem.
As noted in \cite{Hirsch:1986} (see also \cite{Georges:1996}, one
can derive a simple Dyson relation between the Green's functions
of two configurations $g_{S}$ and $g_{S'}$
\begin{multline}\label{eq:QMC:DysonRelation1}
g_{\{ S'\}}^\sigma = g_{\{S \}}^\sigma + (g_{\{S \}}^\sigma-1)
\times
\\
\biggl (
\prod_{i=n}^1
 e^{-\widetilde{{\cal V}}^{i\sigma } (\{S \}) }
\prod_{i=1}^n
e^{ \widetilde{{\cal V}}^{i\sigma } (\{S' \})}
-1
\biggr)
g_{\{S' \}}^\sigma
\end{multline}
with
the notation
\begin{equation}\label{DefTildaNotation1}
\widetilde{{\cal V}}^{i\sigma }_{p\mu l,p'\nu l'} (\{S \}) = \delta_{l,l'}
{\cal V}^{i\sigma }_{p\mu ,p'\nu} (S^{l}_{i})
\end{equation}
(see Appendix \ref{sec:QMC:Appendix}).
Since $\widetilde{{\cal V}}$ acts non--trivially only on the $p=0$
subspace, we can project (\ref{eq:QMC:DysonRelation1}) on it and
get rid of the auxiliary variables. We obtain finally the Dyson
equation for the Green's function $G$ (defined as in
(\ref{eq:QMC:G_qmc_comp}) and considered here as a matrix of size
$L\caln$)
\begin{multline}\label{eq:QMC:DysonRelation}
G^{\{S' \}}_{\sigma } = G^{\{S \}}_{\sigma } + (G^{\{S \}}_{\sigma }-1)
\times\\
\biggl (
\prod_{i=n}^1
e^{-\widetilde{V}^{i\sigma } (\{S \}) }
\prod_{i=1}^n
e^{ \widetilde{V}^{i\sigma } (\{S' \})}
-1
\biggr)
G^{\{S' \}}_{\sigma }
\end{multline}
where
\begin{equation}\label{eq:QMC:DefTildaNotation}
\widetilde{V}^{i\sigma }_{\mu l,\nu l'} (\{S \}) = \delta_{l,l'}
{V}^{i\sigma }_{\mu ,\nu} (S^{l}_{i})
\end{equation}
We used the fact that $G$ and $\widetilde{V}$ are diagonal in the
$\sigma$ index. Eq. (\ref{eq:QMC:DysonRelation}) is a equation
for matrices of size ${\cal N}L$.

We note that (\ref{eq:QMC:DysonRelation}) also holds for a
special case $V (\{S \}) =0$, with $G^{\{S \}}={\cal G}_0$. This
gives a simple way to compute $G_{\{S \}}$ from ${\cal G}_0$,
which requires the inversion of a $L{\cal N}\times L{\cal N}$
matrix $A_{\sigma}$.
\begin{subequations}
\label{eq:QMC:FullUpdate}
\begin{align}
G^{{\{S \}}}_{\sigma } &= A^{-1}_{\sigma } {\cal G}_{0\sigma } \\
A_{\sigma } &\equiv  1 +
\bigl (1- {\cal G}_{0\sigma }\bigr )
\Bigl (
\prod_{i=1}^n
e^{ \widetilde{V}^{i\sigma } (\{S \})}
-1
\Bigr)
\end{align}
\end{subequations}
Eq. (\ref{eq:QMC:FullUpdate}) is often referred to as the ``full update formula''.

Moreover, there is an important simplified relation between two
close configurations   $\{S \}$ and $\{S' \}$ which differ only
for one QMC--spin $S_{i}^{l}$, which allows a faster update of the
Green's function in the algorithm
\begin{align}\label{eq:QMC:FastUpdate1}
G^{{\{S' \}}}_{\sigma } &= A^{-1}_{\sigma } G^{\{S \}}_{\sigma } \\
\nonumber
A_{\sigma } &\equiv  1 +
\bigl (1- G^{\{S \}}_{\sigma }\bigr )
\Bigl (
\prod_{j=n}^i
e^{-\widetilde{V}^{j\sigma } (\{S \}) }
\prod_{j=i}^n
e^{\widetilde{V}^{j\sigma } (\{S' \})}
-1
\Bigr)
\end{align}
can be reduced to
\begin{subequations}
\label{eq:QMC:FastUpdateGeneral}
\begin{align}
p &\equiv
\prod_\sigma
\frac{\det {\cal O_{\sigma }} (S')}{\det {\cal
O_{\sigma }} (S)} = \prod_\sigma\det A_{\sigma } = \prod_\sigma \det A_{ll}^\sigma\\
A_{ll}^\sigma &\equiv   1 +
\bigl (1- G^{\{ S \} }_{\sigma;ll }\bigr )
C_{ll}^\sigma \\
C_{ll}^\sigma &\equiv
\prod_{j=n}^{i}
e^{-V^{j\sigma } (S_{j}^{l}) }
\prod_{j=i}^{n}
e^{ {V}^{j\sigma } (S_{j}^{'l})}
-1
\\
G^{{\{S' \}}}_{\sigma;l_{1} l_{2} } &= G^{{\{S \}}}_{\sigma;l_{1} l_{2} }
 +  \bigl (G^{\{S \}}_{\sigma;l_{1} l } - \delta_{l_1l} \bigr )
C_{ll}^\sigma (A_{ll}^\sigma)^{-1}
G^{{\{S \}}}_{\sigma;l l_{2} }
\end{align}
\end{subequations}
Eq. (\ref{eq:QMC:FastUpdateGeneral}) is often referred to as the
``fast update formula'' \cite{Hirsch:1983}. It is a formula for
matrices of size $\caln$ (compared to $L\caln$ for the full
update). It does not involve a big matrix inversion, therefore it
allows a faster calculation of $G$ than
(\ref{eq:QMC:FullUpdate}). For density--density interactions, the
fast update formula can be further simplified (with no matrix
inversion, see below). These equations are the generalizations of
Eqs.~(130)\footnote{Eq.~(130) in (\onlinecite{Georges:1996}) has
a misprint and should be read as $G'_{l_{1}l_{2}} =
G_{l_{1}l_{2}}+(G-1)_{l_1
l}(e^{V'-V}-1)_{ll}(A_{ll})^{-1}G_{ll_2}$. } and (131) of
(\onlinecite{Georges:1996}).

\paragraph{The Hirsch--Fye algorithm}

In principle, the sum (\ref{eq:QMC:AverageOperator}) could be
done by exact enumeration \cite{Georges:1993:PRB},
\cite{Georges:1996} but in practice one can reach much lower
temperatures by using statistical Monte Carlo sampling. It
consists of the generation of a sample of QMC--spins configuration
$\{S\}$ with probability $\prod_{i,l} w_i(S_i^l)\prod_\sigma \det
\mathcal{O}_\sigma( \{S\})$. If the determinant is not positive,
one needs to take the absolute value of the determinant to define
the probabilities and sample the sign. After computing $G_S$ from
${\cal G}_0$ with the ``full update" formula
(\ref{eq:QMC:FullUpdate}), a Markov chain is constructed by
making local moves, one time slice at a time, selecting a new
value for one QMC--spin and using the ``fast update'' formula
(\ref{eq:QMC:FastUpdateGeneral}) to compute the Green's function
for the new QMC--spin configuration. It may also be convenient to
perform global moves that involve the simultaneous flipping of
many spins in one move (e.g. simultaneous flipping of all the
spins in all the time slices). This can be accomplished directly
using (\ref{eq:QMC:FullUpdate}) or by generating the global move
as a sequence of local moves with
(\ref{eq:QMC:FastUpdateGeneral}) (one has then to keep and restore
the Green's  function to the original configuration in the case
that the proposed global  move is rejected). An interesting
generalization of this global move was  proposed by Grempel and
Rozenberg ~\cite{Grempel:1998}, in which one updates different
Fourier components of the fields. As noted above, the computation
allows, in practice, the computation of any higher order
correlation function since the theory is Gaussian for a fixed
QMC--spin configuration.

It should be noted that for some cases of cluster or multiorbital
problems, this QMC algorithm suffers severely from the sign
problem at low temperature, particularly in the case of
frustrated systems \cite{Parcollet:2004}. In the single--site DMFT
case, this problem is absent: this had been known empirically for
a while and rigorously proved recently \cite{Yoo:cm0412771}.

\paragraph{Remarks on the time discretization}

There is three difficulties coming from the
discretization of the time in the Hirsch-Fye algorithm:

{\it i)} 
One has to take a large enough number of time slices $L$, or in
practice to check that the results are unchanged when $L$ is
increased, which is costly since the computation time increase
approximately like $L^3$.

{\it ii)} 
Since the number of time slices is limited, especially for
multi--orbital or cluster calculation, the evaluation of the
Fourier transform of the Green's function (Matsubara frequencies)
is delicate. In practice, the time Green's function is
constructed from the discrete function resulting from the QMC
calculation using splines, whose Fourier transform can be
computed analytically (see \cite{Georges:1996}). It turns out
however that for this technique to be precise, one needs to
supplement the discrete Green's function by the value of its
derivatives at $\tau=0,\beta$, which can be reduced to a linear
combination of two--particle correlation functions computed by
the QMC calculation \cite{Oudovenko:2002}. Failure to deal with
this problem accurately can lead in some calculations to huge
errors, which can manifest themselves by spurious causality
violations.

{\it iii)} 
When a computation is made far from the particle-hole symmetric
case, the Weiss function ${\cal G}_0$ can be very steep close to
$\tau=0$ or $\tau=\beta$. As a result, it is not well sampled by
the regular mesh time discretization, leading to potentially
large numerical error. A simple practical solution is to replace
${\cal G}_0$ by $\bar{\cal G}_0^{-1}(i \omega_n)  \equiv {\cal
G}_0^{-1} (i \omega_n) - \alpha $ where $\alpha$ is a diagonal
matrix chosen as $\alpha_{\mu\mu} =
\lim_{\omega\rightarrow\infty} ({\cal  G}_0^{-1})_{\mu\mu}
(\omega)$. From Eqs. (\ref{eq:qmc_S}), we see that the new
impurity problem which is equivalent if the $\alpha$ term (which
is quadratic in $d$ and diagonal in the indices) is
simultaneously added to the interaction (or equivalently to the
right hand side of the corresponding decoupling formula). In the
new impurity problem however  $\bar{\cal G}_0$ is less steep than
${\cal G}_0$ close to $\tau=0$ or $\tau=\beta$, so the numerical
error introduced by discretization is less important.

\paragraph{Density--density interactions}
\label{sec:QMC:dens}
\def\spinsigma{{\tilde \sigma}}
\def\spinmu{{\tilde \mu}}
\def\spinnu{{\tilde \nu}}
\def\spinrho{{\tilde \rho}}
The ``fast update''  formula can be further simplified when the
matrix $V$ is diagonal, particularly in the case of
density--density interactions, which is used in most of the
calculations with the Hirsch--Fye algorithm. To be specific, we
concentrate in this paragraph on the first example given above
(the normal state of a Hubbard model). The matrix $V$ of Eq.
(\ref{eq:qmc:DefDecouplingFormula}) is diagonal and given by
(\ref{eq:QMC:V}).
The ``fast update'' formula  (\ref{eq:QMC:FastUpdateGeneral}) for
the flip of the Ising spin $S_{\spinmu \spinnu}^{l \spinsigma
\spinsigma'}$ at the time slice $\tau_{l}$ simplifies. The
non--zero elements of the matrix $C$ are given by
\begin{subequations}
\begin{align}
  C^{\spinsigma}_{ll;\spinmu\spinmu} =& \exp\bigl(  2 \lambda_{\spinmu \spinnu}^{\spinsigma \spinsigma'} S_{\spinmu \spinnu}^{'\spinsigma \spinsigma'}\bigr) -1
\\
C^{\spinsigma'}_{ll;\spinnu\spinnu} =& \exp\bigl(  -2 \lambda_{\spinmu \spinnu}^{\spinsigma \spinsigma'} S_{\spinmu \spinnu}^{'\spinsigma \spinsigma'}\bigr) -1
\end{align}
\end{subequations}
Let us first consider the case $\spinsigma = \spinsigma'$.
$p$ reduces to
\begin{gather}
p = \xi_\spinmu \xi_\spinnu
-  G^{\{S\}}_{\spinsigma;(l,\spinmu); (l,\spinnu)}G^{\{S\}}_{\spinsigma;(l,\spinnu); (l,\spinmu)} C^\spinsigma_{ll;\spinmu \spinmu}
C^\spinsigma_{ll;\spinnu \spinnu}
\\
\xi_\spinrho \equiv
\biggl(1 + \Bigl(1 - G^{\{S\}}_{\spinsigma;(l,\spinrho); (l,\spinrho)}\Bigr) C^\spinsigma_{ll;\spinrho \spinrho}\biggr)
; \spinrho = \spinmu,\spinnu
\end{gather}
Defining $M$ by
\begin{subequations}
\begin{align}
  M_{11} &=  \xi_\spinnu 
   C^{\spinsigma}_{ll;\spinmu \spinmu} /p
\\
  M_{22} &=   \xi_\spinmu 
   C^{\spinsigma}_{ll;\spinnu \spinnu} /p
\\
  M_{12} &=   G^{\{S\}}_{\spinsigma;(l,\spinmu), (l,\spinnu)}  C^{\spinsigma}_{ll;\spinmu \spinmu} C^{\spinsigma}_{ll;\spinnu \spinnu} /p
\\
  M_{21} &=   G^{\{S\}}_{\spinsigma;(l,\spinnu), (l,\spinmu)}  C^{\spinsigma}_{ll;\spinmu \spinmu} C^{\spinsigma}_{ll;\spinnu \spinnu} /p
\end{align}
\end{subequations}
we have the ``fast update'' formula
\begin{widetext}
\begin{multline}
 \nonumber
 G^{\{S'\}}_{\sigma;(l_1,\mu),(l_2,\nu)} =
  G^{\{S\}}_{\sigma;(l_1,\mu),(l_2,\nu)} +
\delta_{\sigma\spinsigma}
 \biggl[
\Bigl(
   G^{\{S\}}_{\sigma;(l_1,\mu),(l,\spinmu)} - \delta_{(l_1,\mu),( l,\spinmu)}
\Bigr)
\Bigl(
M_{11} G^{\{S\}}_{\sigma;(l,\spinmu),(l_2,\nu)}
+
M_{12} G^{\{S\}}_{\sigma;(l,\spinnu),(l_2,\nu)}
\Bigr)
+
\\
\Bigl(
   G^{\{S\}}_{\sigma;(l_1,\mu),(l,\spinnu)} - \delta_{(l_1,\mu),( l,\spinnu)}
\Bigr)
\Bigl(
M_{21} G^{\{S\}}_{\sigma;(l,\spinmu),(l_2,\nu)}
+
M_{22} G^{\{S\}}_{\sigma;(l,\spinnu),(l_2,\nu)}
\Bigr)
\biggr]
\end{multline}
\end{widetext}
This equation can also be obtained by using the fast update
(Sherman--Morrison) formula twice.

The case where $\spinsigma \neq \spinsigma'$ is more
straightforward
\begin{subequations}
\begin{align}
p =& \xi \xi'
\\
\xi \equiv & \biggl(1 + \Bigl(1 -
G^{\{S\}}_{\spinsigma;(l,\spinmu);  (l,\spinmu)}\Bigr)
C^{\spinsigma}_{ll;\spinmu \spinmu}\biggr)
\\
\xi' \equiv &
 \biggl(1 + \Bigl(1 - G^{\{S\}}_{\spinsigma';(l,\spinnu); (l,\spinnu)}\Bigr) C^{\spinsigma'}_{ll;\spinnu
 \spinnu}\biggr)
\end{align}
\end{subequations}
\begin{multline}
  G^{\{S'\}}_{\sigma;(l_1,\mu),(l_2,\nu)} =
  G^{\{S\}}_{\sigma;(l_1,\mu),(l_2,\nu)} +
\\
\delta_{\sigma\spinsigma}
C^{\spinsigma}_{ll;\spinmu \spinmu}
 G^{\{S\}}_{\sigma;(l,\spinmu),(l_2,\nu)}
\Bigl(
   G^{\{S\}}_{\sigma;(l_1,\mu),(l,\spinmu)} - \delta_{(l_1,\mu),( l,\spinmu)}
\Bigr)/\xi
+\\
 \delta_{\sigma\spinsigma'}
 C^{\spinsigma'}_{ll;\spinnu \spinnu} G^{\{S\}}_{\sigma;(l,\spinnu),(l_2,\nu)}
\Bigl(
   G^{\{S\}}_{\sigma;(l_1,\mu),(l,\spinnu)} - \delta_{(l_1,\mu),( l,\spinnu)}
\Bigr)/\xi'
\end{multline}

\paragraph{Analytic continuation}
\label{sec:QMC:MEM}

The quantum Monte Carlo simulation yields the Green's function in
imaginary time $G(\tau)$. For the study of the spectral
properties, transport or optics, Green's function on real axis
are needed and therefore the analytic continuation is necessary.
This in practice amounts to solving the following integral
equation
\begin{equation}
  G(\tau)=\int d\omega f(-\omega)e^{-\tau\omega}A(\omega)
  \label{MEM:Gtau}
\end{equation}
where $A(\omega)$ is the unknown spectral function, and
$f(\omega)$ is the Fermi function. This is a numerically
ill-posed problem because $G(\tau)$ is insensitive to the
spectral density at large frequencies. In other words, the
inverse of the kernel $K(\tau,\omega)=f(-\omega)e^{-\tau\omega}$
is singular and some sort of regularization is necessary to
invert the kernel. Most often, this is done by the maximum
entropy method (MEM)~\cite{Jarrell:1996}.

%

A new functional $Q[A]$, which is to be minimized, is constructed
as follows
\begin{equation}
  Q[A]=\alpha S[A] - {1\over 2}\chi^2[A]
\end{equation}
where $\chi^2$
\begin{equation}
  \chi^2[A] = \sum_{ij=1}^L (\bar{G}_i-G(\tau_i))[C^{-1}]_{ij}(\bar{G}_j-G(\tau_j))
\end{equation}
measures the distance between the QMC data, averaged over many
QMC runs ($\bar{G}_i$) and the Green's function $G(\tau_i)$ that
corresponds to the given spectral function $A(\omega)$ according
to equation Eq.~(\ref{MEM:Gtau}). $C_{ij}$ is the covariant
matrix that needs to be extracted from the QMC data when
measurements are not stochastically independent. The entropy
term, $S[A]$, takes the form
\begin{equation}
  S[A]=\int\left( A(\omega)-m(\omega)-A(\omega)\ln\left[A(\omega)/m(\omega)\right]\right),
\end{equation}
where $m(\omega)$ is the so-called default model, usually chosen
to be a constant, or, alternatively, taken to be the solution of
the same model but calculated by one of the available
approximations.

For each value of the parameter $\alpha$, numeric minimization of
$Q$ gives as the corresponding spectral function
$A^{\alpha}(\omega)$. If $\alpha$ is a large number, the solution
will not move far from the default model, while small $\alpha$
leads to unphysical oscillations caused by over-fitting the noisy
QMC data. In the so-called historic MEM, the parameter $\alpha$
is chosen such that $\chi^2=N$, where $N$ is the total number of
real frequency points at which $A(\omega)$ is being determined.
In many cases, this gives already a reasonable spectral
functions, however, in general the historic method tends to
underfit the data and makes the resulting $A(\omega)$ too smooth.

In the classical MEM, the parameter $\alpha$ is
determined from the following algebraic equation
\begin{equation}
  -2\alpha S(\alpha) = \trace\left\{\Lambda(\alpha)\left[\alpha I+\Lambda(\alpha)\right]^{-1}\right\}
\end{equation}
where $S(\alpha)$ is the value of the entropy in the solution
$A^{\alpha}$, which minimizes $Q$ and $\Lambda(\alpha)$ is
\begin{equation}
\Lambda(\alpha)_{ij}=\sqrt{A^\alpha_i}\left[K^T C^{-1}K\right]_{ij}\sqrt{A^\alpha_j}.
\end{equation}
Here $K_{ij}$ is the discretized kernel $K_{ij} \equiv K(\tau_i,\omega_j)$ and
$A_i$ is the discretized spectral function $A_i=A(\omega_i)d\omega_i$ and
$C_{ij}$ is the above defined covariant matrix.

In applications of DMFT to real materials, the quasiparticle peak
can have a complex structure since at low temperature it tries to
reproduce the LDA bands around the Fermi-level, i.e., the
spectral function approaches the LDA density of states contracted
for the quasiparticle renormalization amplitude $Z$, $A(\omega) =
\rho(\omega/Z+\mu_0)$. The MEM method has a tendency to smear out
this complex structure because of the entropy term. At low
temperature, this can lead to the causality violation of the
impurity self-energy. To avoid this pathology, it is sometimes
useful to directly decompose the singular kernel with the
Singular Value Decomposition (SVD). When constructing the real
frequency data, one needs to take into account only those singular
values, which are larger than precision of the QMC data.

The discretized imaginary time Green's function $G(\tau_i)$ can be SVD
decomposed in the following way
\begin{equation}
G(\tau_i) = \sum_j K_{ij}A_j =
\sum_{j m} V_{i m} S_m U_{m j} A_j
\end{equation}
where $U U^\dagger=1$ and $V^\dagger V=1$ are orthogonal matrices and
$S$ is diagonal matrix of singular values. The inversion is then
\begin{equation}
A_j = \sum_{m<M,i} U_{m j} {1\over S_m} V_{i m} G_i
\end{equation}
where the sum runs only up to $M$ determined by the precision of
the QMC data, for example $S_M > \langle V_{i M}\delta
G_i\rangle$, where $\delta G_i$ is the error estimate for $G_i$.
The magnitude of the singular values drop very rapidly and only of
order $10$ can be kept.


The SVD does not guarantee the spectra to be positive at higher
frequencies nor does it give a renormalized spectral function.
This however does not prevent us from accurately determining
those physical quantities which depend on the low frequency part
of spectra as for example transport or low frequency
photoemission.

\subsection{Mean--field slave boson approach}
\label{sec:IMPsbmf}

In this section we describe a different slave boson
representation, which allows a construction of a mean--field
theory closely related to the Gutzwiller approximation. In this
method we assign a slave boson $\psi_m$ to each atomic state
$|m\rangle$ and slave fermion $f_{\a}$ to each bath channel such
that the creation operator of an electron is given by
\begin{equation}  \label{eq:eqc}
d^\dagger_\alpha = \widetilde{z}^\dagger_{\alpha} f_{\a}^\dagger ,
\end{equation}
with
\begin{equation}
\widetilde{z}_{\alpha}^{\dagger} = \sum_{m,n}
\psi_n^{\dagger}(\Fad)_{nm}\psi_m \equiv \psi^{\dagger}\Fad\psi.
\end{equation}
The matrix elements $\Fa_{nm}$ are closely related to those in
Eq.~(\ref{eq:PH:breakup}) with an important difference: here one
represents electron operator by a product of pseudo--fermion and
two pseudo--bosons. The fermionic presign is completely taken care
of by the pseudo fermion, and therefore the matrix elements that
appear in the definition of pseudo--bosons should be free of the
fermionic presign. In other words, when calculating matrix $F$ in
the direct base, where matrix elements are either $0$, $+1$ or
$-1$, one should take absolute values. In the direct base, the
definition of $\Fa$ is then
\begin{equation}
\Fa_{nm} = |\langle n|d_{\alpha} | m \rangle| .
\end{equation}

The enlarged Hilbert space contains unphysical states that must be
eliminated by imposing the set of constraints
\begin{eqnarray}
  Q &\equiv& \psi^{\dagger} \psi = 1\label{eq:sbmf:c1}\\
  f_{\alpha}^{\dagger} f_{\alpha} &=& \psi^{\dagger} \Fad\Fa \psi .
  \label{eq:sbmf:c2}
\end{eqnarray}
The first constraint merely expresses the completeness relation of
local states, while the second imposes equivalence between the
charge of electrons on the local level and charge of
pseudo--fermions.

Introduction of these type of Bose fields allows one to linearize
interaction term of type $U_{\alpha\beta} n_{\alpha}n_{\beta}$.
For more general type of interaction, one needs to introduce
additional bosonic degrees of freedom that are tensors in the
local Hilbert space instead of vectors. Additional constraints
then can diagonalize a more general interaction term.

Following Ref.~\cite{KotliarR:1986} additional normalization
operators $L_\a$ and $R_\a$ are introduced whose eigenvalues
would be unity if the constraints Eq.~(\ref{eq:sbmf:c1}) are
satisfied exactly but at the same time guarantee the conservation
of probability in the the mean--field type theory
\begin{eqnarray}
R_{\alpha} = (1-\psi^{\dagger} \Fa \Fad \psi)^{-1/2},\\
L_{\alpha} = (1-\psi^{\dagger} \Fad \Fa \psi)^{-1/2} .
\end{eqnarray}
With this modification, the creation operator of an electron is
still expressed by $d^\dagger_\alpha = z^\dagger_{\alpha}
f_{\a}^\dagger$ with projectors equal to
\begin{eqnarray}
z_{\alpha}^{\dagger} = R_{\alpha}\; \psi^{\dagger} \Fad \psi \;
L_{\alpha} .
\end{eqnarray}

The action of the AIM may now be written in terms of pseudo
particles as
\begin{widetext}
\begin{eqnarray}
S &=& \sum_{\alpha}\int_0^{\beta}d\tau\left[
f_{\alpha}^{\dagger}({\partial\over\partial\tau}-\mu+i\lambda_{\alpha})f_{\alpha}
+ \psi^{\dagger}(-i\lambda_{\alpha}\Fad\Fa)\psi
\right] +\int_0^{\beta}d\tau\;\psi^\dagger({\partial\over\partial\tau}+i\Lambda+E)\psi\\
&+&\sum_{\alpha,\beta}\int_0^{\beta}d\tau\int_0^{\beta}d\tau^{\prime}\;
z_{\alpha}^{\dagger}(\tau)f_{\alpha}^{\dagger}(\tau)\Delta_{\alpha\beta}(\tau-\tau^{\prime})
f_{\beta}(\tau^{\prime})z_{\beta}(\tau^{\prime}) - i\Lambda ,
\end{eqnarray}
\end{widetext}
where $i\Lambda$ and $i\lambda_\a$ are introduced for the
constraints Eq.~(\ref{eq:sbmf:c1}) and (\ref{eq:sbmf:c2}),
respectively.

After integrating out pseudo fermions, the following saddle--point
equations can be derived by minimizing free energy with respect to
classical fields $\psi$
\begin{widetext}
\begin{eqnarray}
{1\over\beta}\sum_{i\omega}\sum_{\alpha,\beta}&&\left[ {1\over
2}({G_g}_{\alpha\beta}z_{\beta}^{\dagger}\Delta_{\beta\alpha}z_{\alpha}+
z_{\alpha}^{\dagger}\Delta_{\alpha\beta}z_{\beta}{G_g}_{\beta\alpha})
(L_{\alpha}^2\Fad\Fa+R_{\alpha}^2\Fa\Fad) \right.\nonumber\\
&&\left. +
L_{\alpha}R_{\alpha}({G_g}_{\alpha\beta}z_{\beta}^{\dagger}\Delta_{\beta\alpha}\Fa+
\Delta_{\alpha\beta}z_\beta {G_g}_{\beta\alpha}\Fad)\right]\psi
+(i\Lambda + E - \sum_{\alpha}i\lambda_{\alpha} \Fad\Fa)\psi =0.
\label{eq:sbmf:SaddlePoint}
\end{eqnarray}
\end{widetext}
where $(G_g^{-1})_{\a\b} = ((i\omega+\mu-i\lambda_\a)\delta_{\a\b}
-z^{\dagger}_\a \Delta_{\a\b} z_\b)$. The local electron Green's
function is finally given by
\begin{equation}
  G_{\a\b}=z_\a {G_g}_{\a\b} z_\b .
  \label{eq:sbmf:gf}
\end{equation}
Equations (\ref{eq:sbmf:SaddlePoint}) with constrains
(\ref{eq:sbmf:c1}) and (\ref{eq:sbmf:c2}) constitute a complete
set of non--linear equations that can be solved by iterations.

\begin{figure}[tbh]
\includegraphics[angle=-90,width=0.9\linewidth]{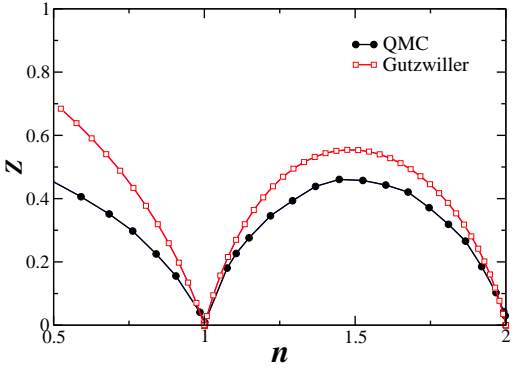}
\caption{The quasiparticle residue $Z$ from the Gutzwiller method
(open squares) is compared to the QMC $Z$ (full circles)
extracted from imaginary axis data. Calculations were performed
for the two--band Hubbard model on Bethe lattice with $U=4D$ for
QMC and $U=5.8D$ for the Gutzwiller. The later  value was chosen
to keep ratio $U/U_{MIT}$ the same in both methods. MIT denotes
metal-insulator transition.} \label{fig:Z_sbmf}
\end{figure}

\begin{figure}[tbh]
\includegraphics[angle=0,width=0.9\linewidth]{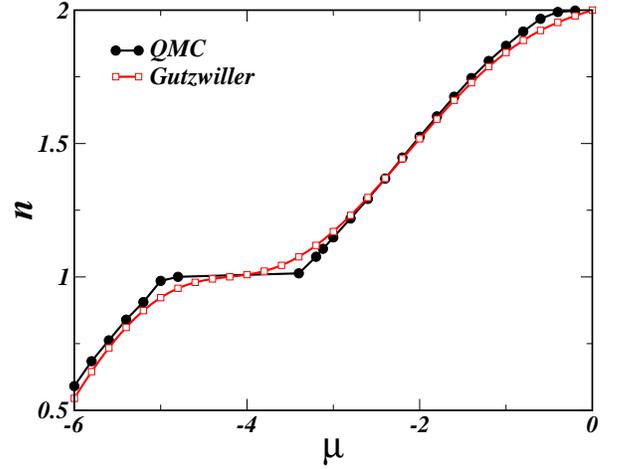}
\caption{
  Doping versus chemical potential extracted from QMC (circles)
  and from the Gutzwiller method (squares) for two--band
  Hubbard model on Bethe lattice with $U=4D$.
} \label{fig:nvmu_sbmf}
\end{figure}

In Fig. \ref{fig:Z_sbmf}, we show a comparison between QMC and
Gutzwiller quasiparticle renormalization amplitude $Z$ for the
two--band Hubbard model on Bethe lattice. We notice that the
Gutzwiller method captures all the basic low--frequency features
of the model and compares very favorably with the QMC results.
Remarkably, the chemical potential also shows a very good
agreement between QMC and the Gutzwiller method as can be seen in
Fig.~\ref{fig:nvmu_sbmf}.

The slave boson technique constructed in
Ref.~\cite{KotliarR:1986} is closely related and inspired by the
famous Gutzwiller approximation, which appears as the saddle
point of the functional integral in terms of the auxiliary boson
fields describing the local collective excitations of the system.
However there is a conceptual difference: rather than attempting
to estimate the total energy, the slave boson approach constructs
an approximation for the Green's function at low energy
Eq.~(\ref{eq:sbmf:gf}).  Fluctuations around the saddle point,
then allow to recover the Hubbard bands as demonstrated in
references \cite{Castellani:1992, Read:1983,Read:1985,
Lavagna:1990}. The Gutzwiller approximation to the Gutzwiller
wave function becomes exact in infinite dimensions
\cite{Metzner:1988} and has been recently evaluated in the
general multiorbital case \cite{Bunemann:1997,Bunemann:1998}. In
the limit of density-density interactions, the form of the
renormalization function $z$ is identical to the one obtained
from the slave boson method. The Gutzwiller approach is an
approach that gives the total energy, and also the Green's
function if one makes the slave boson identification connecting
the Gutzwiller renormalization factor to the Green's function. It
has been applied to Iron and Nickel by B\"unneman and Weber
\cite{Bunemann:2003, Ohm:2002}. Additionally, the slave boson
method gives the exact solution for the Mott transition in a
system with large orbital degeneracy \cite{Florens:2002}.

\subsection{Interpolative schemes}
\label{sec:IMPipt}

This section covers a different type of approximation to the
functional $\Sigma(E_{imp}, \Delta)$. These are not controlled
approximations, in the sense that they are not based on a small
parameter, but instead are attempts to provide approximations
which are valid simultaneously for weak and strong coupling, high
and low frequency, by combining different techniques as well as
additional exact information. By combining various bits of
information and various approaches one can obtain the self--energy
which is more accurate over a broader range of parameters. Their
accuracy has to be tested against more expensive and exact
methods of solution.

\subsubsection{Rational interpolation for the self--energy}
\label{new_ipt}

The iterative perturbation theory (IPT) method was very successful
in unraveling the physics of the Mott transition in the one--band
Hubbard model. Its success is due to the fact that it captures
not only the band limit but also the atomic limit of the problem
at half filling. As we will show in section \ref{IPTold}, the
extensions of the IPT method are possible, but less reliable in
the multiorbital case.  There have been some attempts to ``define"
interpolative methods that are robust enough and give reasonable
results in the whole space of parameters of an multiorbital
impurity model.

In this section, we will review the ideas from
\cite{Savrasov:2004:CM0410410} where a simple rational form for
the self--energy was proposed and unknown coefficient from that
rational expression were determined using the slave boson
mean--field (SBMF) method. This scheme tries to improve upon SBMF,
which gives the low--frequency information of the problem, by
adding Hubbard bands to the solution. For simplicity, only
$SU(N)$ symmetry will be considered.

It is clear that Hubbard bands are damped atomic excitations and
to the lowest order approximation, appear at the position of the
poles of the atomic Green's function. Therefore, a good starting
point to formulate the functional form for the self--energy might
be the atomic self--energy
\begin{equation}
  \Sigma^{(at)}(i\omega)=i\omega-\varepsilon_f
  -\left[G^{(at)}(i\omega)\right]^{-1},
  \label{Eq:Satom}
\end{equation}
where
\begin{equation}
G_{at}(i\omega)=\sum_{n=0}^{N-1}\frac
{C_{n}^{N-1}(\tilde{P}_n+\tilde{P}_{n+1})}
{i\omega+\mu-E_{n+1}+E_n},
  \label{Eq:Gatom}
\end{equation}
and $E_n=\varepsilon_{f}n+\frac{1}{2}Un(n-1)$, $\tilde {P}_n$ is
the probability to find an atom in a configuration with $n$
electrons, and $C_n^{N-1}=\frac{(N-1)!}{n!(N-n-1)!}$ arises due
to the equivalence of all states with $n$ electrons in $SU(N)$.

The atomic self--energy can also be brought into the form
\begin{equation}
  \Sigma^{(at)}(i\omega) = i\omega-\varepsilon_f-
  \frac{\prod\limits_{i=1}^{N}(i\omega-Z_i)}{\prod\limits_{i=1}^{N-1}(i\omega-P_i)},
  \label{Eq:Satomr}
\end{equation}
where $Z_i$ are zeros and $P_i$ are the $N-1$ poles, which can be
calculated from equations (\ref{Eq:Satom}) and (\ref{Eq:Gatom}).
Using the same functional form (\ref{Eq:Satomr}) for the
self--energy at finite $\Delta$ and calculating probabilities
$X_n$ self--consistently results in the famous Hubbard I
approximation.

To add the quasiparticle peak in the metallic state of the
system, one needs to add one zero and one pole to
Eq.~(\ref{Eq:Satomr}). To see this, let us consider the SU(N)
case for the Hubbard model where the local Green's function can
be written by the following Hilbert transform
$G_{loc}(\omega)=H(\omega-\varepsilon_f-\Sigma(\omega))$. If
self--energy lifetime effects are ignored, the local spectral
function becomes $A_{loc}=D(\omega-\varepsilon_f-\Sigma(\omega))$
where $D$ is the non--interacting density of states. The peaks in
spectral function thus appear at zeros $Z_i$ of
Eq.~(\ref{Eq:Satomr}) and to add a quasiparticle peak, one needs
to add one zero $Z_i$. To make the self--energy finite in
infinity, one also needs to add one pole $P_i$ to
Eq.~(\ref{Eq:Satomr}). This pole can control the width of the
quasiparticle peak. By adding one zero and one pole to the
expression (\ref{Eq:Satomr}), the infinite frequency value of the
self--energy is altered and needs to be fixed to its Hartree--Fock
value. The pole which is closest to zero is the obvious candidate
to be changed in order to preserve the right infinite value of
the self--energy. The functional form for the self--energy in the
metallic state of the system can take the following form
\begin{equation}
  \Sigma(i\omega) = i\omega-\varepsilon_f-
  \frac{(i\omega-X_1)\prod\limits_{i=1}^{N}(i\omega-Z_i)}{(i\omega-X_2)(i\omega-X_3)
    \prod\limits_{i=1}^{N-2}(i\omega-P_i)}.
  \label{Eq:Sr}
\end{equation}

To compute the $2N+1$ unknown coefficients in Eq.~(\ref{Eq:Sr}),
the following algorithm was used in \cite{Savrasov:2004:CM0410410}
\begin{itemize}
\item[a)] All $N$ zeros $Z_i$ are computed from the atomic form of
  the self--energy Eq.~(\ref{Eq:Satom}) and probabilities $X_n$ are
  calculated by the SBMF method.

\item[b)] Poles of the atomic self--energy are also computed from
  Eq.~(\ref{Eq:Satom}) with $X_n$ obtained by SBMF. All but one are
  used in constructing self--energy in Eq.(\ref{Eq:Sr}). The one
  closest to Fermi level needs to be changed.

\item[c)] The self--energy at the Fermi level $\Sigma(0)$ is given by the
  Friedel sum--rule
\begin{eqnarray}
\langle n\rangle &=&\frac{1}{2}+\frac{1}{\pi }\mathrm{arctg}\left( \frac{%
\epsilon _{f}+\Re \Sigma (i0^{+})+\Re \Delta (i0^{+})}{\Im \Delta (i0^{+})}%
\right)  \notag \\
&+&\int\limits_{-i\infty }^{+i\infty }\frac{dz}{2\pi i}G_{f}(z)\frac{%
\partial \Delta (z)}{\partial z}e^{z0^{+}}.
\label{FRD}
\end{eqnarray}
This relation is used to determine one
of three unknown coefficients $X_i$.

\item[d)] The slope of the
 self--energy at zero frequency is used to  determine one more unknown
 coefficient. The quasiparticle weight $z$ is
 calculated by the SBMF method and the following relationship is used
\begin{equation}
\frac{{\partial \Re \Sigma }}{{\partial }\omega }\mid _{\omega
=0}=1-z^{-1}. \label{MAS}
\end{equation}

\item[e)] Finally, the infinite frequency Hartree--Fock value of
  $\Sigma$ is used to determine the last coefficient in Eq.~(\ref{Eq:Sr})
\end{itemize}

The $2N+1$ coefficients can be computed very efficiently by
solving a set of linear equations. The method is thus very robust
and gives a unique solution in the whole space of parameters.
It's precision can be improved by adding lifetime effects,
replacing $\omega$ by second order self--energy in the way it is
done in section \ref{IPTold}.

\begin{figure}[tbp]
\includegraphics[height=0.5\textwidth]{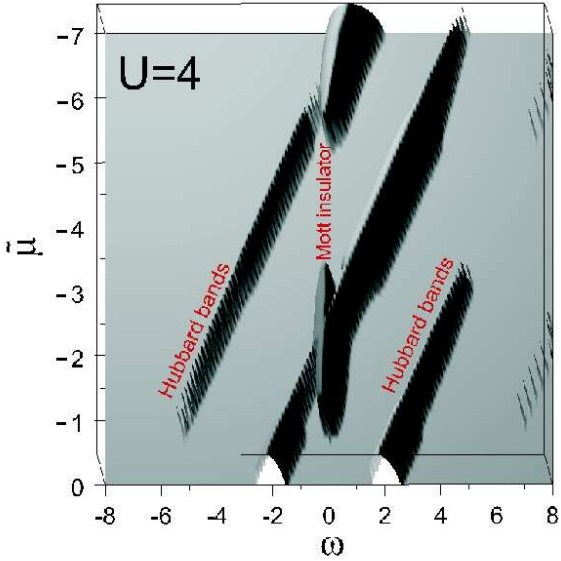}\\[-0.3cm]
\caption{The density of states calculated by the rational interpolative
method plotted as a
function of the chemical potential $\tilde{\mu}=-\epsilon_{f}-(N-1)U/2$
and frequency for the two--band Hubbard
model with $SU(4)$ symmetry and at $U=4D$.}
\label{FigU4trend}
\end{figure}
\begin{figure}[tbp]
\includegraphics[height=0.5\textwidth]{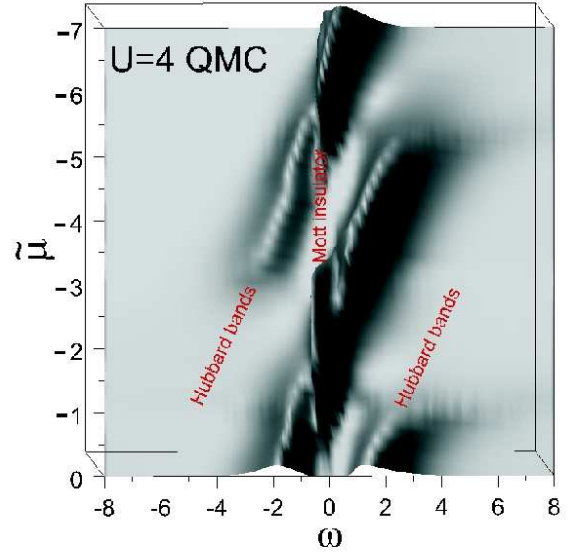}\\[-0.3cm]
\caption{The density of states calculated by the QMC method plotted as a
function of the
chemical potential $\tilde{\mu}=-\protect\epsilon_{f}-(N-1)U/2$ and
frequency for the two--band Hubbard model within $SU(4)$ and at
$U=4D$.
}
\label{FigU4trendQMC}
\end{figure}
A typical accuracy of the method is illustrated in
Fig.~\ref{FigU4trend} by plotting the density of states as a function
of the effective chemical potential $\tilde{\mu}=-\epsilon_f-(N-1)U/2$ and
frequency in the regime of strong correlations. The corresponding QMC
results are shown in \ref{FigU4trendQMC} for comparison.
Several weak satellites can also be seen on this figure
(many atomic excitations are possible) which decay fast at high
frequency.

The semicircular quasiparticle band, which is strongly
renormalized by interactions, is seen at the central part of the
figure. For the doping levels $\tilde{\mu}$ between 0 and -1 and
between -3 and -5, the weight of the quasiparticle band collapses
while lower and upper Hubbard bands acquire all the spectral
weight. In the remaining region of parameters, both strongly
renormalized quasiparticle bands and Hubbard satellites remain.
When the bands are fully filled or emptied, the quasiparticle
band restores its original bandwidth and the Hubbard bands
disappear. From Fig.~\ref{FigU4trend} and \ref{FigU4trendQMC} it
is clear that the rational interpolation for the self--energy in
combination with the SBMF offers a satisfactory qualitative and
quantitative solution of the multiorbital AIM which is useful for
many applications of the LDA+DMFT to realistic systems.

\subsubsection{Iterative perturbation theory}
\label{IPTold}

In this subsection we describe a different iterative perturbation
theory which uses the second order self--energy
Eq.~(\ref{eq:impurity2OSE}) as a main building block and also
achieves correct limits in the large frequency, zero frequency,
band and atomic coupling limit. The idea originates from the fact
that the second order perturbation theory works surprisingly well
in the case of half--filled one--band Hubbard model and captures
all the essential physics of the model. The success of this
approach can be understood by noticing that $\Sigma^{(2)}$ from
Eq.~(\ref{eq:impurity2OSE}) gives the correct atomic limit
although it is expected to work only in the weak coupling limit.
The naive extension away from half--filling or for the multiband
model treatment, however, fails because the latter property holds
only in the special case of the half--filled one--band model. To
circumvent this difficulty, a scheme can be formulated such that
the atomic limit is also captured by the construction. In the
following discussion, only $SU(N)$ symmetry will be considered.

To combine various bits of information in a consistent scheme, an
analytic expression for the self--energy in the form of
continuous fraction expansion
\begin{equation}
{\Sigma }_{\alpha }(i\omega )={\Sigma }_{\alpha}(\infty
)+\frac{A_{\alpha } }{i\omega -B_{\alpha }-\frac{C_{\alpha
}}{i\omega -D_\alpha-\cdots}} , \label{eq:APLimpSIG}
\end{equation}
was set up in Ref.~\cite{Oudovenko:2004:CM0401539}.  All the
necessary coefficients, ${\Sigma _{\alpha }}(\infty ),
A_{\alpha}, B_{\alpha }, C_{\alpha }$, $D_{\alpha }$ ..., can be
determined by imposing the correct limiting behavior at high and
low frequencies.  The basic assumption of this method is that
only a few poles in the continuous fraction expansion
(\ref{eq:APLimpSIG})  are necessary to reproduce the overall
frequency dependence of the self--energy.

Let us continue by examining the atomic limit of the second order
self--energy Eq.~(\ref{eq:impurity2OSE}) when evaluated in terms
of the bare propagator
$G^0_{\beta}(i\omega)=1/(i\omega+\widetilde{\mu}_0-\Delta_\beta)$
\begin{equation}
  \Sigma^{(2)}_{\alpha\;\Delta\rightarrow 0} = {\gamma_{\alpha}\over
    i\omega+\widetilde{\mu}_0},
\label{eq:ipt_al}
\end{equation}
where $\gamma_\alpha = \sum_{\beta}(U_{\a\b})^2 n^0_\b(1-n^0_\b)$
and $n^0_\b$ is a fictitious particle number
\begin{equation}
n^0_\b ={1\over\pi} \int f(\omega) {\rm
Im}G^0_{\beta}(\omega-i0^+)d\omega ,
\end{equation}
associated with bare propagator. The choice of $\tilde{\mu}_0$
will be discussed later.

The continuous fraction expansion in Eq.~(\ref{eq:APLimpSIG}) can
be made exact in the restricted atomic limit, i.e., when the
three significant poles are considered in the Green's function,
and coefficients are calculated from the moments of the
self--energy. By replacing $i\omega$ with
$\gamma_{\a}/\Sigma^{(2)}_\a -\widetilde{\mu}_0$ in expansion
Eq.~(\ref{eq:APLimpSIG}), it is clear from Eq.~(\ref{eq:ipt_al})
that the resulting self--energy functional has the correct atomic
limit and reads
\begin{equation}
  \Sigma_\a(i\omega)=\Sigma_\a(\infty)+{{A_{\a}\over\gamma_\a}\Sigma^{(2)}_\a
  \over 1-{\widetilde{\mu}_0+B_\a\over\gamma_\a}\Sigma^{(2)}_\a -
  {{(C_\a/\gamma_a^2)}\left(\Sigma^{(2)}_\a\right)^2\over1-(\widetilde{\mu}_0+D_\a)
  \Sigma^{(2)}_\a/\gamma_\a - \cdots}}.
  \label{eq:ansatz}
\end{equation}
Coefficients $A$, $B$, $C$ and $D$ can be determined from the
moment expansion
\begin{equation}
A_{\alpha }=\Sigma _{\alpha }^{(1)}, \label{Eq:SE1}
\end{equation}
\begin{equation}
B_{\alpha }=\frac{\Sigma _{\alpha }^{(2)}}{\Sigma _{\alpha
}^{(1)}}, \label{Eq:SE2}
\end{equation}
\begin{equation}
C_{\alpha }=\frac{\Sigma _{\alpha }^{(3)}\Sigma _{\alpha
}^{(1)}-(\Sigma _{\alpha }^{(2)})^{2}}{(\Sigma _{\alpha
}^{(1)})^{2}}, \label{Eq:SE3}
\end{equation}
where the self--energy moments can be expressed in terms of the
density--density correlation functions (see
Ref.~\cite{Oudovenko:2004:CM0401539}).

Finally, the parameter $\widetilde{\mu}_0$ can be determined by
imposing the Friedel sum rule,
%
%
which is a relation between the total density and the real part of
the self--energy at zero frequency, thereby achieving the correct
zero frequency limit. Since the Friedel sum rule is valid only at
zero temperature, the parameter $\widetilde{\mu}_0 $ is
determined at $T=0$, and after having it fixed, equation
(\ref{eq:ansatz}) is used at arbitrary temperatures.

An alternative scheme for determining the temperature--dependent
$\widetilde{\mu}_0$ was proposed by Potthoff et al.
\cite{Potthoff:1997}, and consists of the requirement that the
fictitious occupancy computed from $G^0$ equals the true
occupancy computed from $G$ \cite{Rodero:1982}. A careful
comparison of these approaches was carried out by Potthoff et al.
\cite{Potthoff:1997,Meyer:1999}.

Note that one could continue the expansion in continuous fraction
to the order in which the expansion gives not only restricted but
the true atomic limit. However, in practice this is seldom
necessary because only few poles close to the Fermi energy have a
large weight.

It is essential that the self--energy Eq.~(\ref{eq:ansatz})
remains exact to $U^2$, which can be easily verified by noting
that in the $U\rightarrow 0$ the fictitious occupancy $n^0$
approaches $n$ therefore $A_\a = \gamma_\a (1+O(U))$,
$B\rightarrow -\widetilde{\mu}_0$ and $C\rightarrow 0$. At the
same time, the self--energy Eq.~(\ref{eq:ansatz}) has correct
first moment because expanding $\Sigma^{(2)}$ in the
high--frequency limit yields
$\Sigma_\a^{(2)}=\gamma_\a/(i\omega)+...$ and $A_\a$ is exact
first moment.

Note that in the case of one--band model, the atomic limit
requires only one pole in self--energy therefore the coefficient
$C$ in Eq.~(\ref{eq:ansatz}) can be set to zero and one has
\begin{equation}
\Sigma_\a(i\omega)=\Sigma_\a(\infty)+{{A_{\a}\over\gamma_\a}\Sigma^{(2)}_\a
\over 1-{\widetilde{\mu}_0+B_\a\over\gamma_\a}\Sigma^{(2)}_\a}.
\label{eq:ansatz0}
\end{equation}
Furthermore, double and triple occupancies do not enter the
expression for the moments in the case of one--band model. If one
chooses the functional form for the moments in the atomic limit,
the interpolative self--energy (\ref{eq:ansatz}) has still the
same limiting behavior as discussed above. In this case, no
additional external information is necessary and the system of
equations (\ref{eq:ansatz0}), (\ref{Eq:SE1}), (\ref{Eq:SE2}) and
(\ref{eq:impurity2OSE}) is closed.

\begin{figure}[tbh]
\includegraphics[angle=0,width=0.9\linewidth]{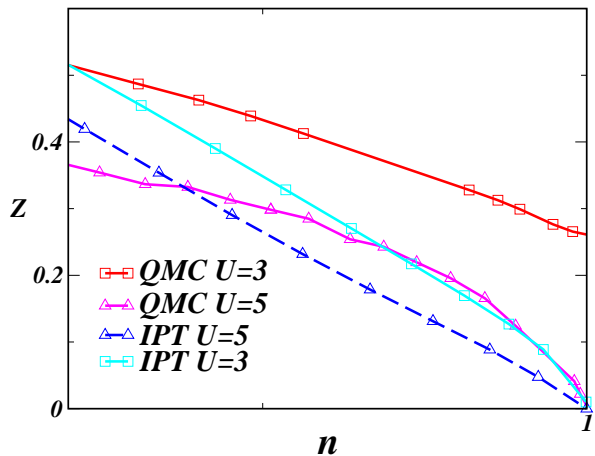}
\caption{Comparison between QMC  and the simplified IPT
Eq.~(\protect\ref{eq:ansatz0}) renormalization amplitude $Z$ for
the three--band Hubbard model on Bethe lattice at $U=3D$ and $5D$.
QMC $Z$ was extracted at temperature $T=1/16$ and IPT at zero
temperature.\label{fig:Z_qmc_ipt}}%
\end{figure}
For the multiband model, an approximate method is needed to
calculate moments which in turn ensure a limiting form consistent
with the simplified atomic limit. Many approaches discussed in
previous sections can be used for that purpose, for example
the Gutzwiller method or SUNCA. In Ref.~\cite{Kajueter:1996} the
coherent potential approximation (CPA) was used to obtain moments
in the functional form consistent with the atomic limit, i.e.,
neglecting last term in Eq.~(\ref{Eq:SE2}).  Another possibility,
also tested by Ref.~\cite{Kajueter:1996}, is to use the ansatz
Eq.~(\ref{eq:ansatz0}) in the case of multiband model. In
Fig.~\ref{fig:Z_qmc_ipt} the quasiparticle renormalization
amplitude $Z$ versus particle number $n$ is displayed for $n$
less than one where this scheme compares favorably with the QMC
method. CPA was used to obtain moments.

\begin{figure}[tbh]
\includegraphics[angle=0,width=0.9\linewidth]{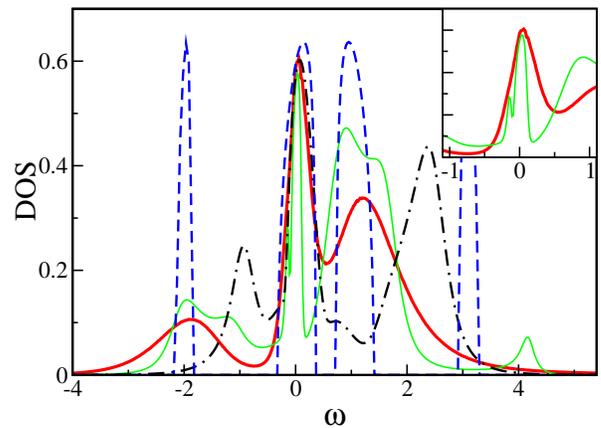}\\
 \caption{Density of states for the two--band
Hubbard model on Bethe lattice at $U=2.5D$ and $n_d=1.1$. The
tick full line marks QMC curve at  temperature $T=1/16D$ while
the rest of the curves correspond to   various IPT schemes at
$T=0$. The thin full line shows the IPT from
Eq.~(\protect\ref{eq:ansatz}) with all four coefficients
determined by high--frequency moments in the functional form of
atomic limit. The inset zooms the region around the chemical
potential where the above mentioned IPT scheme develops a
spurious double peak structure. The dot--dashed line corresponds
to simplified IPT Eq.~(\protect\ref{eq:ansatz0}) with both two
coefficients determined by moments. Finally, the dashed curve
stands for the IPT schemes described in section
\protect\ref{new_ipt}. \label{fig:IPTs} }
\end{figure}
When the particle number slightly exceeds unity, the simplified
IPT scheme Eq.~(\ref{eq:ansatz0}) does not provide an accurate
description of the multiorbital AIM. As shown in
Fig.~\ref{fig:IPTs}, the Hubbard bands are completely misplaced.
Nevertheless, the quasiparticle peak is in good agreement with
QMC result. By taking into account more terms in the continuous
fraction expansion Eq.~(\ref{eq:ansatz}), the high--frequency part
of spectra can be considerably improved since the resulting
approximation obeys more high--frequency moments. Unfortunately,
the quasiparticle peak develops a spurious double peak structure
which severely limits the applicability of the method, as shown
in the inset of Fig.~\ref{fig:IPTs}. Systematically improving
only the high--frequency part of the spectra, by incorporating
more moments into the approximation, can thus spoil the
low--frequency part. This type of unphysical feature can be
avoided using the scheme from subsection \ref{new_ipt} where the
derivative as well as the value of the spectra at zero frequency
was imposed by the information obtained by a more accurate
technique at low frequency, such as the Gutzwiller method.

The most general extension of IPT, and its simplified form
Eq.~(\ref{eq:ansatz0}) was set up in Ref.~\cite{Kajueter:1996}
and in Ref.~\cite{Rodero:1982}. The authors in
Ref.~\cite{Ferrer:1987, Yeyati:1999} tested it in the context of
quantum dots, where it performs satisfactorily. However,
Ref.~\cite{Kajueter:1996} tested it in the DMFT context, and the
difficulties with the spurious double peak structure shown in
Fig.~\ref{fig:IPTs} were found close to integer filling in the
case of occupancies larger than one. When the occupancies are
less than one, the simpler formula Eq.~(\ref{eq:ansatz0}) is
accurate and free from pathologies. It was used to compute the
physical properties of ${\rm La}_{1-x}{\rm Sr}_x{\rm TiO}_3$ in
Refs.~\cite{Anisimov:1997, Kajuter:1997i}. Various materials with
strongly correlated $d$--bands were studied by the group of Craco,
Laad and M\"uller-Hartmann using the IPT method for arbitrary
filling of the correlated bands. The properties of ${\rm CrO_2}$
\cite{Laad:2001,Craco:2003a}, ${\rm Li V_2 O_4}$
\cite{Laad:2003a}, ${\rm V_2 O_3}$ \cite{Laad:2003b} and ${\rm
Ga_{1-x}Mn_x As}$ \cite{Craco:2003b} were explained by the IPT
method.



\section{Application to materials}%
\label{sec:MAT}

In this section we illustrate the realistic dynamical mean--field
methodology with examples taken from various materials.  We chose
situations where correlations effects are primarily responsible
for the behavior of a given physical system. The examples
include: (i) phase transitions between a metal and insulator
where, in the absence of any long--range magnetic order, opening
an energy gap in spectrum cannot be understood within simple--band
theory arguments; (ii) large, isostructural volume collapse
transitions where a localization--delocalization driven change in
lattice parameters of the system is necessary to understand the
transition; (iii) the behavior of systems with local moments which
are not straightforward to study within band theory methods. We
conclude this section  with a brief, non--comprehensive summary
of a few other illustrations of the power of the dynamical
mean--field method that, for lack of space, cannot be covered in
this review.

\subsection{Metal--insulator  transitions}
\label{PDMIT}

\subsubsection{Pressure driven metal--insulator transitions}

\begin{widetext}

\begin{figure}[h]
\includegraphics[width=6.9 in]{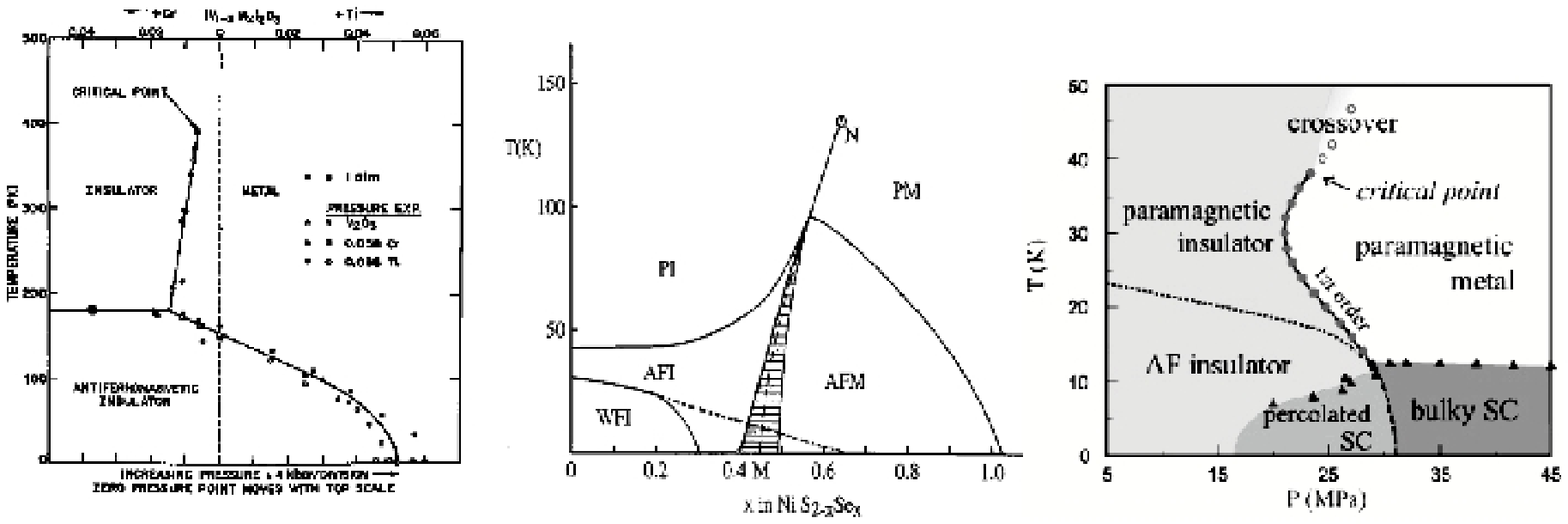}%
\caption{The phase diagrams of V$_{2}$O$_3$, NiS$_{2-x}$Se$_{x}$
and organic materials of the $\kappa $  family (from
\protect\cite{McWhan:1971} (left), \cite{Edwards:1990} (middle),
and  \protect\cite{Kagawa:2004} (right)).
} %
\label{fig:V2O3_NiSSe_org_v}
\end{figure}

\end{widetext}

\begin{figure}[h]
\includegraphics[width=3.5 in]{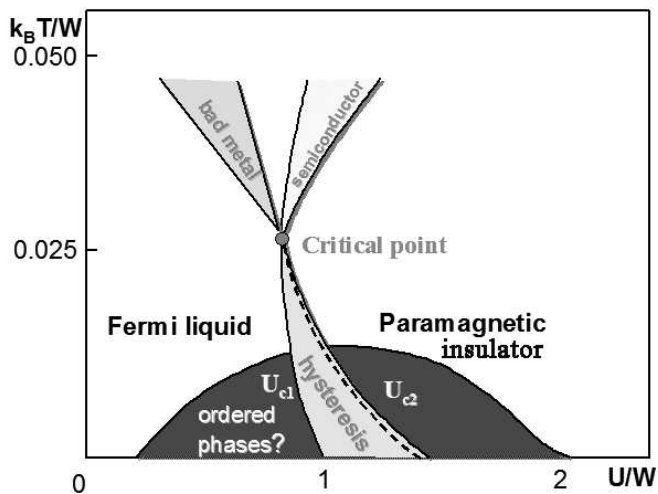}
\caption{Schematic phase diagram of a material undergoing a
Mott metal--insulator transition.} %
\label{fig:dmft_phase_diagram_gk}
\end{figure}

The pressure driven metal--insulator transition (MIT) is one of
the simplest and at the same time most basic problems in the
electronic structure of correlated electrons. It is realized in
many materials such as V$_{2}$O$_3$, where the metal--insulator
transition is induced as function of chemical pressure via Cr
doping, quasi--two--dimensional organic materials of the $\kappa $
family such as (BEDT--TSF)$_{2}$X (X is an anion) \cite{Ito:1996,
Lefebvre:2000}, and Nickel Selenide Sulfide mixtures (for a
review see \cite{Imada:1998} as well as articles of Rosenbaum and
Yao in \cite{Edwards:1990}). The phase diagram of these materials
is described in Fig.~\ref{fig:V2O3_NiSSe_org_v}. It is remarkable
that the high--temperature part of the phase diagram of these
materials, featuring a first order line of metal--insulator
transitions ending in a critical point, is qualitatively similar
in spite of the significant differences in the crystal and
electronic structure of these materials ~\cite{Kotliar:1999:PB,
Chitra:1999, Kotliar:2001:Kluwer}.  This is illustrated in
Fig.~\ref{fig:dmft_phase_diagram_gk} where the schematic phase
diagram of the integer filled Hubbard model is included.

V$_2$O$_3$  has a Corundum structure in which the V ions are
arranged in pairs along the $c$ hexagonal axis, and form a
honeycomb lattice in the basal $ab$ plane. Each V ion has a
$3d^{2}$ configuration. The $d$--electrons occupy two of the
$t_{2g}$ orbitals which split into a non--degenerate $a_{1g}$ and
a doubly degenerate $e_{g}^{\pi }$ orbital. The $e_{g}^{\sigma}$
states lie higher in energy~\cite{Castellani:1978,
Castellani:1978:PRB:18:5001}.  NiSeS  mixtures are charge
transfer insulators, in the Zaanen--Sawatzky--Allen
classification~\cite{Zaanen:1985}, with a pyrite structure. In
this compound the orbital degeneracy is lifted, the configuration
of the $d$--electron in Ni is spin one, $ d^8$, and the effective
frustration arises from the ring exchange  in this lattice
structure. The $\kappa $ -(BEDT--TSF)$_{2}$X (X is an anion)
\cite{Ito:1996, Lefebvre:2000} are formed by stacks of dimers and
the system is   described at low energies by a one--band Hubbard
model with a very anisotropic next nearest neighbor hopping
~\cite{McKenzie:1998, Kino:1996}.

The universality of the Mott phenomena at high temperatures
allowed its description using fairly simple Hamiltonians. One of
the great successes of DMFT applied to simple model Hamiltonians
was the realization that simple electronic models are capable of
producing such phase diagram and many of the observed physical
properties of the materials in question. The qualitative features
related to the Mott transition at finite temperature carry  over
to more general models having other orbital and band degeneracy
as well as coupling to the lattice. The dependence of this phase
diagram on orbital degeneracy has been investigated recently
\cite{Florens:2002, Ono:2002, Kajueter:1997, Ono:2001}.

\vspace*{1cm}

The determination of the  extension of the qualitative phase
diagram away from half filling includes regions of phase
separation near half filling \cite{Kotliar:2002}. Determination
of the low--temperature phases, which are completely different in
the materials in Fig.~\ref{fig:V2O3_NiSSe_org_v}, requires a more
careful and detailed modeling of the material, and the study of
the dependence of the magnetic properties on the properties of
the lattice is only in  the beginning stages~\cite{Chitra:1999,
Zitzler:2002:CM}.

Dynamical mean--field theory~\cite{Georges:1996}, provided a
fairly detailed picture of the evolution of the electronic
structure with temperature and interaction strength or pressure.
Surprising predictions emerged from these studies: %
 a) the observation that for a correlated metal, in the
presence of magnetic frustration, the electronic structure (i.e.
the spectral function) contains both quasiparticle features, and
Hubbard bands~\cite{Georges:1992}.
 b) The idea that the Mott transition is driven by the transfer of
spectral weight from the coherent to the incoherent
features~\cite{Zhang:1993}. This scenario, brought together   the
Brinkman--Rice--Gutzwiller ideas and the Hubbard ideas about  the
Mott transition in a unified framework.
 c) The existence of broad regions of parameters where the
incoherent part of the spectra dominates the transport. The
impurity model subject to the DMFT self--consistency condition is
a minimal model to approach the understanding of the incoherent
or bad metal,  the Fermi liquid state, the Mott insulating state,
and a ``semiconducting" or ``bad insulator" state where thermally
induced states populate the Mott Hubbard gap.
 d) An understanding of the critical behavior near the Mott
transition as an Ising transition \cite{Kotliar:2000}. This
critical behavior had been surmised long ago by Castellani et al.
~\cite{Castellani:1979}.

\begin{figure}[tb]
\epsfig{file=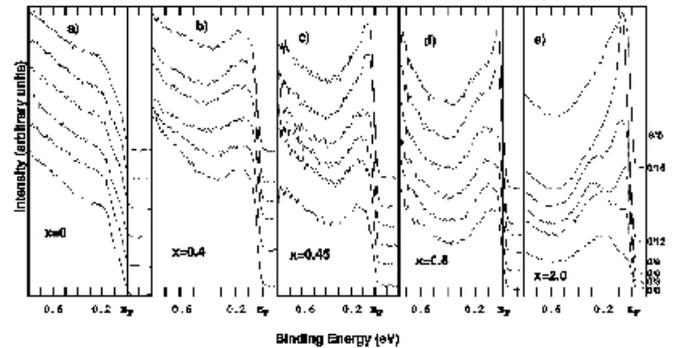,width=4.0in}%
\caption{%
Near--Fermi--level ARPES spectra in NiSeS taken nearly along the
(001) direction for (a) $x=0$ (insulating), (b) $x=0.4$
(insulating), (c) $x=0.45$ (metallic), (d) $x=0.5$ (metallic),
and (e) $x=2.0$ (metallic)  (from
Ref.~\protect\cite{Matsuura:1998}).
}%
\label{fig:NiSSe_PRB58_3690_F2}
\end{figure}

\begin{figure}[tb]
\epsfig{file=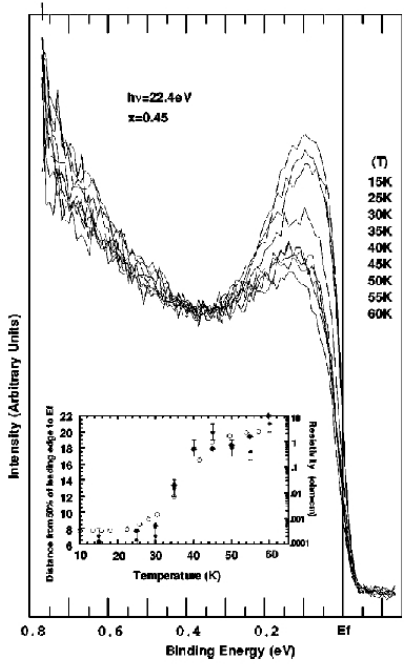,width=2.5in}%
\caption{%
Temperature--dependence obtained in NiSeS of the near-
$\varepsilon_F$ peak for $x =0.45$ at 22.4~eV incident photon
energy. Inset: Distance of the 50\% point of the leading edge from
$\varepsilon_F$ (solid circles); reactivity (open circles,
right--hand scale); area under the near-$\varepsilon_F$ peak
(solid diamonds, scaled in arbitrary units). Analyzer angle:
$0/9$ (from Ref.~\protect\cite{Matsuura:1998}).
}%
\label{fig:NiSSe_PRB58_3690_F3}
\end{figure}

\begin{figure}[tb]
\epsfig{file=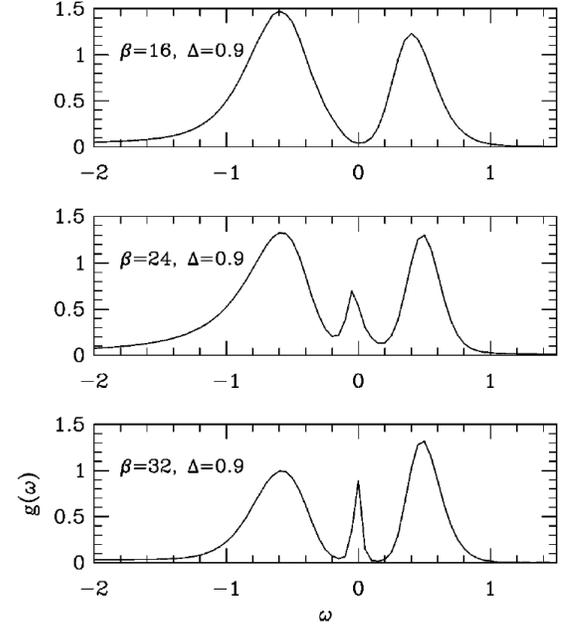,width=3.0in}%
\caption{%
Single--particle Green's functions at half--filling for a fixed
charge--transfer gap and varying temperature. Horizontal axis is
scaled in units of $2t$. Vertical axis has arbitrary units (from
Ref.~\protect\cite{Matsuura:1998}).
}%
\label{fig:NiSSe_PRB58_3690_F7}
\end{figure}

In the last few years experimental developments have confirmed
many of the qualitative predictions of the DMFT approach. For
recent reviews see ~\cite{Kotliar:2004:PT,
Georges:2004:CM,Georges:2004:CM0403123}.

a) Photoemission spectroscopy has provided firm evidence for a
three peak structure of the spectral function in the strongly
correlated metallic regime of various materials and its evolution
near the Mott transition. This was first observed in the
pioneering experiments of A. Fujimori et al.~
\cite{Fujimori:1992a}. The observation of  a quasiparticle peak
near the Mott transition took some additional work. In
Ref.~\cite{Matsuura:1998}, angle resolved photoemission spectra in
NiSeS reveal the presence of a quasiparticle band and a Hubbard
band (see Figs.
~\ref{fig:NiSSe_PRB58_3690_F2},\ref{fig:NiSSe_PRB58_3690_F3}).
These results, together with a DMFT calculation by Watanabe and
Doniach, in the framework of a two--band model are shown in Fig.
\ref{fig:NiSSe_PRB58_3690_F7}.

\begin{figure}[h]
\includegraphics[width=2.9 in]{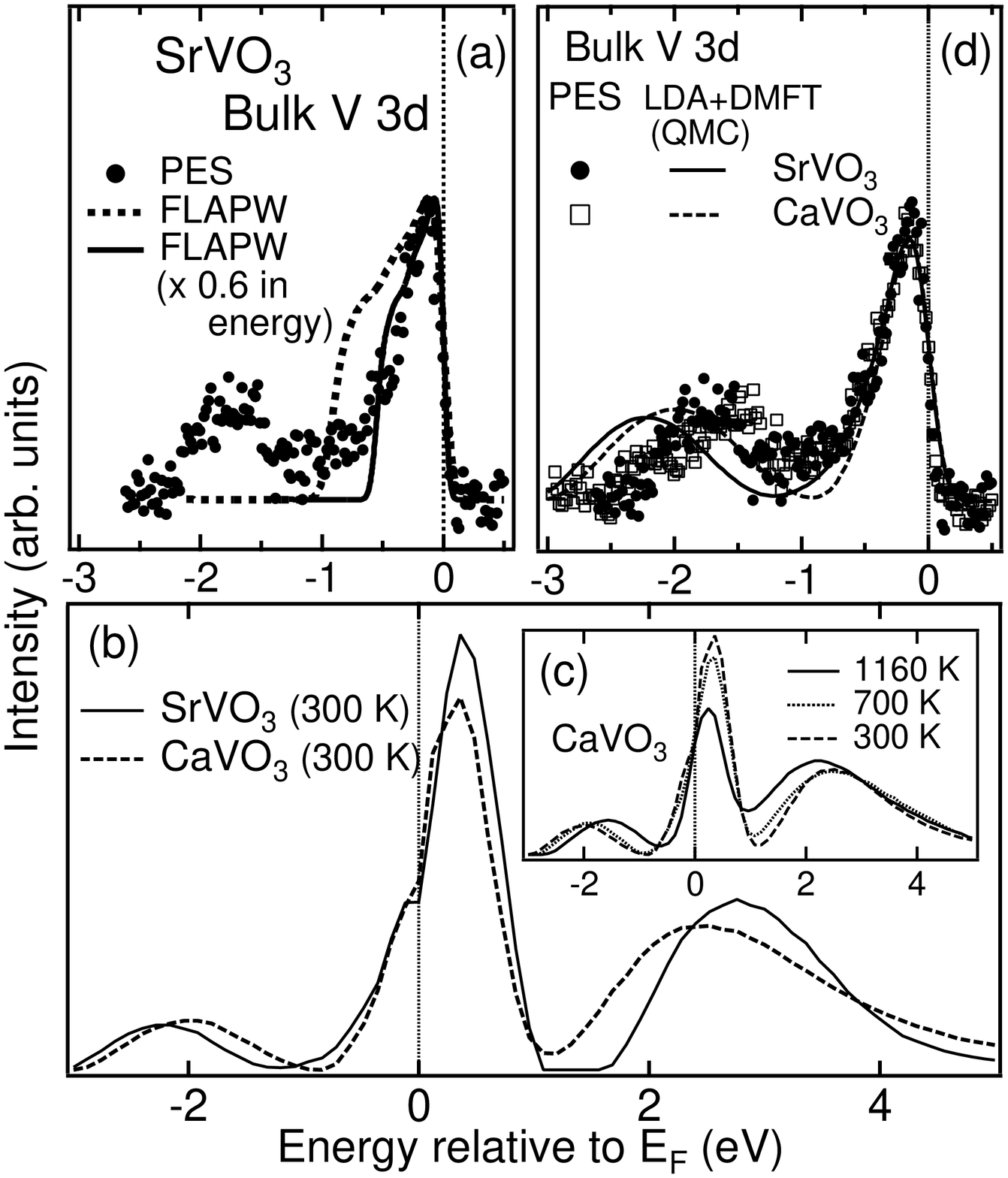}
\caption{ A comparison of LDA (panel $a$), LDA+DMFT (panel $d$), and the photoemission data
for $SrVO_3$ and $CaVO_3$ (after \cite{Sekiyama:2004}).}
\label{fig:srvo3}
\end{figure}

Cubic SrVO$_{3}$ and orthorhombic CaVO$_{3}$ perovskites are
strongly correlated metals. LDA+DMFT calculations
~\cite{Sekiyama:2004} find their spectral
functions to be very similar  in agreement with recent
bulk--sensitive photoemission experiments~\cite{Sekiyama:2002,
Sekiyama:2004}. The comparison of the high--energy LDA+DMFT
photoemission results against the experiments is presented in
Fig.~\ref{fig:srvo3}. LDA qualitatively fails as it cannot produce the
Hubbard band while LDA+DMFT successfully captures this and compares
well with experiment.

\begin{figure}[h]
\includegraphics[width=2.9 in]{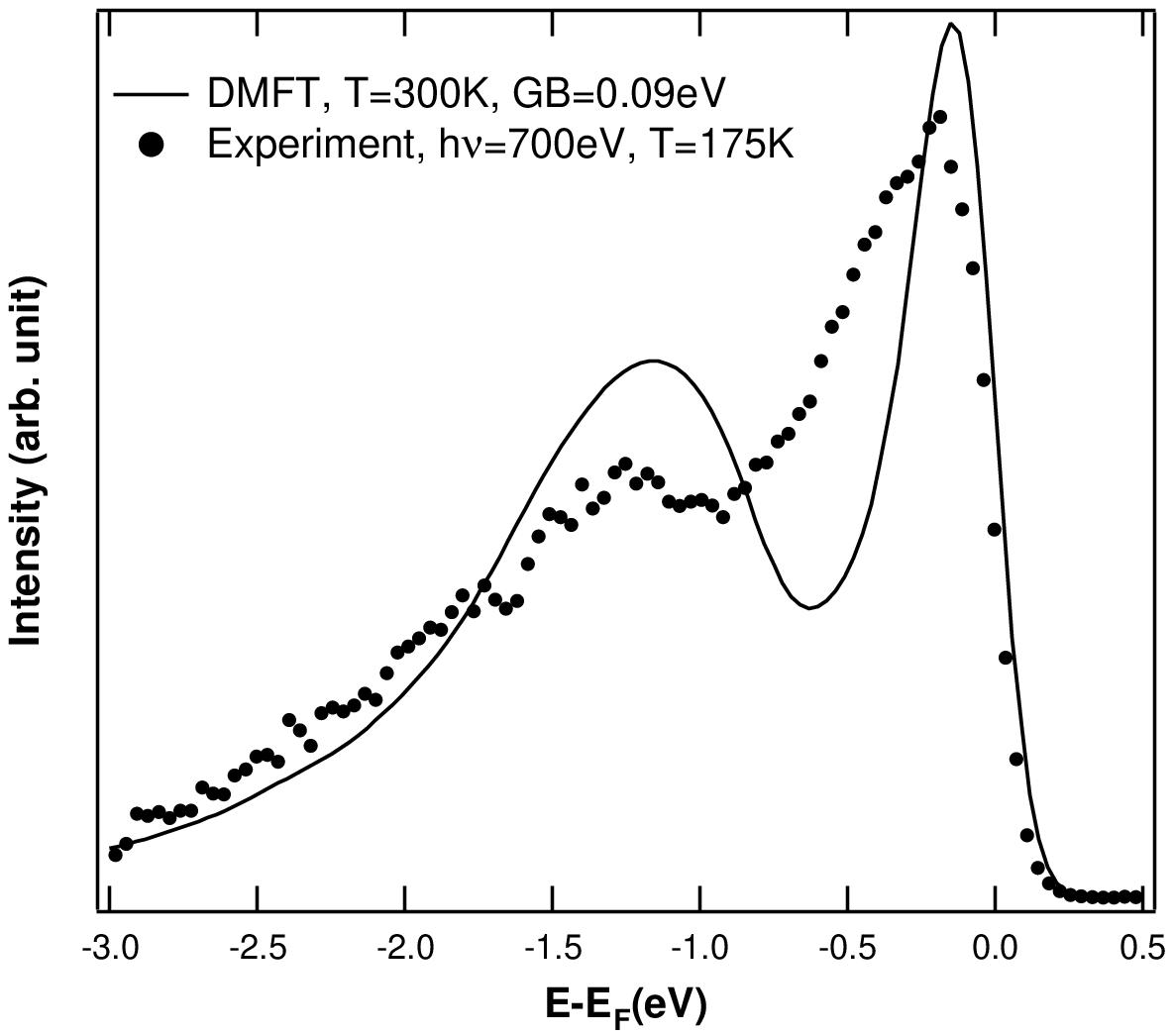}
\caption{Comparison of $h\protect\nu =700$~eV PES spectrum with
LDA+DMFT(QMC) spectrum for $T=300$~K and $U=5.0$~eV in $V_2O_3$. (after
\protect\cite{Mo:2003})} %
\label{fig:V2O3_0212110_F4}
\end{figure}

Using high--energy photoemission spectroscopy, Mo et al. ~\cite{Mo:2003}
studied the V$_{2}$O$_{3}$ system. The spectral function, which
exhibits a quasiparticle peak and a Hubbard band, is displayed in
Fig.~\ref{fig:V2O3_0212110_F4} together with an LDA+DMFT
calculation. The calculation was performed using the LDA density
of states of the Vanadium $t_{2g}$ electrons and a Hubbard $U$ of
5~eV.

\vspace*{1cm}

\begin{figure}[h]
\includegraphics[width=2.9 in]{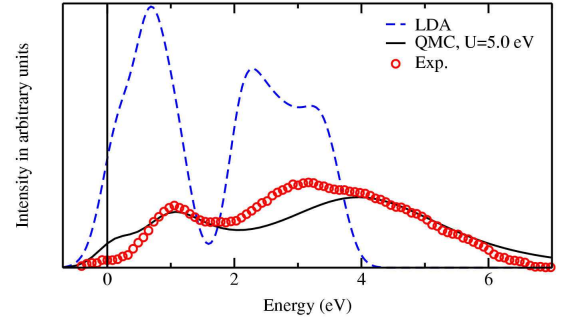}
\caption{Comparison of the LDA and LDA+DMFT(QMC) spectra at
$T=0.1$~eV (Gaussian broadened with 0.2~eV) with the x--ray
absorption data of \protect\cite{Mueller:1997} in $V_2O_3$. The LDA and QMC
curves are normalized differently since the $\protect\epsilon
_{g}^{\protect\sigma }$ states, which are shifted towards higher
energies if the Coulomb interaction is included, are neglected in
our calculation (from Ref.~\protect\cite{Held:2001V2O3}). } %
\label{fig:V2O3_Xray_0011518}
\end{figure}
\begin{figure}[htbp]
\centerline{\includegraphics[width=0.95\hsize]{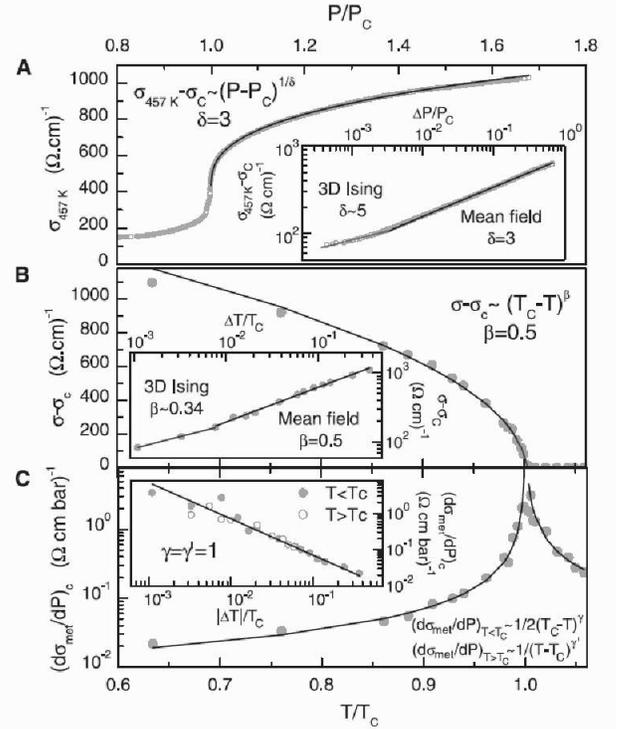}}
\caption{Temperature dependence of the conductivity (a), the
order parameter (b) and derivative of the conductivity (analogous
to a susceptibility) (c) in Cr doped  V$_2$O$_3$ (from
Ref.~\protect\cite{Limelette:2003:science}).} %
\label{fig:V2O3_science_302_89_1}
\end{figure}
\begin{figure}[htbp]
\centerline{\includegraphics[width=0.95\hsize]{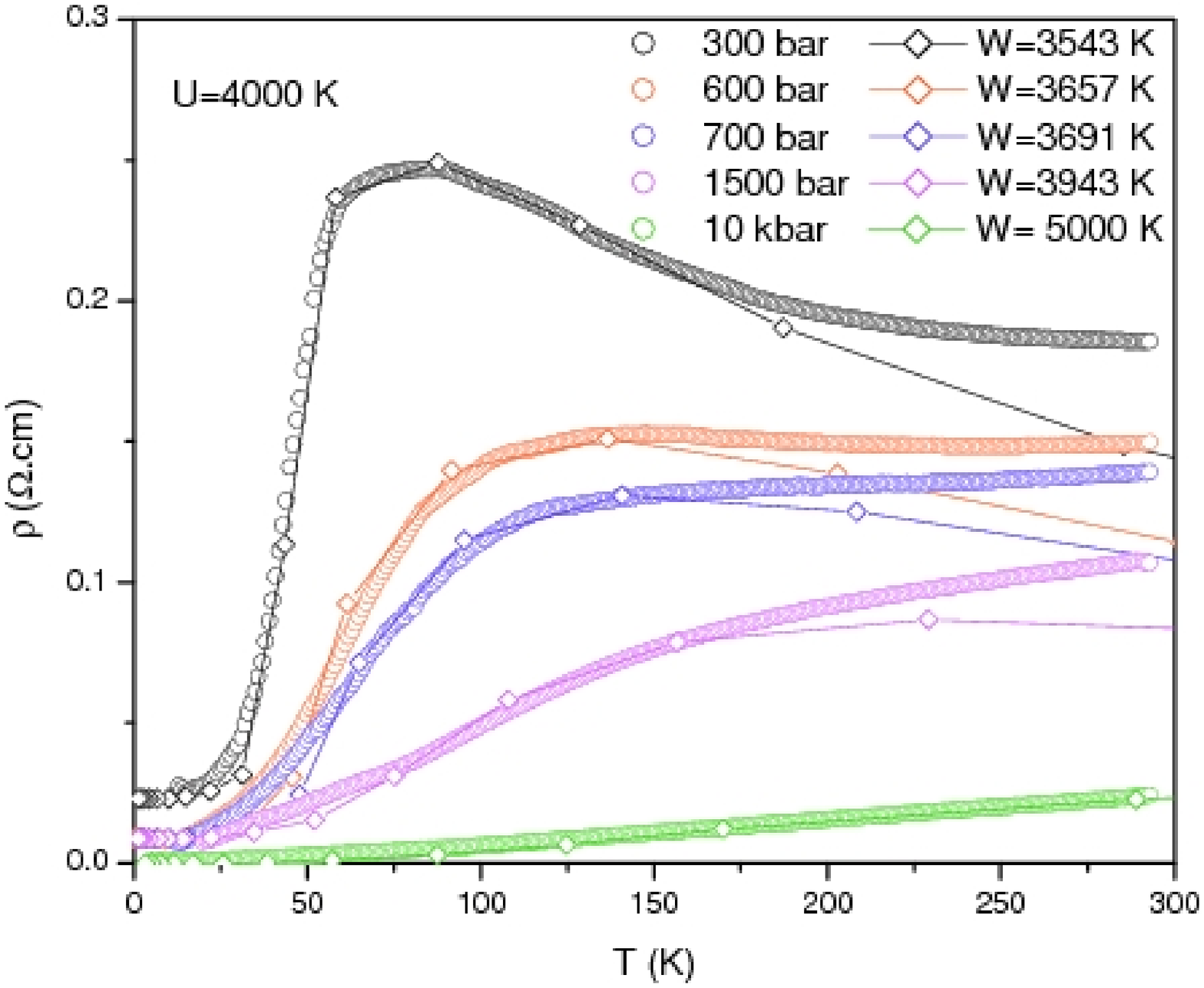}}
\caption{Temperature--dependence of the resistivity in Cr doped
V$_2$O$_3$ at different pressures. The data (circles) are compared
to a DMFT--NRG calculation (diamonds), with a pressure dependence
of the bandwidth as indicated. The measured residual resistivity
$\rho_0$ has been added to the theoretical curves (from
Ref.~\protect\cite{Limelette:2003}).} %
\label{fig:fig6_cm0301478}
\end{figure}


b) Optical spectroscopy  has confirmed the idea of temperature
driven transfer of spectral weight in the vicinity of  the Mott
transition. The first indications  had been obtained in the
V$_2$O$_3$ system~\cite{Rozenberg:1995} where it was found that
as temperature is lowered, optical spectral weight is transferred
from high energies to low energies. Similar observations were
carried out in  NiSeS ~\cite{Miyasaka:2004}, and in the kappa
organics~\cite{Eldridge:1991}, confirming the high--temperature
universal behavior of materials near a Mott transition.

c) The Ising critical behavior predicted by DMFT is now observed
in Cr doped  V$_2$O$_3$ ~\cite{Limelette:2003:science}. The
large  critical region  and the experimental observation  of the
spinodal lines~(Figs.~\ref{fig:V2O3_science_302_89_1} and
\ref{fig:fig6_cm0301478}), was ascribed to the importance of
electron--phonon coupling~\cite{Kotliar:2003:science}. The
situation in organic materials at this point is not
clear~\cite{Kagawa:2003:CM, Kanoda:2004}.

\begin{figure}[htbp]
\centerline{\includegraphics[width=0.85\hsize]{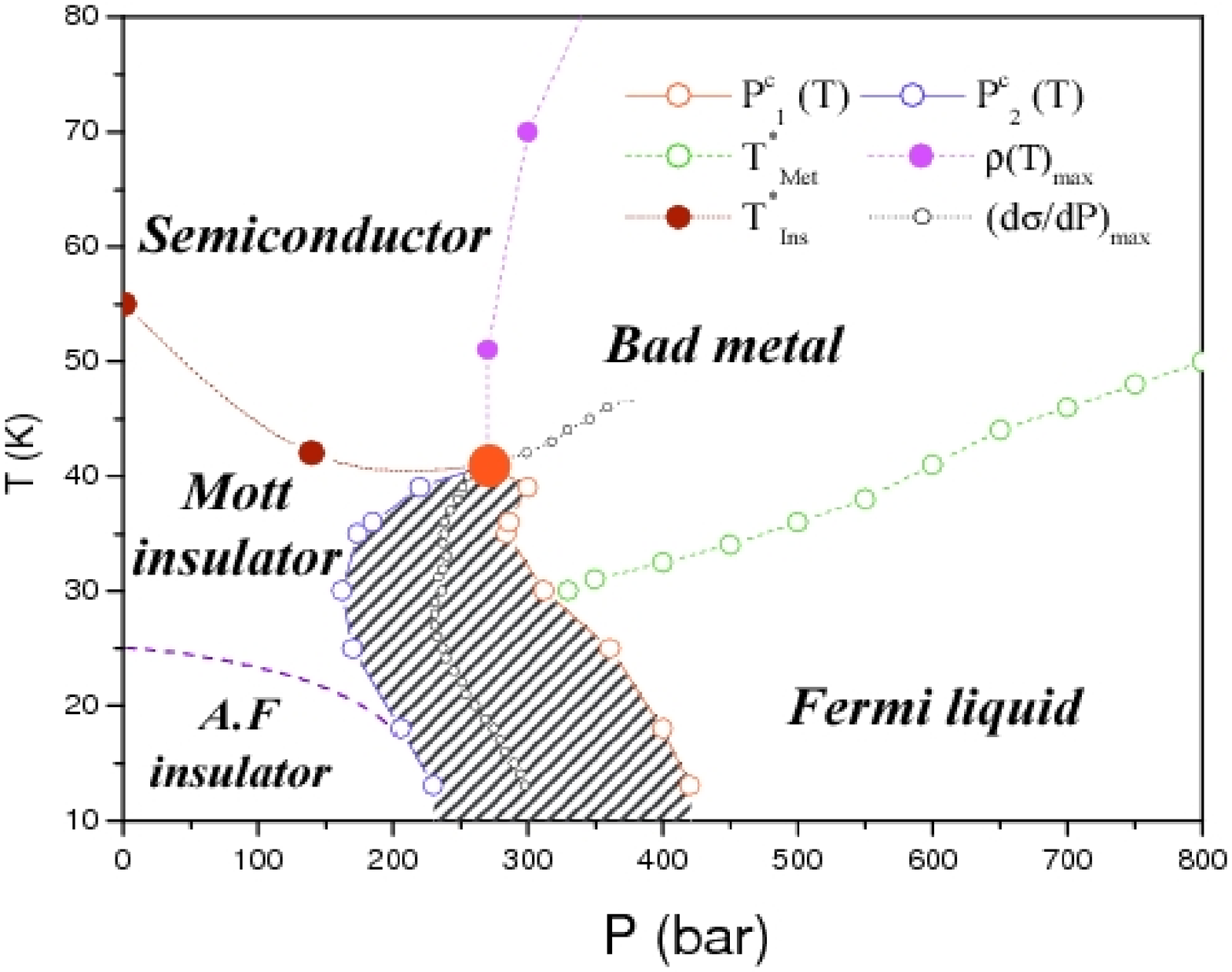}}
\caption{Pressure--Temperature phase diagram of the $\kappa$-Cl
salt (from Ref.~\protect\cite{Limelette:2003}). }
\label{fig:fig1_cm0301478}
\end{figure}

e) Transport studies in V$_2$O$_3$~\cite{Kuwamoto:1980} and in
NiSeS  \cite{Imada:1998} have mapped out the various crossover
regimes of the DMFT phase diagram (see Fig.
~\ref{fig:dmft_phase_diagram_gk}), featuring a bad metal, a bad
insulator, a Fermi liquid and a Mott insulator. More recent
studies in the two--dimensional kappa organics
\cite{Limelette:2003} are consistent with the DMFT picture, and
can be fit quantitatively within single--site DMFT (see Fig.
\ref{fig:fig1_cm0301478}).

The dynamical mean field studies have settled a long standing
question. Is the Mott transition in V$_2$O$_3$, NiSSe and kappa
organics driven by an electronic structure mechanism or 
by the lattice (i.e. the position of the ions ) degrees of
freedom.

This question can only be answered theoretically since lattice
deformations are almost always generically induced by changes in
the electronic structure (see below) and vice--versa. In a
theoretical study one can freeze the lattice while studying a
purely electronic model,
and it is now accepted that the
simple Hubbard model can account for the topology
of the high temperature phase diagram \cite{Georges:1996}.
Hence, lattice
deformations are not needed to account for this effect even
though they necessarily occur in nature.  A cluster study of the
frustrated, two-dimensional Hubbard model using CDMFT (ie. 2x2 plaquette)
~\cite{Parcollet:2004} demonstrated that the  single site DMFT
statement of the existence of a finite temperature Mott
transition survives cluster corrections, even though qualitative
modifications of the single site DMFT results appear at lower
temperatures or very close to the transition.  Finally new
numerical approaches to treat systems directly on the lattice
have further corroborated the qualitative validity of the single
site DMFT results~\cite{Onoda:2003}.

\begin{figure}[htbp]
\centerline{\includegraphics[width=0.85\hsize]{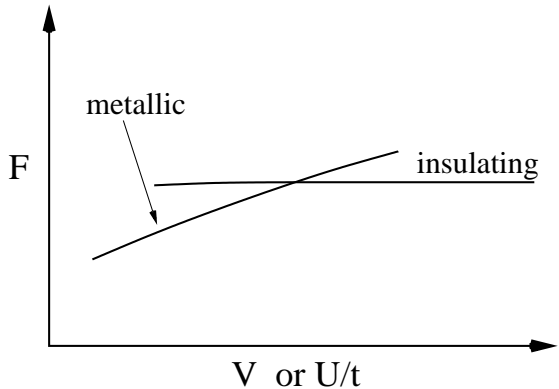}}
\caption{Schematic volume dependence of free energy for a model within
 DMFT.
}
\label{fig:coex}
\end{figure}
The fact that the coupling of the lattice is important near the
electronically driven Mott transition was first pointed out in
the dynamical mean field context in ref
\cite{Majumdar:1994}.
The electronic degrees of freedom are divided
into those described  by the low energy model Hamiltonian
(see section \ref{sec:INTmod}) and the rest, and  the total free energy
of the system is given by the sum of these two contributions
$F_{other}$
and $F_{model}$.
These free energies depend on the volume of the material.
Formally the energy of the model Hamiltonian is a function
of the model Hamiltonian parameters such as
the bandwidth $t$ and
Coulomb interaction $U$, but these parameters
themselves depend on volume.

We have seen that in the absence  of elastic interactions
the Hubbard model has two solutions,
a metallic and an insulating one, in a range
of values of  $U/t$. Hence  $F_{model}(t(V))$ can have
two branches which cross as depicted in Fig.~\ref{fig:coex}.
The  free energy curve obtained by picking at each volume the
lowest of the free energies
has a cusp singularity (an infinitely negative second derivative
at the critical volume)indicating the
formation of a double well structure.

The addition of $F_{other}$, which by construction is smooth,
cannot qualitatively modify this behavior. Furthermore, the
double well structure, which must exist below the Mott transition
temperature, also must persist slightly above the Mott
transition point (given the infinite second derivative at the
critical volume below the transition point of the model
Hamiltonian).  The position where the double well develops
signals the position of the true (i.e. renormalized by
the lattice ) metal to insulator transition.
The exact free energy  is a concave function of
the volume and this concavity which is missed in mean field theory,
is restored through a Maxwell construction.

This qualitative discussion
sketches how the  spectral density functional theory
formalism is used to predict
the volume of materials starting from first principles.
The self-consistent application of LDA+DMFT
determines the energy of  model Hamiltonian
and the one electron Hamiltonian  of both the low
energy and the high energy  degrees of freedom in
a self-consistent fashion.
Results for Ce and Pu are shown in
Figs.~\ref{fig:Ce_etot_0208443} and ~\ref{FigPuTotalEnergy},
respectively. In materials where the model exhibits a transition, the
LDA+DMFT studies produce a double well as will be discussed for Ce and
Pu in sections \ref{subsec:Ce} and \ref{subsec:Pu}, respectively.

It has been recently emphasized  ~\cite{Amadon:2005:CM0504732}
that in Cerium the double well is of purely entropic nature,
while the calculations for Pu only include the energy, but the
qualitative  argument for the existence of a double well applies
to the free energy of both materials.  An analysis of the
influence of the coupling to the lattice on the compressibility
and the electron-phonon coupling has been carried out in
~\cite{Hassan:2004}.

\subsubsection{Doping driven metal--insulator transition}

Doping driven metal--insulator transitions in the
three--dimensional perovskites La$_{1-x}$Sr$_{x}$TiO$_{3-\delta}$
has been extensively explored in the past decade
\cite{Tokura:1993:PRL, Sunstrom:1992, Maeno:1990, Crandles:1992,
Onoda:1997a, Onoda:1997b, Onoda:1998, Hays:1999}. The electronic
properties of the La$_{1-x} $Sr$_{x}$TiO$_{3}$ series is governed
by the $t_{2g}$ subset of the $3d$ orbitals. When $ x =0 $, there
is one electron per   Ti, and the system is a Mott insulator.
Doping with Strontium  or Oxygen introduces  holes in the   Mott
insulator.

In the cubic structure the $t_{2g}$ orbital is threefold
degenerate, but this degeneracy is lifted by an orthorhombic
distortion of the GdFeO$_{3}$ structure resulting in the space
group $Pbnm$. For $x>0.3$ the material is found to transform to
another distorted perovskite structure with space group $Ibmm$.
For larger values of $x>0.8$ the orthorhombic distortion vanishes
and the material assumes the cubic perovskite structure of
SrTiO$_{3}$ with space group $Pm3m$. LaTiO$_3$ is a Mott insulator
which orders antiferromagnetically at $T_{N}\approx 140$~K, with
a Ti magnetic moment of 0.45~$\mu _{B}$ and small energy gap of
approximately 0.2~eV.

The lifting of the degeneracy plays a very important role for
understanding the insulating properties of this compound, and
they have recently been discussed by a  single--site  DMFT study
of this compound~\cite{Pavarini:2004:PRL}. For moderate dopings
La$_{1-x} $Sr$_{x}$TiO$_{3}$ behaves as a canonical doped Mott
insulator. The specific heat and the susceptibility are enhanced,
the Hall coefficient is unrenormalized, and the photoemission
spectral function has a resonance with a weight that decreases as
one approaches half filling. Very near half filling, (for dopings
less than 8~\%) the physics is fairly complicated as there is an
antiferromagnetic metallic phase ~\cite{Kumagai:1993, Okada:1993,
Onoda:1998}. While it is clear that the parent compound is an
antiferromagnetic Mott insulator, the orbital character of the
insulator is not well understood, as recent Raman scattering
~\cite{Reedyk:1997} and neutron scattering investigations reveal
\cite{Furukawa:1997, Furukawa:1999}.

Very near half--filling when  the Fermi energy becomes very small
and comparable with the exchange interactions and   structural
distortion energies,  a treatment beyond single--site DMFT becomes
important in order to treat spin degrees of freedom.
Alternatively, for moderate and large dopings, the Kondo energy
is the dominant energy and DMFT is expected to be accurate. This
was substantiated by a series of papers reporting DMFT
calculations of a single--band or multiband Hubbard model with a
simplified density of states. Ref. \cite{Rozenberg:1994}
addressed the enhancement of the magnetic susceptibility and the
specific heat as the half--filling is approached. The optical
conductivity and the suppression of the charge degrees of freedom
was described in Ref.~\onlinecite{Rozenberg:1996}, while the
observation that the Hall coefficient is not renormalized was
reported in Ref.~\onlinecite{Kotliar:1996, Kajueter:1997}. The
thermoelectric power was investigated by P\'{a}lsson and Kotliar
\cite{Palsson:1998} and the magnetotransport by Lange and
Kotliar~\cite{Lange:1999}.

Given the simplicity of the models used and the various
approximations made in the solution of the DMFT equations, one
should regard the qualitative agreement with experiment as very
satisfactory. The photoemission spectroscopy of this compound as
well as of other transition metal compounds do not completely
reflect the bulk data, and it has been argued that disorder
together with modeling of the specific surface environment is
required to improve the agreement with experiment
~\cite{Sarma:1996, Maiti:2001}. More realistic studies were
carried out using LDA+DMFT.  The results are  weakly sensitive
to  the basis set used while more sensitive to the value of the
parameter $U$ and impurity solver. This  was   discussed in
section ~\ref{sec:IMP}, and it becomes very critical for
materials near the Mott transition since different impurity
solvers give slightly different values of critical $U$, and hence
very different physical spectra for a given value of $U$
\cite{Nekrasov:2000, Held:2001:IJMPB}. However, if we concentrate
on trends, and take as a given that $U$ should be chosen as to
place the material near or above the Mott transition, nice
qualitative agreement with experiment is obtained.

Anisimov \cite{Anisimov:1997}  considered a realistic Hamiltonian
containing Oxygen, Titanium, and Lanthanum bands and solved the
resulting DMFT equations using IPT. Nekrasov et
al.~\cite{Nekrasov:2000} solved the DMFT equations using
$t_{2g}$ density of states obtained from an LMTO calculation. In
this procedure, the bare density of states is rescaled so that it
integrates to one. They solved the DMFT equations using QMC, IPT
and NCA.

\begin{figure}[tbp]
\centering%
\includegraphics[clip=true,width=8cm] {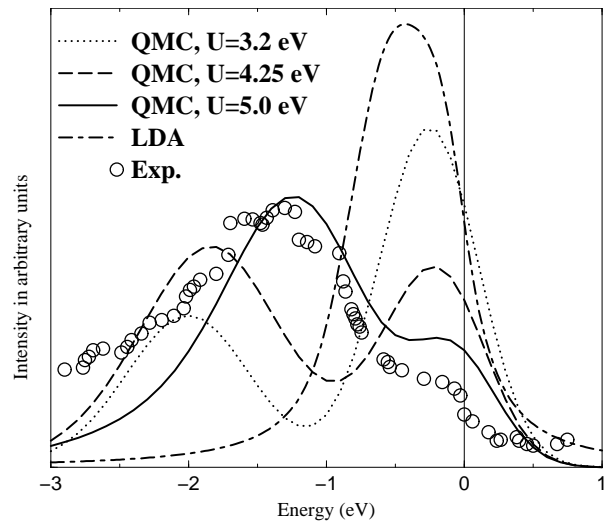}
\caption{Comparison of the experimental photoemission spectrum
\protect~\cite{Fujimori:1992a,Fujimori:1992,Yoshida:2002}, the
LDA result, and the LDA+DMFT(QMC) calculation for LaTiO$_{3}$
with 6\% hole doping and different Coulomb interaction $U=3.2$,
4.25, and 5~eV (from Ref.~\protect\cite{Nekrasov:2000}).}
\label{fig:LaTiO3_cm0005205}
\end{figure}

Comparison of photoemission experiments with results obtained
using the QMC impurity solver for different values of $U$ is presented
in Fig.~\ref{fig:LaTiO3_cm0005205}. One can find favorable
agreement between  experimental and LDA+DMFT results for
$U=5$~eV. LDA+DMFT reproduce the quasiparticle and Hubbard bands
while LDA captures only the spectra around the Fermi level.

\begin{figure}[h]
\centering%
\epsfig{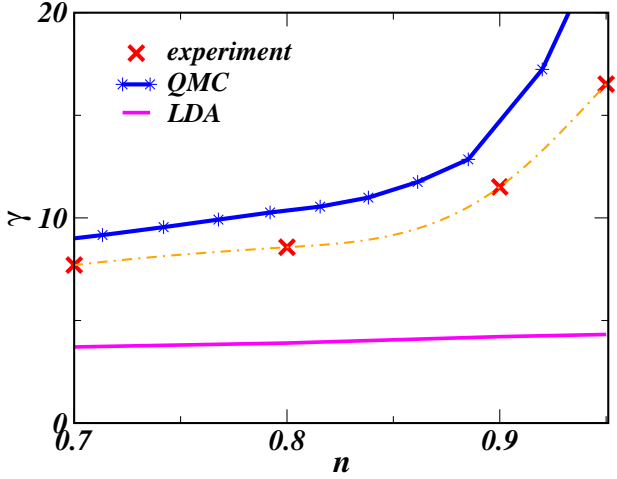}
\caption{Filling dependence of the linear coefficient of specific
heat, $\protect\gamma$, of doped LaTiO$_{3}$  obtained from DMFT
calculations using QMC as an impurity solver (solid line with
stars) with $U=5$, temperature $\protect\beta=16$ and LDA
calculations (solid line). Experimental points are given by
crosses and a dot--dashed line is used as a guide for eye.
Tight--binding density of states was used in the
self--consistency loop of the DMFT procedure. Energy unit is set
to half bandwidth.} \label{fig:Z_qmc_sunca}
\end{figure}

The linear term of the specific heat coefficient was computed by
fitting the $t_{2g}$ density of states to a tight--binding
parameterization. To capture the asymmetry in tight--binding DOS
the next nearest neighbor term, $t'$, on Ti sublattice have to be
taken into account. The dispersion which has been obtained from
the fit is  $\epsilon_{\bf k}=2 t (\cos k_x + \cos k_y) +2 t'
\cos(k_x+k_y) + 2 t_\perp \cos k_z$, where $t=-0.329664$,
$t'=-0.0816$, $t_{\perp}=-0.0205$ in eV units.  Using the
tight--binding DOS and the QMC impurity solver, the Green's
function and the specific heat were calculated. The specific heat
is given in terms of the density of states $N(\mu )$ at the Fermi
level by $\gamma =2.357\left[ \frac{mJ}{molK^{2}}\right]
\frac{N(\mu )[\mbox{states/(eV unit cell)}]}{Z}$ where $Z$ is the
quasiparticle residue or the inverse of the electronic mass
renormalization. In the LDA, the value of $Z$ is equal to one and
the doping dependence can be computed within the rigid band
model. The LDA+DMFT results are plotted against the experiment in
Fig.~\ref{fig:Z_qmc_sunca}. Despite some discrepancies, there is
good semiquantitative agreement.

In general, the LDA data for $\gamma $ are much lower than the
experimental values, indicating a strong mass renormalization.
Also we note that as we get closer to the Mott--Hubbard
transition the effective mass grows significantly. This is
consistent with DMFT description of the Mott--Hubbard transition,
which exhibits divergence of the effective mass at the transition.

\begin{figure}[h]
\centering\epsfig{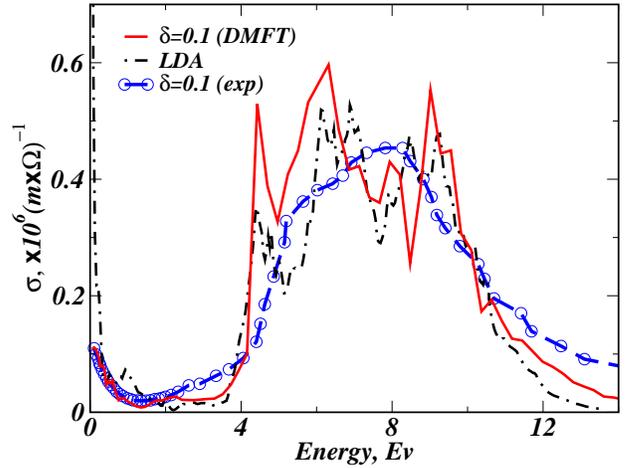}
\newline
\caption{Calculated optical conductivity spectrum for
La$_{x}$Sr$_{1-x}$TiO$_{3}$, $x=0.10$, at large frequency
interval using DMFT method as compared with the
experimental data and results of the corresponding
LDA calculations.} \label{fig:LaTiO3_optics_h4}
\end{figure}

\begin{figure}[h]
\epsfig{file=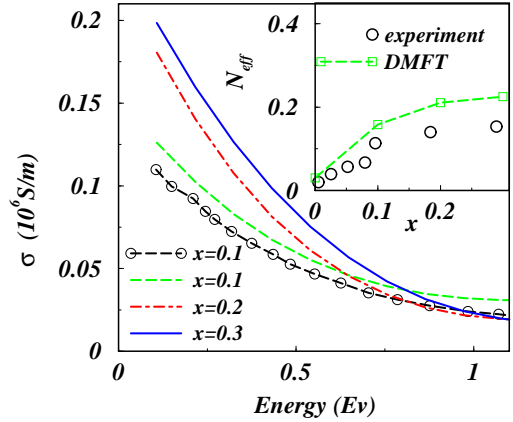,width=2.8in} 
\caption{Low frequency behavior of the optical conductivity for
La$_{1-x}$Sr$%
_{x}$TiO$_{3}$ at $x=0.1, 0.2, 0.3$ calculated using the LDA+DMFT method.
Experimental results~\protect\cite{OpticsLTO2} are shown by symbols for the
case $x=0.1$. In the inset the effective number of carriers is plotted as a
function of doping. Squares show the results of the LDA+DMFT calculations.
Circles denote the experimental data from Ref.~\protect\cite{OpticsLTO2}.}
\label{fig:low_freq}
\end{figure}

Oudovenko et al.~\cite{Oudovenko:2004} considered the optical
properties of La$_{1-x}$Sr$_{x}$TiO$_{3}$. The trends are in
qualitative agreement with those of earlier model studies
\cite{Kajueter:1997} but now the calculations incorporate the
effects of realistic band structures. In
Fig.~\ref{fig:LaTiO3_optics_h4} we plot the calculated optical
conductivity for La$_{x}$Sr$_{1-x}$TiO$_{3}$ at doping $x=0.1$
using the DMFT (solid line) and compare it with the experimental
data (dashed line with open cycles symbols) measured by Fujishima
et al. \cite{Fujishima:1992} and with the LDA calculations
(dot--dashed line).
The low frequency behavior for a range of dopings is
shown in Fig.~\ref{fig:low_freq}.

First we notice that the DMFT result agrees with the experiment
up to the energy of 2~eV. Above 2~eV,  both the LDA and DMFT
optics are quite close and fit the experiment reasonably well.

Its worth  emphasizing that corresponding calculations based on
the local density approximation would completely fail to reproduce
the doping behavior of the optical conductivity due to the lack
of the insulating state of the parent compound LaTiO$_{3}$ within
LDA. As a result, the LDA predicts a very large Drude peak even
for zero doping, which remains \emph{nearly unchanged} as a
function of doping. In view of this data, DMFT captures the
correct trend upon doping as well as the proper frequency
behavior, which is a significant improvement over the LDA.

\subsubsection { Further developments }

Understanding the simplest prototypes of the Mott transition
within DMFT has opened the way to many investigations of modified
and generalized models which are necessary to understand the rich
physics of real materials. Materials with unfilled bands and very
different bandwidth near the Fermi level (examples include
ruthenates Ca$_{2-x}$Sr$_ x$RuO$_4$~\cite{Anisimov:2002}, CrO$_2$
~\cite{Toropova:2005}, cobaltates
~\cite{Ishida:2005,Lechermann:2005:CM0505241}, the classic Mott
insulators VO$_2$~\cite{Goodenough:1971} and V$_2$O$_3$
~\cite{Ezhov:1999}, in layered organic superconductors
~\cite{Lefebvre:2000}, fullerenes ~\cite{Takenobu:2000}, and many
other compounds~\cite{Imada:1998}) %
raise the possibility of an orbitally
selective Mott transition, where
upon increase of the interaction $U$ one band can turn into an
insulator while the other one remains metallic.


This was observed first in connection with the
$\mathrm{Ca_{2-x}Sr_x Ru O_4}$ system
\cite{Nakatsuji:2000,Nakatsuji:2003} and also the
$\mathrm{La_{n+1} Ni_n O_{3n+1}}$ system
\cite{Sreedhar:1994,Zhang:1994,Kobayashi:1994}. The qualitative
idea is that when two bands differ substantially in bandwidth, as
the interaction strength is increased, there should be a sequence
of Mott transitions, whereby first narrow band undergoes a
localization transition with the broader band remaining itinerant
while at large $U$ both bands are localized. The term orbital
selective Mott transition (OSMT) was given for this situation
\cite{Anisimov:2002}.

%
%
%
The Hubbard interaction is separated into $U$ among opposite
spins, $U'$ among different orbitals, and the flipping term,
which is proportional to Hund's coupling $J$.
%

This problem is receiving much attention recently.
%
%
Most of the work was focused on the case of symmetric bands in the
particle hole symmetric point. It has been shown that an OSMT is
possible provided that the difference in bandwidth $t^h/t^l$ is
small enough.  The requirement on this ratio is made less extreme
and therefore the OSMT more clearly visible as $J$ is increased,
in particular if its spin rotationally invariant form is
treated~\cite{Anisimov:2002,Koga:2004,Koga:2005:CM0503651,
Ferrero:2005:CM0503759, Medici:2005:CM0503764,
Arita:2005:CM0504040, Knecht:2005:CM00505106}.

Interestingly, it was shown in
references~\cite{Biermann:2005:CM0505737} that in the regime
where the heavy orbital is localized and the light orbital is
itinerant, the heavy orbital forms a moment which scatters the
light electron resulting in the type of non Fermi liquid behavior
that was first found in the context of the Falikov Kimball model,
and studied extensively in the context of manganites. This type
of non Fermi liquid, containing light electrons scattered by
local collective modes, is then realized in many situations in
transition metal oxides.

The effects of interactions containing three electrons in the
heavy bands and one electron in the light band have not been
studied in detail. They can be eliminated at the expenses of
generating a hybridization term, which at least in some cases,
has been shown to be a relevant perturbation turning the
insulating band into a metallic state via hybridization ~\cite{Medici:2005:CM0502563}.
Finally, we note that understanding this
problem in the context of real materials will require the
interplay of LDA band theory and DMFT many body calculations.

Bands are not necessarily symmetric and their center of gravity may be
shifted relative to each other. The fundamental issue is how
crystal field splittings and spin orbit splittings are
renormalized by many body interactions. This is important not only for
experiments which measure the orbital occupancies but also because
%
%
%
the renormalization of the crystal field splitting
is a  relevant perturbation that can modify dramatically the
nature of the OSMT. This is already seen in the atomic limit,
where shifts in $\epsilon_h - \epsilon_l$ can have dramatic
effect, even in static approaches. For example, in
V$_2$O$_3$~\cite{Held:2005:PB,Held:2001V2O3,Keller:2005} the
``heavy" $a_{1g}$ orbital is quickly renormalized below the
bottom of the LDA conduction band. Notice that  if the heavy
orbital is moved well above the light orbital, one can have an
essentially weakly correlated situation, while if the heavy
orbital moves well below the light orbital and if the number of
electrons is such that the heavy orbital is not full, one
encounters a situation where local collective modes scatter the
light electrons.

A related issue is how interactions renormalize the Fermi surface
beyond the LDA Fermi surface. While in many cases the LDA Fermi
surface provides a good approximation to the true Fermi surface of
many materials such as heavy fermions, there are materials where
this is not the case. The issue was first raised in connection with De
Haas van Alphen and photoemission experiments on CaVO$_3$
\cite{Inoue:2002} and SrRuO$_4$ \cite{Kikugawa:2004} (for a
review see \cite{Mackenzie:2003}).
This problem was first
approached theoretically by Liebsch and Lichtenstein using DMFT
\cite{Liebsch:1999} and then by \cite{Pavarini:2004:PRL,
Okamoto:2004b, Lechermann:2005, Ishida:2005, Zhou:2005}. For a
review and substantial new information on this topic
see~\cite{Lechermann:2005:CM0505241}.
Indeed, the shape of the Fermi surface is easily extracted from the
LDA+DMFT Green's function  from the zeros of the
eigenvalues of the matrix $h^{(LDA)}(\vk) +{\cal M}_{int}(0)-\mu-{\cal M}_{dc}$.
%
%
The self-energy for a multi--orbital system treated within single--site DMFT
cannot in general be absorbed in a chemical potential shift, even if the self-energy
at zero frequency is diagonal, and therefore affects the shape Fermi surface.
Moreover, let us notice that since the form of the double counting enters
explicitly in the equation, a definitive answer to this issue will
require the first principles calculations of this term, for example
using the GW+DMFT technique described in section \ref{sec:Approximations}.

\subsection{Volume collapse transitions}
\label{sec:MATvol}

\begin{figure}[tb]
\epsfig{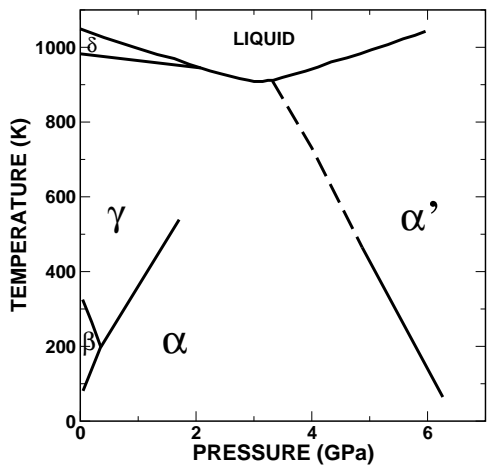}%
\caption{ The phase diagram of Cerium.} %
\label{fig:Ce_PD}
\end{figure}

Several rare earths and actinide materials undergo dramatic phase
transitions as a function of pressure characterized by a first-order
volume decrease upon compression. A classical example of this
behavior is the alpha to gamma ($\alpha \rightarrow \gamma $)
transition in Cerium (see phase diagram in Fig.~\ref{fig:Ce_PD}),
where the volume change is of the order of 15 percent, but
similar behavior is observed in Pr and Gd (for a review see:
\cite{McMahan:1998}).
A volume collapse transition as a function of pressure was also
observed in Americium, at around 15~GPa ~\cite{Lindbaum:2001}.
However, unlike the $\alpha \rightarrow \gamma $ transition which
is believed to be isostructural, or perhaps having a small
symmetry change \cite{Eliashberg:1998, Nikolaev:1999,
Nikolaev:2002}, the volume changing transitions in actinides are
accompanied by changes in the structure.

In the larger volume phase the $f$--electrons are more localized
than in the smaller volume phase, hence, the volume collapse is a
manifestation of the localization--delocalization phenomenon. The
susceptibility measurements indicate that, for example, in Ce the
$\gamma $--phase is paramagnetic with well defined spins while
the $\alpha $--phase is non--magnetic. The challenge is to
understand how small changes in pressure and temperature lead to
phases with different physical properties. A similar challenge is
also posed by the generalized (Smith--Kmetko) phase diagram of
actinides whereby one interpolates between different elements by
alloying. Metallic Plutonium displays a sequence of phase
transitions as a function of temperature between phases with very
different volumes, and the physics of the
localization--delocalization phenomena is believed to be
important for their understanding ~\cite{Johansson:1995,
Savrasov:2001}.
Realistic calculations have been
performed by McMahan and collaborators~\cite{McMahan:2003} for Ce
and Savrasov et al.~\cite{Savrasov:2001} for Pu. Both groups
concluded that while the localized picture of both materials is
important, the delocalized phases ($\alpha $--Pu and $\alpha
$--Ce) are not weakly correlated. This is also in agreement with
recent optical measurements in Ce \cite{Eb:2001}. From the DMFT
point of view, the ``metallic phase" is more correlated than a
naive band picture would suggest, having not only quasiparticles
but some weight in the Hubbard band.

\subsubsection{Cerium}
\label{subsec:Ce}


Johansson proposed a Mott transition scenario
~\cite{Johansson:1974}, where the transition is connected to the
delocalization of the $f$--electron. In the alpha--phase the
$f$--electron is itinerant while in the gamma--phase it is
localized and hence does not participate in the bonding. In the
absence of a theory of a Mott transition, Johansson and
collaborators \cite{Johansson:1995} implemented this model by
performing LDA calculations for the alpha--phase, while treating
the $f$--electrons as core in the gamma--phase.

Allen and Martin~\cite{Allen:1982} proposed the Kondo volume
collapse model for the $\alpha \rightarrow \gamma $ transition.
Their crucial insight was that the transition was connected to
changes in the spectra, resulting from modification in the
effective hybridization of the $spd$- band with the $f$--electron.
In this picture what changes when going from alpha to gamma is the
degree of hybridization and hence the Kondo scale. In a series of
publications ~\cite{Liu:1992, Allen:1992} they implemented this
idea mathematically by estimating free energy differences between
these phases by using the solution of the Anderson--Kondo
impurity model supplemented with elastic energy terms. The modern
dynamical mean--field theory is a more accurate realization of
both the volume collapse model and the Mott transition model. In
fact, these two views are not orthogonal, as it is known that the
Hubbard model is mapped locally to an Anderson model satisfying
the DMFT self--consistency condition. Furthermore, near the Mott
transition this impurity model leads to a local picture which
resembles the Kondo collapse model.


The Cerium problem was recently studied by Z\"{o}lfl et al.
\cite{Zolfl:2001} and by McMahan and collaborators
~\cite{McMahan:2003, Held:2001}. Their approach consists of
deriving a Hamiltonian consisting of an $spd$--band and an
$f$--band, and then solving the resulting Anderson lattice model
using DMFT. McMahan et al. used constrained LDA to evaluate the
position of the $f$--level as well as the value of the
interaction $U$. The hopping integrals are extracted from the LDA
Hamiltonian written in an LMTO basis. Z\"{o}lfl et al. identified
the model Hamiltonian with the Kohn--Sham Hamiltonian of the LDA
calculation in a tight--binding LMTO basis after the $f$--level
energy is lowered by $U(n_{f}-\frac{1}{2})$.

Strong hybridization not only between localized $f$ orbitals
but also between localized $f$ and delocalized $spd$--orbitals
is the main reason to go beyond the standard AIM or PAM and to
consider the Hamiltonian with the full ($s,p,d,f$) basis set. The
starting one--particle LDA Hamiltonian is calculated using the
LMTO method considering $6s$-, $6p$-, $5d$-, and $4f$--shells.
Claiming small exchange and spin--orbit interactions both groups
used $SU(N)$ approximation to treat the $f$--orbitals with the
Coulomb repulsion $U_{f}\approx 6$~eV. McMahan et
al.~\cite{McMahan:2003} used $U_{f}=5.72$ and 5.98~eV for $\alpha
$ and $\gamma $--Ce correspondingly) extracted from the
constrained LDA calculations.

\begin{figure}[tb]
\epsfig{file=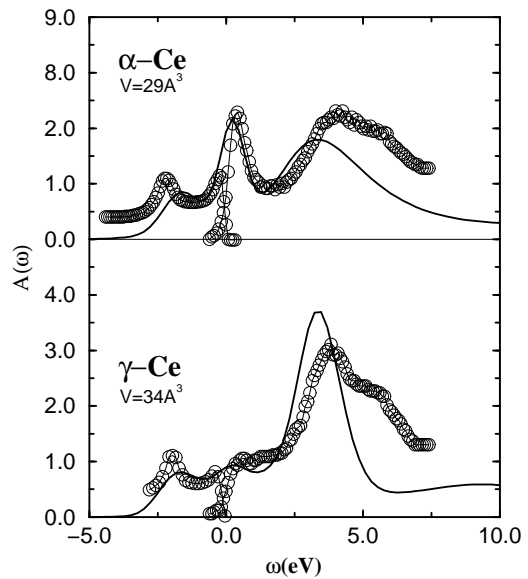,width=2.5in}
\caption{Comparison of the LDA+DMFT(QMC) (solid line) spectra with
experiment (circles) \protect\cite{Liu:1992} (after \protect\cite%
{McMahan:2003}). } \label{fig:Ce_totspec_0208443}
\end{figure}

The differences between these two approaches are attributed to
the impurity solvers used in the DMFT procedure and to the range
of studied physical properties. Z\"{o}lfl et al. used the NCA
impurity solver to calculate the one--particle spectra for
$\alpha $- and $\gamma $--Ce, Kondo temperatures, and
susceptibilities while the McMahan group used QMC and Hubbard~I
methods to address a broader range of physical properties of Ce.
Thermodynamic properties such as the entropy, the specific heat,
and the free energy are studied by McMahan et
al.~\cite{McMahan:2003} in a wide range of volume and
temperatures in search of a signature of the $\alpha -\gamma $
transition. The details of the spectral function obtained in both
publications differ somewhat mostly due to different impurity
solvers used (NCA and QMC) but the qualitative result, a three
peak spectra for $\alpha $--Ce and two peak spectra of $\gamma
$-Ce, is clear for both methods (the spectra from Ref.
~\onlinecite{McMahan:2003} are presented in
Fig.~\ref{fig:Ce_totspec_0208443}). The Kondo temperatures,
$T_{K,\alpha }\approx 1000$~K and $T_{K,\gamma }\approx 30$~K
obtained by Z\"{o}lfl et al. as well as $T_{K,\alpha }\approx
2100$~K and $T_{K,\gamma }<650$~K obtained by McMahan group, are
reasonably close to the experimental estimates of $T_{K,\alpha
}=945$~K and $1800-2000$ ~K as well as $T_{K,\gamma }=95$~K and
60~K extracted from the electronic ~\cite{Liu:1992} and
high--energy neutron spectroscopy ~\cite{Murani:1993},
correspondingly.

\begin{figure}[tb]
\epsfig{file=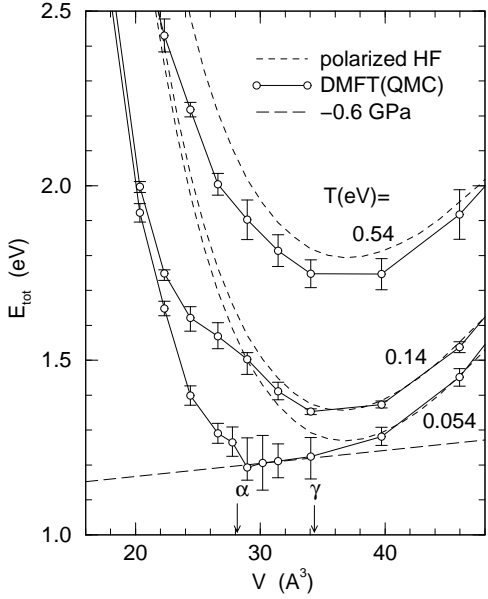,width=2.5in}
\caption{Total LDA+DMFT(QMC) and polarized Hartree--Fock (HF)
energy as a function of volume at three temperatures. While the
polarized HF energy has one pronounced minimum in the
$\protect\gamma$--Ce phase, the LDA+DMFT(QMC) shows a shallowness
($T=0.054,$~eV), which is consistent with the observed $
\protect\alpha$-$\protect\gamma$ transition (arrows) within the
error bars. These results are also consistent with the
experimental pressure given by the negative slope of the dashed
line. (from \protect\cite{McMahan:2003}) }
\label{fig:Ce_etot_0208443}
\end{figure}

To find thermodynamic evidence for the $\alpha \rightarrow \gamma
$ transition, the total energy was calculated. McMahan et al.
compute the total energy which consists of three terms:
all--electron LDA energy, DMFT total energy minus so called
``model LDA" energy which originates from the double counting
term in the DMFT calculations. Volume dependence of the total
energy $E_{tot}(eV)$ is reproduced in
Fig.~\ref{fig:Ce_etot_0208443}. It was found that the DMFT
contribution is the only candidate to create a region of the
negative bulk modulus. In other words, the correlation
contribution is the main reason for the thermodynamic instability
revealing itself in the first order phase transition. As seen
from Fig.~\ref{fig:Ce_etot_0208443} the minimum of the total
energy in the zero--temperature limit corresponds to the volume of
$\alpha $--phase, and for large temperature $T=0.14$~eV the
minimum shifts to higher values of volume roughly corresponding
to the $\gamma $--phase.

With increasing temperature, the contribution to free energy from
the entropy term becomes important. Hence one can look for
another signature of the $\alpha \rightarrow \gamma $ transition:
the behavior of the entropy. The transition was attributed to
rapid increase of the entropy in the region of volumes
28.2-34.4~\AA .\ At large volumes, when spectral weight of the
$4f$--electrons is removed from the Fermi level, the entropy
saturates at value $k_{B}\mathrm{ln}(2J+1)$, and in the case of
the $SU(N)$ approximation assumed in the calculation, the
logarithm tends to $\mathrm{ln}(14)$. For smaller volumes the
quasiparticle peak grows which causes changes in the specific
heat and hence in the entropy through $\int dTC(T)/T$,
substantially reducing it to smaller values.

So, the general qualitative picture which comes out from the
LDA+DMFT calculations is the following. At large volume ($\gamma
$--phase) the $4f$--spectrum is split into Hubbard bands and
therefore a local moment is present in the system. With volume
reduction a quasiparticle (Abrikosov--Suhl resonance) develops in
the vicinity of the Fermi level which causes a drop in the
entropy and disappearance of the local moment. Temperature
dependence of the quasiparticle peak indicated a substantially
larger Kondo temperature in the $\alpha $--phase than in $\gamma
$--Ce phase. The obtained results also suggest that $\gamma $- and
$\alpha $--phases of Ce are both strongly correlated.

Finally the optical properties of Cerium, were computed both in
the alpha and gamma phases by Haule et. al. ~\cite{Haule:2004:CM}.
These authors observed that the Kondo collapse and the Mott
transition scenario can be differentiated by measuring the
optical properties that are controlled by the light electrons, or
by studying theoretically the photoemission spectra of the
$spd$--electrons. In a Mott transition scenario, the
$spd$--electrons are mere spectators, not strongly affected by
the localization of the $f$'s. Alternatively, in the Kondo
collapse scenario a typical hybridization gap should open up in
the $spd$--spectra with a clear optical signature. Their
calculation as well as their interpretation of the  optical data
of Van Der Eb~\cite{Eb:2001} supports the Kondo collapse scenario.


\subsubsection{Plutonium}%
\label{subsec:Pu}

Many properties of Plutonium have been a long standing puzzle
\cite{Freeman:1974}. Pu is known to have six crystallographic
structures with large variation in their
volume~\cite{Hecker:2000}. Pu shows an enormous volume expansion
between $\alpha $- and $\delta $--phases which is about 25\%.
Within the $\delta $--phase, the system is metallic and has a
negative thermal expansion. Transition between $\delta $- and the
higher--temperature $\varepsilon $--phase occurs with a 5\% volume
collapse. Also, Pu shows anomalous resistivity behavior
\cite{Boring:2000} characteristic of a heavy fermion systems, but
neither of its phases are magnetic, displaying small and
relatively temperature independent susceptibility. Photoemission
\cite{Arko:2000} exhibits a strong narrow Kondo--like peak at the
Fermi level consistent with large values of the linear specific
heat coefficient, although inverse photoemission experiments have
not been done up to now.

Given the practical importance of this material and enormous
amount of past experimental and theoretical work, we first review
the results of conventional LDA and GGA approaches to describe
its properties. Electronic structure and equilibrium properties
of Pu were studied earlier by
\onlinecite{Solovyev:1991,Soderlind:1994} as well as recently by
\onlinecite{Jones:2000, Savrasov:2000, Wan:2000, Nordstrom:2000,
Soderlind:2001, Soderlind:2002, Kutepov:2003, Robert:2004}. Using
non--magnetic GGA\ calculations, it was found that the
equilibrium volume of its $\delta$--phase is underestimated by
20-30$\%$. The spread in the obtained values can be attributed to
different treatments of spin orbit coupling for the 6p semicore
states. In particular, the problem was recognized
\cite{Nordstrom:2000} that many electronic structure methods
employ basis sets constructed from scalar--relativistic
Hamiltonians and treat spin--orbit interaction variationally
\cite{Andersen:1975}. Within the Pauli formulation (i.e. when
only terms up to the order $\frac{1}{c^{2}}$ are kept) the
spin--orbit Hamiltonian is given by
\begin{equation}
\frac{2}{rc^{2}}\frac{dV}{dr}\mathbf{\hat{l}\hat{s}} , \label{SO1}
\end{equation}%
whose matrix elements are evaluated on the radial solutions of
the scalar relativistic version of the Schroedinger's equation
$\phi _{l}(r,E)$ carrying no total (spin+orbit) moment
dependence. It has been pointed out \cite{Nordstrom:2000} that in
the absence of proper evaluations of $\phi
_{j=1/2}(r,E_{p^{1/2}})$ $\phi _{j=3/2}(r,E_{p^{3/2}})$ orbitals,
one of the options is to neglect the spin--orbital interaction for
6p states completely. This results in the improvement of volume
which is of the order of 20\% smaller than the experiment as
compared to the relativistic Pauli treatment which gives a 30\%
discrepancy.

One can go beyond the Pauli Hamiltonian and treat the
spin--orbital Hamiltonian as an energy--dependent operator
\cite{Koelling:1977}.
\begin{equation}
\frac{2}{rc^{2}[1+\frac{1}{c^{2}}(E-V)]^{2}}\frac{dV}{dr}\mathbf{\hat{l}\hat{%
s}}  .\label{SO2}
\end{equation}%
For a narrow band, the energy in the denominator can be taken
approximately at the center of the band and the average of the
operator can be evaluated without a problem. Our own simulations
done with the full potential LMTO\ method show that the
discrepancy in atomic volume is improved from 27\% when using Eq.
(\ref{SO1}) to 21\% when using Eq.(\ref{SO2}) and appear to be
close to the results when the spin--orbit coupling for the 6 p
states is neglected. The origin of this improvement lies in a
smaller splitting between $6p^{1/2}$ and $6p^{3/2}$ states when
incorporating the term beyond $ \frac{1}{c^{2}}$.

Models with the assumptions of long range magnetic order have
also been extensively explored in the past
\cite{Solovyev:1991,Soderlind:1994,Savrasov:2000,Wan:2000,Soderlind:2001,
Soderlind:2002,Soderlind:2004,Kutepov:2003,Robert:2004}. While
none of Pu phases are found to be magnetic of either ordered or
disordered type \cite{Lander:2004} all existing density
functional based calculations predict the existence of long range
magnetism. Using GGA and imposing ferromagnetic order, the
predictions in the theoretical volumes for the $\delta $--phase
have ranged from underestimates by as much as 33\%
\cite{Savrasov:2000} to overestimates by 16\%
\cite{Soderlind:2002}. Again, the sensitivity of the results to
the treatment of the spin orbit coupling for 6p semicore needs to
be emphasized. Our most recent investigation of this problem
shows that a 33\% discrepancy found with simulation using the
Pauli Hamiltonian \cite{Savrasov:2000} can be removed if
Eq.(\ref{SO2}) is utilized. This makes the result consistent with
the calculations when spin--orbit coupling for 6p states is
completely omitted \cite{Soderlind:2002,Robert:2004} or when
using fully relativistic calculation \cite{Kutepov:2003}.

Despite the described inconsistencies between the theory and the
experiment for the $\delta $--phase, the volume of the
$\alpha $--phase is found to be predicted correctly by LDA
\cite{Jones:2000,Soderlind:2001,Kutepov:2003}
. Since the transport and thermodynamic properties of $\alpha $- and $%
\delta $--Pu are very similar, the nature of the $\alpha $--phase
and the LDA prediction by itself is another puzzle.

Several approaches beyond standard LDA/GGA schemes have been
implemented to address these puzzles. The LDA+U method was
applied to $\delta $--Pu \cite{Bouchet:2000,Savrasov:2000}. It is
able to produce the correct volume of the $\delta $--phase, for
values of the parameter U$\sim $4~eV consistent with atomic
spectral data and constrained density functional calculations.
However, the LDA+U calculation has converged to the artificial
magnetically ordered state. This method is unable to predict the
correct excitation spectrum and to recover the $\alpha $--phase.
To capture these properties, $U$ must be set to zero. Another
approach proposed  in the past \cite{Eriksson:1999} is the
constrained LDA approach in which some of the 5f--electrons, are
treated as core, while the remaining electrons are allowed to
participate in band formation. Results of the
self--interaction--corrected LDA calculations have been also
discussed \cite{Setty:2003,Svane:1999}. Recent simulations based
on the disordered local moment method \cite{Niklasson:2003} have
emphasized that the volume of the $\delta $--Pu can be recovered
without an assumption of long range magnetic order.

The dynamical mean--field theory offers a way to deal with the
problem using the full frequency resolution. Within DMFT the
magnetic moments do not need to be frozen and can live at short
time scales while giving zero time average values. Also,
performing finite temperature calculation ensures that the
various orientations of moments enter with proper statistical
weights. The LDA+DMFT calculations using Pauli Hamiltonians have
been reported in \cite{Dai:2003,Savrasov:2001}. To illustrate the
importance of correlations, the authors
\cite{Savrasov:2003,Savrasov:2001} discussed the results for
various strengths of the on--site Coulomb interaction U. The
total energy as a function of volume of the FCC lattice is
computed for $T=600$~K using the self--consistent determination
of the density in a double iteration loop as described in
Section~\ref{sec:SDF}. The total energy is found to be
dramatically different for non--zero $U$ with the possibility of
a double minimum for $U^{\prime }s \cong 4$~eV which can be
associated with the low volume $\alpha $- and high--volume $\delta
$--phases.

\begin{figure}[tbh]
\includegraphics[height=3.0in]{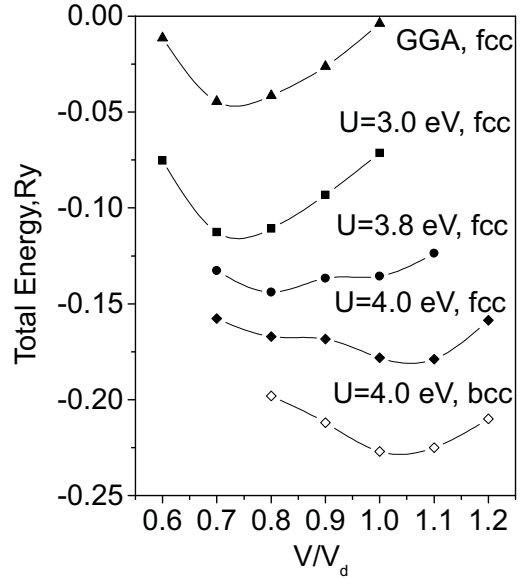}%
\caption{Total energy as a function of volume in Pu for different
values of $U$ calculated using the LDA+DMFT approach. Data for
the FCC lattice are computed at $T=600$~K, while data for the BCC
lattice are given for $T=900$~K.}
\label{FigPuTotalEnergy}
\end{figure}

The calculations for the BCC structure using the temperature
$T=900$~K have  been also reported \cite{Savrasov:2003:CM0308053}.
Fig.~\ref{FigPuTotalEnergy} shows these results for $U= 4$~eV
with a location of the minimum around $V/V_{\delta } = 1.03$.
While the theory has a residual inaccuracy in determining the
$\delta $- and $\varepsilon $--phase volumes by a few percent, a
hint of volume decrease with the $\delta \rightarrow \varepsilon $
transition was clearly reproduced.

The values of $U\sim 4$~eV, which are needed in these simulations
to describe the $\alpha \rightarrow \delta $ transition, were
found to be in good agreement with the values of on--site Coulomb
repulsion between $f$--electrons estimated by atomic spectral
data \cite{Desclaux:1984}, constrained density functional studies
\cite{Turchi:1999}, and the LDA+U studies~\cite{Savrasov:2000}.

The double--well behavior in the total energy curve is
unprecedented in LDA or GGA based calculations but it is a
natural consequence of the proximity to a Mott transition.
Indeed, recent studies of model Hamiltonian systems
\cite{Kotliar:2002} have shown that when the $f$--orbital
occupancy is an integer and the electron--electron interaction is
strong, two DMFT solutions which differ in their spectral
distributions can coexist. It is very natural that allowing the
density to relax in these conditions can give rise to the double
minima as seen in Fig.~\ref{FigPuTotalEnergy}.

\begin{figure}[tbh]
\includegraphics*[height=2.8in]{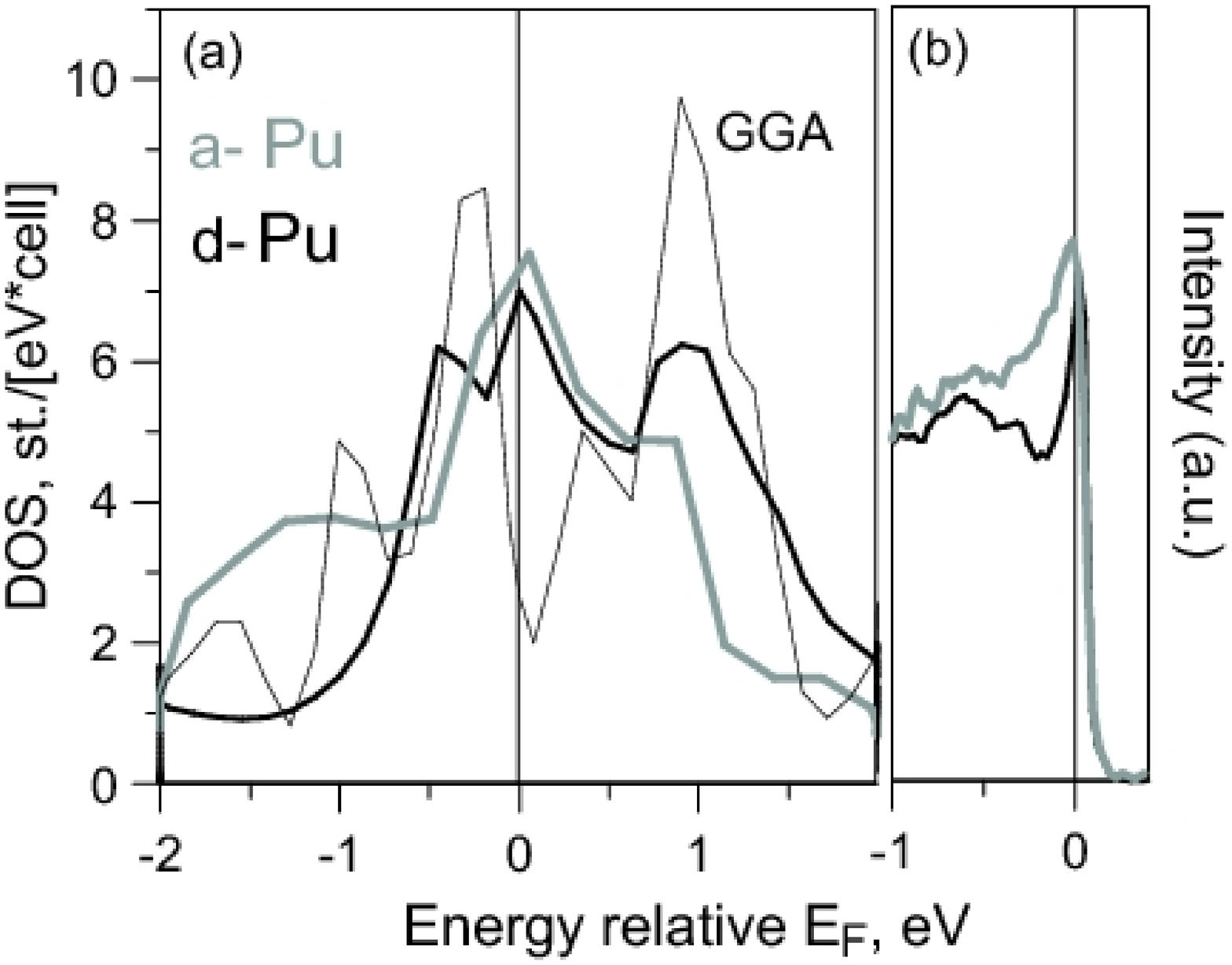}
\caption{a) Comparison between calculated density of states using
the LDA+DMFT approach for FCC Pu: the data for
$V/V_{\protect\delta }=1.05,\,U=4.0$~eV (thick black solid line),
the data for $V/V_{\protect\delta }=0.80,\,\,U=3.8$~eV (thick gray
line) which correspond to the volumes of the $\protect\delta $ and
$\protect\alpha $--phases respectively. The result of the GGA\
calculation (thin solid line) at $V/V_{\protect\delta
}=1\,\,(U=0)$ is also given. b) Measured photoemission spectrum of
$\protect\delta $ (black line) and $\protect\alpha $ (grey line)
Pu at the scale from -1.0 to 0.4~eV (after Ref.
\onlinecite{Arko:2000}).} \label{FigPuSpectra}
\end{figure}

The calculated spectral density of states for the FCC structure
using the volume $V/V_{\delta }=0.8$ and $V/V_{\delta }=1.05$
corresponding to the $ \alpha $- and $\delta $--phases have been
reported ~\cite{Savrasov:2001, Savrasov:2003:CM0308053}.
Fig.~\ref{FigPuSpectra} compares the results of these dynamical
mean--field calculations with the LDA method as well as with the
experiment. Fig.~\ref{FigPuSpectra}~(a) shows density of states
calculated using LDA+DMFT method in the vicinity of the Fermi
level. The solid black line corresponds to the $\delta $--phase
and solid grey line corresponds to the $\alpha $--phase. The
appearance of a strong quasiparticle peak near the Fermi level
was predicted in both phases. Also, the lower and upper Hubbard
bands can be clearly distinguished in this plot. The width of the
quasiparticle peak in the $\alpha $--phase is found to be larger
by 30~percent compared to the width in the $\delta $--phase. This
indicates that the low--temperature phase is more metallic, i.e.
it has larger spectral weight in the quasiparticle peak and
smaller weight in the Hubbard bands. Recent advances have allowed
the experimental determination of these spectra, and these
calculations are consistent with these measurements
\cite{Arko:2000}. Fig.~\ref{FigPuSpectra} (b) shows the measured
photoemission spectrum for $\delta $- (black line) and $\alpha
$--Pu (gray line). A strong quasiparticle peak can clearly be
seen. Also a smaller peak located at 0.8~eV is interpreted as the
lower Hubbard band. The result of the local density approximation
within the generalized gradient approximation is shown in
Fig.~\ref{FigPuSpectra} (a) as a thin solid line. The LDA produces
two peaks near the Fermi level corresponding to 5$f^{5/2}$ and
5$f^{7/2}$ states separated by the spin--orbit coupling. The
Fermi level falls into a dip between these states and cannot
reproduce the features seen in photoemission.

\begin{figure}[tbh]
\includegraphics[height=2.4in]{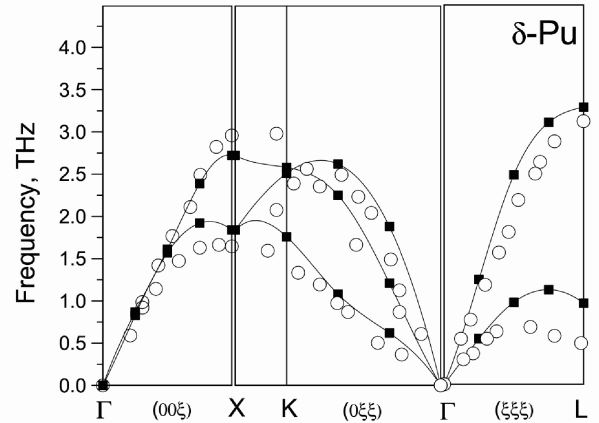}
\caption{Calculated phonon spectrum of $\protect\delta$--Pu
(squares connected by full lines) in  comparison with experiment
(open circles \protect\cite{Wong:2003})} \label{FigPuPhonons}
\end{figure}

A newly developed dynamical mean--field based linear response
technique ~\cite{Savrasov:2003} has been applied to calculate the
phonon spectra in $\delta $- and $\varepsilon
$--Pu~\cite{Dai:2003}. Self--energy effects in the calculation of
the dynamical matrix have been included using the Hubbard~I
approximation \cite{Hubbard:1963}. A considerable softening of
the transverse phonons is observed around the $L$ point in the
calculated frequencies as a function of the wave vector along
high--symmetry directions in the Brillouin zone for the $\delta
$--phase (see Fig.~\ref{FigPuPhonons}) This indicates that the
$\delta $--phase may be close to an instability with a doubling of
the unit cell. Another anomaly is seen for the transverse
acoustic mode along (011) which is connected to the non--linear
behavior of the lowest branch at small $q$. Overall, the phonon
frequencies are positive showing the internal stability of the
positions of the nuclear coordinates in $\delta $--Pu.
Remarkably, the experiment~\cite{Wong:2003} which came out after
the publication ~\cite{Dai:2003} has confirmed these theoretical
predictions. The measured points are shown on top of the
calculated curves in Fig.~\ref{FigPuPhonons}.

The presented results allow us to stress that, in spite of the
various approximations and their respective shortcomings
discussed in the text, it is clear that LDA+DMFT is a most
promising technique for studying volume collapse transition
phenomena. Besides the satisfying agreement with experiment
(given the crudeness of various approximations) both the studies
of Plutonium and Cerium have brought in a somewhat unexpected
view point for the electronic structure community. The
delocalized phase ($\alpha$--Plutonium and $\alpha$--Cerium) are
not weakly correlated, in spite of the success of density
functional theory in predicting their volume and elastic
properties.


\subsection{Systems with local moments}

\label{subsec:MATloc}

The magnetism of metallic systems has been studied intensively
\cite{Moriya:1985}. Metallic ferromagnets range from being very
weak with a small magnetization to very strong with a saturated
magnetization close to the atomic value. For a review of early
theories see e.g. Refs. \cite{Herring:1966, Vonsovsky:1974,
Moriya:1985}. Weak ferromagnets are well described by spin
density wave theory, where spin fluctuations are localized in a
small region of momentum space. Quantitatively they are well
described by LSDA. The ferromagnetic to paramagnetic transition
is driven by amplitude fluctuations. In strong ferromagnets,
there is a separation of time scales. $\hbar /t$ is the time
scale for an electron to hop from site to site with hopping
integral $t$, which is much shorter than $\hbar /J,$ the time
scale for the moment to flip in the paramagnetic state. The spin
fluctuations are localized in real space and the transition to
the paramagnetic state is driven by orientation fluctuations of
the spin. The exchange splitting $J$ is much larger than the
critical temperature.

Obtaining a quantitative theory of magnetic materials valid both
in the weak and strong coupling regime, both above and below the
Curie temperature, has been a theoretical challenge for many
years. It has been particularly difficult to describe the regime
above $T_{c} $ in strong ferromagnets when the moments are well
formed but their orientation fluctuates. A related problem arises
in magnetic insulators above their ordering temperature, when this
ordering temperature is small compared to electronic scales. This
is a situation that arises in transition metal monoxides (NiO and
MnO) and led to the concept of a Mott insulator. In these
materials the insulating gap is much larger than the N\'{e}el
temperature. Above the ordering temperature, we have a collection
of atoms with an open shell interacting via superexchange. This
is again a local moment regime which cannot be accessed easily
with traditional electronic structure methods.

Two important approaches were designed to access the disordered
local moment (DLM) regime. One approach \cite{Hubbard:1979:I,
Hubbard:1979:II, Hubbard:1981} starts form a Hubbard like
Hamiltonian and introduces spin fluctuations via the
Hubbard--Stratonovich transformation \cite{Stratonovich:1958:sov,
Wang:1969, Evenson:1970, Cyrot:1970} which is then evaluated
using a static coherent potential approximation (CPA) and
improvements of this technique. A dynamical CPA
\cite{Al-Attar:1999} was developed by Kakehashi
\cite{Kakehashi:1992, Kakehashi:1998, Kakehashi:2002} and is
closely related to the DMFT ideas. A second approach begins with
solutions of the Kohn--Sham equations of a constrained LDA
approximation in which the local moments point in random
directions, and averages over their orientation using the
KKR--CPA approach \cite{Gyorffy:1979, Faulkner:1982}. The average
of the Kohn--Sham Green's functions then can be taken as the first
approximation to the true Green's functions, and information about
angle resolved photoemission spectra can be extracted
\cite{Gyorffy:1985, Satunton:1985}. There are approaches that are
based on a picture where there is no short range order to large
degree. The opposite point of view, where the spin fluctuations
far away form the critical temperature are still relatively long
ranged was put forward in the fluctuation local band
picture~\cite{Capellmann:1974, Korenman:1977:I, Korenman:1977:II,
Korenman:1977:III, Prange:1979:IV, Prange:1979:V}.

To describe the behavior near the critical point requires
renormalization group methods, and the low--temperature treatment
of this problem is still a subject of intensive research
\cite{Belitz:2002}. There is also a large literature on
describing ferromagnetic metals using more standard many--body
methods \cite{Liebsch:1981, Treglia:1982, Manghi:1997,
Manghi:1999, Nolting:1987, Steiner:1992}.

While the density functional theory can in principle provide a
rigorous description of the thermodynamic properties, at present
there is no accurate practical implementation available. As a
result, the finite--temperature properties of magnetic materials
are estimated following a simple suggestion
\cite{Liechtenstein:1987}. Constrained DFT at $T=0$ is used to
extract exchange constants for a \textit{classical} Heisenberg
model, which in turn is solved using approximate methods (e.g.
RPA, mean--field) from classical statistical mechanics of spin
systems \cite{Liechtenstein:1987, Rosengaard:1997, Halilov:1998,
Antropov:1996}. The most recent implementation of this approach
gives good values for the transition temperature of Iron but not
of Nickel \cite{Pajda:2001}. However it is possible that this is
the result of not extracting the exchange constants correctly,
and a different algorithm for carrying out this procedure was
proposed \cite{Bruno:2003}.

DMFT can be used to improve the existing treatments of DLM to
include dynamical fluctuations beyond the static approximation.
Notice that single--site DMFT includes some degree of short range
correlations. Cluster methods can be used to go beyond the
single--site DMFT to improve the description of short range order
on the quasiparticle spectrum. DMFT also allows us to incorporate
the effects of the electron--electron interaction on the
electronic degrees of freedom. This is relatively important in
metallic systems such as Fe and Ni and absolutely essential to
obtain the Mott--Hubbard gap in transition metal monoxides.

The dynamical mean--field theory offers a very clear description
of the local moment regime. Mathematically, it is given by an
effective action of a degenerate impurity model in a bath which is
sufficiently weak at a given temperature to quench the local
moment. This bath obeys the DMFT self--consistency condition. If
one treats the impurity model by introducing the
Hubbard--Stratonovich field and treats it in a static
approximation, one obtains very simple equations as those
previously used to substantiate the DLM picture.


\subsubsection{Iron and Nickel}
\label{subsec:FeNi}

Iron and Nickel were studied in Refs.
\onlinecite{Lichtenstein:2000, Lichtenstein:2001}. The values
$U=2.3$ $(3.0)$~eV for Fe (Ni) and interatomic exchange of
$J=0.9$~eV for both Fe and Ni were used, as obtained from the
constrained LDA calculations \cite{Bandyopadhyay:1989,
Anisimov:1997, Lichtenstein:1997, Lichtenstein:1998}. These
parameters are consistent with those of many earlier studies and
resulted in a good description of the physical properties of Fe
and Ni. In Ref. \onlinecite{Lichtenstein:2001} the general form
of the double counting correction $V_{\sigma
}^{DC}=\frac{1}{2}Tr_{\sigma }\mathcal{M}_{\sigma }(0)$ was
taken. Notice that because of the different self--energies in the
$e_{g}$ and $t_{2g}$ blocks the DMFT Fermi surface does not
coincide with the LDA Fermi surface.

The LDA+U method, which is the static limit of the LDA+DMFT
approach, was applied to the calculation of the magnetic
anisotropy energies~\cite{Imseok:2001}. This study revealed that
the double counting correction induces shifts in the Fermi
surface which brings it in closer agreement with the De Haas Van
Alphen experiments. The values of $U$ used in this LDA+U work are
slightly lower than in the DMFT work, which is consistent with
the idea that DMFT contains additional screening mechanisms, not
present in LDA+U. This can be mimicked by a smaller value of the
interaction $U$ in the LDA+U calculation. However, the overall
consistency of the trends found in the LDA+U and the DMFT studies
are very satisfactory.

More accurate solutions of the LDA+DMFT equations have been
presented as well. The impurity model was solved by QMC in Ref.~
\onlinecite{Lichtenstein:2001} and by the FLEX scheme in Ref.
\onlinecite{Katsnelson:2002}. It is clear that Nickel is more
itinerant than Iron (the spin--spin autocorrelation decays
faster), which has longer lived spin fluctuations. On the other
hand, the one--particle density of states of Iron closely
resembles the LSDA density of states while the DOS of Nickel,
below $T_{c}$, has additional features which are not present in
the LSDA spectra \cite{Iwan:1979, Eberhardt:1980, Altmann:2000}:
the presence of the famous 6~eV satellite, the 30\% narrowing of
the occupied part of $d$--band and the 50\% decrease of exchange
splittings compared to the LDA results. Note that the satellite in
Ni has substantially more spin--up contributions in agreement with
photoemission spectra \cite{Altmann:2000}. The exchange splitting
of the $d$--band depends very weakly on temperature from
$T=0.6T_{C}$ to $T=0.9T_{C}$. Correlation effects in Fe are less
pronounced than in Ni due to its large spin splitting and the
characteristic BCC structural dip in the density of states for
the spin--down states near the Fermi level, which reduces the
density of states for particle--hole excitations.

\begin{figure}[tbp]
\centerline{\epsfig{file=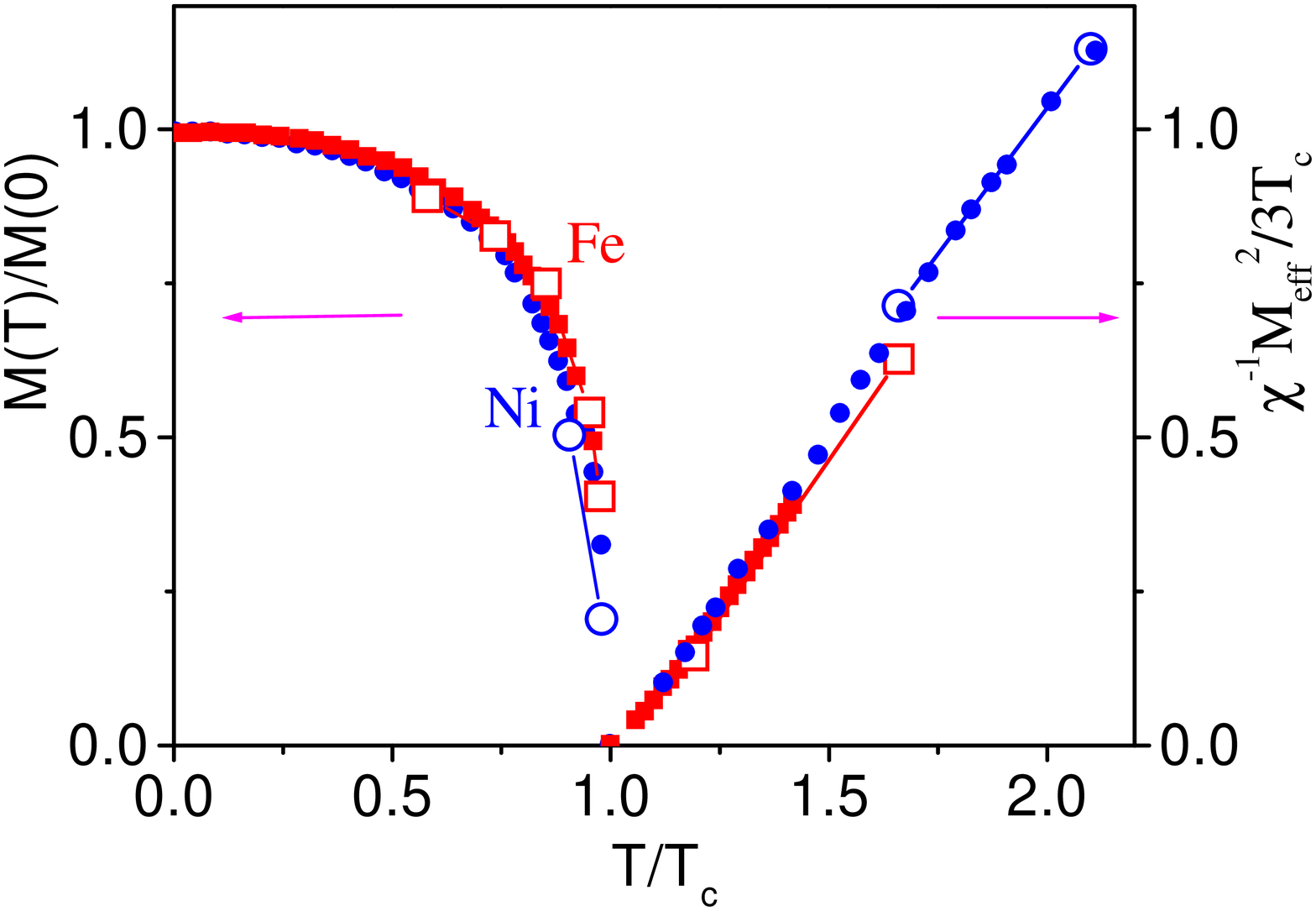,
width=9cm,height=7cm} } \caption{Temperature dependence of
ordered moment and the inverse ferromagnetic susceptibility for
Fe (open square) and Ni (open circle) compared with experimental
results for Fe (square) and Ni (circle) (from
Ref.~\onlinecite{Stocks:1998}). The calculated moments were
normalized to the LDA ground state magnetization (2.2
$\protect\mu_B$ for Fe and 0.6 $ \protect\mu_B$ for Ni). }
\label{fig:Mchi}
\end{figure}

The uniform spin susceptibility in the paramagnetic state, ${\chi
}_{q=0}=dM/dH$, was extracted from the QMC simulations by
measuring the induced magnetic moment in a small external
magnetic field. It includes the polarization of the impurity
Weiss field by the external field \cite{Georges:1996}. The
dynamical mean--field results account for the Curie--Weiss law
which is observed experimentally in Fe and Ni. As the temperature
increases above $T_{c}$, the atomic character of the system is
partially restored resulting in an atomic like susceptibility
with an effective moment $\mu_{eff}$. The temperature dependence
of the ordered magnetic moment below the Curie temperature and
the inverse of the uniform susceptibility above the Curie point
are plotted in Fig.~\ref{fig:Mchi} together with the
corresponding experimental data for Iron and Nickel
\cite{Wolfarth:1986}. The LDA+DMFT calculation describes the
magnetization curve and the slope of the high--temperature
Curie--Weiss susceptibility remarkably well. The calculated
values of high--temperature magnetic moments extracted from the
uniform spin susceptibility are $\mu _{eff}=3.09$ $(1.50)\mu
_{B}$ for Fe (Ni), in good agreement with the experimental data
$\mu _{eff}=3.13$ $ (1.62)\mu _{B}$ for Fe
(Ni)~\cite{Wolfarth:1986}.

\vskip 0.5cm

The Curie temperatures of Fe and Ni were estimated from the
disappearance of spin polarization in the self--consistent
solution of the DMFT problem and from the Curie--Weiss law. The
estimates for $T_{C}=1900$ $(700)$~K are in reasonable agreement
with experimental values of $1043$ $(631)$~K for Fe (Ni)
respectively \cite{Wolfarth:1986}, considering the single--site
nature of the DMFT approach, which is not able to capture the
reduction of $T_{C}$ due to long--wavelength spin waves. These
effects are governed by the spin--wave stiffness. Since the ratio
of the spin--wave stiffness $D$ to $T_{C}$, $T_{C}$/$a^{2}D,$ is
nearly a factor of 3 larger for Fe than for Ni
\cite{Wolfarth:1986} ($a$ is the lattice constant), we expect the
$T_{C}$ in DMFT to be much higher than the observed Curie
temperature in Fe than in Ni. Quantitative calculations
demonstrating the sizeable reduction of $T_{C}$ due to spin waves
in Fe in the framework of a Heisenberg model were performed in
Ref. \onlinecite{Pajda:2001}. This physics whereby the long
wavelength fluctuations renormalize the critical temperature
would be reintroduced in the DMFT using E--DMFT. Alternatively,
the reduction of the critical temperature due to spatial
fluctuations can be investigated with cluster DMFT methods.

The local susceptibility is easily computed within the DMFT--QMC.
Its behavior as a function of temperature gives a very intuitive
picture of the degree of correlations in the system. In a weakly
correlated regime we expect the local susceptibility to be nearly
temperature independent, while in a strongly correlated regime we
expect a leading Curie--Weiss behavior at high temperatures $\chi
_{local}=\mu _{loc}^{2}/({3}T+const.)$ where $\mu _{loc}$ is an
effective local magnetic moment. In the Heisenberg model with
spin $S$, $\mu _{loc}^{2}=S(S+1)g_{s}^{2}$ and for the
well--defined local magnetic moments (e.g., for rare--earth
magnets) this quantity should be temperature independent. For the
itinerant electron magnets, $\mu _{loc}$ is temperature--dependent
due to a variety of competing many--body effects such as Kondo
screening, the induction of local magnetic moment by temperature
\cite{Moriya:1985} and thermal fluctuations which disorder the
moments \cite{Irkhin:1994}. All these effects are included in the
DMFT calculations.

The comparison of the values of the local and the $q=0$
susceptibility gives a crude measure of the degree of
short--range order which is present above $ T_{C}$. As expected,
the moments extracted from the local susceptibility are a bit
smaller (2.8 \ $\mu _{B}$ for Iron and 1.3 \ $ \mu _{B}$ for
Nickel) than those extracted from the uniform magnetic
susceptibility. This reflects the small degree of the
short--range correlations which remain well above $T_{C}$
\cite{Mook:1985}. The high--temperature LDA+DMFT clearly shows
the presence of a local moment above $T_{C}$. This moment is
correlated with the presence of high--energy features (of the
order of the Coulomb energies) in the photoemission. This is also
true below $T_{C}$, where the spin dependence of the spectra is
more pronounced for the satellite region in Nickel than for that
of the quasiparticle bands near the Fermi level. This can explain
the apparent discrepancies between different experimental
determinations of the high--temperature magnetic splittings
\cite{Kisker:1984, Kakizaki:1994, Sinkovic:1997, Kreutz:1998} as
being the results of probing different energy regions. The
resonant photoemission experiments \cite{Sinkovic:1997} reflect
the presence of local--moment polarization in the high--energy
spectrum above the Curie temperature in Nickel, while the
low--energy ARPES investigations \cite{Kreutz:1998} results in
non--magnetic bands near the Fermi level. This is exactly the
DMFT view on the electronic structure of transition metals above
$T_{C}$. Fluctuating moments and atomic--like configurations are
large at short times, which results in correlation effects in the
high--energy spectra such as spin--multiplet splittings. The
moment is reduced at longer time scales, corresponding to a more
band--like, less correlated electronic structure near the Fermi
level.


\subsubsection{Classical Mott insulators}
\label{subsec:CMI}

NiO and MnO represent two classical Mott--Hubbard systems (in
this section we shall not distinguish between Mott--Hubbard
insulators and charge transfer insulators \cite{Zaanen:1985}).
Both materials are insulators with the energy gap of a few eV
regardless whether they are antiferromagnetic or paramagnetic.
The spin--dependent LSDA theory strongly underestimates the energy
gap in the ordered phase. This can be corrected by the use of the
LDA+U\ method. Both theories however fail completely to describe
the local moment regime reflecting a general drawback of band
theory to reproduce the atomic limit. Therefore the real
challenge is to describe the paramagnetic insulating state where
the self--energy effects are crucial both for the electronic
structure and for recovering the correct phonon dispersions in
these materials. The DMFT calculations have been performed
\cite{Savrasov:2003} by taking into account correlations among
$d$--electrons. In the regime of large $U$, adequate for both for
NiO and MnO in the paramagnetic phase, the correlations were
treated within the well--known Hubbard~I approximation.

The calculated densities of states using the LDA+DMFT method for
the paramagnetic state of NiO and MnO \cite{Savrasov:2003} have
revealed the presence of both lower and upper Hubbard sub--bands.
These were found in agreement with the LDA+U calculations of
Anisimov \cite{Anisimov:1991} which have been performed for the
ordered states of these oxides. Clearly, spin integrated spectral
functions do not show an appreciable dependence with temperature
and look similar below and above phase transition point.

\begin{figure}[htb]
\includegraphics[width=2.9 in]{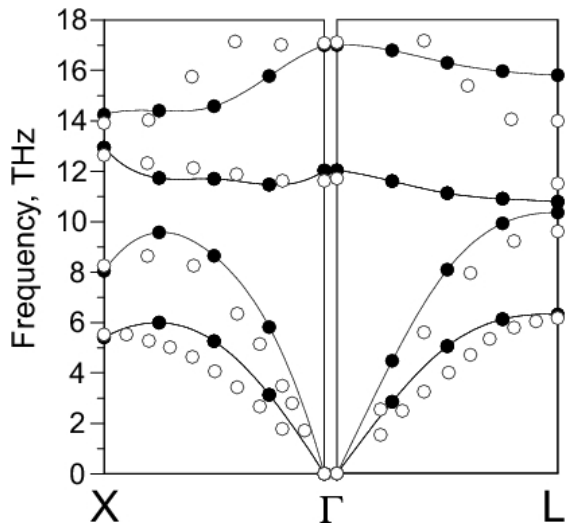}
\caption{Comparison between calculated using the DMFT method
(filled circles) and experimental (open circles) phonon
dispersion curves for NiO.} \label{fig:NiO_phonons}
\end{figure}

The same trend is known to be true for the phonon spectra which
do not depend dramatically on magnetic ordering\ since the
N\'{e}el temperatures in these materials are much lower of their
energy gaps. Fig.~\ref{fig:NiO_phonons} shows phonon dispersions
for NiO along major symmetry directions. A good agreement with
experiment \cite{Roy:1976} can be found for both acoustic and
transverse modes. A pronounced softening of the longitudinal
optical mode along both $\Gamma X$ and $\Gamma L$ lines is seen
at the measured data which is in part captured by the theoretical
DMFT\ calculation: the agreement is somewhat better along the
$\Gamma X$ direction while the detailed $q$--dependence of these
branches shows some residual discrepancies.

The results of these calculations have been compared with the
paramagnetic LDA, as well as with the antiferromagnetic LSDA and
LSDA+U\ solutions in Ref.~\onlinecite{Savrasov:2003}. The
paramagnetic LDA did not reproduce the insulating behavior and
therefore fails to predict the splitting between the LO and TO
modes. Due to metallic screening, it underestimates the
vibrations for NiO and predicts them to be unstable for MnO. The
spin resolved LSDA solution imposes the existence of long--range
magnetic order and is an improvement, but strongly underestimates
the energy gap. As a result, an apparent underestimation of the
longitudinal optical modes has been
detected~\cite{Savrasov:2003}. On the other hand, the
calculations with correlations produce much better results. This
is found both for the LSDA+U and the LDA+Hubbard~I calculations
which can be interpreted as good approximations to the DMFT
solutions for the ordered and disordered magnetic states. Such an
agreement can be related with the fact that the direct $d-d$ gap
is fixed by $U$, and the charge transfer gap comes out better in
the theory. Thus, the local screening of charge fluctuations are
treated more appropriately.


\subsection{Other applications}
\label{subsec:MAToth}

DMFT concepts and techniques are currently being applied to
investigate a broad range of materials and a wide variety of
strong correlation problems. This is a very active research
frontier, comprising topics as diverse as manganites, ruthenates,
vanadates, actinides, lanthanides, Buckminster fullerenes,
quantum criticality in heavy fermion systems, magnetic
semiconductors, actinides, lanthanides, Bechgaard salts,
high--temperature superconductors, as well as surfaces
heterostructures alloys and many other types of materials. We
mention below a small subset of the systems under investigation
using the techniques described in this review, just in order to
give the reader a glimpse of the breadth of this rapidly
developing field and the great potential of DMFT methods  to
investigate strongly correlated materials. For earlier reviews see
Ref.~\onlinecite{Held:2001:IJMPB} and also \cite{Held:2003:PSIK},
\cite{Lichtenstein:2002:CM} and \cite{Lichtenstein:2004:KLUWER,
Kotliar:2001:Tsvelik, Georges:2004:CM, Georges:2004:CM0403123,
Kotliar:2004:PT,Kotliar:2005}.

{\large Transition Metal Oxides}\\
$\bullet $ A large body of DMFT studies focused on the
\textit{Manganites} with the perovskite structure, like
La$_{1-x}$Ca$_{x}$MnO$_{3}$ or La$_{1-x}$Sr$_{x}$MnO$_{3}$. These 
materials  attracted attention because of their ``colossal"
magnetoresistance, which is an extreme sensitivity of resistance
to an applied magnetic field~\cite{Tokura:2003:PT, Tokura:2000,
Dagotto:2002}. The phase diagram of these materials in the
temperature composition plane is very rich, displaying
ferromagnetism, antiferromagnetism, and charge and orbital ordering.
Several physical mechanism and various interactions are important
in these materials such as the double exchange mechanism (i.e.
the  gain in kinetic energy of the $e_{g}$ electrons when the
$t_{2g}$ electrons are ferromagnetically aligned), the coupling
of the $e_g$ electrons to Jahn--Teller modes (i.e. distortions of
the Oxygen octahedra which lift the cubic degeneracy of the $e_g$
orbitals) and breathing Oxygen phonon modes. These materials, as
well as the copper oxides,  also spurred new studies on strong
electron-phonon coupling problems and their interplay with the
electron electron interactions. Semiclassical treatments of the
core spins and the phonons, help to dramatically simplify the
solution of the DMFT equations.

For various  DMFT studies of  the electron phonon
coupling problem in the manganites see  \cite{Furukawa:1994, Millis:1996,
Phan:2003, Fishman:2002, Fishman:2003, Fishman:2003:PRB,
Chernyshev:2003, Izyumov:2001,Michaelis:2002,Held:2000,
Imai:2000,Tran:2003, Ciuchi:1999, Blawid:2003, Pankov:2002,
Deppeler:2002,Deppeler:2002:b, Fratini:2001, Fratini:2000,
Ciuchi:1997, Benedetti:1999, Venketeswara:2003:CM,
Ramakrishnan:2003:CM,  Ramakrishnan:2003:CM:2}.

$\bullet $ \textit{High-temperature superconductivity} The
discovery of high-temperature superconductivity in the cuprates
posed the great theoretical and computational challenge of
uncovering the mechanism of this phenomena, which is still not
sufficiently understood to this day. The cluster extensions of
DMFT are actively being applied in an effort to further unravel
the mystery of the high-temperature superconductivity
\cite{Civelli:2005,Lichtenstein:2000,Dahnken:2005, Dahnken:2004,
Potthoff:2003:PRL, Maier:2005:CM0504529, Macridin:2005,
Maier:2004, Macridin:2004, Maier:2002:PRB66, Aryanpour:2002,
Huscroft:2001, Maier:2000:PRL}.

$\bullet $ \textit{Other transition-metal oxides} have recently
been studied with DMFT.
Na$_x$CoO$_2$ has received much interest due to anomalous
thermoelectric properties in addition to superconductivity upon
hydration.
~\cite{Ishida:2005,Lechermann:2005:CM0505241,Saha-Dasgupta:2005}.
LiVO$_2$ displays an unusually large effective mass
for a $d$-electron system, which gave rise to the idea that this
may be an example of a heavy-fermion $d$-electron material ~\cite{Nekra:2003}. TiOCl
displays $1d$ orbital ordering at low temperatures, and 
exhibits a spin-Peierls transition ~\cite{Seidel:2003,Saha:2005,Hoinkis:2005}.
Ca$_{2-x}$Sr$_{x}$RuO$_{4}$ attracted a lot of attention as it
exhibits unconventional $p$--wave superconductivity at low
temperatures, a Mott transition, and a surface-sensitive
spectra~\cite{Liebsch:2000, Anisimov:2002,
Liebsch:2003:CM0306312,Liebsch:2003:CM0301536,
Lichtenstein:2001:Sr2RuO4}.

{\large Organic Materials and Supramolecular Structures}\\
 $\bullet $
{\textit Fullerines} K$_{n}$\textit{C}$_{60}$. The doped
Buckminster fullerenes are solids formed from $C_{60}$ -- a
molecule shaped like a soccer ball -- with the alkali metal
sitting in the middle. Their proximity to the Mott transition was
pointed out by \onlinecite{Gunnarsson:1997}. At low temperatures
K$_{3}$C$_{60}$ is an $s$--wave superconductor, where both the
strong electron--phonon interaction and the Coulomb repulsion
need to be taken into account. DMFT has been helpful in
understanding the transition to superconductivity
~\cite{Capone:2002:SCIENCE, Han:2003, Capone:2000, Capone:2002}.
A Mott insulating state has also been realized recently in
another nanostructured supercrystal family, that of potassium
loaded zeolites, and realistic DMFT calculations have been
carried out ~\cite{Arita:2003}.

$\bullet $ \textit{Bechgaard salts} \cite{Vescoli:2000,
Biermann:2001}. In addition to the quasi-two dimensional organic
compounds  of the kappa and theta families mentioned in
section~\ref{sec:MAT}, there have been chain-DMFT studies of
$(TMTTF)_{2}X$ and $(TMTSF)_{2}X$. These  are strongly
anisotropic materials made of stacks of organic molecules. Their
optical properties are very unusual. At high temperatures the
electrons move mainly along the chains, before undergoing
inelastic collisions forming a ``Luttinger liquid" -- one of the
possible states of the one--dimensional electron gas. By
contrast, at sufficiently low temperatures the electronic
structure becomes effectively three--dimensional. For example,
they exhibit a very narrow Drude peak carrying only one percent
of the total optical weight ~\cite{Vescoli:2000,Biermann:2001}.
Investigations of the unusual properties of these materials, and
in particular of the crossover between the low- and
high--temperature regimes, using cluster generalizations of DMFT
have been performed in Refs. ~\onlinecite{Vescoli:2000,
Biermann:2001}. Other aspects have recently been addressed in
Refs.~\onlinecite{Georges:2000, Giamarchi:2004:CM}.

{\large Heavy Fermion Systems and Silicides}\\
$\bullet $ \textit{Heavy fermion materials} are compounds
containing both $f$--electrons and lighter $s,p,d$--electrons.
They can form a ``heavy" Fermi liquid state at low temperatures
where the quasiparticles are composites of $f$--electron spins and
conduction electron charges, or can order antiferromagnetically
at low temperatures (for recent review see \cite{Stewart:2001}).

The boundary between the Fermi liquid and the itinerant
antiferromagnet has been  a subject of intensive theoretical and
experimental study. Recent neutron scattering experiments are
consistent with a local spin self--energy, and have motivated an
extended  dynamical mean--field descriptions of the Kondo lattice
model ~\cite{Si:2001,Si:1999, Si:2001:NATURE, Jian-Xin:2003,
Grempel:2003, Ingersent:2002, Si:2003:PRB} which reproduces many
features of the experiment at zero temperature. For a recent
discussion see ~\cite{Si:2003}. A corresponding study of Anderson
lattice model~\cite{Sun:2003} suggests that nonlocal effects,
which require a cluster dynamical mean--field study, play an
important role.

An interesting issue is whether the optical sum rule, integrated
up to some cutoff of the order of one eV, can be a strong
function of temperature in heavy fermion insulators. The $f$--sum
rule states that if the integration is performed up to infinite
frequency, the result is temperature independent. In most
materials, this sum rule is obeyed even when a finite upper
limit   of the order of an electronic energy is used. This was
found in Ce$_3$Bi$_4$Pt$_3$~\cite{Bucher:1994} and in FeSi
~\cite{Schlesinger:1993})~\cite{Urasaki:2000, Smith:2003,
Marel:2003:TRIESTE}. In Ref. \onlinecite{Damascelli:1999}, it has
been shown that the integrated optical weight up to 0.5~eV is a
strong function of temperature, and if an insulating gap much
smaller than the cutoff is open, spectral weight is transferred to
very high frequencies. This problem was theoretically addressed
using single--site  DMFT applied to the Anderson lattice model,
and the theory supports a gradual filling of the gap without area
conservation ~\cite{Rozenberg:1995}. More recent studies applied
to a multiband Hubbard model \cite{Urasaki:2000, Smith:2003,
Marel:2003:TRIESTE} and to the Anderson
lattice~\cite{Vidhyadhiraja:2003} yield excellent quantitative
agreement with the most recent experiments.
%
%
Of great interest is the behavior of these materials in an
external field, which is easily incorporated into the DMFT
equations ~\cite{Meyer:2001,Medici:2005:CM0502563}.%
One interesting phenomena is the possibility of metamagnetism,
namely   the anomalous increase of the magnetization and the
concomitant changes in electronic structure as a function of
external field, known as a metamagnetic transition. This is
displayed in many heavy fermion systems such as
CeRu$_{2}$Si$_{2}$. An important issue is whether a transition
between a state with a large Fermi surface and a small Fermi
surface takes place as a function of magnetic field.

{\large Inhomogeneous Systems}\\
$\bullet $ \textit{Magnetic semiconductors} are materials where
the magnetization is strongly tied to the carrier concentration.
They offer the possibility of controlling the charge conductivity
(as in the usual semiconductors) and the spin conductivity (by
controlling the magnetization), by varying the carrier
concentration. Great excitement in this field has been generated
by the discovery of high--temperature ferromagnetism in these
materials~\cite{Vollhardt:1999, Kubler:2000, Kubler:2002}. A main
challenge is to understand the strong dependence of the
magnetization on the carrier concentration of the magnetic atoms
and on the concentration of the conduction electrons or holes.
This problem is closely related to the Anderson lattice model and
several DMFT studies of this problem have
appeared~\cite{Chattopadhyay:2001:PRL, Das:2003}. 
DMFT has been successfully applied to half magnets such as NiMnSb
~\cite{Chioncel:2003:PRB68, Chioncel:2005:PRB71,
Irkhin:2004:CM0406487} as well as magnetic multilayers
\cite{Irkhin:2004:CM0403685}. For a recent DMFT study of the
dependence of the critical temperature on various physical
parameters see~\cite{Moreno:2005:CM0507487}.

$\bullet $ \textit{Strongly inhomogeneous systems: systems near
an Anderson transition.} The dynamical mean--field theory has
been formulated to accommodate strongly inhomogeneous situations
such as systems near an Anderson transition, by allowing an
arbitrary site dependence of the Weiss
field~\cite{Dobrosavljevic:1997}. Recent progress in simplifying
the analysis and solution of these equations was achieved using
the typical medium approach of Dobrosavljevic et
al.~\cite{Dobrosavljec:2003}. The statistical DMFT approach can
also be used to study the interplay of disorder and the
electron--phonon coupling~\cite{Bronold:2003}.

$\bullet $ \textit{Heterostructures  surfaces and interfaces}

Another application of DMFT is to study correlation effects on
surfaces, which likely to be more pronounced than in the
bulk~\cite{Okamoto:2004, Okamoto:2004:CM, Okamoto:2004:PRB,
Freericks:2004:CM}.
For an early discussion of these effects see
Refs.~\onlinecite{Sawatzky:1995, Hesper:2000}. DMFT equations for
the study of correlation effects on surfaces and surfaces phase
transitions were written down by Potthoff and collaborators
~\cite{Potthoff:1999, Potthoff:1999:FILMS, Potthoff:1999:MASS}.
and successfully applied to study the Mott transition on surfaces
by~\cite{Perfetti:2003, Liebsch:2003}. DMFT has been applied to
inhomogeneous situations involving a much longer periodicity.
These studies were motivated by the discovery of stripes in
high--temperature superconductors.


\section{Outlook}
\label{sec:otlook}

Dynamical mean--field methods represent a new advance in
many--body physics. They provide an excellent description of the
strongly correlated regime of many three--dimensional
transition--metal oxides, which had not been accessible to other
techniques. Additionally, DMFT has given new insights into
strongly correlated electron systems near a metal--insulator
transition, or localization--delocalization boundary.

The combination  of advanced electronic structure  methods with
the dynamical mean--field technique  has already resulted in new
powerful methods for modeling correlated materials. Further
improvements are currently being pursued, as the implementations
of GW  methods and dynamical mean--field ideas
\cite{Aryasetiawan:1998,Zein:2002,Sun:2002,Sun:2004,Biermann:2003}.

In the field of statistical mechanics, the development of
mean--field theories was followed up  by the development of
renormalization group approaches  incorporating the physics of
long wavelength fluctuations which become dominant near critical
points. The development of effective  renormalization techniques
for correlated electrons and electronic structure applications is
a major challenge  ahead. It will  allow for more  accurate
derivation of low--energy  Hamiltonians, and improve the solution
of model Hamiltonians beyond  the dynamical mean--field theory
based on small clusters.

The forces acting on the atoms have been recently evaluated in
the realistic DMFT treatment of phonons in correlated electron
systems~\cite{Savrasov:2003, Dai:2003}. The indications that DMFT
captures correctly the forces on the atoms in correlated materials
bodes well for combining this development with molecular dynamics
to treat the motion of ions and electrons simultaneously. This
remains one of the great challenges for the future.

In conclusion, DMFT is a theory which can accurately capture
\emph{local} physics. We emphasize that the notion of
\emph{local} is flexible, and generically refers to some
predefined region in which correlations are treated directly
(e.g. a single--site, or a cluster of sites). Current
computational limitations restrict the local region to be a
relatively small number of sites for lattice models. Despite this
restriction, DMFT and its cluster extensions have been very
successful in describing a wide variety of materials properties
where conventional techniques such as LDA have failed. Therefore,
it seems that there finally exists a general tool which can
accurately treat many of the problems posed by strong correlation
in realistic materials. With the increasing number of realistic
DMFT implementations and studies of materials, more detailed
comparisons with experiments will emerge. Ultimately, this
experience will allow us to understand which aspects of the
strong correlation problem lie within the scope of the method,
and which aspects require the treatment of non--Gaussian, long
wavelength fluctuations of collective modes not included in the
approach.

\acknowledgements

The content of this review has been greatly influenced
by numerous colleagues
and collaborators, theorists and experimentalists,
that have lead to establish Dynamical Mean Field Theory as
a practical tool for the study of the physical
properties of strongly correlated materials.
Particular thanks go to the KITP, and to my coorganizers of the program
on realistic studies of correlated materials, O. K Andersen, A. Georges and
A. Lichtenstein, that catalyzed many efforts in the field including this review.
We thank 
E.  Abrahams, A.  Anisimov, V.  Antropov, A.  Arko, F.  Aryasetiawan, S.  Biermann, 
G.  Biroli, C.  Bolech, M  Capone, C.  Castellani, R.  Chitra, M.  Civelli, 
T.A.  Costi, X.  Dai, L.  de Medici, S.  Florens, T.  Giamarchi, K.  Held, 
D.  Hess, M.  Imada, J.  Joyce, V.  Kancharla, M.  Katsnelson, 
D.  Koelling, A.  Liebsch, W.  Metzner, A.  Migrliori, A.  Millis,  
S.  Murthy, G.  Palsson, S.  Pankov, I.  Paul, A.  Poteryaev, 
M.  Rozenberg, A.  Ruckenstein, S.  Sachdev, P.  Sun, J.  Thompson, D.  Vollhardt, 
J.  Wong, N.  Zein, and X.Y.  Zhang.

The work at Rutgers (K. Haule, G. Kotliar , C.A. Marianetti, and V. Oudovenko )
was supported   by an award of the NSF-ITR program, under grant NSF-ITR-0312478.
Gabriel Kotliar is a beneficiary of a "Chaire Internationale
de Recherche Blaise Pascal de l'Etat et de la R\'egion d'Ille
de France, g\'er\'ee par la Fondation de l'Ecole Normale
Sup\'erieure".
O. Parcollet was supported by the ACI of the French ministry of research.
S. Savrasov acknowledges NSF DMR grants 0238188, 0342290
and DOE Computational Material Science
Network.
The work was supported by the NSF DMR Grants No.
0096462, 02382188, 0312478, 0342290,  US DOE division of Basic
Energy Sciences Grant No DE-FG02-99ER45761, and by Los Alamos
National Laboratory subcontract No 44047-001-0237. Kavli Institute
for Theoretical Physics is supported by NSF grant No. PHY99-07949.

\appendix

\section{Derivations for the QMC section}
\label{sec:QMC:Appendix}

\paragraph{Derivation of Eq. (\ref{eq:QMC:ZasdetO})}

First, we show that
\begin{equation}\label{eq:QMC:formula.lemma.1}
\Tr_{{d^{\dagger},d}} \left(\prod_{1\leq k\leq K}e^{d^{\dagger }\ak d} \right)
= \det \left(1 + \prod_{1\leq k\leq K}e^{\ak} \right)
\end{equation}
where $d_{i}$ ($1\leq i\leq n$) are fermionic operators, $\ak$  ($1\leq k\leq K$)
 $n\times n$ matrices, and the notation $d^{\dagger }\ak d \equiv \sum_{1\leq
i,j\leq n}d^{\dagger }_{i}\ak_{ij}d_{j}$.
Indeed using
\begin{equation}\label{lem1.f1}
\left[ d^{\dagger }A d ,  d^{\dagger }B d  \right] = d^{\dagger }
\left[A,B \right] d
\end{equation}
and Baker--Campbell--Hausdorff formula $e^{A} e^{B} = e^{M}$ with
\begin{equation}
M\equiv A + B + \frac{1}{2}[A,B] + a_{2} [A,[A,B]] + \dots
\end{equation}
we have
\[
\exp (d^{\dagger }A d) \exp (d^{\dagger }Bd) = \exp (d^{\dagger }M d)
\]
By recursion, this generalizes to $K$ matrices, so we just have
to prove the result for $K=1$, $A^{(1)}=M$. If $M$ is diagonal,
the result is straightforward. For a general matrix $M$, by
directly expanding the exponential of the left hand side and
using Wick theorem, we see that $\Tr _{{d^{\dagger},d}}
e^{d^{\dagger }M d}$ is a series in $\Tr M^{k}$ ($k\geq 0$) and
is therefore invariant under any change of basis, and therefore
$\det (1 + e^M)$. Hence the result follows by diagonalizing $M$.

Second, we use the  determinant formula
\begin{equation*}
\begin{vmatrix}
1      & 0    &\dots  &   0   &   B_{L} \\
-B_{1} & 1    & \dots   &  \dots&   0  \\
0     &-B_{2} & 1     & \dots     &   \dots     \\
\dots  & \dots  &\dots  & 1 & 0 \\
\dots  & \dots  &\dots  & -B_{L-1} & 1 \\
\end{vmatrix}
= \det (1 + B_{L}B_{L-1} \dots B_{1})
\end{equation*}
where $B_{1},\dots,B_{L}$ are $n\times n$ matrices. This formula
results by recursion from the general formula for block matrices:
\begin{align}
\label{eq:QMC:APPblockdet}
\det
\begin{pmatrix}
A & B \\
C & D
\end{pmatrix}
=&
\det
\left[
  \begin{pmatrix}
1 & B \\
0 & D
\end{pmatrix}
\begin{pmatrix}
A-BD^{-1}C & 0 \\
D^{-1}C & 1
\end{pmatrix}
\right]
\nonumber
\\
=&
 \det  D
\det (A -B D^{-1}C)
\end{align}
\\[0cm]

\paragraph{Derivation of Eq.(\ref{eq:QMC:gisOinverse})}

The first step is to obtain an explicit formula for $g_{\{S\}}$
\cite{Blankenbecler:1981}. A quick way is to replace $a_{p\mu}$
and  $a_{p\nu}^\dagger$ in the trace by $\exp(\psi^\dagger
\Lambda^\dagger_\mu a)$
 and $\exp(a^\dagger \Lambda_\nu \psi)$, where $\psi$ is an auxiliary fermion and
$(\Lambda_\mu)_{ij} = \lambda \delta_{j0}\delta_{i\mu}$ where $0$
is the index of $\psi$. Starting from the explicit trace
expression of the Green's function, distinguishing the cases $l_1
\geq l_2$ and $l_1<l_2$, using Eq. (\ref{eq:QMC:formula.lemma.1})
and  Eq. (\ref{eq:QMC:APPblockdet}), we finally expand to second
order in $\lambda$ and obtain (see also  \cite{Georges:1996} ):
\begin{widetext}
\begin{equation}
g_{\{S\}}^\sigma(l_1,l_2) =
  \begin{cases}
      B^\sigma_{l_1 -1} \dots B^\sigma_{l_2} \bigl( 1 +  B^\sigma_{l_2 -1} \dots B^\sigma_1 B^\sigma_L \dots B^\sigma_{l_2}
\bigr)^{-1}
& \text{ for } l_1 \geq l_2
\\
    -  B^\sigma_{l_1 -1}\dots B^\sigma_1 B^\sigma_L \dots B^\sigma_{l_2} \bigl( 1 + B^\sigma_{l_2 -1} \dots B^\sigma_1 B^\sigma_L
\dots B^\sigma_{l_2} \bigr)^{-1}
& \text{ for } l_1 < l_2
  \end{cases}
\end{equation}
\end{widetext}
A straightforward  calculation then shows that
$g_{{\{S \}}}^\sigma  {\cal O}_\sigma (\{S \}) = 1$.

\paragraph{Derivation of Eq.(\ref{eq:QMC:DysonRelation1})}

Equation (\ref{eq:QMC:DysonRelation1}) follows from the observation that
${\cal O}_\sigma (\{S \}) \prod_{i=n}^1
e^{-\widetilde{{\cal V}}^{i\sigma } (\{S \}) } $
depends on the configuration $\{S \}$ only on its diagonal blocks, which leads to
\begin{multline}
  \label{eq:QMC:deriveDyson}
  {\cal O}_\sigma (\{S \})
\prod_{i=n}^1
e^{-\widetilde{{\cal V}}^{i\sigma } (\{S \}) }
-
{\cal O}_\sigma (\{S' \})
\prod_{i=n}^1
e^{-\widetilde{{\cal V}}^{i\sigma } (\{S' \}) }
 =
\\
\prod_{i=n}^1
e^{-\widetilde{{\cal V}}^{i\sigma } (\{S \}) }
-
\prod_{i=n}^1
e^{-\widetilde{{\cal V}}^{i\sigma } (\{S' \}) }
\end{multline}
which yields Eq. (\ref{eq:QMC:DysonRelation1}).

\paragraph{Derivation of the fast update formula (\ref{eq:QMC:FastUpdate1})}

We present here the steps to go from
Eq. (\ref{eq:QMC:FastUpdate1}) to Eq. (\ref{eq:QMC:FastUpdateGeneral}).
Since the difference between the two $V$ is in the $l$ block,
$A_\sigma$ has the form
\[
A_{\sigma } =\begin{pmatrix}
1      & 0    &\dots  & A^\sigma_{1l} &  \dots    & 0    \\
0      & 1    &\dots  & A^\sigma_{2l} &  \dots    & 0  \\
\vdots &\vdots &\vdots &\vdots &\vdots &\vdots \\
0     & 0 &  \dots   & A^\sigma_{ll} & \dots     &  0     \\
\vdots &\vdots &\vdots &\vdots &\vdots &\vdots \\
0     & 0 &  \dots   & A^\sigma_{Ll} & \dots     &  1     \\
\end{pmatrix}
\]
Using (\ref{eq:QMC:APPblockdet}), we have $\det A_\sigma = \det
A^\sigma_{ll}$. If $ \det A^\sigma_{ll} \neq 0$, we use the
Woodbury formula where $M$ is a $N\times N$ matrix and $U$ and
$V$ are $N\times P$ matrices \cite{Golub:1996}:
\[
(M + U^{t}V)^{-1} = M^{-1} - M^{-1} U  \bigl (1 + {}^{t}V M^{-1}U
\bigr )^{-1}{}^{t}V M^{-1}
\]
with
$N=L\caln$, $P = \caln$, $M=1$, ${}^{t}U =
\bigl(A_{il}\bigr), {}^{t}V =  \bigl( \delta_{il} \bigr), 1\leq i \leq L$).
to get
\begin{equation}
A_\sigma^{-1}=
\begin{pmatrix}
1      & 0    &\dots  & -A^\sigma_{1l}(A^\sigma_{ll})^{-1} &  \dots    & 0    \\
0      & 1    &\dots  & -A^\sigma_{2l}(A^\sigma_{ll})^{-1} &  \dots    & 0  \\
\vdots &\vdots &\vdots &\vdots &\vdots &\vdots \\
0     & 0 &  \dots   & (A^\sigma_{ll})^{-1} & \dots     &  0     \\
\vdots &\vdots &\vdots &\vdots &\vdots &\vdots \\
0     & 0 &  \dots   & -A^\sigma_{Ll}(A^\sigma_{ll})^{-1} & \dots     &  1     \\
\end{pmatrix}
\end{equation}
which leads to Eq. (\ref{eq:QMC:FastUpdateGeneral}).




\section{Software for carrying out realistic DMFT studies.}%
\label{sec:app}

There is a growing interest to apply DMFT to realistic models of
strongly--correlated materials. In conjunction with this review,
we provide a suite of DMFT codes which implement some of the ideas
outlined in the review (http://dmftreview.rutgers.edu). These
codes should serve as a practical illustration of the method, in
addition to lowering the barrier to newcomers in the field who
wish to apply DMFT. A strong effort was made to isolate the
various aspects of the DMFT calculation into distinct subroutines
and programs. This is a necessity both for conceptual clarity,
and due to the fact that various pieces of the code are under
constant development. Additionally, we hope that this will
increase the ability of others to borrow different aspects of our
codes and apply them in future codes or applications. Each of the
codes performs some task or set of tasks outlined in the LDA+DMFT
flow chart (see Fig. \ref{FigLDA+DMFT}) or the simpler DMFT flow
chart (see Fig. \ref{Fig:DMFT}).

\subsubsection {Impurity solvers}
DMFT is a mapping of a lattice problem onto an impurity problem.
Therefore, at the heart of \emph{every} DMFT calculation is the
solution of the Anderson impurity model (section \ref{sec:IMP}).
Solving the AIM is the most computationally demanding aspect of
DMFT. No one solver is optimum for all of parameter space when
considering both accuracy and computational cost. Therefore, we
provide a variety of impurity solvers with this review.
Additionally, more impurity solvers will be added to the webpage
with time, and existing impurity solvers will be generalized. It
should be noted that some solvers are more general than others,
and some of the solvers are already embedded in codes which will
perform DMFT on simple Hubbard models. The following solvers are
currently available: QMC, FLEX, NCA, Hubbard I, and interpolative
solver. (See sections
\ref{sec:IMPqmc},\ref{sec:expandinU},\ref{sec:expandinV},
\ref{sec:IMPatm}, and \ref{sec:IMPipt}, respectively.)

\subsubsection{Density functional theory}
Density functional theory (see section \ref{sec:DFT}) is the
primary tool used to study realistic materials, and in practice
it is usually the starting point for the study of realistic
materials with strong correlations. Furthermore, current
implementations of DMFT require the definition of local orbitals.
Therefore, DFT performed using an LMTO basis set is ideal match
for DMFT. However, we should emphasize that any basis set may be
used (i.e. plane waves, etc) as long as local orbitals are
defined.

Sergej Savrasov's full--potential LMTO code (i.e. LMTART) is
provided to perform both DFT and DFT+U (i.e. LDA+U or GGA+U)
calculations (see sections \ref{sec:DFT} and \ref{sec:SDFldu},
respectively). This code possesses a high degree of automation,
and only requires a few user inputs such as the unit cell and the
atomic species. The code outputs a variety of quantities such as
the total ground state energy, bands, density of states, optical
properties, and real--space hopping parameters. The code
additionally calculates forces, but no automatic relaxation
scheme is currently implemented.

A Microsoft windows based graphical interface for LMTART,
Mindlab, is also provided. This allows an unfamiliar user an
intuitive interface to construct the input files for LMTART, run
LMTART, and analyze the results in a point--and--click
environment. This code is especially helpful for plotting and
visualization of various results ranging from the projected
density of states to the Fermi surface.

\subsubsection{DFT+DMFT}

As stressed in this review, the ultimate goal of our research is
a fully first--principles electronic structure method which can
treat strongly--correlated systems (i.e. see section
\ref{sec:ScreenedInt} \ref{sec:SDFrsp}). Because this ambitious
methodology is still under development, we continue to rely on
the simplified approach which is DFT+DMFT (section
\ref{sec:DFT}). Although simplified, DFT+DMFT is still
technically difficult to implement, and currently we only provide
codes which work within the atomic--sphere approximation (ASA) for
the DFT portion of the calculation. One of the great merits of
DFT+DMFT is that it is a nearly first--principles method. The user
only needs to input the structure, the atomic species, and the
interactions (i.e. U). The DFT+DMFT code suite is broken into
three codes.

The first part is the DFT code, which is simply a modified
version of LMTART. It has nearly identical input files, with
minor differences in how the correlated orbitals are specified.
Therefore, the main inputs of this code are the unit cell and the
atomic species. The main role of this code is to generate and
export the converged DFT Hamiltonian matrix in an orthogonalized
local basis for each k--point.  Therefore, this code essentially
generates the parameters of the unperturbed Hamiltonian
automatically. This information is needed to construct the local
Green's function.

The second part is the code which implements the DMFT
self--consistency condition Eq.~(\ref{DMFdel}), which requires a
choice of correlated orbitals. This code takes the Hamiltonian
matrix and the self--energy as input, and provides the bath
function as output.

The third part is the various codes which solve the Anderson
impurity model, and these have been described above in the first
section. These codes take the bath function as input and provide
the self--energy, which is used in the self--consistency condition
in the preceding step.

These three pieces allow one to perform a non--self--consistent
DFT+DMFT calculation as follows. First, the DFT code is used to
generate the local, orthogonalized Hamiltonian matrix at each
k--point. Second, one starts with a guess for the self--energy and
uses the DMFT self--consistency condition code to find the bath
function. Third, the bath function is fed into the impurity solver
producing a new self--energy. The second and third steps are then
repeated until DMFT self--consistency is achieved. This is
considered a non--self--consistent DFT+DMFT calculation. In order
to be fully self-- consistent, one should recompute the total
density after DMFT self--consistency is achieved and use this as
input for the initial DFT calculation. This process should be
continued until both the total density and the local Green's
function have converged. At the present time, we have not
provided a routine to recompute the total density following the
self--consistent DMFT calculation, and therefore only
non--self--consistent DFT+DMFT calculations are currently
supported.

One should note that the above pieces which compose the DFT+DMFT
suite are three separate codes. Therefore, one must write a
simple script to iterate the above algorithm until
self--consistency is reached (ie. the self-energy converges to
within some tolerance). Additionally, the DFT portion of this
code suite (i.e. the first part) can in principle be replaced by
any DFT code as long as a local basis set is generated.

\subsubsection{Tight--binding cluster DMFT code (LISA)}
The LISA (local Impurity Self-Consistent Approximation) project is
designed to provide a set of numerical tools to solve the quantum
many-body problem of solid state physics using Dynamical Mean Field
Theory methods (single site or clusters).  The input to the program
can be either model Hamiltonians, or
the output of other ab-initio calculations (in the form of tight
binding parameters and interaction matrices). This should greatly
facilitate the development of realistic implementations of dynamical
mean field theory in electronic structure codes using arbitrary basis
sets.

This tool is provided to allow non-DMFT specialists to make DMFT
calculations with a reasonable investment.  However, DMFT methods are
still in development and undergoing constant improvements.  In
particular, new impurity solvers need to be developed and new cluster
schemes will possibly be explored.  Therefore numerical tools have to
be flexible to accommodate foreseeable extensions of the methods.  In
particular, one needs to be able to switch the solver easily while keeping
the same self-consistency condition.  This can be achieved most
efficiently with modern programming techniques (e.g. object orientation,
generic programming without sacrificing speed since intensive parts of
the program are quite localized and can be easily optimized).  These
techniques allow for a standardization of DMFT solvers by
using an abstract solver class such that any new solver can be used
immediately in various DMFT calculations. The use of an abstract Lattice
class allows for programs designed for tight-binding models like the
Hubbard model to also be used for realistic calculations.  A
decomposition of the self-consistency conditions into small classes is beneficial in 
that various summation techniques on the Brillouin zone
can be used or new cluster schemes can be tested.

With LISA, we hope to achieve flexible, reusable, and efficient
software that is general enough to solve a variety of models and to
serve as a basis for future developments.  Documentation, including
examples, is provided with the web page. At present, a library and a
self--contained DMFT program are provided to solve a generalized
tight--binding Hamiltonian with single--site or many variants of
cluster DMFT described in section \ref{subsec:CDMFT} with the Hirsch-Fye
QMC method.  The tight--binding Hamiltonian may be very simple, such
as the traditional Hubbard model or the $p-d$ model of the cuprates,
or very complex, such as a real material with longer range hoppings.
This is markedly different than the DFT+DMFT code which takes the structure
as input and generates the Hamiltonian.  The
tight--binding Hamiltonian may be generated by a variety of different
electronic structure methods and codes, or trivially specified in the
case of a model Hamiltonian.


\section{Basics of the Baym--Kadanoff functional}
\label{sec:pedag}

The aim of these  online notes is to provide  a  more pedagogical
description of
the use of functionals by using the Baym--Kadanoff functional as
an example, and to derive, step--by--step, a few simple relations
and formulae
which are used in the main text.

In the Baym--Kadanoff theory, the observable of interest is the
following operator
\begin{equation}
 \psi^\dagger(x)\psi(x'),
\end{equation}
and its average is the electron Green's function $G(x',x)=-\langle
T_{\tau}\psi(x')\psi^\dagger(x)\rangle$.  As in the main text
of the review we use the notation $x=(r,\tau)$. The aim of the theory is to
construct a functional that expresses the free energy of the system
when the Green's function is constrained to have a given value.

First, we  modify the action of the system so that it gives
rise to the observable of our choice.  This is achieved by
adding a source term to the action in the following way

\begin{widetext}
\begin{equation}
  e^{-F[J]}=\int D[\psi^\dagger\psi]\exp\left({-S-\int
  dxdx^\prime\psi^\dagger(x)J( x,x^\prime)\psi(x^\prime)}\right),
  \label{Eq:freeEBK}
\end{equation}
\end{widetext}
where the action $S$ is
given by $ S = S{_0} + \lambda  S_1 $ where $S_0$ is the free part
of the action and $S_1$ the interacting part.  
In electronic structure calculations   
\begin{equation}
S_1 = \frac{1}{2}\lambda \int dxdx'\psi^\dagger(x)\psi^\dagger(x')
v_C(x-x')\psi(x')\psi(x).
\end{equation}
and $ v_C$ is the Coulomb interaction.

$\lambda $ is a coupling constant that allows us to "turn on"
the interaction. When  
$\lambda =0$  we have a non interacting problem and
we have   the interacting
problem of interest when $\lambda =1 $.

The modified free energy Eq.~(\ref{Eq:freeEBK}) is a functional of
the source field $J$ .
$F=F[J]$. By varying modified free energy Eq.~(\ref{Eq:freeEBK}),
$F$ with respect to  $J$  we get
\begin{equation}
{  { \delta F[J, \lambda ]} \over { \delta J }} = G, \label{Eq:deltaF}
\end{equation}
 The solution of this equation, gives $J= J(\lambda, G )$.

Its meaning is the source  that  results for a given  
Green's function G  when the interaction is $\lambda $.
Notice that when  $\lambda$ is set to unity and G is the
true Greens function of the original problem (i.e.  $ J=0, \lambda=1 $ )
J vanishes by definition.

When G is the true Greens function of the original problem and  
$\lambda=0$, $J (\lambda=0, G ) $ is non--zero and
is equal to the interacting self--energy $\Sigma_{int}$. This is
because the interacting self--energy is the quantity that needs
to be added to the non--interacting action to get the
interacting Green's function. We will show how this works
mathematically below.

We now  make a Legendre
transform from source the  $J$ to Green's function $G$, to get a
functional of Green's function only
\begin{equation}
  \Gamma_{BK}[G, \lambda ]=F[J[\lambda, G ] , \lambda ]-\trace[J G]
  \label{GBK}
\end{equation}

 with the differential
\begin{equation}
  \delta\Gamma_{BK}=-\trace[J\,\delta G].
  \label{Eq:deltaG}
\end{equation}

$\Gamma_{BK}[G]$ is  the functional which, as we will show
below, gives the  free--energy  and the Greens function
of the interacting system at its saddle point.  It is 
very useful for constructing numerous approximations. The
Legendre transform is used extensively in statistical mechanics,
and the above procedure parallels transforming from the canonical
to the grand--canonical ensemble where the chemical potential
replaces the density as the independent variable.

Now we want to connect the solution of the interacting system
$\lambda=1$ with the corresponding non--interacting $\lambda=0$
problem and split functional $\Gamma_{BK}[G]$ into the simple
non--interacting part and  a more complicated interacting part.

\subsection{Baym--Kadanoff functional at $\lambda=0$}

If $\lambda$ is set to zero, the functional integral
(\ref{Eq:freeEBK}) can readily be computed
\begin{widetext}
\begin{equation}
e^{-F_0[J_0]}=\int D[\psi^\dagger\psi]\exp\left({-\int
dxdx^\prime\psi^\dagger(x)(\frac{\partial}{\partial\tau}-\mu+H_0+J_0)\psi(x^\prime)}\right)
=\mathrm{Det}(\frac{\partial}{\partial\tau}-\mu+H_0+J_0),
\end{equation}
\end{widetext}
and the free energy becomes
\begin{equation}
  F_0[J_0]=-\trace\ln\left(G_0^{-1}-J_0\right),
\end{equation}
where we neglected a constant term $\trace\ln(-1)$.
Here $J_0$ is $J(\lambda=0)$ and $F_0$ is $F(\lambda=0)$, while
$G_0=(\omega+\mu-H_0)^{-1}$ is the usual non--interacting Green's
function. Taking into account Eq.~(\ref{Eq:deltaF}), the Green's
function at $\lambda=0$ is
\begin{equation}
G=\frac{\delta F_0[J_0]}{\delta J_0}=\left(G_0^{-1}-J_0\right)^{-1}.
\label{Eq:GS0}
\end{equation}
Since the Green's function $G$ is fixed at the interacting
Green's function, it is clear that the source field $J_0$ is the
interacting self--energy, viewed as a function of the Greens function G  i.e.,
\begin{equation}
{J_0} \equiv \Sigma_{int}[G] \equiv  G_0^{-1} - G^{-1}
 \label{J0}
\end{equation}
viewed as a functional of $G$ (since  $G_0$ fixed and given from
the very beginning).
 In general, $J_0$ is the constraining field that needs to be
added to the non--interacting action $S_0$ to get the interacting
Green's function. Finally, the Baym--Kadanoff functional at
$\lambda=0$, being the Legendre transform of $F_0[J_0]$, takes the
form
\begin{equation}
\Gamma_0[G]=-\trace\ln\left[G_0^{-1}-\Sigma_{int} [G ] \right]-\trace[\Sigma_{int}[G]G]
\end{equation}

\subsection{Baym--Kadanoff functional at $\lambda=1$}

When the interaction is switched on, the functional is altered
and in general we do not know its form. We will write it as
\begin{equation}
\Gamma_{BK}[G]=-\trace\ln\left[G_0^{-1}-\Sigma_{int}[G]\right]-\trace[\Sigma_{int}[G]G]
+ {\Phi}_{BK}[G]
\label{defBK}
\end{equation}
where $\Phi_{BK}$ is a non--trivial functional of $\lambda $ and
G. We are interested in  $\lambda =1$  but it is useful sometime
to retain its dependence of $\lambda $ for theoretical
considerations.  It will be shown (see section \ref{sec:BKint})
that $\Phi_{BK}$ can be represented as the sum of  all
two--particle irreducible skeleton diagrams.

We have seen in the previous subsection that $J$ vanishes at
$\lambda=1$, $J (\lambda =1, G ) =0 $.  This has the important consequence that the
Baym--Kadanoff functional is stationary at $\lambda=1$ (see
Eq.~(\ref{Eq:deltaG})) and is equal to free energy of the system
(see Eq.~(\ref{GBK})).

Stationarity of $\Gamma_{BK}$ means that the saddle point equations
determine the relationship between the quantities that appear in the
functional, i.e.,
\begin{eqnarray}
\frac{\delta\Gamma_{BK}[G]}{\delta
G}&=&\trace\left\{\frac{\delta\Sigma_{int}}{\delta
G}\left[(G_0^{-1}-\Sigma_{int})^{-1}-G\right]\right\}\nonumber\\
&-&\Sigma_{int}+\frac{\delta\Phi_{BK}[G]}{\delta G}=0.
\end{eqnarray}
The first term in parenthesis vanishes from Eq. ($\ref{J0}$). 
Therefore  {\it at the stationary point },
which determines the Greens function of interest $G_{sp}$,  the 
constraining field, denoted by $\Sigma_{int}[G]$ in Baym--Kadanoff
theory, is equal to the derivative of the interacting part of
functional, i.e.,
\begin{equation}
\Sigma_{int}[G_{sp}] =\left. \frac{\delta\Phi_{BK}[G]}{\delta G}\right|_{G_{sp}}
\label{Eq:dPhidG}
\end{equation}

Using the definition of $\Sigma_{int}$ in  eq. (\ref{J0} ) 
we see that this is nothing but the standard
Dyson equation, a non linear equation that determines the Greens function
of interest, $G_{sp}$, at the saddle point of the BK functional:
\begin{equation}
\left(G_0^{-1} - G^{-1}_{sp}
\right) =\left. \frac{\delta\Phi_{BK}[G]}{\delta G} \right|_{G_{sp}}
\label{dysn}
\end{equation}

Eq.~(\ref{Eq:dPhidG}) offers a  the diagrammatic interpretation
of $\Phi_{BK}$ as a sum of all two particle irreducible skeleton
graphs.  Namely, a functional derivative amounts to opening or erasing
one Green's function line and since the self energy by definition
contains all one particle irreducible graphs, ${\Phi}_{BK}$ must
contain all two particle irreducible graphs (skeleton graphs, see 
\cite{Dominicis:1964,Dominicis:1964b}).

Note that the  functional $\Gamma_{BK}$ can also be regarded as a stationary
functional of  two  independent variables, $G$ and $\Sigma_{int}$.
\begin{equation}
\Gamma_{BK}[G,\Sigma_{int}]=-\trace\ln\left[G_0^{-1}-\Sigma_{int}\right]-
\trace[\Sigma_{int} G] +\Phi_{BK}[G]
\end{equation}
The derivative with respect to $\Sigma_{int}$ gives
Eq.~(\ref{J0}), while derivative with respect to $G$ leads to
Eq.~(\ref{Eq:dPhidG}).

Finally,  by construction, the free energy of the interacting
system at ($\lambda=1$) is simply obtained by evaluating
$\Gamma_{BK}$ at its stationary point,  which is the true Greens
function of the system, which by abuse of notation we will  still refer
to as G (instead of $G_{sp}$):
\begin{equation}
F_{BK}=\trace\ln G-\trace[\left( G_0^{-1} - G^{-1} ]G \right)
 +\Phi_{BK}[G]. \label{Eq:FEBK}
\end{equation}

\subsection{Interacting part of Baym--Kadanoff functional}
\label{sec:BKint}



In this subsection, we want to give an alternative proof that the
interacting part of the Baym--Kadanoff functional $\Phi_{BK}$ is
the sum of all two--particle irreducible skeleton diagrams.


To prove this we  go back and reintroduce the coupling constant
$\lambda $  which multiplies the interacting part of the
Hamiltonian $H_{int}=\lambda V$ and  the interacting part of the
action which was used to  define the path between the
non--interacting $\lambda=0$ and interacting $\lambda=1$ system.

We  first  evaluate the derivative of the Baym Kadanoff functional
with respect to $\lambda$ at fixed G, namely
\begin{equation}
 \frac{\partial {\Gamma}_{BK} [G,\lambda ] }{\partial \lambda}=
 \frac{\partial {\Phi}_{BK} [G,\lambda ] }{\partial \lambda}
\end{equation}
Using Eq.~(\ref{GBK}) and the relation between J and G,
Eq.~(\ref{Eq:deltaF}) valid at any given $\lambda$,
($\Gamma_{BK}[G]=F[J_\lambda , \lambda ]-\trace[J_{\lambda} , G]$), we
obtain
\begin{equation}
 \frac{\partial {\Phi}_{BK} [G,\lambda ] }{\partial \lambda}=
\left. {{\partial F[J,\lambda ] }  \over    {\partial \lambda} } \right|_{J=J(\lambda,G)}
\end{equation}
Here $J$ is a function of both $\lambda$ and $G$, i.e., $J(\lambda,G)$.

The derivative of the free energy  functional with respect to the
coupling constant  (at fixed source ) is readily obtained
\begin{eqnarray}
  \frac{\partial F}{\partial \lambda} = \frac{1}{Z}\int D[\psi^\dagger\psi] V
  e^{-S}=\frac{1}{\lambda}\langle H_{int}\rangle.
  \label{hint}
\end{eqnarray}
and 
\begin{equation}
 \frac{\partial {\Phi}_{BK} [G,\lambda ] }{\partial \lambda}=\frac{1}{\lambda}
\langle H_{int}\rangle [\lambda, J (\lambda, G )]
\label{target}
\end{equation}
Notice that $H_{int}$ or $S_{1}$ is independent of of J, but the average $\langle\quad\rangle$
is carried out with respect to a weight which contains the source explicitly.
Integrating equation (\ref{target}) : 
\begin{equation}
\Phi_{BK}[G] = {\int_0}^1 \frac{1}{\lambda}\langle H_{int}\rangle
[\lambda, J (\lambda, G )] d\lambda 
\label{phibk}
\end{equation}

This is another example
of  the coupling constant integration formula for the interaction
energy of the effective actions constructed in our review. Equation (\ref{phibk})
can be expanded in a standard perturbation theory. We  take   as the
inverse unperturbed propagator ${G_{0}}^{-1} - J(\lambda=0) $ ( 
namely $G$ )  and  as interaction vertices $H_{int}$ and $[J(\lambda ) - J(\lambda=0)]
\psi^{\dagger} \psi $ as interaction vertices.
This means that the perturbation theory contains two kinds of vertices
, the first carries 4 legs and the second denoted by a cross carries only two legs
and represents  $[J(\lambda ) - J(\lambda=0)]
\psi^{\dagger} \psi $. 
The role of the second
vertex is to eliminate the graphs which are two particle
reducible, this cancellation is illustrated in figure \ref{Fig:DFT_2roads}.
which demonstrates that for each   reducible graph (i.e. one having
a self energy insertion ), there is also a cross, their sum is zero
as a result of the equation 
\begin{equation}
 J[\lambda]-J[\lambda=0]+\Sigma[\lambda]=0
\label{oequestion}
\end{equation}
 
This equation is proved by noticing
that at a given value of $\lambda$ the Greens function
of the problem is by definition ${{G_{0}}^{-1} - J(\lambda=0)-\Sigma[\lambda]}^{-1}=G$
and since by definition ${{G_{0}}^{-1} - J(\lambda=0)}^{-1}=G$ combining these two equations we
obtain Eq. (\ref{oequestion}).
\begin{figure}[tbh]
\includegraphics*[width=5cm]{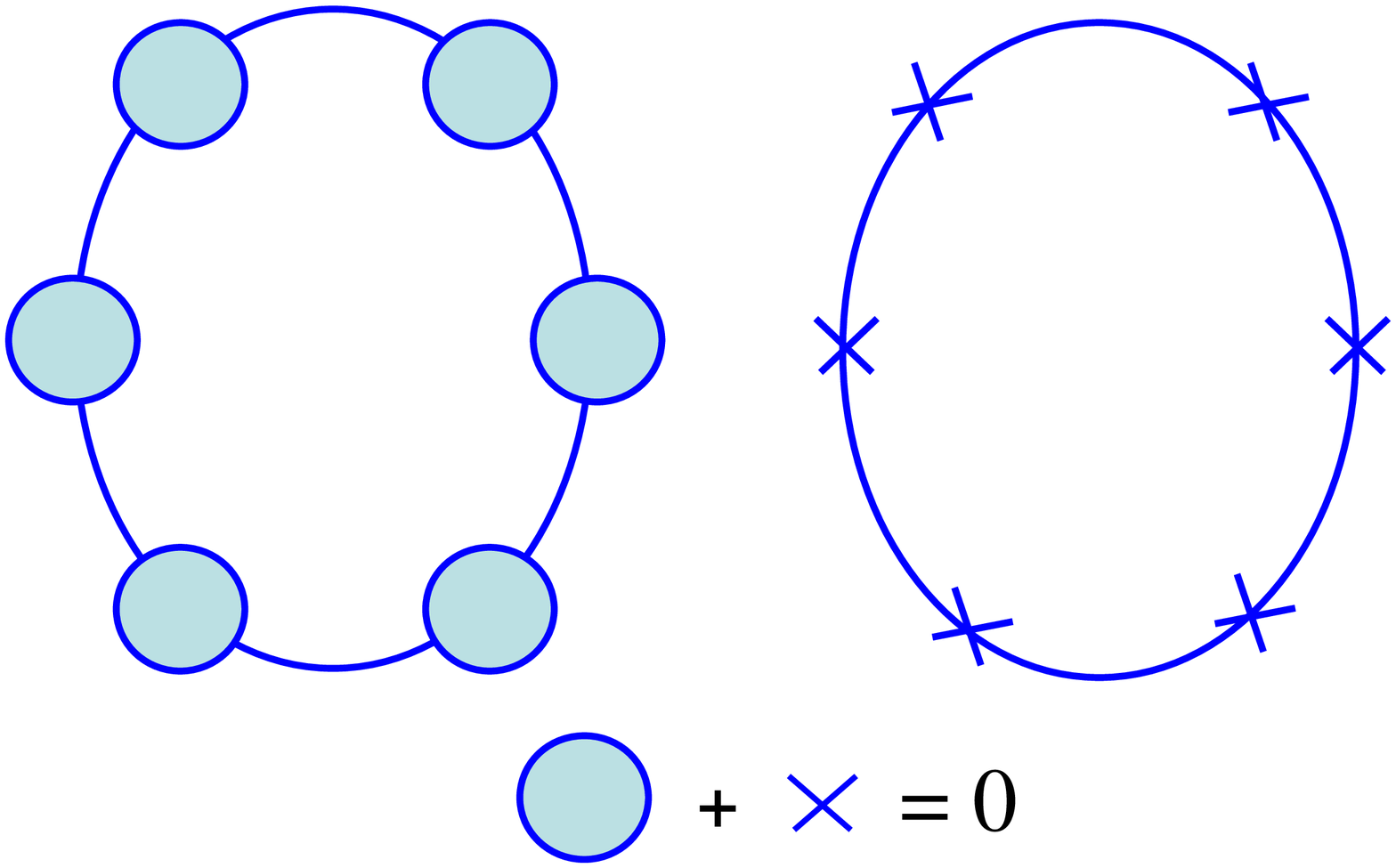}
\caption{Mechanism of cancellation of reducible graphs. The  "X"
denotes $J(\lambda)-J(\lambda=0)$ and the circles in blue are the
self-energy insertions that make the graph reducible.} 
\label{Fig:DFT_2roads}
\end{figure}
The role of the
coupling constant integration is to provide the standard symmetry
factors in the free energy graphs.

\subsection{The total energy}

In this subsection, we want to derive the relationship between the
total energy of the system and the corresponding Green's
function. Let us start by the definition of the Green's function
\begin{equation}
  G(x_1,x_2)=-\langle T_\tau \psi(x_1)\psi^\dagger(x_2)\rangle.
  \label{G_def}
\end{equation}
The non--interacting (quadratic) part of the Hamiltonian
\begin{eqnarray}
\beta   H_0=\int dx_1 dx_2 \psi^\dagger(x_1) H0_{x_1 x_2}\delta(\tau_1-\tau_2)
  \psi(x_2),
\end{eqnarray}
can be expressed by the Green's function in the following way
\begin{eqnarray}
\beta   \langle H_0\rangle = \int dx_1 dx_2 H0_{x_1
  x_2}\delta(\tau_1-\tau_2) \langle \psi^\dagger(x_1)\psi(x_2)\rangle\\
  =\int dx_1 dx_2 H0_{x_1 x_2}G(x_2,x_1)\delta(\tau_2-\tau_1+0^+).
\end{eqnarray}
To get the interacting part of the total energy, we are going to
examine the time derivative of the Green's function which follows
directly from the definition Eq.~(\ref{G_def}) and takes the form
\begin{eqnarray}
\left(\frac{\partial
  G(x_1,x_2)}{\partial\tau_1}\right)_
  {
    \begin{array}{l}
      \tau_1\rightarrow\tau_2-0^+\\
      \vr_1\rightarrow \vr_2
    \end{array}
  }
  = \langle \psi^\dagger(x_1)\left[H-\mu n,\psi(x_1)\right]\rangle
  .\ \ \ \ \
\label{tmd}
\end{eqnarray}
The resulting commutator can be simplified by noting that the
following two commutators take a very simple form
\begin{eqnarray}
&  \int dx \psi^\dagger(x)\left[\psi(x),V\right]&=2V\\
&  \int dx \psi^\dagger(x)\left[\psi(x),H_0\right]&=H_0 .
\end{eqnarray}
where $V$ is the normal--ordered electron--electron interaction.
The factor two in the above equation follows from the fact that
the interaction term is quartic in $\psi$ while $H_0$ is
quadratic.

It is more convenient to express the equations in imaginary
frequency than the imaginary time. Using the transformation
\begin{equation}
  G(x_1,x_2)=
  T\sum_{i\omega}e^{-i\omega(\tau_1-\tau_2)}G_{i\omega}(\vr_1,\vr_2),
\end{equation}
one obtains for the non--interacting part
\begin{equation}
  T \int d\vr_1 d\vr_2 H0_{\vr_1 \vr_2}\sum_{i\omega}
  G_{i\omega}(\vr_2,\vr_1)e^{i\omega 0^+}=\langle H_0\rangle,
  \label{h0_1aa}
\end{equation}
while the time derivative from Eq.~(\ref{tmd}) gives
\begin{equation}
  T\int d\vr_1 \sum_{i\omega}(i\omega)e^{i\omega 0^+}G_{i\omega}(\vr_1,\vr_1) =
  \langle H_0+2 V-\mu n\rangle .
  \label{v0_1}
\end{equation}
Combining equations (\ref{h0_1aa}) and (\ref{v0_1}) leads to the
following expression for the interaction energy
\begin{widetext}
\begin{equation}
 \langle V\rangle =\frac{1}{2}T\int d\vr_1
  d\vr_2\sum_{i\omega}e^{i\omega 0+}\left[(i\omega+\mu)\delta(\vr_1-\vr_2)-H0_{\vr_1
  \vr_2}\right]G_{i\omega}(\vr_2,\vr_1)=
  \frac{1}{2}\trace[e^{i\omega 0+}\;G_0^{-1} G]=\frac{1}{2}\trace[\Sigma G].
\label{Eq:intE}
\end{equation}
Here we took into account that $\sum_{i\omega}e^{i\omega 0+}=0$.
Finally, the total energy becomes
\begin{equation}
  \langle H\rangle =\frac{1}{2}T\int d\vr_1
  d\vr_2\sum_{i\omega}e^{i\omega 0+}\left[(i\omega+\mu)\delta(\vr_1-\vr_2)+H0_{\vr_1
  \vr_2}\right]G_{i\omega}(\vr_2,\vr_1)=
  \trace[H_0 G+\frac{1}{2}\Sigma G].
\end{equation}
\end{widetext}


\bibliographystyle{apsrmp}
\bibliography{bib/journals,bib/review,bib/impurity,bib/articles,bib/books}


\end{document}